# ILYA MIKHAILOVICH KAPCHINSKY

To the 90th birthday

Collection of memories

*(English translation by R.A. Jameson, dedicated in appreciation to Dr. N.V. Lazarev*
*OCR using convertio.co & Adobe Acrobat XI, Google translate.)*

# И.М. КАПЧИНСКИЙ

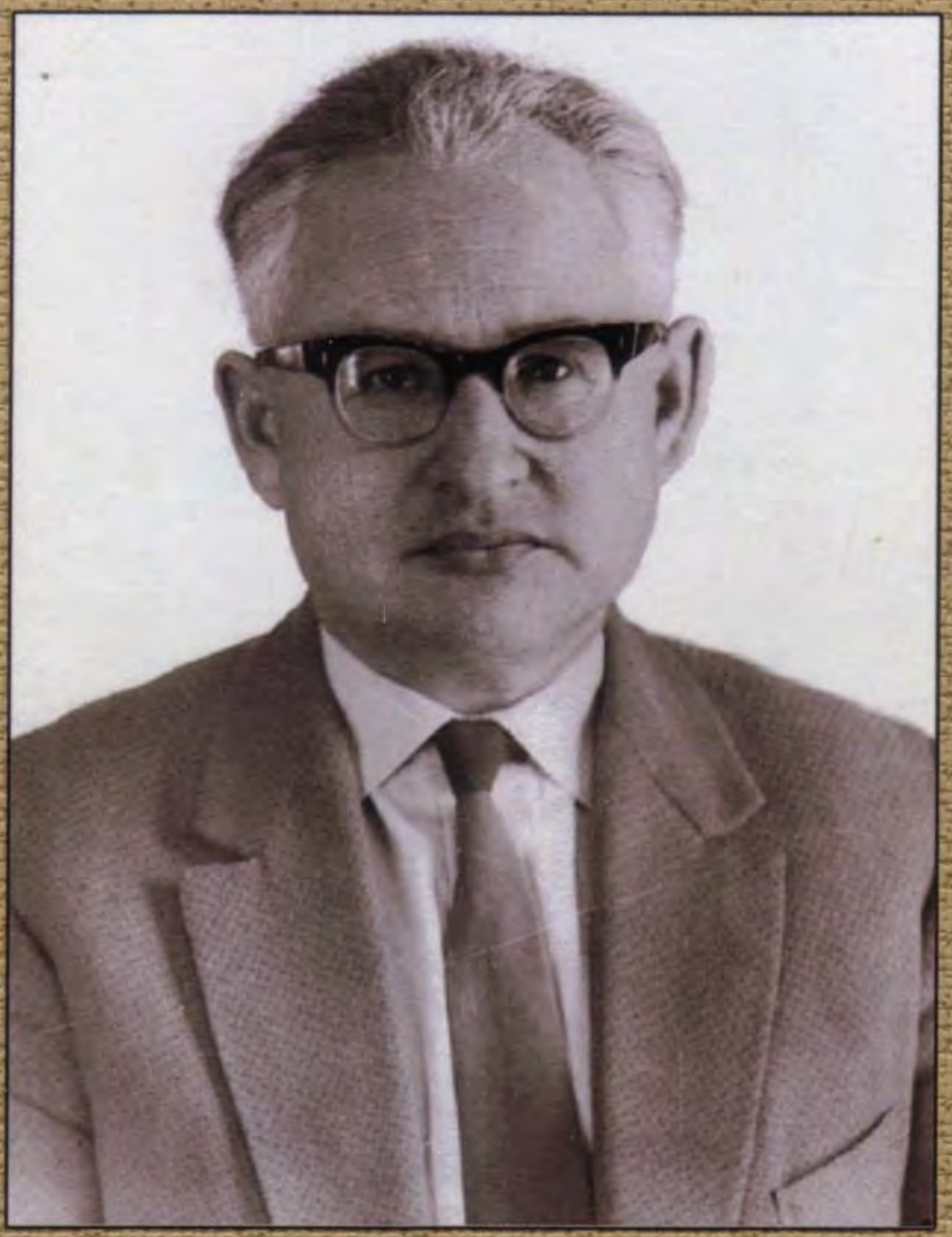

*к 90-летию со дня рождения*

## СБОРНИК ВОСПОМИНАНИЙ

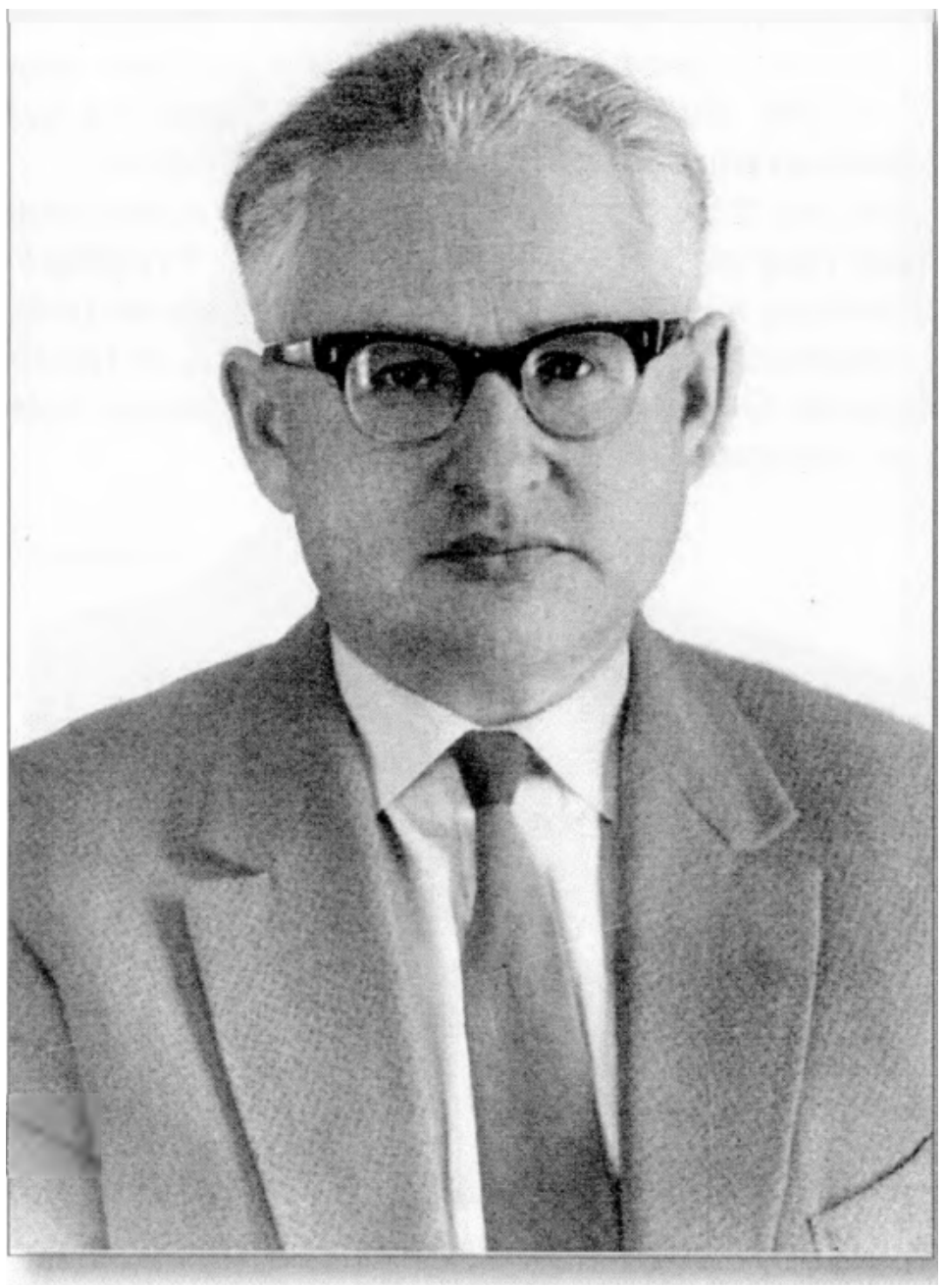

**ILYA MIKHAILOVICH KAPCHINSKY**

To the 90th birthday
Collection of memories

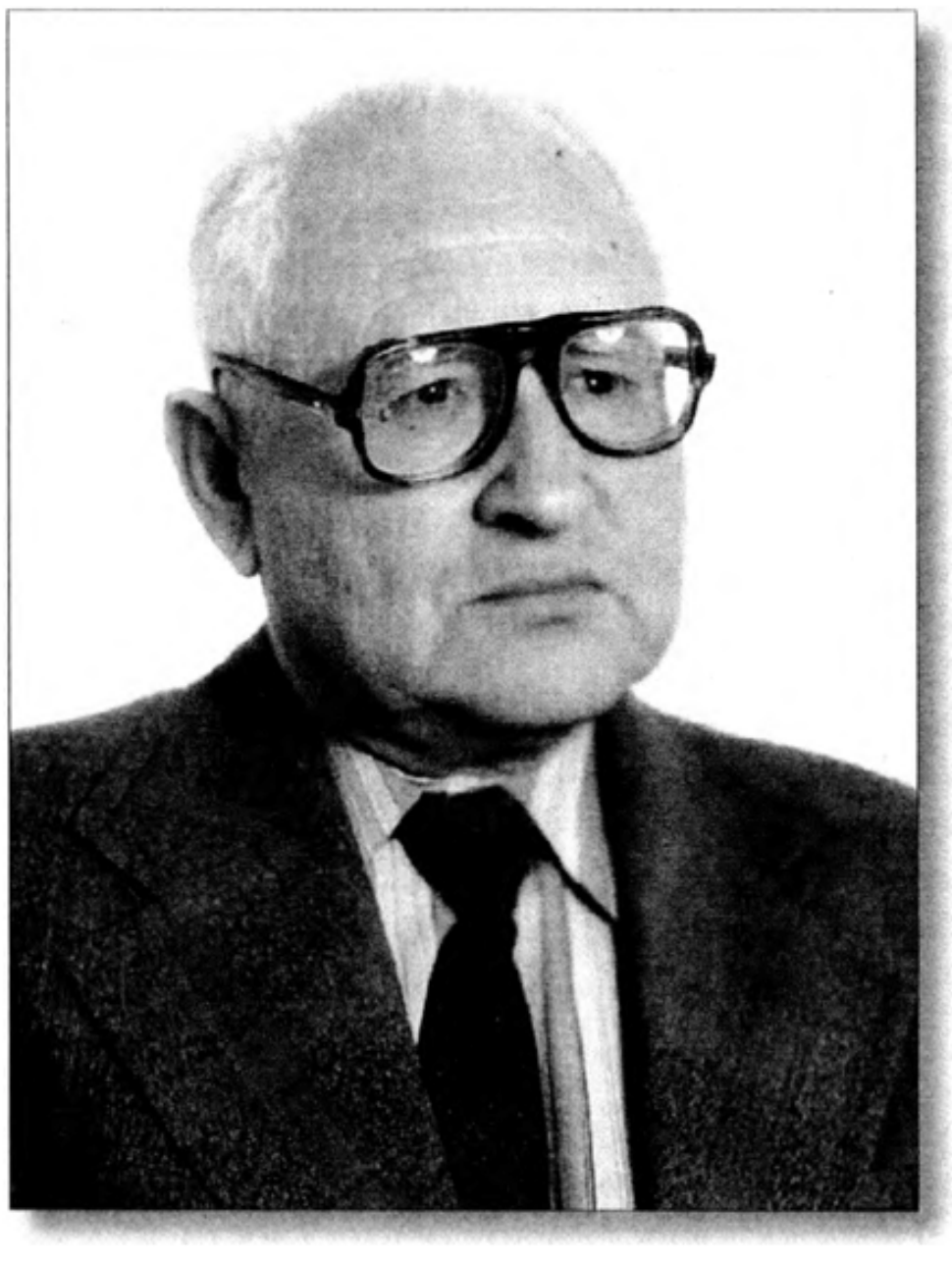

Ilya Mikhailovich KAPCHINSKY
- Laureate of the Lenin and State Prizes,
Honorary Doctor of the Frankfurt Goethe University.
Recipient of the American Physical Society Honorary Award,
Professor, Doctor of Engineering
(1919-1993)

The idea for this publication belongs to Nikolai Vladimirovich Lazarev, a close collaborator of Ilya Mikhailovich, head of one of the laboratories in the ITEP department he heads. It was through the efforts of N.V. Lazarev that most of the materials in the collection were gathered. We are deeply grateful to him and all his assistants for the publication of the preprint "Scientist, Teacher, Leader", released at ITEP in June 2009. Special thanks to the ITEP administration, represented by Deputy Director of the Institute Boris Yurievich Sharkov for support and assistance in publishing the preprint.

## FOREWORD – B.Yu. Sharkov, IN MEMORY OF PROFESSOR ILYA MIKHAILOVICH KAPCHINSKY

On July 5, 2009, Professor Ilya Mikhailovich Kapchinsky, a prominent specialist in the field of the theory of oscillations, radio engineering, physics of charged particle beams, and accelerator technology, would have turned 90 years old. Unfortunately, his heart stopped on May 2, 1993.

After graduating from the Faculty of Physics of Moscow State University, for the first half of his life, Ilya Mikhailovich was engaged in experimental development and theoretical studies of pulsed electronic circuits and the practical application of the results of these works in the creation of complex radio engineering devices. While working at the research institute, he creatively participated in the creation of the first domestic radio location stations. He summarized the results of research of that period in the monograph "Methods of the Theory of Oscillations in Radio Technology" (1954), which was republished soon abroad. The most complete and vivid talent of I.M. Kapchinsky opened up at the Institute of Theoretical and Experimental Physics, where he worked from 1958, first heading the laboratory he founded, and then the department of linear accelerators.

The most important scientific works of I.M. Kapchinsky are associated with the development of linear ion accelerators. At ITEP, under his leadership, the physics projects of proton linear accelerators (CLs) with hard-focusing on energies of 25 and 100 MeV were created. I.M. Kapchinsky supervised the activities of scientific teams in coordinating the design work, commissioning and commissioning of these accelerators at ITEP and IHEP.

Being a brilliant theoretician-analyst, Ilya Mikhailovich, thanks to his previous experience in industry, rarely combined these qualities with a deep understanding of all the subtleties of engineering and the difficulties that had to be overcome in the development and implementation of projects.

At the I-2 LU at ITEP, shortly after the launch, the world's highest pulsed intensity of the proton beam (-250 mA) was achieved, and the I-100 LU at IHEP in 1967-1970 was the largest linear proton accelerator in the world. Thanks to the correct understanding of the most complex issues of beam dynamics, the accuracy of calculations and the firmness in observing all the calculated tolerances, both machines soon after launch exceeded the design parameters of the beam with downtimes of less than 1%.

Construction and launch in 1966-1967. hard-focusing LU of protons at 25 and 100 MeV has advanced domestic technology to the forefront in the world. Works by I.M. Kapchinsky, connected with the creation of high current ion linear accelerators, are widely known in Russia and abroad. In this area, he became a recognized leader.

Kapchinsky obtained important results in the theory of intense beams of charged particles. He carried out theoretical studies on the effect of space charge on transverse and longitudinal vibrations of particles in accelerated beams, developed the theory of collective interactions of particles with self consistent beam eigenfields, together with V.V. Vladimirskii proposed to use in the theory of high-current beams the microcanonical phase distribution (which in English terminology received the name “KV Distribution” in honor of our authors, which made it possible for the first time to obtain analytically valuable, previously inaccessible results.

Kapchinsky studied the effect of longitudinal Coulomb repulsion on the autophasing process at an extremely high phase current density, determined the theoretical limitations of the beam current in linear accelerators, and formulated requirements for the phase density of injected beams and matching devices for the input optics of an accelerator.

Kapchinsky (together with V.V. Vladimirsky and V.A. Teplyakov) discovered the effect of spatially uniform quadrupole focusing, on the basis of which a new method for accelerating ion beams was developed. This method made it possible to significantly reduce the injection energy and increase the limiting value of the beam current, which opened up prospects for the development of ion linear accelerators with a high average current. Accelerators with spatially homogeneous quadrupole focusing (our abbreviation is POKF, foreign abbreviation is RFQ) have received all general recognition and are used in most Russian institutes and in all foreign scientific centers where ion linear accelerators are built or operated.

On the initiative of I.M. Kapchinsky began the introduction of hard magnetic quadrupole lenses into linear accelerators, this direction is associated with the development at ITEP of new frequency ranges for resonant accelerators and the search for a solution to the problem of radiation resistance of high current machines.

Kapchinsky was the first to propose a two-frequency scheme of ion accelerators, developed a physical justification for a prototype of a high-current accelerator with a high average beam current for applications under the ISTRA project for an energy of 56 MeV. The main parts of this LU were created and studied at ITEP. Started at the suggestion of I.M. Kapchinsky's development of LU of heavy low-charge ions on the problem of inertial thermonuclear synthesis.

Since 1958, I.M. Kapchinsky published about 150 scientific papers on the subject of accelerators, including one discovery and 10 inventions. He published three monographs (1954, 1966, 1982), one of which (Theory of Resonant Linear Accelerators) was published in the USA in 1985 in a revised and supplemented form.

For work in the field of accelerator physics, I.M. Kapchinsky was awarded the State and Lenin Prizes of the USSR. General recognition of the scientific merits of I.M. Kapchinsky was expressed by the award of an honorary prize of the American Physical Society and the high title of Honoris Causis, Honorary Professor of the University of Frankfurt.*

I.M. Kapchinsky was a remarkable organizer of science. He created large research teams at the branch scientific research institute, at ITEP and IHEP, trained a large number of young scientists, candidates and doctors of science. He gave lectures on the theory of linear accelerators at the ITEP, IHEP, MRTI, and the Institute of Linguistics (Italy). His last fourth monograph, "Selected Topics in Ion Linacs Theory" contains lectures given by him at the University of Maryland (USA) in 1993. This book was then published at the same time (with drawings made by him with his own hand) in Los Alamos (LA-UR-93-4192).

Many scientific ideas and valuable groundwork of I.M. Kapchinsky, which he did not realize during his lifetime, continue to be developed by specialists educated by him.

ITEP Deputy Director for Research, Corresponding Member RAS B.Yu. Sharkov

* Only a few of the most outstanding scientists were previously awarded this title, including Maria Sklodowska-Curie, the founder of the GSI Institute Professor Schmelzer, and the last (in 2004) one of the greatest physicists, the Belevsky laureate Professor Hans Bethe, who died in 2005.

## I. LITTLE KNOWN HERITAGE OF I.M. KAPCHINSKY

### AUTOBIOGRAPHY*

I, Kapchinsky Ilya Mikhailovich, was born on July 5, 1919 in Odessa. Father - film director, died in 1981. Mother - a nurse, died in 1966. Brother, born in 1923, head of a laboratory at a scientific research institute (Moscow).

In 1937 I graduated from secondary school No. 45 in Kyiv, and in 1942 I graduated from the physics department of Moscow State University. Since November 1942, I worked as an engineer in the recording workshop of the Tashkent film studio. In October 1943 I returned to Moscow and until November 1945 worked as an engineer in the laboratory of plant No. 465 of the NKEP. From November 1945 to July 1958, I worked at NII-20 of the Committee for Defense Equipment, where I successively held the positions of senior engineer, senior researcher, head of a laboratory, and head of a department. I published my first scientific work in 1946. In April 1948 I was awarded the degree of candidate of technical sciences. During my work at NII-20, I was repeatedly the deputy chief designer of objects and the supervisor of scientific research. I published 17 scientific papers on the theory of oscillations and impulse technology, was a co-author of eight special developments.

From July 1958 to the present, I have been working at the Institute of Theoretical and Experimental Physics of the SSAE. Organized a laboratory, which was in April 1974 transformed into a department. The department conducts research on the physics of beams in linear accelerators of protons and heavy ions. In this department, the physical projects of two large proton accelerators at 25 and 100 MeV were developed and, in addition, a number of engineering developments were carried out for these accelerators. I was entrusted with the scientific supervision of the construction and commissioning of these accelerators in ITEP (1966) and IHEP (1967). From April 1964 to March 1971, I worked part-time as the head of the injector department at the Institute of High Energy Physics of the State Committee for Astronomy.

From 1958 to 1989 I published 137 scientific papers on the physics of charged particle beams and accelerator technology.** Published three monographs. The monograph "Methods of the Theory of Oscillations in Radio Engineering" (1954) has been translated into Japanese and Chinese, and the monograph "Theory of Linear Resonant Accelerators" (1982) has been translated into English.

I gave courses of lectures on the theory of oscillations, pulse technology, and the theory of linear accelerators to students of the Academy of the Defense Industry, graduate students and engineers of a number of scientific institutes in Moscow.

In May 1962, I was awarded the degree of Doctor of Engineering. In 1967 I received the academic title of professor. I was awarded the State Prize of the USSR (1970) and the Lenin Prize of the USSR (1988) for my work in the field of accelerator physics. In 1989, I received an honorary doctorate from the University of Frankfurt. Goethe (Germany).

In 1953 I married Lyubov Mikhailovna Kon, a doctor who graduated from the Saratov Medical Institute in 1952. I have two children and three grandchildren. The son works as a researcher at MRTI, he is married. The daughter is a doctor and is married.

* Written in 1990
** Total I.M. Kapchinsky published about 150 scientific papers.

## STUDENT CONTACTS

### Ashgabat*

* These memoirs of I. M., with minor revisions, are published in the collection of memoirs about A. D. Sakharov "He lived between us ...". Ed. "Practice", Moscow, 1996, p. 310-314.

It is very difficult to write about Andrei Dmitrievich Sakharov now, in the second half of 1990: you are afraid to pile up beautiful tales. I will try to speak only about what I remember distinctly. Unfortunately, the memory was preserved in fragments, often accidental.

We studied together at the Faculty of Physics of Moscow State University in 1938-42. He was our classmate. The same as all of us, but not much and not the same. Andrei was very friendly, relations were even and comradely. But something still made me treat him in a special way. One episode stuck in my memory. We had a physics club in our second year. The circle was led by Sergei Grigoryevich Kalashnikov, at that time an assistant professor in the Faculty of Physics. We made presentations. Kalashnikov called all of us by our first or last names. And only to Andrey Kalashnikov addressed by name and patronymic. Why? Andrei's father was a prominent physicist, the author of a problem book that we used to study. But to imagine that it was precisely for this reason that Kalashnikov addressed Andrei so respectfully, of course, is not possible. There must be another reason, unknown to me. I did not think about this reason; nevertheless, he took such an appeal to Andrey without surprise. Indeed, Andrei inspired us with special respect.

Before the war, we finished the third year. On June 23, 1941, having passed the last exam of the spring session, I went to the Krasnopresnensky District Committee of the Party to sign up as a volunteer. I was not sent to the front, but assigned to the Krasnopresnensky fighter battalion. There were many students of Moscow State University in the battalion, mostly historians and biologists. We put on military uniforms and settled in an empty school. At night, they were secretly located among the graves of the Vagankovsky cemetery: it was believed that enemy agents from this cemetery would signal to German pilots. During the day they slept for several hours, and then they learned the charter and cleaned their rifles. Rarely appeared in the city. After some time, I was demobilized. As the battalion's chief of staff, Captain Lukyanov, explained, there is an order signed by Stalin - to send senior physics students for further education. Now I understand that in the confusion of those days, the order (if it existed) could not have been fully carried out. Several of our fellow students, very good guys, died in October 1941 in the ranks of the people's militia (among them I remember Lenya Sokolov, Petya Vasiliev-Dvoretsky).

When I ended up at the faculty in September, I found out that all our students who had stayed in Moscow in August and were suitable for health reasons (perhaps according to personal data) were taken to the Zhukovsky Military Academy. There was no further enrollment of physics students. For one reason or another, many children were not in Moscow in August. They are in the majority and ended up in Ashgabat at the end of the year, where Moscow University was evacuated. In August-September, I did not meet Andrei in Moscow.

I left Moscow on October 16, the day of the well-known general urban panic. After some ordeals in mid-December 1941, he ended up in Ashgabat. There were professors, associate professors and only a few students in Ashgabat at that moment. However, a large train arrived a week later. A lot has arrived students of our course, boys and girls. I would like to name Kot Tumanov, Yura Jordan, Petya Kunin, Leon Bell, my closest comrades. Andrey Sakharov arrived in the same train.

In the hostel, our beds - Andrei and mine - stood side by side. Probably, for this reason, we had a lot of contact with him in Ashgabat.

How were the classes in our last, fourth year? The university was located in the building of the Ashgabat Pedagogical Institute, in the suburb of Keshi. The curriculum was redrawn in a military way. We were offered to graduate from university in one of two specialties: “Defensive telecommunications” or “Defense materials science”. The special course in the specialty “electrocommunications” was the theory of vibrations, and in “materials science” (if my memory serves me) special courses - magnetism, x-ray diffraction analysis "Theoretical physicists, of course, were not trained. Andrey formally graduated with a degree in materials science. They taught us two general theoretical courses - electrodynamics and quantum mechanics. Associate Professor V. S. Fursov read electrodynamics very intelligibly. Quantum mechanics, somewhat theatrically - Associate Professor AA Vlasov. We attended lectures carefully.

I remember that after classes, Andrey would come to the hostel, sit on his bed and, fixing his eyes on infinity, would think. Andrei and I talked only about physics. On other topics, everyday or military-political, Andrey did not resonate. It was difficult to talk to Andrew. He spoke slowly and abruptly. I did not always catch the connection between his statements. Nevertheless, communication with Andrei gave, as far as it turned out to be accessible to me, a lot in understanding physics. In particular, this affected quantum mechanics. We had practically no books, and it was possible to comprehend physics only on the basis of lecture material. A discussion with Andrey of certain quantum effects (including, I remember, the tunneling effect) clarified a lot in quantum mechanics for me. Andrew knew how to think through to the end.

As an exercise, somehow A.A.Vlasov suggested that I consider the propagation of radio waves along a waveguide. In those years, this task was not known to us. I stubbornly worked out that the phase velocity of wave propagation depends on frequency.

The result seemed wrong to me. I returned to the calculations many times and could not get what I thought was the correct answer. Finally, I shared my doubts with Andrey. In his style, Andrey thought quietly, looking at infinity, and confidently said that dispersion must take place. After that, I reported my calculations to A.A. Vlasov. Anatoly Alexandrovich said to me: “But it will work for you!”. But I understood that it would “go” not with me, but with Andrei.

Our life in Ashgabat was difficult. Food, to put it mildly, was not enough. Officially, we had a daily coupon for 400 g of bread and a plate of mash. (Zatirukha was a dish that was flour stirred in hot water). With their scholarship, they could also buy a bunch of green onions and a glass of sour milk, which we called matsoni, in the market. Sometimes potatoes gave a drop, but there was no fat. It was Andrey who managed in this situation to calculate the available source of fats: castor oil was sold in the pharmacy. By his own example, Andrei showed that you can fry potatoes with castor oil. We quickly got used to the smell and many took advantage of Andrey's discovery.

In the atmosphere of hard life that surrounded us, Andrei and Petya Kunin at one time developed the idea of organizing a seminar on the general theory of relativity. But among the hungry guys, the idea did not meet with a response and gradually died out.

In July 1942, we graduated from the Physics Department of Moscow University in Ashgabat, having four courses under our belt. The so-called special works were not carried out due to the complete absence of laboratories. The heat was unaccustomed to us, at times a dusty hot wind blew, called in Ashgabat “Afghan”. In the sands right under the city, a student, Professor Gelfand's sister, Delya, could not stand it and died of heat stroke. At night, the breath of the Kara-Kums was felt, the stuffiness did not subside. The elderly professors suffered. I still remember how heavily the ill Professor Teodorchik breathed.

We passed state exams, received diplomas and were distributed. I don’t know who went on distribution that year. Everyone aspired to Moscow or to reunite with their families. I only firmly remember that Andrey (perhaps the only one) went to the factory in the city of Kovrov in accordance with a ticket. Andrei always seemed to be an idealist.

In the post-war years, I rarely met Andrei. Most often, such a case turned out to be a seminar at FIAN. After one seminar, Andrey and I started talking about poetry. Previously, we did not touch on such topics. Andrey said with some kindness that the jacket he was wearing was a gift from Galich. Galich had already been expelled from the country. It turned out that we have the same attitude towards Galich's poems.

I remember that at one of the seminars, Yakov Borisovich Zel'dovich, in response to a remark from his seat, cheerfully said: “Andrey Dmitritch is here. He won't let me lie." There was something serious in this playful remark.

I remember the seminars that took place during a difficult period for Andrey. And Gorky was ahead.

In a telephone conversation at the end of 1988, I congratulated Andrei on his election to the Presidium of the Academy of Sciences, but I realized that he was not happy with his election. Andrei Dmitrievich was burdened by the need to participate in discussions of academic organizational issues, such as the distribution of funding.

The students of our course first got together in September 1938. But the war scattered us, we finished at different times and in different conditions in the period 1942-1947. Therefore, at our periodic meetings, we celebrate the anniversary not of graduation from the University, but of admission to the University. Andrei was also present at our anniversary gathering in 1988. Elena

Georgievna was with him. It's hard to say what Andrew looked like. Time, apparently, has no power over us, real age is no longer perceived. We see in front of us only the same faces that have become familiar, which we saw in student times.

On a slushy December day in 1989, our course said goodbye to Andrey. We made up our full changing of the guard of honor. It was at a memorial service at FIAN.

In conclusion, I want to say on behalf of my fellow students: no matter where we are, no matter what work we do, at all times we were proud that we were comrades and classmates of Andrei Dmitrievich Sakharov. These are not just words.

## HERITAGE

## M.L. LEVIN*

We met Misha Levin for the first time on September 1, 1938, at the Faculty of Physics of Moscow State University, in the long corridor leading to the large Lenin Auditorium. And we immediately found each other. We agreed on a common love of poetry.

I remember that at some lecture we sat in the last rows of the Lenin audience, and I read my poems to Misha. Fifty-five years ago I really liked my poems. Misha had an absolute poetic taste, but his love for poetry was so great that he listened to me and, as it turned out, remembered the poems.

The same thing happened in subsequent years. I really appreciated Misha's condescending attitude towards my poems. Out of old friendship, he never made a joke about my lines.

And Misha knew how to make fun and immediately grasped the humorous side of unsuccessful lines. Misha often responded with light verses to the surrounding serious events and found a funny side in these events. I remember that in 1972, at a seminar on collective acceleration methods, Misha sent me a note:
“Forward, the rings are compressed, along the Veksler path ... "

Misha had an extraordinary memory for poetry. He memorized poetry immediately and for the rest of his life. I remember there was such a case. In 1990, the volume Foreign Poetry in B.L. Pasternak. Misha called me: “Ilya, this volume contains a translation of Gerweg's “Hate Song”. You've been translating this poem since school. Send me your translation. To my shame, I must confess that I was able to remember only half of the translation.

I really liked to visit Misha's house on Bolshaya Kaluga. Most often they gathered with Kot Tumanov and Kika Yaglom. But if they stayed with Misha alone, then they read poetry. The cult of Pasternak reigned. Do not forget the warm atmosphere created in the family by Revekka Saulovna and Lev Naumovich, Misha's parents.

In 1942, Kot Tumanov and I finished our studies at the Physics Department of Moscow State University in Ashgabat, Misha got stuck in Tashkent. He was very bored. He asked that we each write to him separately so that he could receive more letters from friends. But he promised to write to us only to the general address "letters of redoubled intellectual power."

In August 1943 I returned to Moscow from evacuation. The train arrived at night. The city was darkened. On the platform of the station - fellows. They stole my suitcase, where all my property was. Misha was already in Moscow and found out about it on the same day. An hour later, he and the Cat appeared at my place. The time was military, severe. Misha brought me panties, and the Cat brought me a T-shirt. So friends helped me start my life in science.

* Published in the collection "Mikhail Lvovich Levin. Life, Memories, Creativity. Nizhny Novgorod, 1995, p. 206

## PLANT No. 465 and NII-20. FIFTEEN YEARS OF LIFE

It gives me pleasure to return to my youth and remember how in October 1943 I came to the 465th plant. After graduating from the Faculty of Physics at Moscow University, I returned to Moscow from evacuation and looked for applications for my strength. My university comrade Sasha Starobinsky temperamentally persuaded me to go to work at the 465th plant, where he had been working for several months.

A very strong scientific team was assembled at the plant, which, in my opinion, laid the foundations of domestic radar. The Antenna Laboratory was headed by Mikhail Alexandrovich Leontovich, one of the greatest Soviet physicists, later an academician. The outstanding radio engineer Anton Yakovlevich Breitbart played a leading role in the laboratory of markers. I started working in the laboratory of Professor Semyon Emmanuilovich Khaikin. In the laboratory of S.E. Khaikin then developed phase aircraft guidance systems. The principles of these systems still go back to the ideas of Academicians Mandelstam and Papaleksi. Meeting with S.E. Khaikin determined my scientific interests. From that time to the present, my work has always been based on the theory of oscillations.

The group in which I worked was led by senior engineer Vasily Vasilievich Vladimirsky, now a corresponding member of the Academy of Sciences of the USSR * I was an engineer in this group, and Borya Samoilov, Misha Khaikin and Sasha Starobinsky worked as technicians. Sasha, unfortunately, died early, and Borya and Misha were well-known physicists in later years, Misha was a corresponding member of the USSR Academy of Sciences.

* V.V. Vladimirsky (1915 - 2008) since 1946 was ITEP Deputy Director for Research

Common interests connected me in life and at the plant with employees of the plant laboratories Irina Borisovna Andreeva and Ilya Solomonovich Abramson. I.B. Andreeva then for many years was at NII-20 the head of the laboratory in which I worked. Later, Irina Borisovna became one of the leaders of work in the field of marine physics in the Soviet Union.

I did not have to deal with the director of the plant, Boris Isaakovich Forshter, but I fell in love with the chief engineer of the plant, Mikhail Lvovich Sliozberg, immediately and for the rest of my life. M.L. Sliozberg was an extraordinary person both in his purely human qualities and in the breadth of scientific, engineering and artistic erudition. The character and all the activities of Mikhail Lvovich left a decisive imprint on the formation of the plant staff and on the development of the Institute staff.

I would like to mention the mobilization of laboratory staff at the end of 1943 for urgent work in the 5th workshop of the plant. The shop manager, if my memory serves me right, was Boris Alexandrovich Tomilin, a very good man, with whom we subsequently worked together at the training ground in Donguz. It was cold in the workshop, and then we worked without a break, without leaving the workshop for more than a day. That's where I started smoking. I painfully got rid of smoking in twenty years.

Two years after my arrival, the plant was transformed into the Central Design Bureau, soon into the Research Institute, where I worked until 1958.

We worked a lot. It was only over the years that we managed to figure out that there are times in winter when we didn't see the sun at all: we arrived at work before dawn, returned home late at night. It takes a long time to write about the content of works and business relations. However, many episodes were deposited in my memory that were not directly related to work, but remained milestones in my life and, to some extent, signs of the times. Some of them I would like to bring.

I firmly remember the case when Anatoly Prokofievich Belousov and I went to the first tests of the SON-4 station in Donguz. The post-war times were still difficult, and we were only able to purchase train tickets for a free-sleeper car.

Anatoly Prokofievich, with his characteristic enterprise and foresight, managed to get into the car before the start of the official boarding and occupied the two upper luggage racks - for me and for himself. We tied ourselves with belts to the heating pipes running under the ceiling of the car, and thus reached Orenburg in safety. The rest was already easier.

I remember how, on another trip to the training ground, Mikhail Lvovich Sliozberg and I walked around the steppe at night and found out who was better acquainted with Mayakovsky's poems. I considered myself an expert on Mayakovsky's poetry and was sure of victory. However, I was quickly put to shame. I still remember the voice of Mikhail Lvovich: "Loving Mayakovsky, this is a dynasty

We talked not only about poetry. When the SON-4 station stubbornly failed to provide the most important parameter, we discussed in desperation whether the antenna column was made of the right material.

The peculiar life of our small group in a lonely steppe house, about fifty kilometers from the village of Kapustin Yar, will not be erased from memory. Then rocket technology took the first steps. The head of the institute's expedition was Vadim Mikhailovich Taranovsky, scientific director of the work, a talented scientist. A description of our life and activities there could take up many pages, but in order to save space, I will not fill them out. I just want to mention the name of Isaac Mikhailovich Golovchiner, who brought a powerful stream of cheerfulness into our existence.

Comrade Moyne left a very instructive memory (unfortunately, I do not remember his name and patronymic). He was the head of the Institute's planning department, and he pressured us into strict planning and reporting. But after some transformations, Moin turned out to be an employee of one of the scientific laboratories (in my opinion, the laboratory of Naum Adolfovich Barshai). A short time passed, and Comrade Moyne began to energetically explain why scientific work could not be planned. But it was already too late.

I have great respect for the memory of Konstantin Mikhailovich Gerasimov, a major business executive on a national scale, who for a short time became the director of our Institute. Until now, I quote his words at an opportunity: “If the laboratory in the absence of the chief works no worse than with him, then this is a good boss.”

Of course, I fondly remember Grigory Vasilyevich Balakov, Boris Efimovich Vander and many other comrades at the Institute of that ancient time. I am very close to the staff of Department No. 8, which I was fortunate enough to head for a number of years.

Thinking about my work at plant No. 465, and then at NII-20, I involuntarily repeat the words from Bunin's poem: "... the voice of an old life, from which only beauty remains."

I would like to wish the staff of the Institute, which is dear to me, prosperity and new outstanding engineering achievements.

## TO THE DEVELOPMENT OF THE THEORY OF INTENSE BEAMS AND LINEAR ACCELERATORS WITH POKF IN THE USSR*

* Speech by I.M. Kapczynski at the solemn meeting at the University of Frankfurt, dedicated to the awarding of the scientific degree of Honorary Doctor of Natural Philosophy to him.

(Review of my life and work in science)
Ladies and gentlemen! Dear colleagues!
I hope you will forgive my bad English. Unfortunately for me, my German is even worse than my English.

It is a great honor for me to receive an Honorary Doctorate degree from such a world-renowned university as the University Johann Wolfgang Goethe in Frankfurt am Main.

Now there is an opportunity to recall and present some aspects of my life in science.

The theory of oscillations was my first youthful love. My work in the forties and partly in the fifties was connected with the theory of low-frequency generators of sinusoidal oscillations that do not contain inductances. Such circuits are known as PC oscillators. The tasks were to stabilize the frequency and amplitude of oscillations, as well as to suppress higher harmonics. The results of this work formed part of my book, published in 1954. The book was also published in Japan. I have the happy opportunity to say that thanks to the scientific activity of my teachers, professors Andronov and Khaikin, the level of development of the theory of oscillations in the Soviet Union was very high.

In the mid-fifties, my interests moved to the area of proton linear accelerators. In 1966-1967, two of the largest linear proton accelerators in the Soviet Union were built. The first of them - in Moscow at the Institute of Theoretical and Experimental Physics; the proton energy was 25 MeV, the pulsed current was 250 mA. The second one is in Protvino (near Serpukhov) at the Institute of High Energy Physics. This last accelerator produced protons with an energy of 100 MeV and a pulsed current of 150 mA. The values of the pulsed current were large for that time. With the light hand of Colin Taylor, the laboratories that took part in the "design and construction of these two accelerators" were referred to at CERN as the "Kapchinsky team".

When designing linear accelerators in the fifties, it became clear that the theory of intense proton beams should be developed. It is obvious that at present the theory of intense beams, thanks to the work of American and Western European scientists, is well developed. However, in the fifties there was a bare field. Our preliminary results were presented at the International Conference in Geneva in 1959. For the first time, the problem of the self-consistent Coulomb field of a particle beam in a smooth focusing channel and in a channel with an alternating gradient was solved. The proposed phase distribution is currently widely used in the theory of intense beams. In the American and Western European literature, this phase distribution is known as the "KV-distribution" (Kapchinsky-Vladimirsky KV-distribution).
Let us pay some attention to the problem of a self-consistent field. In a general setting, this problem was considered for the first time by the Russian scientist Vlasov.
So let's assume that all acting forces, including external forces and Coulomb forces, are linear. Then the first integral of motion can be expressed in the following form:

$$I(x, y, p_x, p_y, z) = a_{11}(z)\cdot x^2 + a_{12}(z)\cdot y^2 + a_{21}(z)\cdot p_x^2 + a_{22}(z)\cdot p_y^2,$$

where 2 is the independent variable along the axis. In accordance with the well-known Liouville theorem, the distribution of the phase density must be a function of $I$: $n = n_0 \cdot f(I)$

The space charge density is

$$\rho(x, y, z) = n_0 \cdot \iint f(x, y, p_x, p_y, z)\, dp_x dp_y \,.$$

With a HF distribution, all particles have the same integral of motion: $n = n_0 \cdot \delta(I - I_0)$. As can be shown, in this case the beam has a cylindrical shape and the space charge density distribution is constant in each section.

The Coulomb potential is

$$U(x, y, z) = -\frac{\rho(z)}{4\varepsilon_0}\cdot\left[x^2 + y^2 + \frac{a_x - a_y}{a_x + a_y}\cdot(x^2 - y^2)\right],$$

where $a_x(z)$, $a_y(z)$ are the semiaxes of the sections. The function $U(x,y,z)$ leads to linear Coulomb forces, as suggested. Thus, the HF distribution gives the simplest and most rigorous solution for the self-consistent beam field.

Now I would like to dwell on the development of linear accelerators with spatially uniform quadrupole focusing (SPQF).

It is well known to focus particles in a linear accelerator with drift tubes using a spatially periodic structure consisting of static quadrupole lenses with an alternating gradient. Until the seventies, this was the best way to focus. However, this method requires a high voltage pre-injector. The capture of particles in the acceleration mode is relatively low.

In 1968-70. Dr. Teplyakov and I published the first works devoted to the substantiation of high-frequency electric focusing in a spatially homogeneous quadrupole structure (SHQS). Ion linear accelerators with SHQS are currently widely used and widely described. More than 200 papers on the theory and technique of SHQS have been published. A brilliant review of the work on SHQS was made in 1983 by Professor Klein.

In principle, an accelerator with SHQS is a four-wire line with quadrupole symmetry of high-frequency power supply. Charged particles moving along the longitudinal axis of symmetry experience the action of an electric field with a variable sign of the gradient. This leads to the effect of quadrupole focusing in a spatially homogeneous structure.

The SHQS is of interest primarily because the focusing rigidity does not depend on either the particle energy or the particle phase with respect to the high-frequency field. The structure of the accelerating electrodes makes it possible to vary the acceleration efficiency and the equilibrium phase over a wide range. The capture of particles in the acceleration mode can be increased to 95-97%. This is twice as high as the best values for other structures. The injection energy can be small. However, at a low injection energy, the limiting value of the beam current remains high.

SHQS is expedient as the initial part of a linear accelerator. As the particle speed increases, the rate of acceleration decreases.

Acceleration of particles in the SHQS section with almost no losses, starting from a low input energy, makes it possible to create a high-current cw accelerator.

The use of the SHQS effect in the initial part of the linear accelerator makes it possible to simplify and reduce the cost of the formation of the input beam for injector accelerators. However, the main impetus for the development of structures with SHQS appeared when new directions for the use of linear ion accelerators became clear: the creation of high-flux neutron generators for radiation materials science associated with the problems of TNR; formation of high-current proton beams for the electronuclear method of producing atomic fuel; creation of linear accelerators of heavy ions for various experimental installations of nuclear physics and linear accelerators of superheavy low-charge ions for inertial thermonuclear fusion; creation of light generators of neutral particle beams. The development of linear accelerators for these problems has previously been associated with insurmountable difficulties. For the initial part of the linear accelerator, the main difficulties were associated with a low coefficient of particle capture into the acceleration regime and with a high injection energy.

The use of a beam from an ion source is especially important, since it is difficult to obtain beams of high intensity in such sources.

The practical development of structures with SHQS began in the Soviet Union in 1970, and abroad has developed on a wide front since 1979. Let us present some experimental data on the SHQS section of the Istra-56 accelerator. Up to a beam current of 100 mA, particle losses in the channel are practically absent, and at a maximum beam current of 250 mA they reach 25%. Under matched initial conditions and at an initial normalized phase current density of 1 A/cm.mrad, the output beam emittance increases in proportion to the beam current. As the beam current is increased to 100 mA, the normalized emittance increases by about a factor of two. The possibilities of reducing the emittance growth require further study.

Thank you for attention.

## EVENING SPEECH AT THE FRANKFURT GOETHE UNIVERSITY

Ladies and gentlemen!

When a person is over sixty years old or has come close to seventy, it is important for him to know whether his life was useful for other people. As I mentioned a few hours ago, I hope I have made some progress during my lifetime - the development of low frequency PS oscillators; design and construction of two largest proton linear accelerators in the Soviet Union; the idea of a special phase distribution of accelerated particles, which is known to specialists in accelerators as the KV distribution; the idea of high-frequency quadrupole focusing in linear accelerators.

I am pleased to say here that the theoretical and experimental developments of the SHQS at the Institute of Applied Physics of the Goethe University, Frankfurt am Main are brilliant achievements in this field. Professor Klein directs these works. I would also like to acknowledge Dr. Schempp's outstanding ideas in the field of resonant structures for linear accelerators with SHQS.

There are other reasons for me to be proud of receiving an Honorary Doctorate from your University. I love the poetry of the great German poet Johann Wolfgang Goethe. Of course, I knew his poems by heart as a child. Please forgive me, but I will try to say a few lines in my bad German. We all build our lives by remembering the words of Goethe:

"We nurture and cherish
Flowers of the sky in earthly life"

I would like to thank you again.

## FROM THE HISTORY OF THE LAUNCH OF THE I-100 LINEAR ACCELERATOR IN SERPUKHOV

The linear accelerator, consisting of three sections, each approximately 30 m long, was launched sequentially for a month - section by section. The achievement of the design proton energy at the end of each section was fixed by measuring the beam current with a transient induction sensor and a Faraday cup installed behind it. The Faraday cylinder was covered with a copper plate. The thickness of the plate was chosen such that protons with the design energy penetrated through it. Protons of lower energy were retained by the plate. When the accelerating field in the section exceeded the critical level, the readings of both sensors were abruptly equalized. The path length was determined from nomograms from Segre's book "Experimental Methods of Nuclear Physics". Since the error in determining the thickness of the plate could reach ±3% in this case, the plate was taken 3% thinner.

So the first and second sections were successfully launched.

On the night of July 28, the last, third section was to be launched. Everything was ready for launch. On this night, we were honored by her presence and inspired to a feat by a certain lady - a leading worker of the department of accelerator and thermonuclear facilities of the SSAE, Comrade S. We helpfully explained to Comrade. S. - what two pointer devices she needs to look at and what the movement of the arrows will mean.

At two o'clock in the morning, the accelerating voltage began to rise. The device, operating from an individual sensor, showed the presence of a beam current. Here the voltage is raised above the calculated critical level. The device operating from the Faraday cylinder is silent. Higher. No more - electrical breakdowns. The beam passes, but is not accelerated. A gross error in the calculation of the third section of the accelerator? Tov. S. muttered something about the fact that you need to be able to launch accelerators.

There was no successful launch.

At three o'clock the analyzing magnet was turned on. Set up the installation. The energy spectrum is narrow. So there is acceleration. What's the matter? And then one of the young theorists says:
- How so? After all, I counted the last target with an accuracy much better than 1%!
- But Segre's nomograms do not give such accuracy!
- This year, a theoretical work appeared, significantly refining the formulas from Segre's book. I counted according to new formulas.
- Have you checked in the second hand?
- No, we didn't.
This is getting suspicious.
- Bring the nomograms!
And there is. Big mistake: the plate is made 17 mm thick. The nomograms give a thickness of 14 mm for an energy of 100 MeV.

The next day, a 14 mm thick plate was placed and the desired effect was obtained. The matter was simple: the new exact formulas turned out to be very cumbersome, the calculation on them took three days of work on a Mercedes calculating machine. And an arithmetic error of 21% was made.

After this Comrade S. said:
- It turns out that launching accelerators is easy. You just need to press the button.

During the very first studies of the linear accelerator, which followed after the physical launch, it turned out that a significant part of the beam was lost somewhere in the first section. Dismantling the accelerator threatened with a long stop without a guarantee of an answer. And then an experiment was proposed. It was possible to try to localize the location of the losses by measuring the beam current at the output of the first section when the focusing cells were successively switched off. The curves were carefully taken, after which, in fact, a discussion broke out - which focusing cell is to blame.

- Cell 81 is to blame ...

The doctor of science objected:
- Cell 83 is to blame ...

The candidate of science expressed his opinion:
- Cell 82 is to blame!

The dispute was long, fundamental and could only be resolved by some more subtle experiment. Such an experiment was proposed and carried out by the candidate. It became obvious that one of the drift tubes, namely cell 82, was to blame. Scientific doubts were already resolved when the head of the injector vacuum group entered the room and said:
- I heard you look in the first section for a place where the particles are lost. I suggest looking at the 82 cell drift tubes.

Everyone looked at each other and asked curiously:
- Why?

The leader of the group pulled out a thick notebook.
- I found an entry in my notebook: during the installation of the accelerator on May 19, 1966, in the region of 82 cells, a heavy flange was dropped inside the resonator. Maybe he hit the drift tube?
- Who knew about it?
- Me and mechanic B.
Why didn't they say then?
- After all, the pipes are firmly fixed. And this would delay the installation.

Resonators are activated. However, the hatch was opened. An engineer climbed inside the resonator. The first tube 82 of the cell was knocked off by more than 1 mm at a tolerance of 50 microns.

## OBJECTIVES OF LABORATORIES

Note N.V. Lazarev: Among the countless folders with various completed or just begun scientific works of I.M. there are notebooks with concise, sometimes understandable only to himself, records of issues discussed at the operatives. Even a cursory glance at these brief lines shows how extensive was the range of issues that aroused his interest and which he kept under his control. All entries have their own dates, in general I.M. had an internal discipline that allowed him not to forget the details of the planned and important and smaller (many tens, if not hundreds) cases simultaneously carried out under his leadership both at ITEP and at IHEP. There is no need to give examples of everyday work, the results of I.M. as a leader and scientific leader speak for themselves. For large periods of time of several years for each subdivision of the I.M. made detailed lists of work priority. Such was the style of his leadership of the team of the department.

Perhaps it would not be superfluous to show how, somewhere in 1981-1982. I.M. outlined tasks for the laboratories included in the department.

Laboratory No. 121
1. Operation of I-2. With the allocation of employees of laboratory No. 123 to maintain a number of technological systems - with the obligatory implementation of decisions for them (under the responsibility of the head of laboratory No. 123).
2. Design work, creation and maintenance of technological systems of technological systems of new accelerators in building No. 119 (mechanics, temperature control, power supply, vacuum, RF systems, diagnostics).

3. Development and provision of MTC (hard magnetic quadrupoles).
4. Electrical maintenance of the premises of laboratories Nos. 121, 123, 124.
5. Maintenance of a mechanical workshop for all works of laboratories Nos. 121, 123, 124.
6. ACS LU I-2 and new boosters.
7. Research on electron cooling.

(For laboratory No. 122 - separately from others)

Laboratory No. 123
1. Development and creation of electronics and automation systems for new accelerators.
2. Maintenance of electronics and automation systems of LU I-2.
3. Development and creation of pre-injectors for new accelerators.
4. Development and creation of ion sources for new accelerators.
5. Development and creation of stabilized systems of high-current and 6. high-voltage power supply for all laboratories of the department.
6. Supervision of work on the construction of new resonant LU.

Launch and experimental work with a beam on new resonant LA.
To paragraphs 6 and 7 - with the allocation of laboratory staff 123 and 124 for the maintenance of technological systems and accelerating systems with the obligatory implementation of decisions for them.

Laboratory No. 124
1. Development of RF accelerating structures for new resonant
LU.
2. Development and creation of RF generators for new resonant LANs.
3. Calculation and theoretical work.

A hierarchy of priorities introduced for a specific period of time.
1. Operation of the I-2 (Ne) installation. 2. LF IPP (NCHU-1 and NCHU-2). 3. II B|+ (Xe2+). 4. Alvarez 10 MeV (MTK). 5. LIU-5; 6. LF TYP VR \ 7. AIIS I-2. 8. Operation of the CMU installation. 9. 2 sections TYPr VR+. 10. LF continuous mode 0.5 mA. 11. Alvarez 40 MeV (MTK). 12. Electronic cooling.

## TO THE HISTORY OF I-100 CREATION

Note N.V. Lazarev: Ten years after the creation of his main brainchild - the largest I-100 linear accelerator in the world at that time, Ilya Mikhailovich summed up some results, what was the situation with proton accelerators in the world at the beginning of our work, and also what (of the most important cases) , by which institution and by whom specifically it was done. Below is a summary written by him for himself, apparently for some kind of report.
October 12, 1977

What was available at the start of work on the I-100 in 1958.
1. The work of KIPT and IHF was still classified.

2. The USA and England began to open from 1955 and detailed descriptions and theory of machines were known in Berkeley and in Minnesota. 1952 - Blewett's proposal and the theory of phase oscillations. Theoretical work of Smith-Glückstern. One-Particle Approximation.
3. Working LU - with grid focusing. Beam current 5-50 μA (Berkeley, Minnesota, Kharkov for KM).
4. RTI work since the mid-1950s - the MLK accelerator with hard focusing at 35 MeV for the LIPAN. HELL. Vlasov is the most complete development of the one-particle theory with hard focusing.

II. Division of work in a triangle.
1. ITEP - coordination; theory; accelerator calculations (dynamics); development of a focusing system; PHYSICAL SUBSTANTIATION.
2. RTI - chief technologist - resonators; RF power; AR; vacuum and water; PROJECT.
3. NIIEFA-FORINJECTOR; working design of drift tubes and power system.
About the work of RTI: bold decisions, successfully managed: resonators of large electrical length for TT idle transitions; powerful RF generators; multicavity power supply system with new requirements for AR systems; the rigidity of the resonators (there was no pgshshz1et!) and the cascade vacuum.
Contacts: I.Kh. Nevyazhsky, N.K. Titov.

III. The main task is high intensity in the ring.
Diagram energy - intensity. Increase output current
up to 100 mA. The limits weren't clear as there was no theory of collective processes. Status: there were no concepts of “phase volume”, “capacity”, “matching”, etc. in the theory of LUs (The work of A. D. Vlasov on longitudinal repulsion was pioneering and would be interesting if he had not received a very low current limit 20 mA and would not defend it). The first statement of the question and the first theoretical solution - in the report of V.V. Vladimirsky (Geneva, 1959).

IV. PATHWAYS: increasing LU acceptance; increase in the phase density of the preinjector beam.
NIIEFA assignment: preinjector beam current 400 mA; phase volume 1 cm * mrad. NIIEFA took up these conditions (A.I. Solnyshkov). This determined the direction of NIIEFA work on ion optics (V.S. Kuznetsov, Yu.P. Sivkov).

V. Accelerator-injection problems.
1. In short-wave injectors (A, - 1.5 m) Uk is less than Uk PS.
2. The RF energy margin is small.
3. Poor use of the LU beam due to longitudinal capture in the PS.

VI. Consideration at ITEP of a number of acceleration systems.
1. NUMBER OF BASS: B = 1 km; 250 accelerating spirals, length A,/2.
A = 115 m D = 2.6 MHz). The radius of the spiral is 15 cm with a length of 1.5 m. Advantages: grouping of particles around f5 PS and no extra particles; UK LU = UK rings; necessary energy reserve for acceleration. Disadvantages (causes of failure) - inconsistency of longitudinal volumes due to a large difference in the rate of acceleration; large ungrouping path; huge - function because of the turns. (Perhaps they did it now: achromatic turns, ungrouping in the ring.)
2. FOUR-CRANKSHAFT SYSTEM. Spatially homogeneous focusing (the idea of V.V. Vladimirsky, my calculations, discussions).

H. DISCUSSION WITH I.Kh. NEVYAZHSKII. [His] REFUSAL. The revival is now with the ideas of V.A. Teplyakov: non-stationary bunches, double H-resonator.
VII. To resolve the issue with acceptance and with No., Alvarez was considered with A from 2 to 8 m. A = 4 m turned out to be optimal for our currents, but we did not dare, since the machine was very lengthened. We stopped at A = 2 m, which agreed with the RTI - as the optimal wave for the [lamp] GI-27A. Other things being equal, the current is like A2.

VIII. Development of a focusing system
1 For the first time - suppression of the 5th harmonic.
2.Methodology for taking into account the edge effects of lenses.
3 Choice of FODO structure. Discussion on FOFDOD, which is better in terms of tolerances.
4. Theorem on 4 parameters and difficulties in calculating the matching channel: the surface of minima. Exit to I-2. (Points 1-4 - V.K. Plotnikov).
5. The first samples of impulse lenses. (Points 5-6 - N.V. Lazarev and A.M. Kozodaev).
6. Development of a pulsed power supply system.
IX. The first report in Dubna in 1963 [On the project of the I-100 injector].
(And M. Kapchinsky, V.G. Kulman, N.V. Lazarev, B.P. Murin, I.Kh. Nevyazhsky, V.K. Plotnikov, B.I. Polyakov).

## PROSPECTS FOR THE USE OF LINEAR ACCELERATORS OF CHARGED PARTICLES

Linear ion and electron accelerators have long been an indispensable part of large experimental facilities for high-energy particle physics. The improvement and modernization of linear accelerators was largely dictated by the operating conditions of physical installations. These tasks are still the most important in the creation of new accelerators. At the same time, in recent years a large number of national economic fields of technology have appeared, the development of which is closely connected with the development of the theory and technology of linear accelerators.

Among the areas of technology, the further development of which cannot be ensured without the creation of appropriate linear accelerators for protons and deuterons, the following should be mentioned:
1) installations for the production of medical radionuclides, both for the purposes of diagnostics and for the purposes of therapy;
2) plants for the destruction of radioactive waste from nuclear power plants;
3) high-power neutron generators for radiation testing of fusion reactor materials;
4) powerful generators of directed beam energy;

5) installations for the inspection of space objects;
6) light installations for screening luggage at airports;
7) plants for the development of nuclear fuel for nuclear power plants.
Linear induction accelerators are currently the only known sources of high-power electron beams, making them the most suitable exciters for high-energy free electron lasers (FELs). It is possible to use electronic LIAs to create directed beam energy. Experiments are being made on the use of powerful electron beams of LIAs for driving rocks. FELs excited by linear induction accelerators are effective devices for the production of unique solid-state elements of electronic technology.

A separate area of development is linear accelerators of heavy ions for inertial thermonuclear fusion plants and a number of other applications (for example, ion implantation).

The development of the fields of technology enumerated above is proceeding unevenly. However, it is obvious that continuity in the development of the physics and technology of linear accelerators must be ensured. Physical and technological research must proceed at a faster pace than the development of applied applications of linear accelerators, while at the same time meeting the priority needs of experimental physics.

## RULES FOR YOURSELF*

1. Whatever you hear - DO NOT WORRY - will not help ... DO NOT WORRY - everything passes! ...
2. Regardless of the characters and business qualities - TO EVERYONE - WITHOUT INTERNAL IRRITATION!
3. To any situations - CALMLY! Everything is formed...

And further:

1. Behavior at the reports: speak when you need to, and not when you can.
2. Do not point out the shortcomings of others and do not look for them without instructions.
3. Avoid "I", do not point to your merits.
4. Do not point out the shortcomings of others, just give the right conclusions and results.
5. Do not skimp on co-authors.
6. Enter yourself (as a co-author) when the responsible share is yours and you wrote.
7. Thanks - do not forget and coordinate.
8. In any papers that you write yourself, do not indicate your merits, but only a list of what has been done.

* The sheet with this entry was always lying on the desktop under glass

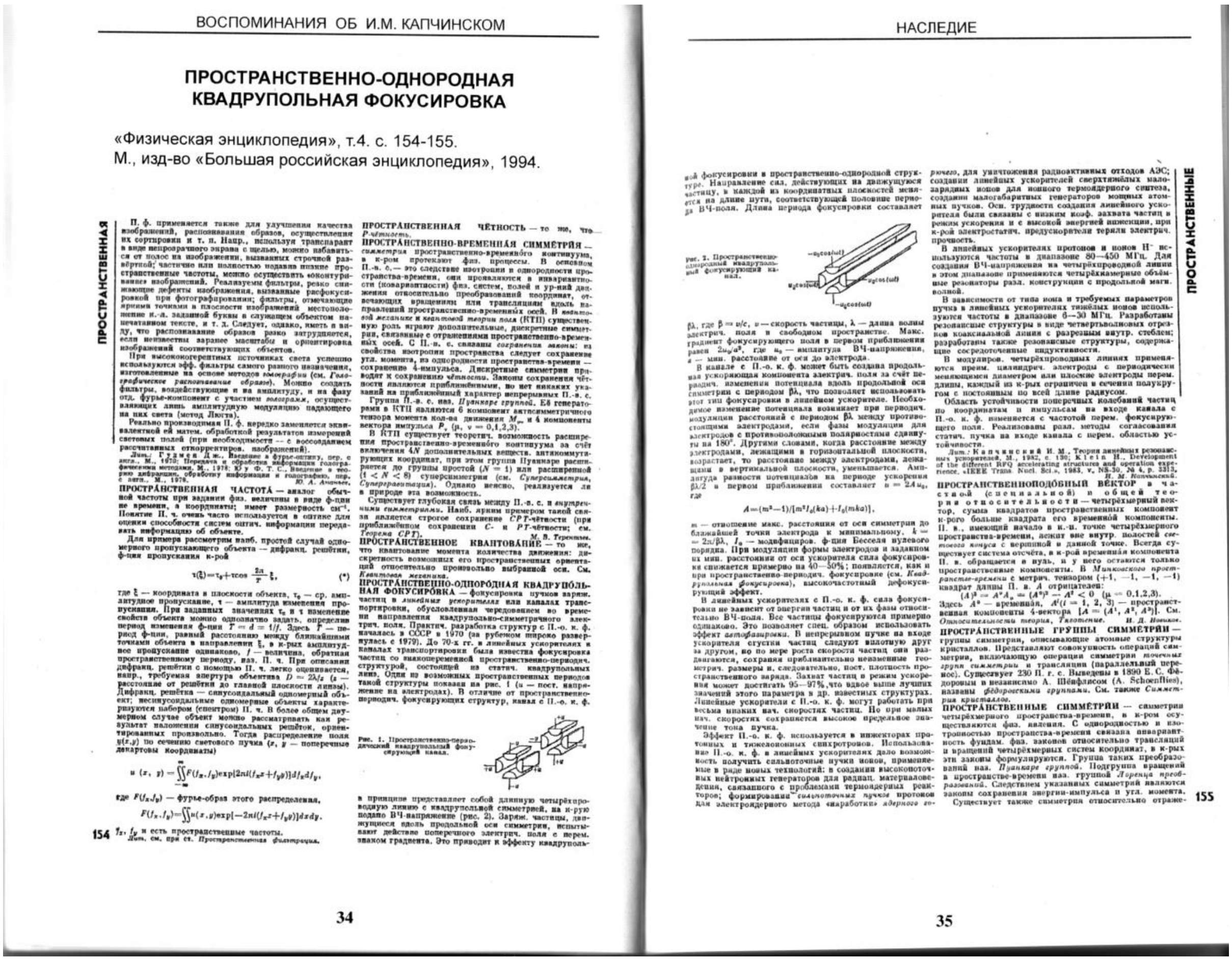

ВОСПОМИНАНИЯ ОБ И.М. КАПЧИНСКОМ

НАСЛЕДИЕ

ПРОСТРАНСТВЕННО-ОДНОРОДНАЯ КВАДРУПОЛЬНАЯ ФОКУСИРОВКА

«Физическая энциклопедия», т.4. с. 154-155.
М., изд-во «Большая российская энциклопедия», 1994.

34

35

**SPATIALLY HOMOGENEOUS QUADRUPOL FOCUSING**

"Physical Encyclopedia", v.4, p. 154-155.

M., publishing house "Big Russian Encyclopedia", 1994.

## POEMS*

*** Several poems by I. M. from a large cycle of 1935 - 1946.**

VIOLIN

Below, a wave came ashore with a groan, Spitting wounded water in the face...
A little higher - on a stone emaciated An old man sat - bony and gray-haired.
At the very feet - the sand spilled over the beach (Sunset was already bloody yellow).
Everything was silent. Even the stones were silent, Whispering welcoming the wave!
Pressing the thin body of the violin to his shoulder, Squinting his eyes attentively at the threads of the strings, The old man played a languid and unsteady motif fanned by legends...
And he took off, obediently echoing the bow; Eaten into the blood, woke up and worried; Perhaps this is how the sea rings and smells, When the many-pood squall subsides!
He fell on the limestone, warmed by Steep splashes and colors of dawn; Did the old man weave the tunes of the abandoned ghetto into his melodies?
He sang about the days - cold and harsh ... Through the towns scattered around - At evening parties, at local weddings - A wandering orchestra played everywhere ...
Even then he dreamed at night - And this day, boiling hot,

And the fabulous glory of Lisa Gilels,
And Oistrakh's swift bow!
Is it not because - like the sea after a squall, Like a distant dream that came into life, Its melody disturbed and called,
Feelings tremble...
I felt longing in my chest (Maybe it's the spring years?)
It was a pity that I will not grab the heart,
And I won't give it to him for the songs.
Odessa, 1938

THEODOSIA. AIVAZOVSKY
The waves are rippling,
The sea, shuddering, sleeps;
Ripe disc, full of yolks,
Down, squinting, looking.
City handful of diamonds All scattered under the scythe.
Moonlight, smearing the expanses,
He lay down on the water in a stripe.
Outline of a sailboat Shimmers with silver;
Capel slender infantry Quietly beats overboard.
There are no sounds. Only the waves from below are splashing against the granite;
A full month shines from above;
The sea, shuddering, sleeps.
Feodosia, 1936

The night was cold, mute...
Dark cars on the way.
Without raising your head
I stood, and did not know how to leave!
This hour was not even scary to me - In the rush of cash registers, in the platform crush ... So you left, Natasha,
Flickering through the stained window.

Behind him are empty sidewalks.
Apart from the slow years
Under my temporal blows I wander into someone else's night alone...
And around - unkind cold,
The wind, bitter and damp,
In rainbow puddles smelling of smoke,
Blue clouds tattered cut ...
I believe in the days - the time will come!
But I can't twist my memory.
I will remember both longing and darkness,
And the car on the path of life.
Tashkent, 1942

THOUGHTS ON LECTURES ON PHYSICS
Time is a scalar, -

preaching lecturer;
But I do not believe
his words:
Time -
also a giant vector, with an arrow resting on death,
into nothing.
Who will count his units?
Is there any capriciousness in the world?
Each of us,
like this chalk
rushes
By black board
human life.
I want a minute
moments
didn't steal!
I will break the vector with a stubborn hand,
So that the arrow of time
not to death
rested
And in life
young
like a downpour in spring!
Moscow, 1937

Who was not on the Dnieper at sunset?
The breath of autumn is uniquely fresh,
And clouds like gold in the sky
Like lights of unfulfilled hopes.
But at this hour, solemn as glory, A stream stream along the old channel,
The river was calmly majestic - It seemed to her that youth had come.
Kyiv, 1936

When the night embraces the streets, The constellations will be knitted into a bouquet, Filled with their sounds The pulse on the hand;
The two-horned Hump of the crescent moon will snap out into the sky, It will fill with furious anxiety,
And the body will be crushed by emptiness, - In the brain, the twists of thoughts tremble, I can’t find a corner for myself;
I wander hot and dreary And my heart is cold! My breasts feel tight
And lips - fresh as night!
And the shoulders are so warm
Eyes - favorite for a long time!
The rage of the meeting burns me,
It's like she's here with me!
I dig into my lips, into my neck, into my shoulders, I am an unbridled dream ...
When the night hides the streets,

Will spin in a medley of feelings - I wander dreary and hot And I whisper stuffy lines ....
Moscow, 1937

ODESSA

I am the best of songs
Of those that I still have to sing, I will present to my native Odessa, the Best of the cities in the country!
The Odessa spirit was infused with poetry,
And I'm proud that hangs over me Like a Black Sea storm, a simple Gangster nickname - Odessa!
Swallowing salt, all bleeding with a word,
With a knapsack wandering between the rocks, For the first time here with the Odessa fisherman High Til walked over the stones! ..
But our life was not simple in vain... And it began...
We lived to the day
How soldiers pass along your tormented boulevards of Alien banners.
What do they need?
All this expanse at the pier,
And our suns are lemon balls,
And our songs are a cheerful roll They do not understand in the Odessa fire! The Romanians walked, kissing the earth in their hearts, Not out of love - under the machine-gun heat!
And even the sea, not accepting prose,
They were covered with a cloudy yellowness ...
But the Odessa consonances will flourish!
All my life in slaves to wash - you eh?
Be by the sea, even brighter and better,
My fate is a big cradle!
Tashkent, 1942

May it be sung hundreds of times!
Let this voice, let the hands sing of thousands of poets!
I am not ashamed to write about them.
About what kingdoms Her huge eyes give,
That even the old Stradivari would not be able to tell everything.
And revealing something new
I'll shout the news to the world
About what a good
How much I want her!
We were not afraid of old words with her. So let love, and let the moons Play a thousand pianos,
I'm not ashamed to write about them!
Tashkent, 1943

We did not glue wax wings in Crete,
But they lived in a difficult, stressful age.
It started here! From here, a whirlwind of discoveries Led a rapid run into the centuries.
We were trained to touch the spaces,
And penetrate to the nuclear beginnings ...
It is not for nothing that romance swirls on the tables of laboratories at night!
Squat, calm and quiet,

We sucked hard with milk,
And left paints and poems,
To follow the instrument arrows.
Here every step was hard for us!
But thoughts over the centuries of ray,
Only to catch their breath they went from time to time Into the quiet world of the art gallery ...
We started. Launching a rocket
Or sailing off to the moon again,
Remember: in this oblique movement - All our youth, and greedy love!
Moscow, 1946

Her eyes - among the constellations closely. - Such depths and the ocean did not know.
She was strangely wonderful
Unexpectedly tender to the pain!
Suddenly immersed in her pupils,
I was heavy, and went to the bottom like a stone ...
I could wander around Lanzheron all day, Dream about her, and think about one thing.
I called her eyes mine
Their dark world, like a fairy tale, having memorized ...
And still her simple name sounds like a symphony in my ears!
Odessa, 1939

I could weave a feeling into a verse, rhyme a heartbeat,
And, burning, bring a Burning poem ...
It’s not the first time for me to pour in verses Everything that has boiled in my heart;
But I'm afraid to scorch Your eyes with a bold line.
Odessa, 1939

## OMAR KHAYYAM

I want to touch you. And if the Heart, wandering through the old streets,
Will not find you, answer at least a song,
Even with a sly smile, even with the ringing of a dutar ... Look how the moon is circling over the city, Pouring with its honey power,
And the southern sky above us is no worse,
Than that which carried your consonances!
You loved to wander under the honey moon,
Where the Milky One wrapped you in a starry ribbon!
So get out now! Meet me! Let's go, as friends, along the roads of Tashkent!
I am also rushing about... Tell me what you did when geometry called...
Tell me how by dawn, having finished the calculations,
You were looking for new rhymes, it used to be! ..
How accurate thought you mixed up on wine,
As rhyme did not spoil the language of the theorem,
How deftly he knew how to wrap the delicate sting of the poem with the petals of greeting!
Look, through the lunar stream - the crossing!
From the dark centuries I'm reaching out to you...
So come out, poet! Confirm my right To life and love - to poetry and science!

Tashkent, 1942- 1943

I'm here again, I'm passing through Moscow In the inky haze, among the night stones; But for some reason, with a new longing, A blind shadow crawled up to me here.
Is it not because I wander in the powder, Night winds drive the snowy trash,
And there is no girl - that same, that good one, With whom life is like evening - in half? .. Isn't it because with a lonely heart I reached out for a gentle hand? ..
I didn't leave anything in the East!
I am here again! But I wander in anguish...
Moscow, 1943

The wind of time blows me. The further I go, the stronger, the more persistently he hits me in the back. He freaks out, he turns into a hurricane. And now I can't stand up anymore. The whirlwind picks me up and carries me swiftly through space.
I can't come to my senses.
Days and nights pass with such speed that weeks - day after day - merge into a solid gray twilight - as for a Welsh Traveler.
The sun rotates at a wild speed, leaving yellow-white stripes in the sky, like a match smoldering in the dark.
I want to linger, I want to look around, think, concentrate a little, but I can’t: time grabbed me by the scruff of the neck, the whirlwind twists me and drags me along.
Below, in the semi-darkness, something is moving back at breakneck speed. Someone is screaming at me through the howl of the hurricane that this something is called Life. I try to make out individual outlines, but I can’t, I don’t have time. I can't even look back. I was knocked down, a blizzard drags me; sometimes the terrible simum of the desert burns my back; but constantly - forward, forward, in the gray twilight of weeks.
The wind of time drives me.
Moscow, 1939

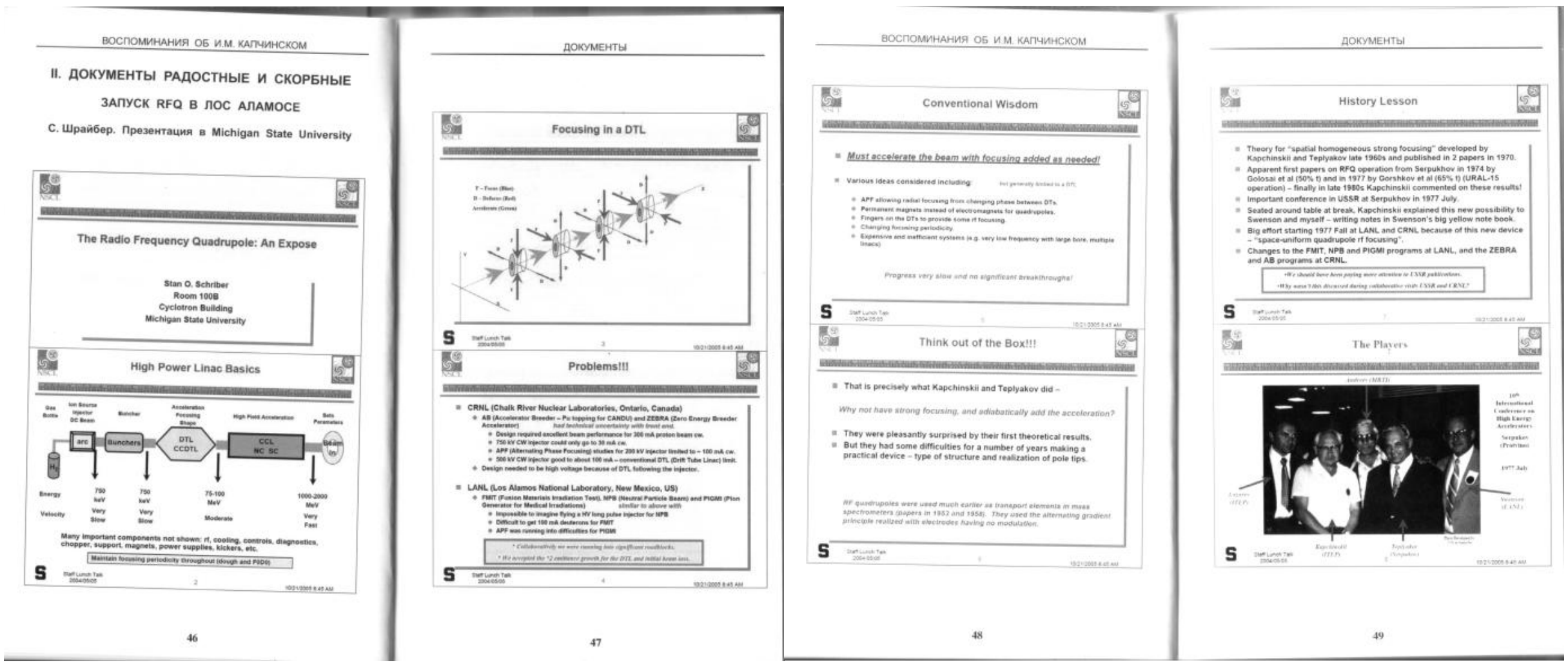

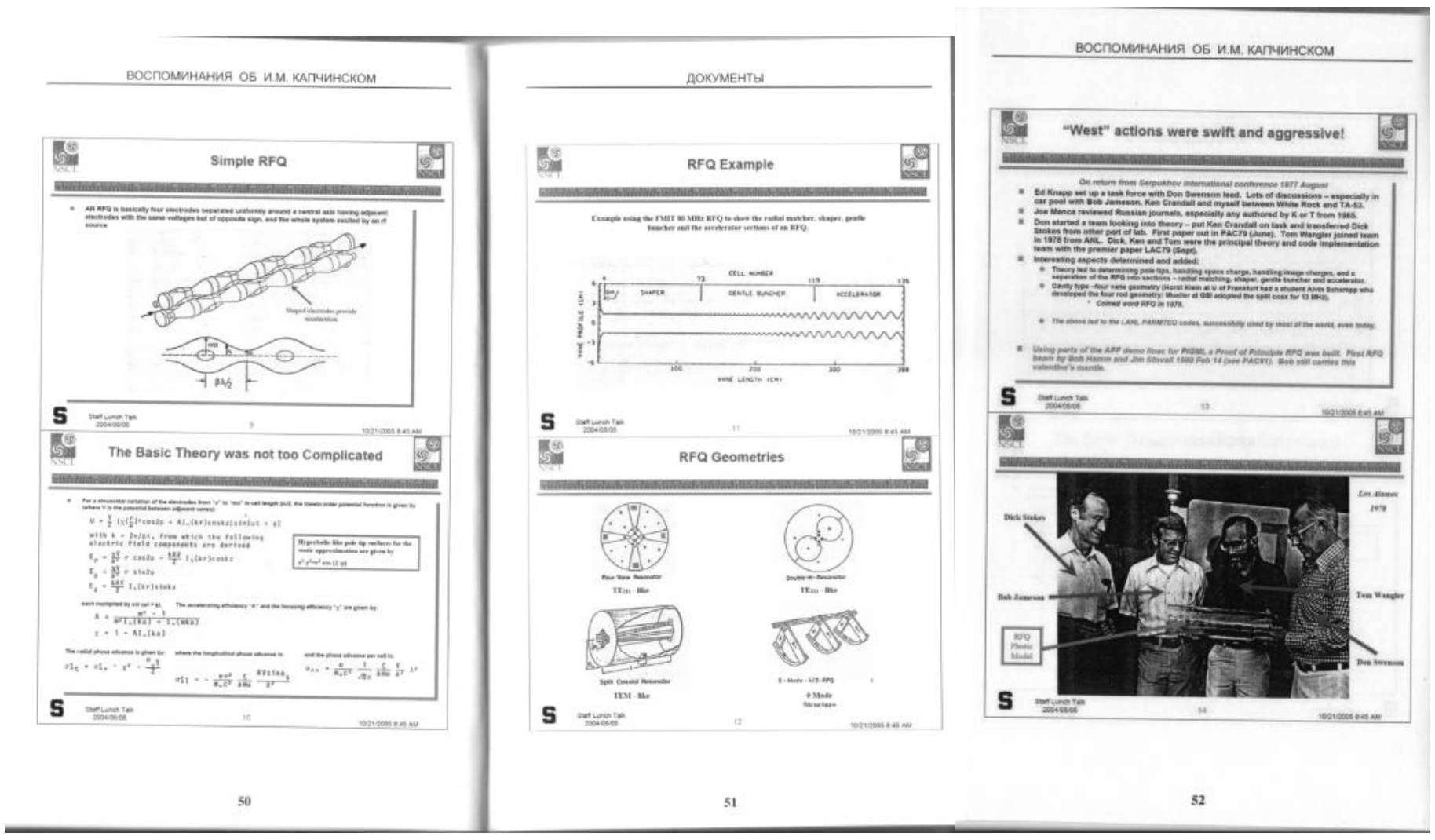

## II. DOCUMENTS JOYFUL AND MOURNFUL

### SUCCESSFUL START MESSAGE – RFQ launch at Los Alamos

Immediately after the launch of the first RFQ model, the leader of the MS-811 LASL Accelerator Division, Edward Knapp (with colleagues) sent telexes of the same content to ITEP (and IHEP).
The text of the telex received at ITEP from Los Aolamos Scientific Laboratory (LASL):
15 February 1980
Copy to I.N. Kapchinsky
Copy to Chuvilo Director ITEP
Edward Knapp MS-811 University of Caalifornia, LASL   UNCLASS
The RFQ is alive and well in Los Alamos
Fifteen mA of protons were accelerated from 0.1 MeV to 0.64 MeV in 425 MHz RFQ structure with a length of 1.1 m and diameter 0.15 m during the first attempt on February 14, 1980.
Signed by: R. Jameson, D. Swenson, R. Stokes, J.Potter, E.L.Kemp, M. Machalek, E.A. Knapp

In response, a telex was sent from ITEP:
E.A. Knapp, LASL, N.M., USA
Our congratulations to you and your colleagues upon your remarkable realization of the RFQ structure. We wish you more success, hope to discuss the details.

I.M. Kapchinsky
I.V. Chuvilo

### REVIEW on the scientific and social activities of Professor Ilya Mikhailovich Kapchinsky

Professor I.M. Kapchinsky since 1958 has been heading the department of linear accelerators he created at the Institute of Theoretical and Experimental Physics. From 1964 to 1971, he simultaneously worked at the Institute for High Energy Physics, where he headed the injector department he had created.

Tov. Kapchinsky is a scientist with a wide range of scientific interests and a prominent specialist in the theory of oscillations, radio engineering, physics of charged particle beams and accelerator technology.

Until 1958, I.M. Kapchinsky headed the department of the branch research institute. The first scientific work of I.M. Kapchinsky, published in 1946, was devoted to research on the nonlinear theory of harmonic generators that do not contain inductances. For the first time, oscillatory systems with a large number of degenerate degrees of freedom and with "small nonlinearity" were studied. Subsequently, regenerators have found wide application in technology. The development of the theory of narrow-band amplifiers for ultralow frequencies, which have found application in radio astronomy, was also of great importance. Research by I.M. Kapchinsky on shock excitation of quartz oscillators were successfully used in radar technology.

The most important scientific works of I.M. Kapchinsky are associated with the development of linear ion accelerators. At ITEP under the leadership of I.M. Kapchinsky, with his direct participation, created physical projects for large linear proton accelerators at 25 and 100 MeV. As the head of the ITEP department, and then the head of the IHEP department, comrade. I.M. Kapchinsky led the activities of scientific teams in coordinating scientific work, commissioning and commissioning of both linear accelerators. At the time (1967-1968) the ITEP and IHEP linear accelerators achieved the world's highest pulse intensity of proton beams among machines of this class.

The construction of 25 and 100 MeV hard-focusing linear accelerators in the Soviet Union certainly brought forward the domestic technology of linear accelerators to the first places in the world. The works of Professor Kapchinsky in this area are widely known in the Soviet Union and abroad.

And I.M. Kapchinsky obtained important results on the theory of intense beams of charged particles. Theoretical investigations have been carried out on the effect of space charge on transverse and longitudinal vibrations of particles in accelerated beams. He developed the theory of collective interactions of particles in self-consistent eigenfields of beams. Together with V.V. Vladimirskii, a phase distribution was proposed that is widely used in the theory of high-current beams (according to the English terminology, KV distribution). The effect of longitudinal Coulomb repulsion on the autophasing process at extremely high phase current densities has been studied. Theoretical restrictions on the phase density of beams in ion sources and on the matching devices of the input optics of the accelerator are determined.

And I.M. Kapchinsky together with V.V. Vladimirsky and V.A. Teplyakov discovered the effect of spatially homogeneous quadrupole focusing, on the basis of which a new method for accelerating ion beams was developed. This method made it possible to significantly reduce the injection energy into the linear accelerator and raise the limiting value of the beam current, which opened up prospects for the creation of linear accelerators with a high average current. Accelerators with spatially homogeneous quadrupole focusing (foreign abbreviation RFQ) have received universal recognition and are used in most Soviet institutes and in all foreign scientific centers where linear accelerators are built or operated.

On the initiative of I.M. Kapchinsky began the introduction of hard magnetic quadrupole lenses into linear accelerators; this direction is associated with the development at ITEP of new frequency ranges for linear resonant accelerators and with the solution of the problem of radiation resistance of high-current machines.

I.M. Kapchinsky was the first to propose a two-frequency scheme for a high-current linear proton accelerator for energies up to 100 MeV. A linear accelerator with an energy of 56 MeV is being built in the department according to this scheme. In parallel, the development of a linear accelerator for heavy low-charge ions is underway on the problem of inertial thermonuclear fusion.

During the period from 1958 to 1989. comrade Kapchinsky published 137 scientific papers on the theory and technology of linear accelerators. Published three monographs (1954, 1966, 1982) and received eight patents for inventions. Monograph 1954 reprinted in Japan and China. 1982 monograph on the theory of linear accelerators - in the USA.

For work in the field of accelerator physics, I.M. Kapchinsky was awarded the State and Lenin Prizes of the USSR.

Tov. Kapchinsky is a good organizer of science. He created large scientific teams in the branch scientific research institute, at the Institute of Theoretical and Experimental Physics and at the Institute of High Energy Physics. A large number of young scientists were trained, fruitfully working in the field of linear accelerators. In the teams headed by I.M. Kapchinsky, candidates and doctors of sciences have grown.

As head of the department, comrade Kapchinsky skillfully combines high demands and great sensitivity to his employees. He pays a lot of attention to educational work.

Director of the Institute for Theoretical and Experimental Physics. Doctor of Physics and Mathematics Sciences, Professor I.V. Chuvilo

Chairman of the Council of the Labor Collective of the Institute of Theoretical and Experimental Physics, Doctor of Physics and Mathematics. Sciences D.G. Koshkarev

**OBITUARY**

On May 2, at the age of 74, a member of the Scientific and Technical ITEF Council
Chief Researcher of Department No. 120 Laureate of the Lenin and State Prizes
Honorary Doctor of the University of Frankfurt
Professor Ilya Mikhailovich KAPCHINSKY

A remarkable person, a brilliant accelerator physicist, a prominent engineer, and an experienced organizer of scientific research, has passed away.

After graduating from Moscow State University, I.M. Kapchinsky fruitfully worked in the field of the theory of oscillations, radio engineering, with his active participation the first samples of domestic radars were developed.

In 1958 I.M. Kapchinsky organized the Laboratory of Linear Accelerators at ITEP, which became a department headed by him for more than three decades.

The talent of I.M. Kapchinsky fully manifested himself in the development of the theory and overcoming all engineering difficulties in the construction of proton linear accelerators-injectors I-2 at ITEP and I-100 in Serpukhov.

Works by I.M. Kapchinsky on new structures for the acceleration of intense ion beams implemented in the ISTRA-36 accelerator received the widest recognition in our country and in all accelerator centers of the world.

Kapchinsky is the author of an internationally accepted discovery, 12 inventions, about 200 scientific papers, three of his monographs have been published in many countries and have become reference books for several generations of scientists. He created the "Kapchinsky School", highly valued in the accelerator community.

Ilya Mikhailovich has always been a sensitive and caring person, his absence will be keenly felt by all of us.

The memory of the outstanding scientist I.M. Kapczynski will always be in our hearts.

His employees, friends, management and trade union committee of the institute.

**EUROPEAN ORGANIZATION FOR NUCLEAR RESEARCH (CERN)**

Dear Professor Chuvilo,

We would like to express our sympathy to the colleagues of the late Professor I.M. Kapchinsky, this great man of the entire community of specialists working in the field of linear accelerators.

His contacts with CERN lasted for many years, beginning in 1959, when the formulation of the now generally accepted K-V distribution of particles was the specific guideline of the early period. Since then at CERN, we have consistently received fruitful support from Professor Kapczynski in discussions on how to move our linac projects (linac 1, linac 2, and experimental linac at 3 MeV) towards the particle current limits predicted in his pioneering work.

However, his invention of the RFQ principle in 1970 has most of all influenced modern linear accelerator designs. Since the late 1970s, all linear accelerator designs have been based on RFQ structures in order to solve the inherent problems that arise at low energies. The range of applications was astonishing - from 2.5 keV/n to 2.5 MeV, from protons to uranium ions, from zero current to 200 mA, and from low cycle duty cycle to continuous operation - all of which have been successfully achieved.

In recent years, Professor Kapchinsky has become more free to travel to Western Europe and the USA, so that the younger generation of accelerator physicists working abroad have been able to directly appreciate the exceptional height of his genius.

Many of us will feel deep sadness for him.

Sincerely yours: Kurt Hübner, Head of the Proton Synchrotron Department; H. Hazeroth; D. Warner; M. Weiss.

## FOREWORD BY PROFESSORS R. GLUCKSTERN AND M. REISER TO THE LECTURES OF I.M. KAPCHINSKY AT THE UNIVERSITY OF MARYLAND (published at Los Alamos after his death)

In February 1993, Professor Ilya Mikhailovich Kapchinsky arrived at the University of Maryland as a professor of physics and electrical engineering. During his visit, he gave 10 lectures on selected sections of the theory of linear ion accelerators. Unfortunately, Professor Kapczynski died on May 2, 1993, before he could deliver his final lectures. As is customary, he wrote and submitted the full texts of his lectures in advance. Since we feel that the lectures given by one of the world's largest experts in linear ion accelerators will be of great interest to all those working in the field of accelerators, we received permission from Ms. Lyubov Kapchinskaya to publish them. Due to the long history of successful development of linear accelerators at Los Alamos, we felt that publishing these lectures here would allow them to reach the most appropriate audience. Unfortunately, Professor Kapczynski's failing health prevented him from fulfilling his wish to visit Los Alamos.

In order to preserve for readers the specific style of presentation of Professor Kapczynski's lectures, we have edited them only in relation to the common English language. The chosen topics and the expressiveness of their disclosure are strictly his.

Robert L. Gluckstern, Martin P. Reiser, College Park Maryland, October 1993.

## III. MEMORIES OF FAMILY AND FRIENDS FRAGMENTS OF OUR LIFE

**L.M. Kapchinskaya**

Life gave me 40 interesting and happy years - these are the years lived together with Ilya Mikhailovich. Our meeting might not have happened. And “it’s scary to imagine now that I would have opened the wrong door ...”.

As a senior student at the Saratov Medical Institute, during the winter holidays, having passed the exam ahead of schedule, I came to Moscow to visit theaters and museums. I have been successful with both. But, returning in the evening to Bolshaya Yakimanka, where I stayed with my aunt, I was already “nodding” from fatigue in the subway, because. behind was a busy day.

My aunt, an exceptionally generous soul, lived with her 6-year-old son in an 11-meter communal apartment that looked like a pencil case, and her husband, my mother's brother, spent his 10-year term in the Taishet camps. He was in the same camp with Lidia Ruslanova. The story of his arrest is banal, like the stories of millions of other prisoners.

In 1948, after the formation of the state of Israel, the first ambassador of this country to the Soviet Union was Golda Meir. She met with fans of her young country more than once near the synagogue on Arkhipov Street ... My uncle was at one of these meetings. After Golda's speech, he made his way through the crowd to her and shook her hand. After a short time, on the night of November 8, 1948, they came for him. Well, during the investigation, which lasted more than a year, he already became a spy for all existing and non-existent countries in the world. The standard "three" gave him a standard ten, and there was one more prisoner at the logging site. We met him from the camp after 7 years on the platform of the Kazan railway station on August 11, 1955, after N.S. Khrushchev opened the gates of the camps. Instead of the once handsome man, a hunched gray-haired man in a worn and patched padded jacket, with a skinny bag behind his back, stepped out of the carriage. He was 51 years old.

On my first visit to Moscow, in January 1951, there was such an episode: one morning there was a knock on my room, and unknown woman. She introduced herself, but her last name Kapchinskaya didn't tell me anything. She turned out to be a neighbor from the top, 6th floor. We talked about something and she left. This meeting is not imprinted in my memory. Two days later I went to a friend in Leningrad, and from there to Saratov. And only a year later, when I again arrived in Moscow in the winter, I was reminded of this acquaintance. It turned out that mother I.M., having heard from my aunt that a provincial student girl had come to visit her, came to see me. Her eldest son, who was already 32 years old, was in no hurry to marry. And the youngest son Lenya was already married. A year later, it turned out that she liked me then, but she did not introduce us to I.M. I knew from my aunt that my dad's parents and sisters live in Israel. Sofya Grigoryevna did not want to risk the career of her son, who worked at the numbering institute in Kuntsevo.

My father's parents and sisters moved to Palestine in 1935 from Poland, fleeing the regime of Jozef Pilsudski. One sister remained in Poland, who in 1939, after the annexation of Western Belarus to the Soviet Union, moved to Brest with her husband and four-year-old daughter. Her husband was a history teacher at one of the schools in Brest. At the end of the academic year of 1941, he was awarded a ticket to the Caucasus. He doubted for a long time to go - not to go, because. The BBC radio station warned of an impending German attack on the Soviet Union. The last warning was issued on June 12, 1941: “Citizens of the Soviet Union! Germany is preparing a surprise attack on you!” On June 14th was our next refutation. And he left. It is easy to imagine the further tragedy of this family: at the end of June, in a white suit and white shoes, he reached Saratov, and his wife, the father's sister, and their daughter died in the Warsaw ghetto in 1943.

When I arrived in Moscow in the winter of 1952, I asked my aunt to take me to someone where I could see such a miracle as television. There were two possibilities: one - to friends in a neighboring house, the second - to the neighbors on the floor above. Of course, going up to the floor is much easier than walking across the street in winter. Television at that time worked 2 times a week, on Monday and Thursday. Showed mainly performances and films. And from the news one could learn about a record harvest in some region, about the winners of the socialist competition, stories about the impressive victories of socialism over capitalism, and about the weather. And I also wanted to look at another miracle - a refrigerator.

My aunt and I went up to the 6th floor and entered the apartment, where I recognized my last year's acquaintance as the hostess. The apartment was a large room of 34 sq. m. in a communal apartment, divided into 3 parts by plywood partitions. That evening, on January 23, 1952, Gulak-Artemovsky's opera Zaporozhets beyond the Danube was performed. The main parts were sung by Ukrainian opera stars Litvinenko-Wolgemut and Patorzhinsky. Now few people remember these names. TV KVN-49 with a lens filled with distilled water to enlarge the image on the screen made a huge impression on me! And the screen size, as I remember now, was approximately 14-15 by 10 centimeters. In the same family, I also saw a refrigerator: this is the Gas Apparatus - 45. Volume - 45 liters and without a freezer. But at that time it was also a miracle of technology.

At the end of the opera, the three of us (Sofya Grigorievna, her husband Mikhail Yakovlevich and I) sat down to drink tea with a cake "Leningradsky" or "Leningrad", on which sat a chocolate beetle. This beetle went, of course, to me. I describe all these details because now they have become signs of that distant time.

At 11 o'clock in the evening a young man entered the room. It was their son who came home from work. We were introduced to each other. He sat down to dinner without taking off his jacket, and I continued to watch TV. I didn’t care what was happening around, the main thing was the TV. There was some general conversation, but I have not yet joined the situation. After dinner, my new friend suggested that we go to his room. For some reason, I was sitting on his desk, and he was standing next to me. We talked for an hour or an hour and a half, about what - I don’t remember. Then Ilya (I already knew his name) took me to my aunt on the 5th floor. We agreed to meet tomorrow.

The next day we went to the Savoy restaurant, later renamed Berlin. It was the first trip to a restaurant in my life. Fountains beat there, stucco angels “flyed” under the ceiling, and fresh flowers stood on beautiful tablecloths. All this struck my provincial imagination. She sang, not yet running around the stage, an elegantly dressed singer, there were no speakers that would deprive visitors of the opportunity to talk.

I stayed in Moscow for a few more days, the holidays were ending, and being late for the beginning of the semester was punished by the deprivation of the scholarship. I could not allow this, because. our family lived very modestly, and my scholarship for an excellent student was as much as 35 rubles. I remember that 1 meter of crepe de chine (there was such a silk fabric) is also worth 62 or 35 rubles. Dad was a mechanical engineer, he created and managed the Repair and Surgical Workshop. The workshop created by dad became a factory. In 2008, the 70th anniversary of the workshop was celebrated. Mom worked as a dentist very far from home, with poor transport. Both studied and received a specialty when the family already had two children - me and my only and beloved brother. He is 6 years younger than me, lives with his family in Saratov, and for 40 years he has been the head of the emergency surgery department at the clinic of the Medical Institute. Many generations of Saratov doctors passed through his department and literally through his hands. It seems that all the removed gallbladders in Saratov and the Saratov region are the work of his hands. Lev Mikhailovich Kon - surgeon by the grace of God, Honored Doctor of Russia.

During the remaining 3-4 days in Moscow, Ilya and I visited the Udarnik cinema, at a concert in the Hall of Columns and just walked around Moscow. Then I gladly left for my Saratov. I understood that I had met a very interesting person, that there might not be such a second meeting in my life, but it would be so difficult for me to part with my parents and Saratov that I didn't think about anything serious.

And then the correspondence began, which was difficult for me, because the level of my answers had to correspond to the level of letters received. I received interesting letters, but somehow I had nothing to write about. I keep two thick bundles of letters even now, but in more than 15 years I have never been able to open them.

Somehow I once again visited Moscow, somehow And M. was 1-2 days in Saratov on his way to a business trip to Donguz (Orenburg region). He said that in 1947 he also went to Donguz through Saratov, and for some reason 2-3 hours of free time formed. He decided to see the city and walked along Lenin Avenue. This is a long street that stretches across the city from the station to the Volga (now it has been renamed Moskovskaya). He was passing by the Medical Institute, and a bunch of beautiful girls ran out to meet him. Then he thought that he should look for his wife in Saratov. He later decided that I was one of those girls. At that time I was already studying at the institute. It was, of course, a fantasy. Although, how do you know?

In my opinion, it was interesting for both of us to communicate, but there were no serious conversations about anything. And in a few months I had to be assigned to a job. And I received it - Kazakhstan, in Kokchetav. I cried well from such a prospect - Kokchetav shone for me for 3 years. I didn't even write to Moscow about this, so that it wouldn't, God forbid, be taken as a hint at marriage, which gave free distribution. I didn't even fully understand what scared me more - after all, after 3 years from Kokchetav, I will still return to my Saratov, but from Moscow?

My parents, seeing my mood, decided to do something to change the situation. As a result of great efforts and my mother's trip to Moscow, I received the Saratov region. And the closest place in the Saratov region is the city of Engels, formerly Pokrovsk. Lev Kassil wrote a lot about Pokrovsk, because. it was his hometown. It is located on the other side of the Volga. When the Germans lived in this city, it was a very clean green city, but the Germans were completely expelled in September 1941 from the Volga German Republic to Kazakhstan and Siberia. In September 1952, when I got to Engels, this city was already completely different - dusty and dirty.

In the 1st City Polyclinic, I started working as a teenage doctor. This is a very uninteresting job, but I always ran away from the district work, but it was still better. House calls also had to be made, and then a skinny filly harnessed to a cart was sent for me. I easily jumped on the cart (now it would be so!) and drove around the city with the coachman. The main time of the working day was occupied by the road from Saratov and back. It was necessary to take a trolley bus to the Volga early in the morning, then for about an hour to sail along the river on a steamboat or the Record barge, and from the Volga to the clinic, go for about 30 minutes through the picturesque island of Osokoriy, and then through the city. In the evening - everything is reversed. But it's summer. There was no bridge across the Volga then. There was a single railway bridge just below Saratov. During the war, the Germans tried in vain to bomb it in order to interrupt communication between the right and left banks of the Volga. If this happened, it would be a terrible catastrophe. This did not happen, but a barge was sunk, on which there were one and a half thousand children, women and the elderly, ready to sail the next day for evacuation somewhere up the Volga. The German army then stood between Stalingrad and Saratov, near Kamyshin, this is 200 km from Saratov. Our family was already thinking about evacuation.

When navigation ended, but there was no freeze-up yet, the crossing of the Volga was a daily risk to life.

People, including me, plunged into private motor boats and sailed in dense fog, not seeing the shores. To determine the direction of the flow of the river, the minder threw pieces of paper into the water and watched where he carried them. In addition, the boats were always overloaded. Cars were allowed to cross the Volga when the ice reached a thickness of 40 cm. And until that moment they walked 5-6 km on the ice, falling and tumbling. In the spring, I walked with great risk on the ice with cracks through which the water was already flooding the ice floes. It was possible to go to Engels by train, but the trip took 3! 4 hours one way, because. the train stood for 30-40 minutes at each crossing. Anything is possible when you're young! When traveling by train, it was saved by the fact that an interesting company formed for me on the train.

So I worked for almost a year, but I did not write to Moscow about my difficulties. And even when I was in Moscow for 2 weeks in February 1953, I did not talk about it so as not to complain about life. Time passed, the correspondence continued.

And at that time, trouble happened in the Kapchinsky family: on the night of January 21, 1953, I.M.'s father, Mikhail Yakovlevich, was arrested. I must say here a few words about Mikhail Yakovlevich. He lived a long, very rich and ambiguous life, he died almost at the age of 92, in 1981.

In 1918 M.Ya. joined the Communist Party. During the Civil War, he fought against Denikin, being underground in Odessa. Participated in CHONs. From that time until the end of his life, he wore fragments of a shell in the lower leg of his left leg. Why he didn't delete it, I don't know, he was probably afraid. In the early 1920s, he was "transferred" to culture, and it so happened that in 1922 he became the founder, organizer and first director of the Odessa film studio. There is still a memorial room dedicated to him.

For half a century of work in Soviet cinematography, M.Ya. directed 5 feature films and 34 popular science films. In 1925, he was transferred to Moscow and appointed director of the newly opened 1st State Film Factory on Zhitnaya Street, which later became the Mosfilm film studio.

In 1925, filmmakers were ordered to make a film for the 20th anniversary of the 1905 revolution. This is how the legendary "Battleship Potemkin" was born. The film was directed by S.M. Eisenstein, his assistant - G.V. Alexandrov. Difficulties in filming were huge, because. there was no production base for filming on Zhitnaya. M.Ya. provided all possible assistance and support to the filming of this film, personally selected workers film crew. The picture was taken in 3 months. Record time! On December 24, 1925, the painting "Battleship Potemkin" was shown to the government, delegates and guests of the 14th Party Congress at the Bolshoi Theater. The last shots of the picture - a red flag, a symbol of the revolution, is flying on the mast of the ship, went to a long standing ovation from the audience, who rose from their seats (according to the published memoirs of G.V. Aleksandrov, the flag on a black-and-white film was painted by hand). From that night, the triumphal procession of the film "Battleship Potemkin" began on the screens of the whole world.

M. Ya directed the Moscow film studio for only a year, before his first arrest. He (and the entire management of the film factory) was arrested as part of the Dzerzhinsky-led campaign for economy. The accusation is an overrun of film stock. Filmmakers during filming do several takes. It turned out that you need to do only one, everything else is an overrun. Another article of the charge was the payment of bonuses to film studio employees for successful work. So for the "Battleship" Eisenstein was given 50 rubles, the operator Tisse - 25. In 1926, Felix Edmundovich died, the next spring all the accused (including Mikhail Yakovlevich) received short sentences and were immediately amnestied.

The second arrest was under Yagoda in 1935. Some economic article was also incriminated. In 1936, Yagoda ceased to be People's Commissar of Internal Affairs and headed the People's Commissariat of Communications of the USSR. Mikhail Yakovlevich was released.

The last arrest was the most terrible. On the night of January 21, 1953, 5 film directors from the Popular Science Film Studio were arrested, all Jews. They were: M.Ya Kapchinsky, L.B. Sheffer, Ozerskaya, Levit-Gurevich. I forgot my fifth surname, because 55 years have passed since those events. M.Ya. was taken out of bed sick, he had an attack of Meniere's disease.
They had to fabricate the case of the film directors. Until March 5, the official day of Stalin's death, the least accommodating of those arrested managed to learn about torture and sit in a punishment cell. Since March 5, physical sanctions have been discontinued. M.Ya. told me what a punishment cell was: it was a cold room, meter by meter, with cement walls and floors. In the middle is a stool for the prisoner. Water on the floor, water running down the walls, water dripping from the ceiling. I don't remember how many days he spent there. This depended on the behavior of the arrested person and the nature of the investigator.

Sofya Grigorievna later told me that I.M. all night, when the apartment was being searched, he sat on a chair, resting his head on his hands, which lay on the table. During that night he turned gray. In prison, the investigator showed Mikhail Yakovlevich a warrant for the arrest of his son. Apparently, this was an element of pressure on the person under investigation, and perhaps it was in essence, but Stalin's death prevented further arrests.

Fortunately, nothing has changed at YM's work.

Looking ahead, I'll tell you that Mikhail Yakovlevich remained in prison for the entire 1953 year. M. I later said that the investigator once asked him during interrogation: "Why did you, such an intelligent person, become a Jew?" And in the course of the investigation, he was first led to a Japanese spy, but ignorance of the language let him down, and then he was promoted to resident, but they could not build this accusation, because torture had already been canceled.

Since the end of 1953, the so-called open courts began. Their openness consisted in the fact that the guards led the prisoners to the courtroom on the third floor along the common staircase, and relatives could see them. It was already a lot. There were trials at Kalanchevka, in the building of the City Court. Of course, no one was allowed into the meeting room. The trial of M.Ya. took place on February 10, 1954. He was tried last, and his four "accomplices" have already received their dozens under Art. 58-10, counter-revolutionary agitation, and M.Ya., as an unconfessed person, was presented with Art. 59 - inciting ethnic hatred. It was difficult to understand where this article came from, and what they had scraped together over the whole year of the investigation.

The whole family was waiting in the hallway. And suddenly, a few hours after the start of the court session, we were all invited into the courtroom. M.Ya was found guilty, given a term that actually served - 1 year and 21 days, and released in the courtroom. From that moment Mikhail Yakovlevich was free!!!

We were all on our way home together. At home, the sons scraped him off everything prison, laid him in a clean, soft bed. After a couple of months, when he recovered a little, coped with the exacerbation of bronchial asthma, with which he was released from prison, he returned to work at the Studio of Popular Science Films.

Regular arrests almost completely erased the name of Mikhail Yakovlevich Kapchinsky from reference books and encyclopedias. During searches and requisitions, its richest archive, photographs and library were destroyed.

Let's go back. In Saratov, in the summer, on June 14, 1953, there was a knock at our window at night. We lived then on the 1st floor. The postman brought a telegram marked "night". Previously, there were. We were all pretty scared. It turned out that this was I.M. announced that he was coming in the morning. And he arrived in a very resolute mood and with a diamond ring (it is still on my finger): he came to get married! He had three and a half days at his disposal. There was no agreement between us. In the registry office, the deadline for registration of marriage is 2 weeks.

The first day, June 14th, was spent on our conversations. They were difficult, I had grievances. In the end, everything was decided amicably. I.M. I was so unsure of the results of my trip that I didn’t even tell my mother at home why he was going. When she received our telegram in Moscow, she fainted.

In the registry office, my cousin had familiar girls, because. 2 months before that, he divorced his wife. Therefore, on June 16, our marriage was registered. There was no white dress that every girl dreams of, there was no accompanying crowd of loved ones with flowers. There were only two witnesses necessary for the ceremony - my own brother and cousin. There was no wedding, because father I.M. was in prison, only in the evening three pairs of friends of my parents came to us. But the wedding was promised to me sometime later ... And it took place! On our tenth day, June 16, 1963. In the Mirror Hall of the Prague restaurant, we gathered 50 relatives and friends. And it was a wonderful wedding! And behind was already a 10-year trial period and two children.

A day after registration, on the 18th, I.M. already left. The timing of my arrival was not specified. Then - again correspondence, however, in a different tone. There were no home phones either in Moscow or in Saratov. My mother's friends scared me that I was going to live with my mother-in-law, but, knowing my character, I was not afraid of this. Sofya Grigoryevna was a difficult person, but she turned her best side to me. We lived together for 13 years without a single quarrel. She was a good hostess, and I tried to learn a lot from her.
In Engels, at work, I issued a dismissal, but left for Moscow only on July 22. The first months of my life in Moscow, I did not work, because. again she did not want to go to the site, but she could not find anything else. There were no acquaintances in Moscow. But later, through Saratov, I found a job as a teenage doctor.

My mother recalled that when I was born, my grandmother, looking at the baby, said: "This nose will require a lot of dowry." From Saratov to Moscow, as a dowry, I carried one pillow.
In July 1954, we were expecting our first child. I went to Saratov for childbirth, so that at first I could have help from my mother. It all ended tragically: 3 weeks after birth, the boy Lyonechka died. The injury was terrible for us, but even in this situation, I.M. could find some soothing words for me. The forensic medical examination concluded that the cause of death was a heart defect incompatible with life. Now such defects are operated on in utero. Some time later, once again analyzing this situation, I realized that the conclusion was deceptive: it was a birth traumatic brain injury, which was the result of improper delivery with a large fetus. For some reason, there was not even a doctor during childbirth, only midwives.

A year later, a boy, Misha, was born in Saratov. He grew up healthy and smart, retaining these qualities to this day. And after 6 years, the girl Olya was born. It was great luck - a boy and a girl. In addition, the girl was remarkably white and curly. Unfortunately, in the process of evolution, these qualities have disappeared. When Misha was just a baby, I.M. composed a lullaby that began with the words "far from the big highways" and came up with a melody without hearing. He hummed it when he had to put the children to bed. So I can quite consider him the first bard.

Life went on with its ups and downs and difficulties. I.M. worked very hard. Came from Kuntsevo at 10-11 pm. The metro had not yet been built there, and, most importantly, in 1953, the Stalinist late, and even night style of work was still preserved in the country.
Once, in the summer of 1957, a fateful event took place - Vasily Vasilyevich Vladimirsky came to us, at Bolshaya Yakimanka, house 39! He had to walk up to the very high 6th floor. There was no prior agreement, because we lived without a phone. I didn't know V.V.V. and could not immediately appreciate the significance of this visit. And he came as a demon - a tempter - to invite Ilya Mikhailovich to work at TTL, the current ITEP.

Their acquaintance goes back to 1943, when they met at plant number 465, part of which later became NII-20. Irina Borisovna Andreeva, a smart woman, also worked there, and strong, they say - a woman with a masculine mind. If I am not mistaken, I met N.V. Lazarev and R.M. Vengrov.

Registration lasted 9-10 months. More than once they called me to the Lubyanka, which did not cheer me up. Once there I.M. they asked: “Why do you write in the questionnaire that your wife in Israel has two aunts?” - "Because there are two of them" - "No, there are four of them!". Our alert eye knew everything and even more than us. V.V.V. took some steps as a catalyst. And on July 1, 1958, when I was walking home along B. Yakimanka, a Moskvich stopped near me. Remir Moiseevich Vengrov was driving, I.M. was sitting next to him, and the back seat was completely littered with flowers. I immediately understood and burst into tears. It was the day of farewell at NII-20.

TTL for I.M. - This, of course, is not only a new team and a different cabinet. This is a change of SCIENTIFIC DIRECTION! And already in 1959, Vasily Vasilyevich took them to Geneva together with I.M. report.

The first foreign trip to the socialist country (as it was supposed at first) took place in 1967. It was a whole event. Let's go to Poland I.M., V.K. Plotnikov and someone else. Maybe it was a "specialist in hydrodynamics", or maybe someone in essence, I don't remember.

On the visiting commission (there was such) I.M. asked how many square meters of housing was built in Moscow last year, some more abracadabra. Of course, this was “important” for the Polish physicists, and they could ask their guests about it.

The next trip took place in March 1968 already in the capitalist country, in Switzerland. I.M., S.A. went to Geneva, to CERN. Ilyevsky and, it seems, A.M. Stolov from Leningrad. And then, as a result of the six-day Arab-Israeli war, the exit gate slammed shut for I.M. for 21 years! Somewhere in the mid-80s, already under M.S. Gorbachev, one day Ivan Vasilyevich Chuvilo, director of the ITEP, said to I.M.: “We will soon let the Jews out.” But this “soon” took place oh, how not soon.

In 1989, the University of Frankfurt am Main. Goethe awarded the title (or degree) of Honorary Doctor to three physicists: Ilya Mikhailovich, a physicist from Geneva and a physicist from South Africa. For all the years of awarding this title, it was given to a Soviet scientist for the first time. But then the competent authorities faced the question of leaving I.M. in Germany (then it was West Germany), and even with his wife. Normal people from normal countries do not go to such a solemn event alone. A large positive role here was played by the head of the 1st department of the institute V.F. Radchenko. He spoke about I.M. that he was a real Russian intellectual. This, apparently, determined the degree of trust in him. Good will always be remembered. I was summoned to Staromonetny Lane for a preventive conversation. to the State Committee for Atomic Energy, and two young men gave me instructions on how to behave abroad and on my return to Moscow. For them, as they said, this was the first time a scientist and his wife had traveled abroad. They dragged me out with their muzzles on the table, but I dutifully listened to everything.

The trip took place in May 1989. Everything was interesting and wonderful! The graduation ceremony was unusual for me. The recipient was not just called to the table, shook hands and presented a diploma. No, there was lyrical music performed by a quartet, obviously informal performances, a lot of humor, large bouquets of flowers for wives, a hall decorated and fragrant with lilacs. And all this was slow and long. Physicists arrived from other German cities. And then there was a banquet in some town near Frankfurt, in the Rothschild castle. The castle was built by Rothschild for his daughter, but when the Nazis came to power, its owners left for America. Everything was organized exceptionally. I know that Horst Klein showed his great organizational skills. In relation to I.M. many showed so much interest and so much attention that I was very happy for him. We were taken around the country a lot there, we saw a lot, and Klein and Schempp took care of us. I will forever be grateful to them. The first years after our stay in Germany, we exchanged New Year's messages, and then everything died out by itself.

We were in Frankfurt am Main on May 9th, and we were wondering how our Victory Day is going there. No processions or demonstrations. Just in the evening there was a meeting in the mayor's office. We were invited there, but we could not go.

About the work of I.M. in essence I will not write: firstly, I will not cope, and secondly, I hope others will write about it. I can only say that he worked hard and with pleasure. At home, he never had his own office, his desk was in the common room, where there was a TV, and we all smoked. I called this room "red corner", "skhodnyak", "komsomole headquarters”, “suite”, because visiting guests often lived there. All definitions were the same. I.M. knew how to abstract from everything around him and sit and work, not seeing or hearing anything happening around him. Everyone tried to be quiet, but the TV was on.

What happened before we met?

I studied in Kyiv at school number 45. By a funny coincidence, our children, Misha and Olya, studied at the Moscow special school number 45. In school and student years, I.M. wrote poetry. I had never seen them before and only now decided to touch them. Some of them I will include in this book. These are either lyrics or patriotic poems, typical for a young man of those years.

In the last grade of the school, he won a prize at the all-Ukrainian competition of young poets and, in the form of an award, got the opportunity to go to Moscow for some further review (another winner of the competition, Yuriy Timoshenko, the future Tarapunka, was traveling with him on the same train) and a scholarship for the entire period of study at an institute or university. As an excellent student, I.M. had the right to enter without exams in any university in the country. And he was seriously inclined to enter the IFLI (Institute of Philosophy, Literature, Art) and engage in literature. But the love of physics won, the popular science books on physics read by M.P. Bronstein and Ya.I. Perelman. While he hesitated between physics and literature, admission to Moscow State University ended. I.M. took the documents to the Mining Institute. He studied at Gorny for one year, and then moved to the Faculty of Physics of Moscow State University. They still didn’t want to let him go, his excellent students were needed there too.

I.M. received a poetry scholarship one year, and then suddenly the payments stopped. He complained about the problem to his parents, Mikhail Yakovlevich decided to find out the reason for what happened, went to talk to the Union of Writers of Ukraine. The official who received him was laconic: “Do you remember who signed the decision of the jury of the competition? So-and-so. And he is an enemy of the people! Be glad your son is not arrested."

By the beginning of the war, I.M. completed the 3rd year of the physics department. The guys who did not have health restrictions and the correct questionnaire were sent to study at the Air Force Academy. Many, mostly girls, were sent near Smolensk to dig trenches, from there they returned on foot, which is 400 km. I.M. went to the draft board, but he was not taken to the front because of his eyesight. Then he went to ask in the people's militia, there were no health restrictions. He was enlisted and sent on night duty to put out lighters at the Vagankovsky cemetery. With a sense of patriotism, he did this for about a month, and at the end of the summer Stalin issued a decree ordering to gather all senior physics students for finishing their studies. So he ended up in Ashgabat. And his family, on the terrible days for Moscow on October 15-16, 1941, left for Tashkent, where Mikhail Yakovlevich lived with his sister and family.

In 1943 I.M. returned to the capital. To get into military Moscow, I had to cheat. He applied for admission to graduate school, but when he arrived in the city, he went to work at a military enterprise. Pure science was far from the immediate needs of defense, but it was necessary to work for the common victory. So he ended up in the future NII-20. The theme of the work at first was sound pickups for tracking enemy aircraft, and later the direction of the institute's work naturally switched to radar.

Fast forward to our lives.

There was a lot in common in our characters, we lived all the years without a single quarrel, but two moments were just a happy coincidence: we both loved poetry and both supported Dynamo Moscow. Well, still I.M. played a little bit for Dynamo Kiev and Chernomorets Odessa, and I for Sokol Saratov, but this was already secondary and without great emotions. How many hours, and in total, maybe days or months, we spent at the stadiums! What matches we saw and what players! I envy myself right now. We saw Tommy Lawton and Stanley Matthews (Arsenal London), Pelé and Garrincha (Brazil) and many other great players and the team in general. I'm not talking about our players, such as Khomich, Yashin, Netto, Bobrov, Maslachenko, Streltsov. I.M. carefully filled in the small cells of football calendars. From the age of 5, they began to introduce their son Misha to football. At first he was bored, and he could stand the whole match with his back to the football field, and then he began to get involved in the vicissitudes of the game. I remember he was with us at the last match of Lev Yashin, when after the first half he symbolically handed over his gloves to the goalkeeper Pilguy. Unfortunately, Pilgui did not become a worthy successor to Yashin's skill, although for some time he was in the main team at the Dynamo goal. In winter, they went to hockey, which then was not the private, under the roof of the Palaces of Sports, but it was necessary to defend in the cold and in the snow for 2.5 hours on the North Stand of the Dynamo stadium. In 1955, the European Speed Skating Championship was held in Moscow. They also froze, but the spectacle was breathtaking. We had brilliant skaters, such as Grishin, Goncharenko, but the Swede Erik Erickson became the champion.

We with I.M. both loved poetry, and, I'm afraid to seem like a braggart, they knew it. Everyone had their own preferences, for example, I.M. he loved Pushkin very much, knew by heart almost all of his poems and all of Onegin, loved Pasternak, Gumilyov, Mayakovsky, Samoilov. Lermontov, Blok, Akhmatova, Simonov were closer to me. Both loved and knew Yevtushenko, Galich, Korzhavin, Okudzhava.

We had such a game: when we were on a long car trip, one of us began to read some quatrain, the other continued and so on until the end of the poem. It brightened up a bad beaten road. It is already the seventeenth year of my loneliness, and I still involuntarily turn to I.M. if I need to remember some poetic line.

Since the autumn of 1961, poetry evenings began to be held in concert halls, sports palaces, and the Polytechnic Museum. What a pleasure it was! I don't remember who we threw the children at, Olenka was still quite a baby, but we ran endlessly to these concerts. We tried to go to those halls where in one concert you could hear and see Voznesensky, Akhmadulina, Okudzhava, Yevtushenko, Rozhdestvensky, N. Matveeva. After the performances, I still had to approach the poets and take autographs for their collections. For some reason he didn't pull me away, he just stood on the sidelines, embarrassed and patiently waiting for me to finish my round. Now I don't know why I needed it, but then I definitely needed it!

We went to the spectacle a lot. Going to the theater was determined by the ability to get tickets in addition to the box office. And the opportunities at different times were different. At one time we reviewed all the plays in Sovremennik, then the possibility of trips to the Taganka appeared. Then came the turn of the Operetta, which at that time was on the rise. In the "Big", unfortunately, they were rare, but they were. Watching many plays in a row in one theater is not very good, but there was no other way out.

I remember that in Sovremennik G. Volchek staged "The Ballad of a Sad Tavern". There Tabakov wonderfully played the role of brother Layman. He played the role of this bastard and scoundrel, which was impossible to look at without disgust, so well that for us it cast a shadow on all his subsequent roles. Even now, not remembering the essence of the play, I remember this vile feeling that he later evoked in the most positive roles in other plays. Then the trips to the Obraztsov Theater began. Well, and besides - concerts in the Great Hall of the Conservatory. In September, I bought subscriptions to concerts of our best vocalists, instrumentalists and symphony orchestras. That was great! Olenka went with us.

In the summer, on weekends, we sometimes went with the children to the dacha of the Vengrovs, taking advantage, but, I hope, not abusing their wide hospitality. It is always a pleasure to be a guest in a friendly family. Inna and Rema met in the third grade of the school and kept warmth and high relations for the rest of their lives. We met with employees of NII-20 at their place.

In 1953, when I moved to Moscow, Ph.D. theses were being defended at NII-20. Not a month went by that we didn't go to a post-protection banquet. The circle of people on them was basically the same. All this was interesting to me, because smart people gathered. The toasts were beautiful and interesting, and among the guests there was one (I won't give his last name), who ended each toast with the words: "And to dear Anatoly Prokofievich." He added these same words to all other people's toasts. Anatoly Prokofievich Belousov was the chief engineer of the institute and was also present here. We sang our lyrical songs at these gatherings, involving other guests in this activity. More than other songs, they loved the song that appeared, "Distant bonfires burn, the moon bathes in the river ..." Those were good times.

I remember many pleasant moments, here are some of them. In the summer of 1967, during the pre-launch period on the I-100, many ITEP employees worked in Protvino. I remember how we celebrated the 35th anniversary of Vladimir Konstantinovich Plotnikov at the Orbita restaurant. But the restaurant closed quite early, and the whole big company of us moved to finish drinking and eating up. We had a one-room service apartment there, where oval metal numbers were nailed to the furniture. There were very pleasant gatherings.

I remember how once we went for mushrooms with Vasily Vasilyevich and Anna Nikolaevna Vladimirsky. And this region abounds in mushrooms. Let's go foxes. They collected a large plastic blue bucket of chanterelles. I made them with potatoes and fried onions. How delicious everyone was!

Ilya Mikhailovich's closest associates at ITEP were interesting, intelligent, highly professional, decent, well-read, and well-educated people.

On the territory of the Donskoy crematorium, the eldest daughter of A.S. Pushkin, Natalya Alexandrovna Gartung-Pushkina. When we visited the graves of I.M.'s parents, we always approached the burial place of Natalia Alexandrovna. She, born in 1832, died in March 1919 and I.M. considered himself a contemporary.

One of our close acquaintances, Nina German, studied at VGIK with A. Galich's wife and maintained good relations with this family. And then one day she invited Alexander Arkadievich to her house for us. He gladly accepted the invitation. It was January 5th, 1965. Galich came with his niece, brought a guitar. I remember that I brought a bottle of Belovezhskaya for him. I don't know much about vodkas, but I liked this one too. The bottle was successfully drunk. Galich sang a lot. How much we enjoyed - and not to say. I.M. brought with him a heavy Yauza-5 tape recorder. The cassette was well recorded, it was subsequently rewritten many times and, one might say, went to the people. I still can't understand why neither we nor anyone else took a camera with us.

In September 1979, I.M. was invited to the Physics School on Accelerators to give several lectures. It was near Minsk, it seems, in the youth complex "Youth". On the way we stopped to see Khatyn. The impression is terrible and unusually strong. We got to Yunost in the middle of the day. I.M. went to take shape, I was left to wait in the car. After 10-15 minutes he came up to me somewhat embarrassed. He showed the received papers and said where to go for accommodation. It was not the closest building. In this building, I was placed in a 4-bed room, and he was placed in another 4-bed room.

The mood was completely ruined. Having settled, i.e. pushing the suitcase under my bed, we went out for a walk and look around. Near the main building, a young man almost ran up to him with a quick step with the words: “Ilya Mikhailovich! We've been waiting for you, let's go, I'll show you your suite on the 9th floor." Justice has been restored. But I.M. I didn't even think that he could claim something more than a communal apartment.

Recently, I was looking through the addresses received by Ilya Mikhailovich for his seventieth birthday and found that it turned out to be I.M. was awarded to this date by a higher organization "valuable gift". It is a pity that now it is impossible to find out what kind of "valuable gift" it was and who enjoyed it. Here are some interesting points that are being revealed 20 years later.

Our family has always had a problem with vacations. Starting from 1962, shortly after defending his doctorate, I.M. The vacation was 2 months, and I, an otolaryngologist, have only 3 weeks. At their own expense, they did not give a day. During these 3 weeks I had to manage to visit my parents in Saratov and relax somewhere with my family. In order to somehow extend my meager vacation, I was a gratuitous donor 3-4 times during the year. Each blood donation gave 2 days to release. I.M. didn't know about this, of course, I only sometimes wondered why I wear a long-sleeved sweater at home. Due to bad veins, I always had hematomas in the elbow bend area. Rest in a civilized manner for the four of us, i.e. we didn't have the opportunity, so we always drove “savages”, and when I got behind the wheel in 1965, we drove somewhere by car.

Now, when I see how the people of Canada travel here lightly on long journeys, I involuntarily recall our fees when the trunk had to be filled with canisters of gasoline and semi-smoked Moskovskaya sausage for 4 rubles per kg., All other provisions and water. I'm not talking about the difference in the quality of the roads. But then there was one indisputable advantage over today - we were young and were together.

I.M. loved sports. In his single years, he was engaged in mountaineering and was proud of the certificate and badge of the mountaineer. Then for this sport there was no need for medical certificates. Then there were skates. And there were always skis. Every year on February 17, as long as his heart allowed, he went skiing for 2 weeks. These were Ershovo near Moscow, Moszhinka, or some other rest home. And on February 17, shifts began in all these health resorts. And almost always from these trips he brought interesting acquaintances. Sometimes they were married couples, and more often they were women. Warm relations with some remained for a long time.

In the early sixties, a large ski jump was put into operation on the Leninsky, now Sparrow Hills. One winter we gathered there to watch some international competitions. While I arranged the children, prepared food for them, while they were waiting for the buses and while we approached the springboard, the last skier was already flying from it. I had a good time then. Some compensation was that we met acquaintances from NII-20 there- M.L. Sliozberg and I.M. Golovchiner.

Odessa is the hometown of I.M. The house where the Kapchinsky family lived in the 1920s, Deribasovskaya 21, was destroyed by an aerial bomb during the war. When we came to Odessa in the seventies, the Komsomolets cinema already stood in this place.

Ilya Mikhailovich said that because of my powers of observation, he lives under an x-ray machine.

I.M. in the military ticket in the column "military rank" was - “private - untrained”, and I was a reserve captain. This often served as an occasion for jokes.

We have rested many times in Sukhumi and never in Sochi. I think that B. Pasternak's line played a role here: "And the vulgarity of Sochi informs the nature of modest Kobulets."
A frequent phrase used by Ilya Mikhailovich in a conversation was this: "Explain to me using logic ..."

Favorite fruit of I.M. was a white cherry. His father, Mikhail Yakovlevich, made white cherry jam with lemon every summer. Very beautiful in appearance, amber in color with visible interspersed with pieces of lemon. And very tasty.

In 1982, the third monograph by I.M. "Theory of linear accelerators". The Americans became interested in this book and decided to translate it into English. They themselves were engaged in this work, but many points had to be coordinated, and some papers had to be signed by I.M. This was, perhaps, the most difficult for the Americans in this work, because I.M. had no right to contact with foreigners. In the summer of 1986, the work was completed and the last, final signature of Ilya Mikhailovich was required.

At the beginning of September every year, the International Book Fair takes place on the territory of VDNKh. A representative of an American publishing house came to this fair, and I.M. They were in different rooms so as not to see each other. Some of our trusted representatives took the brought book from the American woman, brought it to the room where I.M. went on, received the necessary signature and took the book back. In my opinion, we were the only ones to be ashamed of it. And how difficult it was for I.M. come up with a reason why he cannot come to an international conference this time. Corresponding member, who worked at ITEP USSR Academy of Sciences Iosif Solomonovich Shapiro in a similar situation once said in his hearts: “I can’t once again write that I’m pregnant.”

At the age of 50, Ilya Mikhailovich began to have health problems. Once he went to the medical unit (or polyclinic) to Susanna Nikolaevna for help for a trip to a rest home. She intended to put some decent pressure on him and refused a certificate, adding: "You will fall on the track and die, and I will answer for you." Unfortunately, we doctors have such "tactful" attacks. I wrote a certificate to him myself, and the problem was eliminated. On September 29, 1972, he returned from an international conference from Dubna in a serious condition. There, after the banquet, in the hotel, he began to have pain in his heart. Remir Moiseevich Vengrov said that he would get validol from someone and call a doctor. Rema had a sad experience in matters of the heart with his father. I.M. answered: "You don't understand anything, I need a pyramidon." Medical assistance was limited to this. On the train and on the bus, he still carried a heavy briefcase with the proceedings of the conference.

In the following days, everything became clear - it was a heart attack. After lying at home for the prescribed weeks, he left for a cardiological sanatorium in Peredelkino, and then went to work. In subsequent years, there were similar conditions, he ended up in the hospital, but the diagnosis sounded like an exacerbation of coronary disease. Then I realized that sometimes we missed microinfarcts. Unfortunately, we were not lucky with the doctor at the Academic Polyclinic. She was a sweet person, but a weak therapist.

I still regret that at one time I did not ask V. Kurakin to transfer I.M. from the 3rd floor down. The steps on those stairs are high and I.M., having come to work, did not leave the building during the working day. He was worried that he could not walk around the units, he stopped going to the canteen, he took food with him.

And yet, besides many other things, I regret that I did not delve into official conflicts enough, considering it indecent. But he listened to my advice. Purely women's advice could mitigate some situations.

In recent years, he lived with almost constant pain in his heart. Sometimes this state was reflected in the intonation of conversation with employees. In the evening, at home, this was the subject of great experiences. Analyzing some working moments already in a calm state, he understood that people were not at all obliged to enter into his difficult affairs of the heart, but he often could not cope with himself. At home, I always had syringes and ampoules ready. With this baggage, I accompanied him on business trips abroad. Unfortunately, the moments were different, sometimes I had to use them. There was no peace, but one could not even talk about retirement even in jest, he was always all at work, there were always unrealized ideas.

If there were mobile phones then, how much easier it would be. I.M. liked to tinker in the garage on the weekend, I also liked to take a walk in the Neskuchny Garden at the weekend for 3 hours. I could not participate in these campaigns and lived in constant anxiety.

Ilya Mikhailovich always had many friends. They were, as it were, in three dimensions: school friends, university friends and colleagues from ITEP and NII-20. The most numerous was the group of school friends. I was always struck by their warm and respectful attitude towards each other, it was always interesting for them to meet. Gathered together often. They finished school in Kyiv in 1937. Many died at the front. And the survivors communicated with each other. There was a class day - March 24th. On this day, those living in Moscow always gathered. They gathered either at our place, or at the classmates of Nadia and Grisha Herman-Rashba, who live in Zhukovsky, or at the Praga restaurant. In "Prague" there are always rooms for any number of visitors. In addition, there were fees for birthdays. And in May 1977, we went to Kyiv, where there was a large gathering on the occasion of the 40th anniversary of graduation from school. Graduates of the school came from other cities, there was even one of their teachers. In March 1979 we celebrated their collective 60th anniversary. All this has always been interesting. I was friendly with everyone and at such moments I also felt like a graduate of the 45th Kyiv school.

In October 1992, at the invitation of CERN physicist Mario Weiss, we flew to Geneva. I am far from a physicist, but it seems to me that visiting and working at CERN is the dream of every accelerator physicist. Ilya Mikhailovich was given a separate office and a computer for work. Help me minds that this computer occupied the whole wall. (But maybe it's my fantasy - I don't remember exactly now). And there I.M. first entered the world space, i.e. in Internet. There was a lot of joy.

We lived there not far from the institute, in a building intended for those arriving on a business trip or for temporary work - Bob Gluckstern (Robert Gluckstern), an accelerator physicist and Chancellor of the University of Maryland (USA), also lived there at the same time. He invited Ilya Mikhailovich to come to Maryland for 3 months to give a course of lectures on accelerators. The issue was not resolved immediately, because I.M. refused, citing insufficient English for lecturing. In the end, Bob said that his, Bob, English arranges for I.M.. This settled the matter.

Once, during our stay in Geneva, Mario Weiss gathered five accelerator physicists of the first wave in a restaurant. They were Ilya Mikhailovich, Mario Weiss, Gluckstern, Lapostolle, who came from France, and I don't remember who else. I did not understand anything of what they were talking about, but how interesting it was to look at these already enthusiastic middle-aged people!

We flew to Washington on January 31, 1993. The 20-degree frosts were left behind, and in Washington it was +14. We were greeted by Bob Gluckstern, cousin of I.M. with his wife, Edik and Bella and their daughter Lenochka. Lena had recently graduated from the University of Maryland with a degree in accelerators. She said that when she studied, and someone came to the University, associated with linear accelerators, they brought him to the class or auditorium where she studied, and introduced her as I.M.'s niece.

We were settled on the territory of the University, for some reason in the building of excellent students, in a two-room apartment. Living conditions were very good. I.M. at 10 in the morning the car took away and brought home at 5-6 o'clock. And at that time I was leaving by bus somewhere to shop or relatives came for me. In Maryland, I.M.'s 3 cousins live with their families, all are from Kiev. I.M. lectured once a week. Judging by the number of listeners and the questions asked, the lectures aroused interest.
\
For me, the main purpose of the trip to America was to show I.M. cardiologist and take the necessary medicines with you. Edik found us cardiologist. A visit to him did not bring joy. After the examination, he said: "You can't leave America without doing anything." It was, of course, about the operation. But he understood that we had no money for the operation. He didn't charge us for the visit - it's 250 or 300 dollars, he said that what he would do would be free of charge for us, but other staff had to pay for staying in the hospital. The amount was not named, but it certainly was unbearable. With that we left. They turned to the Jewish community of Maryland for us, there was no such money there. The university also could not pay such (I don't know what) money. But they found a way out: 70 km from the University in the city of Baltimore there are clinics of the Medical Faculty of the University. There is a department of cardiac surgery. The issue has been resolved.

Before hospitalization I.M. electrocardiography was done. According to its results, he was left in the hospital, the operation was scheduled for the next day, April 20th. Our stay in the States ended on April 30th and I..M. I was very worried whether the institute would think that we had stayed in America at all, because for observation it would have been necessary to stay for another month. I spent that night in the ward, they put me a cot. The next day at one in the afternoon he was taken to the preoperative room, I was allowed to stand next to the gurney on which he lay. A few minutes before I.M. I was taken to the operating room, he was reading something from Poltava to me.

I waited until the end of the operation, my relative Maya Krakovskaya, who had been with me all this time, talked to the Nicaraguan surgeon, he told me about the progress of the operation and assured me that everything would be fine. In the evening we left for home.

The next morning, Maya called the hospital to inquire about I.M.'s condition. and ask permission to come. The surgeon said that everything corresponds to the time that has passed since the operation. Then he said that the patient would still sleep all day, but if you want to come to stroke him, then come. It surprised and moved me. We arrived, he was no longer sleeping, you could sit next to him and talk, and kiss, and not just stroke.

In the hospital I.M. stayed 10 days. Everything was as it should be, strength was added. Every day someone from my family took me there. On April 30, Edik brought us home. Everyone had a sense of relief - such an epic was left behind.

On May 1 and 2, I felt good, we walked around the house, our relatives visited us.
i.M. said that for the first time in a long time, his heart did not hurt, and that if he had some more time, then there were ideas. He was mentally already at work.

On May 2, at 9 pm, when nothing foreshadowed trouble, suddenly I.M. ran into the bedroom from the living room and threw himself on the bed. He could not say anything, but I realized that a disaster had occurred, and I was losing him. It was a Sunday evening, the university was empty, the students had left before Monday morning. Only in the next apartment I found a guy. On my fingers, I explained something to him, he called the doctors. They quickly arrived, did something, fussed, But everything was already clear to me. Near me appeared two girls from the Russian department of the university, with whom I was on friendly terms. Who found them, I don't know. They called Edik and Bela, Prof. Gluckstern, well, but everyone was visiting. Ilya Mikhailovich, already lifeless, was taken to the hospital, I went there in a police car with student girls. There he continued to carry out rehabilitation activities, but everything was already useless. It was a pulmonary embolism.

Soon Gluckstern and his wife Liz and Edik and Bela came to see me at the hospital. We spent some time in the hospital. I asked permission to say goodbye to Ilya Mikhailovich. There has never been a more bitter moment in my life.

I stayed in the States for another week. I was very supported by my relatives and friends of the University staff. I was touched by the attitude towards me of complete strangers who came, said something, brought flowers, postcards. A farewell was held in the Funeral Home, where professors, teachers, students of the University came.

I flew home with an urn. On the plane, I was put in a business class, where for some reason there was not a single passenger except for me, only the stewardesses sometimes rested. After 5 years, I brought the urn with me to Canada.

My life is broken. For many years I lived with the physical feeling that I was cut vertically and the remaining half of me was constantly bleeding. They say that time is the best healer, it is true, but the healer is very, very slow.

3 weeks after the death of Ilya Mikhailovich, on May 23, 1993, Misha defended his doctoral thesis at MEPhI. Ilya Mihailovich has been waiting for this day so much! In my subjective opinion, the defense was brilliant. But I cried all the time. Friends were afraid to come up to me to congratulate me.

I.M. planned, after the approval of his doctoral thesis, to transfer Misha to ITEP, but not to his own department. Misha worked at MRTI, things were going badly there, the institute was practically not funded. But, it didn't happen. Misha's wife, Masha, also worked for pennies. A family with two children lived hard. There was nothing I could do to help.

And then Misha thought about emigration. The last argument to him was the fighting in Moscow in October 1993. Of the three countries - Australia, Canada and South Africa, willingly accepting as immigrants people in professions that are scarce for them, they chose Canada. In February 1995, the family left. I took their departure hard. I could not imagine then that in 3 years we would be together again, in the same city and even in the same house - we are on the 34th floor, and Misha and his family are on the 24th.

Misha's development in Canada was long and difficult, he started by selling flowers. There was no real work for eleven months. But it's all right now. He works as a mathematician-analyst in a bank. Creative work, he is satisfied. Mashenka works as an office worker in a small engineering business. Tanya, the eldest girl, graduated from the university with a degree in statistics, works in a financial company, her husband is trying to establish his own computer business. The youngest, Sonya, graduated from the University of Toronto last year, then improved for a year in Denmark, in Copenhagen. Now she has gone to study in Montreal for two years in order to return from there as a master. Her future plans are a doctoral degree. Specialty - sports medicine and physiology.

After Misha and his family left, Olya also thought about emigration, she wanted to take her two boys away from the army. The process of registration was difficult, Canada does not take doctors, although there are very few of them here, but, in the end, everything worked out. We left in March 1998.

Now Olya is engaged in ultrasound examination. Her husband, Volodya, was a truck driver, he traveled around America 5 days a week. After a serious accident 2 years ago (he was taken out of the cab of a truck with an autogen), when he accidentally survived, and inexplicably saved his right leg, after three operations on his knee, he will study in will be free, in addition, all the time from the moment of the accident until the start of work in a new specialty, he receives 80% of his salary. The boys are fine. The eldest, Sanya, is a 4th year student at the Faculty of Applied Mathematics at the University of Waterloo, 120 km from Toronto. This university enjoys a reputation as the best in Canada in mathematics and computers. Lives in a hostel. The youngest, Ilyusha, is studying in the 7th grade of a school for gifted children, he was selected there for testing. Studying well. The main subject there is mathematics. All his free time belongs to the computer and music, he plays viola and piano.

Well, I, of course, went for the children. Go, go, said the parrot as the cat dragged him out of the cage. Here I am the same parrot. I can't complain about life, I get a decent pension, I have my own subsidized apartment. But every 2 years I visit Russia. I want to see relatives, friends, walk around Moscow and Saratov. I have a large social circle, consisting of our Russian intelligentsia, reaching out here for the children. Without this circumstance, none of us would begin to move around the planet.

Here in the apartments of Olya and I hang paintings brought from Moscow. They were given to Ilya Mikhailovich on various occasions - one by Volodya Plotnikov and four by Ida Egorkina-Batalina.

For all 40 years, there was a reliable and loving person around me. Our daily communication was not ordinary and boring, there was always humor in our conversations. Ilya Mikhailovich said that his marriage is a one in a million chance. And he loved the words of Mikhail Svetlov: "I will put on dozens of bald heads, just be young."

In our life there was respect, and love, and tenderness, and the desire to always be together.
I wrote everything that I could and how I could.

## MY FATHER

### M.I. Kapchinsky

My memory of life begins with the memory of my parents. I am very small. I know this because everyone else is very big, much bigger than me. Around - unusual nature, trees with strange hairy trunks and long narrow leaves. Next to me are mom and dad. Mom is tall and beautiful. Dad is tall and strong and knows the answers to all questions. High gate. At the gate is a tall, strict boy in a red tie. "Boy, are you a sentry?" - Mom asks ... That's it, there's a cliff, the answer of the sentry boy is no longer fit in the memory. Unwinding the events back, I can assume that it probably happened in the Crimea at the gates of Artek.

I have very few memories of my father from my childhood. He came home from work late, when I was already driven to bed. I don't know why, but even on Sunday, mom and dad didn't have much time for me. (Looking back, my sister was small). Every case when dad went for a walk with me or just talked for a long time about "serious things" was remembered for a long time.

Dad taught me to ride a two-wheeled bicycle. It didn't turn out very well. Postponed until next Sunday. On the following Sunday, something did not work out for dad, they postponed it for another week. During these two weeks, my boy friends managed to learn me. When dad offered to go to school, I refused, boasting that I already knew how. Dad was disappointed and reminded me reproachfully several times.

A car appeared in the family, and now dad began to disappear in the garage all weekend. Rare joint walks stopped altogether. But by that time I had already grown up, and the boyish company replaced the family. Yes, and the passion for reading seriously rolled up, there was no time left for doing nothing at all. To take care of the car, a mechanic from the institute's garage, named Matveich, was hired. For a couple of years, Matveich repaired and lubricated the family unit, then they parted ways with dad, Matveich was fired, and since then dad has been working on the car only himself.

But holidays were completely family time. At first it was the Crimea or the Caucasus, certainly in August or September, later they mastered the Baltic coast near Kaliningrad. South Holidays were, as I remember, always savage - they rented rooms from private owners. Traveled only by train, never by plane. And they already went to the Baltic States in their car.

Dad could not drive a car, poor eyesight did not allow. Mom was always behind the wheel. And dad is nearby, with maps and a car atlas. The trip was always planned in advance - the daily route, places of gas stations, lunches, overnight stays were signed. We usually spent the night in a campsite in tents, sometimes right in the car on the unfolded seats, having moved a little into the forest from the road.

The toilet bowl is broken. Dad armed himself with a hammer, pliers, strings, wires, sticks, and retired with a faience appliance. Earned. But not for long, again several times it broke, repaired, broke, repaired. My grandfather, my dad's dad, got tired of it, he brought a plumber. Dad found out, was terribly dissatisfied: "Why did he do this? I asked! I was more proud of this than the accelerator!" This was even before the launch of Protvino.

I remember from my childhood that my dad smoked. There were always packs of Belomor in the house. Dad quit smoking. Apparently, there was a medical recommendation. The cigarettes disappeared, and Papa never smoked again.

Even in its "smoking" period, I discovered the White Sea left unattended and decided to find out why adults love it so much. I put it in my mouth (bad luck, I got the tobacco end) and began to chew. The impression was so strong that a sharp aversion to tobacco and tobacco smoke remained for life. And for that I am also grateful to my father. But he never knew about it. I could not then admit that I had stolen his cigarette. And then never had to say a word.

A Yauza tape recorder appeared in the house. It was dad who began to study English, and the purchase of a tape recorder was an indispensable condition for learning. I don't know if the tape recorder helped him in English, but the coils with bard songs formed immediately and never disappeared.

Papa was very fond of Galich and Okudzhava. He did not like Vysotsky's songs, he considered him a bad poet. No matter how hard I tried, I couldn't convince him. My father's acquaintance came, his name was Volodya, he brought a guitar and sang in his extremely pleasant voice to his own accompaniment of the songs of Galich and Okudzhava. Dad hid the tape recorder behind the closet and secretly recorded Volodino singing. The room was locked with a key, I had to eavesdrop under the door. Once the parents managed to get an invitation to Galich's home concert, and they went with a tape recorder. The tape from the concert then sounded non-stop for several months in a row.

The artist Stepanova came to us and painted my father's portrait. Dad liked the portrait very much. He said that his main essences were correctly captured in the portrait: "the high forehead of a thinker, the thick cheeks of an inhabitant and the narrow eyes of the head of the laboratory." The portrait hung in my parents' bedroom, and later in my father's office space.

The fact that my dad is not simple, but famous, I learned in a pioneer camp. It was a departmental ITEP camp. Once, after lights out, a counselor came into the bedroom, not our squad leader, but one of the senior counselors, and started a conversation about how we, unworthy children, behave badly and dishonor our fathers. And he cited two parents as an example, from whom one should take an example. I don't remember the name of the first one, I never met it again, and the second example was my own father! Then in the dark I almost burst with pride, it's good that no one saw.

I do not remember that my father's colleagues were often at our house. But dad talked about them very often, so the names of his leading employees were on my hearing. He always called them by their first and last names and spoke with unfailing respect for everyone. It is obvious that the disrespectful employees next to him did not linger. There was no talk about work at home, but he regularly discussed official intrigues with his mother. Dad was very worried about these troubles, he was literally seething. Once, when, apparently, it was completely hot, he complained to a guest who happened at that moment at our house. She asked him: "Ilyusha, if the situation is so serious, why don't you quit?" I don't remember the answer. Two months later, the situation completely collapsed and dad never remembered it again.

My father rarely took part in my school life, but each time very intensively. His passive participation consisted in calls to the school. For some reason, my math teacher called not just parents, but purposefully - dad, for every my four. And he dutifully went to her and listened that nothing good would come of his son.

Active participation in my memory happened three times. I couldn't draw at all. No matter how hard I tried, my tags came out ugly dirty, crooked, lopsided. It was disgusting for me to look at them. And the drawing mark went to the certificate. And so dad came up to me, asked what I was supposed to draw for the final lesson, took away my drawing tools and drew a magnificent Bunsen burner in three projections and in section. It turned out amazingly beautiful! The drawing teacher also liked the drawing very much, he gave (dad) five plus and apologized that he could not bring (me) five in a year, because there were many previous twos and threes. He did not suspect a forgery.

Another time I had to write an essay on Mayakovsky. Mayakovsky is one of my father's favorite poets. He looked at my creative suffering (school essays were difficult for me) and volunteered to write it himself. I was terribly surprised because nothing like this had ever happened before. Dad wrote the full text, with dates and quotes (all from memory). I carefully rewrote it with a little editing (Dad's vocabulary was slightly outdated, for example, he wrote "greedy American capitalists" instead of the usual stamp "greedy American imperialists") and gave it to the teacher. There was no five for the content, dad was terribly offended. As a joke, of course, not seriously.

The third participation was the brightest. Somehow, the school organized a physics evening for students in the 9th and 10th grades, and dad was invited to give a lecture. He approached the matter very seriously, and all the abstracts of the lecture with the allotted timing and graphic illustrations were written by him in advance. He chose the main ideas and history of the development of charged particle accelerators as the subject of the lecture. The lecture was extremely successful, he was listened to very attentively, there were many questions. Dad later recalled with pleasure the receptive audience. He highlighted a question from a tenth grader about whether focusing and phasing would help if the accelerating chamber were filled with hydrogen. It was the only time I heard my dad speak in public.

During the construction and initial work of the complex in Protvino, Dad went there every week, spent two days there - Wednesday and Thursday. He had a one-room apartment in Protvino, we all often went there at different times in winter and summer. In winter there were skiing, in summer - walks in the green surroundings, swimming in the river, picking mushrooms. I remember I always got up before a walk the question is which direction to head today - to go to Protva or go to the Oka. Lilies of the valley grew in the woods on the way to the institute entrance in the spring. Later, the forest was cut down, and houses were built in place of the lilies of the valley.

Shortly before the launch of the accelerator, French President Pompidou came to Protvino. On this occasion, a decent road was built from Protvino to Serpukhov. Pompidou was skeptical of a launch date on the anniversary of the revolution, saying "Russians will regret making the revolution so early." The launch, however, took place right on time. The Pope suggested sending a telegram to Pompidou: "The revolution was made on time." The idea was wrapped up, the telegram was not sent, Pompidou remained out of the loop. This whole story is from dad's words.

On the instructions of the editor, I asked my father to write some funny or amusing story for the Physics and Technology newspaper. He wrote how, during the launch of the synchrotron at Protvino, a poster was prepared in advance that a record beam energy of 70 GeV had been reached. In fact, the energy turned out to be higher, and the poster had to be transported. The number 70 can be easily translated to 76 or 79. The measured energy was 78 GeV. They decided not to embellish reality, and as a result, the poster showed off 76. So it went to the newspapers that a record energy of 76 GeV had been reached.

There was never any discussion about my becoming a physicist. It was something taken for granted. Understandable, first of all, by me. Physics was so in tune with my soul that it would take a fair amount of parental effort to change my educational and career vector. My admission to the Fiztekh was prevented by the fifth point (in the Fiztekh of those years there was an unspoken percentage quota for Jews). Dad strained his acquaintances. Dating helped, I was accepted. Well, I did everything so that later neither Dad nor the Phystech would regret it. Dad certainly did not regret it, he sometimes jokingly said that my admission to the Physicotechnical Institute was the biggest achievement in his scientific career.

My father took an active part in my physical and technical education only once. In my freshman year, for my first semester physics exam, I chose "Kepler's Laws" as an elective question. In the midst of preparations, somewhere before the New Year, Dad found me among a pile of books and sheets of paper, asked what the problem was, got an explanation and suddenly caught fire. He went to his office (the part of the common room immediately around the desk served as his office) and, a few hours later, he brought me eleven sheets of paper covered with his neat small handwriting, with a complete derivation of Kepler's laws from Newton's laws. The text (with short inserts between the integrals) was written in black ink, the formulas in red, and the pictures in blue.

I passed that exam not only successfully, but triumphantly. The teacher who took the exam remembered me, and many years later, when I asked him to become an opponent in my dissertation (he himself was fluffy by profession, so coincidentally), he immediately agreed, called his graduate students and set them an example of me and mine " good reputation."

But that was the only time. By that time, my father was already very ill. The first heart attack knocked him out, as far as I remember, outside the house, he was returning from Dubna, from the accelerating conference; he managed to get home by himself. He felt bad, he sadly told his mother that that was all, and he would never see his grandchildren. When they say that dad died very early, I both agree with this and disagree. Because all the years allotted to him after that heart attack were a gift of fate and one should be grateful for them.

When my first child was supposed to be born, dad asked very much that if there was a girl, then to be named Sophia, in honor of his mother. A girl was born, but they did not call her Sonya, her name had long been chosen. And my second daughter became Sonya.

After graduating from the Physicotechnical Institute, I got a job in the theoretical department of MRTI, having received as a topic for work the collective acceleration of ions in direct electron beams. The Pope did not approve of this direction, considering it unpromising. He urged me to change the subject many times. Every time I refused, because the work was insanely interesting, and in the theoretical department I was surrounded by wonderful people; I went to work every day like a holiday. The years have shown that the pope was right, and the direction really turned out to be a dead end. But, firstly, who could have known it then, and secondly, it was very beautiful physics.

In those years, I had, so to speak, on duty, to study my father's book "Particle Dynamics in Linear Resonant Accelerators", its first edition, which I myself called "green", according to the color of the cover. I really liked the book. It successfully combined a scientific monograph, a reference book, a textbook, and at the same time turned out to be a methodological guide - it called for applying its ideas in related fields. I have come across very few books that successfully combine these qualities. More precisely, only one more - “Electromagnetic Waves” by L.A. Weinstein. The conclusion of the microcanonical distribution was read in one breath. Like a poem. Or a good detective, who likes what comparison. The second edition, "red", turned out, in my opinion, not so successful. Maybe because it was not written from scratch. Dad armed himself with glue and scissors and began to shred the green copy - cut, paste, cross out paragraphs, write in paragraphs ... As a result, the manuscript was prepared very quickly, but something elusive was gone from it.

My dad didn’t closely follow my career, but he tracked it. After two years of work, he asked me if I had a candidate plan. I honestly said no. And immediately received instructions: "Now sit down and write." At first I was at a dead end, not imagining how one could take and write a dissertation plan without leaving the spot. However, the impetus was given, and a month later the plan in my head somehow formed by itself. Ten-something years later, my dad asked in exactly the same way if my doctoral thesis was ready. I honestly answered that no, and not expected. “How is it not,” he wondered, “for a theoretician to work for ten years and not gain a doctorate? You are probably lazy.

Now sit down and write a plan." This time it did not help, there was not enough material, and I begged myself for another two years. Dad remembered the deadline, but when two years later I received his direct question, I already had an answer prepared.

Dad didn't make it. The mournful call from Washington came in the afternoon in early May. In recent years, when the iron curtain has rusted and crumbled, dad began to travel abroad on invitations. I remember he set a strict condition for his superiors that he would not go without his wife, and he and my mother traveled together to Germany, Switzerland, Italy, and America. And we were looking forward to returning them, looking forward to souvenirs, photographs and interesting stories. The trip to America was the last, another heart attack turned out to be fatal, despite the just undergone surgery. And in Sheremetyevo we already met my mother alone, with a funeral urn in her hands.

My dad's photograph is on the shelf next to my desk. We often talk in the evenings. At first, I felt awkward, as if I changed professions, left science. I lost this feeling when I managed to find my vein, I began to work in the field of financial mathematics, and it turned out that, according to the research methods, financial mathematics is not mathematics at all, but pure applied theoretical physics. We observe the financial market and try to understand how it can be described mathematically. I had no one to share my ideas with, the only listener of the treatise "What is an exchange from the point of view of physics" was my father's photograph.

## A LETTER TO A FRIEND

**N.O. Rashba**

Dear Ilyusha!

For twenty years now, we have been living in different worlds, but my affection for you and friendly disposition have not diminished.

In half a year I will turn ninety, and even if I turn out to be a long-liver, only a few years will remain before meeting you.

I live calmly. I walk, I sew different clothes for myself, I knit for my grandchildren, I go to art museums, I read a lot. I myself don't understand how this happened, but I wrote a book "Experienced" about our life in the Union; and people, not only friends, read it and respond approvingly. I told in it about the 45th school, about the outstanding 86th class, and you appear in the book on many pages. You were always the best student of all of us, and although there were six other excellent students in the group, none of them could compare with you, but you never boasted of your superiority. Football, poetry, chess were also among your interests.

I treasured our friendship and it never broke. I remember how you and Lyuba came to us in Zhukovsky, loaded with food and household items, in order to ease our household chores. You have bags of vegetables in both hands, Lyuba has a necklace made of packs of toilet paper around her neck. For your arrival, I am preparing my signature dishes - mushrooms and pilaf. You love this food, and I'm glad I did.

Grisha is studying proverbs and sayings. He has many publications. He acquaints you with them even before the publication - in proofs. Your assessment of an article or a new book for Grisha is the most authoritative.

Once in your youth, on your initiative, it was decided to gather once every five years with the whole class and tell each other about your family, work, successes and possible failures. On your own initiative, these gatherings became annual and remained a significant event in our lives.

Grisha and I ended up in Alma-Ata during the war and returned to Moscow only in 1953, without having the slightest idea about life and events in the country and in the world. You and Lyuba were shocked by our ignorance, and at every meeting you explained something to us. The receiver finally acquired in Moscow helped, and the opportunity to listen to the BBC at night.

And how happy you were when we showed up with Grisha's sister Nina and her husband Volodya Vishnyak! Volodya has a wonderful memory. He knows all the songs of Russian bards and sings them with the guitar in his pleasant voice. From him we heard first of all Okudzhava, and then Galich. Volodya sang with pleasure, but he was very afraid to perform these forbidden songs in your apartment building. We begged him together, he gave up and continued to speak. I still have a cassette with one of Volodya's recordings made on the sly. All our voices are heard, Volodino: “No, no, no!”, And then his own singing. Again Grishin’s exhorting voice, Lyubino: “Volodya, everything is off, all doors and windows are locked!”, And yours: “Well, please, we all really ask you!” - and, finally, Volodya's singing again.

Then we parted ways to our own countries, and neither Grisha nor Volodya are with us...

I know that Luba has your portrait, painted by my son's first wife, and I remember that you are pleased with it, although it is made in the manner of super-avant-garde, so to speak. I also consider the portrait very successful and remember all the details well.

This concludes our conversation. I was very pleased to remember with you some of the events of our lives. Take care of your family.

Kiss, Nadia.

## IV. MEMORIES OF COLLEAGUES I. M. KAPCHINSKY

### WORLD LEADER IN ION LINEAR ACCELERATORS

**N.V. Lazarev**

<u>Background</u>

Ilya Mikhailovich Kapchinsky passed away at the height of his scientific fame, alas, unacceptably early. Achievements of I.M. in the field of accelerator science and technology, especially on the problem of accelerating high-intensity ion beams, the proposal of new ideas and the creation of linear proton accelerators that were advanced for their time, are known throughout the world. With him and with V.V. Vladimirsky (during his work on accelerators) connected the worldwide fame of ITEP as an advanced scientific center for the development of both ring and linear accelerators for the highest energies of the proton beam at that time.

However, not everyone knows that at the beginning of his scientific career, I.M. engaged in scientific methods and techniques of pulsed radio electronics used in radar. In this area, he then achieved great success and recognition: he defended his Ph.D. thesis, published a valuable monograph [1], which was soon republished abroad. Therefore, many could not explain his decision to drastically change the area of his scientific interests to an initially unknown to him accelerator subject. But subsequent life showed the foresight and fidelity of the decision.

Perhaps, I had the opportunity to work with Ilya Mikhailovich in close contact and under his leadership longer than others. For the first time, fate brought us to NII-20, where in the laboratory of Irina Borisovna Andreeva, the development of pulse systems of the first domestic early warning radar station, an analogue of the American station 5SN-584, was carried out. I apologize in advance for the fact that, having undertaken to write about I.M., I will have to say a lot about myself. I ended up in a small friendly team of I.B. after 1% of the technical school course, which had to be abandoned due to the fact that there was nothing to live on. I already had the experience of a beginner radio amateur, so they hired me as a technician, and after six months of intensive work under the guidance of I.B. and older comrades, I became quite well versed in tube electronics. I must say that less than others I communicated with Ilya Mikhailovich Kapchinsky, who was continuously immersed in some kind of calculations, a laboratory researcher who was not sitting like the rest of us, each at his own circuit and oscilloscope, but at a table with a stack of scribbled sheets of paper, a slide rule and a reference literature.

Although we all worked regardless of the time, our salaries were small. Life was harsh then, neither I nor the other laboratory staff had enough money. Therefore, when I.M. asked me if I would like to participate with him in additional work for the biophysical department of Moscow State University, I gladly agreed. According to the calculated I.M. In the circuit, I assembled a highly sensitive DC amplifier, checked its gain, stability and adjustments on the stand. Our demonstration of the operation of this amplifier from signals from different points of the body of a crucified and twitching frog, both on the screen of an oscilloscope and on a large pointer instrument, delighted biophysicists. Apart from the common work at NII-20, this amplifier was our first joint work with I.M., thanks to which we got to know each other better, and I learned something new from him that you don't learn from books. When I came into close contact with I.M., I saw that he not only writes reports on the work of the laboratory and performs some mysterious calculations, but also perfectly understands all the intricacies of the operation of individual circuits that we established on our stands, as well as in their interactions, as was the design of the station.

Thanks to the exceptionally clear filigree writing of I.M. letters of the Greek alphabet, and in consonance with his last name, we called him "Kapa" behind his back. But, it must be said that even then I.M. he knew how to put himself in such a way that he did not allow familiar relations with all of us, members of the IB team.

Then our paths diverged for a long time: I passed the secondary school exams externally, on the advice of I.B. enrolled in MMI (MEPhI), and after graduating from the institute was admitted to the ITEP graduate school. However, in my dissertation work on y-spectroscopy, which was carried out at a nuclear reactor, due to the strong background of scattered radiation, which led to an overload of detectors with Na1 crystals and coincidence circuits, which masked the desired effect, insurmountable difficulties arose. Yes, and I myself was overexposed in these conditions and ended up in the hospital. After recovery, I had to look for a new field of activity. And then, unexpectedly, I met I.M., who was invited by ITEP Deputy Director V.V. Vladimirsky, who knew well the high scientific potential of I.M. for a long-term joint work at plant No. 465, to lead the work on the creation of high-current linear accelerators (LU) of protons with hard focusing - injectors of proton synchrotrons. I accepted the offer of I.M. and since then - since 1957, 36 years of continuous work under his leadership until his unexpected death in 1993.

Development and creation of high-current injectors for proton synchrotrons

In those distant times, when the ITEP (U-7) and IHEP (PS-70) proton synchrotrons were still being designed, the project manager V.V. Vladimirsky, Academicians A.I. Alikhanov and A.L. Mintz was well aware of the need to develop high-current pulsed injectors for these machines. However, the experience available in the country in the development of grid-focused accelerators with an output current of only about 1 mA (at KIPT and JINR) could not be used to create injectors with an output beam current of about 100 mA. The construction of the LU-injector then seemed to be an exceptionally new and complicated matter, therefore, in order not to delay the launch of the U-7 at ITEP, it was decided to establish the initial injection from a standard 4 MV Van Graaff direct-acting accelerator. This decision fully justified itself, and from 1961 to 1967 injection into the U-7 was provided thanks to the work of the EG-5. To increase the intensity, it was necessary to significantly increase both the beam current and the injection energy, and it was also necessary to increase the reliability of the injector.

The design, development and construction of LU-injectors for the U-7 and PS-70 synchrotrons was carried out under the general supervision of Corr. USSR Academy of Sciences V.V. Vladimirsky and Academician A.L. Mintz by teams of employees of ITEP, RYAN (later MRTI) and NIIEFA with the involvement of several dozen scientific development of new injectors at ITEP in July 1958, a laboratory of linear accelerators was organized, headed by I.M. Kapchinsky, its first employees were E.N. Daniltsev and V.K. Plotnikov, I and V.A. were soon accepted. Batalin, then appeared V.I. Edemsky and other employees. The selection of personnel I.M. paid great attention to the conversations with the engineers who offered their services, at first he went out himself and took me or someone from his closest circle for training. Later, he began to constantly control this function and trust us, but the admission to scientific positions remained exclusively with him.

From the very first months, detailed plans were drawn up and a wide front of computational, theoretical and experimental work was launched, which made it possible to obtain much-needed data in a short time to determine the parameters of the main components and technological systems of the I-2 LU (i.e., the second injector) for energy 25 MeV, which was also a model of the world's largest proton LU I-100 for an energy of 100 MeV, the injector of the PS-70 synchrotron in Protvino, which was being developed at the same time.

According to the nature of his character, I.M. He took his work very seriously and tried to do everything fundamentally. Fluency in the mathematical apparatus, excellent knowledge of physics and radio engineering, reading English technical literature without a dictionary and difficulties - all this allowed I.M. in a short time to brilliantly master a new rapidly developing field of science and technology. Moreover, putting in order the fragmentary works of other authors and developing his own theoretical studies, I.M. developed a coherent theory of resonant ion accelerators with hard focusing. There was no such literature at that time, and the execution of tasks for the design of two such accelerator-injectors for proton synchrotrons at ITEP and IHEP required a good understanding of the physical foundations of the new technology from a large number of performers. Therefore, I.M. prepared and delivered a full course of lectures on the theory of linear ion accelerators for engineers and scientists at seminars at RYAN (MRTI) and at ITEP. The recording of these lectures later served as the basis for the publication of a very valuable book.

It was extremely important that I.M. simultaneously combined knowledge and theory and the whole complex engineering issues and problems arising in the creation of new technology It was a rare combination of these qualities, developed as a result of previous work on the development of radar stations, that allowed him to become the undisputed leader, whose word was decisive at meetings regularly held with us, in RYAN and NIIEFA.

The development of injectors was carried out extremely intensively, although everything had to be done for the first time. To ensure the correct calculation of channels capable of transmitting intense pulsed beams, I.M. Kapchinsky and under his leadership at ITEP developed a theory of focusing taking into account the collective interaction of particles, a method for calculating the focusing and accelerating systems was developed, which made it possible to perform detailed calculations on the dynamics of particles, determine the parameters of the longitudinal and transverse motion of particles in the accelerator, all its main physical parameters . A system of tolerances for the manufacture of the main elements of beam acceleration and focusing was also developed. At the conducted by I.M. in the operative laboratory, each team leader reported on what had been done on their subject during the week, on the difficulties that had arisen. By the joint efforts directed by I.M., measures were outlined to overcome them. This order was maintained in subsequent years in the educated department. The abstract notes of I.M. issues that were dealt with on the operatives during the creation of the I-2 and I-100 LUs, which show that he delved into everything, both large and small, in detail, "kept his finger on the pulse" of all the work that was carried out both in our country and in related organizations.

The results obtained in the laboratory by calculation and on test benches allowed I.M. in 1960 - 1961 to develop physical substantiations of hard-focusing injectors I-2 and I-100. Under the guidance and with the direct participation of I.M. Kapchinsky published (first in the form of reports and then preprints) complete collections of the main physical parameters for the I-2 - in 1965 [2]; for I-100 - in 1967 [3]. And the new one that I.M. into the theory of such machines, was presented by him in his brilliantly defended doctoral dissertation.

It is difficult to enumerate all the scientific and technical achievements and discoveries made on the way to the launch on November 2, 1966 of the country's first hard-focusing LU I-2. On that day, an accelerated beam was obtained with a pulsed current of 30 mA, which gave a flash on a quartz plate installed at a distance of 10 m from the second resonator.

The correspondent of the Pravda newspaper the next day took pictures of this luminous plate and the general view of the LU, placed in the pre-holiday (by November 7) issue with explanations by I.M. Kapchinsky and ITEP founder Academician A.I. Alikhanov. The beam current after 3 months was increased to 135 mA, which at that time exceeded the output currents of foreign proton injectors, as well as the design current value of 120 mA. After a number of improvements to the ion source and the matching channel, the world's highest pulsed proton beam current of over 230 mA was achieved at the output of the I-2 injector at a design energy of 24.6 MeV.

The photos shown here show pictures of: an RF resonator (Fig. 1), a pre-injector for a voltage of 750 kV (Fig. 2), the process of installing the last drift tube into the I-2 resonator (Fig. 3), and adjusting the drift tubes (Fig. 4 ), general view of LU I-2 in the year of launch (Fig. 5).

The commissioning of a proton accelerator with a pulsed beam current more than 100 times greater than the grid focusing accelerator currents was noted first by our brief note in the journal Atomic Energy [4] and by two reports at a conference in Cambridge, USA [5,6]. Then, in the journal PTE, prepared under the guidance of I.M. a set of 12 articles by 48 authors devoted to the physical foundations of the I-2 LU project, the engineering part of the project, all technological systems, tuning methods and machine start-up. Of these, in five works by I.M. took a decisive part [7-11].

It should be noted that the I-2 LU turned out to be designed and tuned in full accordance with the theory; during its installation, adjustment and adjustment, the design tolerances were observed extremely strictly. There are many examples of I.M. in the struggle for strict observance of all mechanical, radio engineering and other tolerances. But for over 40 years, the I-2 LU has been operating without deteriorating the parameters of the output beam, and not once did it have to open the vacuum casing and the resonators themselves.

Immediately after the launch of the LU, ITEP began work on the construction of an ion duct to transfer the synchrotron to injection from I-2, which was done exactly one year later in November 1967 and was also noted by the Pravda newspaper in the article “The second birth of the accelerated calf." The intensity of U-7 increased to 6*1011 protons/pulse, i.e. more than 5 times, which means that the time for collecting statistics by experimental physicists has been reduced to the same extent, and, consequently, the cost of their work.

The capabilities of the I-2 LU were not limited to beam injection into the synchrotron. I.M. supported the development of the multi-purpose use of I-2 for research on the radiation resistance of electronic products, for the production of short-lived isotopes, for research in the field of radiation chemistry, etc. [12]. These works were carried out on additional pulses of the accelerated beam in the pauses between injection pulses and did not limit the operation of the synchrotron in any way.

And M. continuously worked much more than the allotted time, however, everyone could always call him and arrange a meeting to talk about work or about their personal needs and problems. For himself, he wrote the memo cited here, the first paragraph of which was “to treat all people and situations calmly ...”, but if this could be observed! The character of I.M. he was emotional, with the inevitable sharp conversations outwardly he tried to restrain himself, but he could not calm his excitement, the pressure rose, his face turned red.

Simultaneously with the studies of the beam acceleration regimes and the operation of all I-2 technological systems, the construction of the I-100 LU was being completed at an increasing pace. I.M. put the matter in such a way that each independent worker (there were no others, however, at that time) conducted parallel developments on his own subject for both the I-2 and the I-100.

Life has shown the correctness of several decisions, new in comparison with the I-2 design, adopted in the project of the I-100 injector. It is also worth noting the unique RF generators developed at RYAN for the LU I-100 with a frequency of 150 MHz and a pulsed power of 5 MW. Heavy-duty flash tubes - GI-27A triodes, produced by the Svetlana Design Bureau under a special Government Decree, adopted thanks to the enormous authority and perseverance of Academician A.L. Mints, are still the best in the country (and maybe in the world) in this frequency range. Below we will talk about the modernization of these triodes on Svetlana according to our ideas and technical specifications for a frequency of 300 MHz.

The I-100 injector was created by the combined efforts of so many teams that it is impossible to name all of them here.

ITEP, represented by I.M. Kapchinsky with his closest collaborators, fully represented the physical side of the project and carried out the scientific management of its implementation, i.e. in addition to conducting his own developments, he coordinated with the ultimate goal - obtaining a high-intensity beam - all technical solutions, tasks, requirements, and real characteristics of technological systems, equipment units, measuring equipment. The first (and, several years before the launch, the only) publication of the main authors of the I-100 project was made at the International Conference on Accelerators in Dubna in 1963 [13].

The work was carried out intensively and amicably. Of particular note is the exceptionally great help from our head office in placing heterogeneous and sometimes unique orders in industry throughout the country, decisions on difficult issues were made at quickly convened technical meetings, where the word of I.M. Kapchinsky was decisive.

At the final stage of the creation of the I-100, the staff of the specially organized department of the IHEP injector, headed by I.M. Kapchinsky, who was sent there for architectural supervision and management of installation, commissioning and commissioning.

The design energy of 100 MeV - twice as high as at the then largest accelerators at CERN and AI1_ (USA) - was reached on July 28, 1967 at a beam current of 5 mA. In the launch of the I-100 from ITEP, in addition to I.M. Kapchinsky, V.K. Plotnikov, V.A. Batalin, R.P. Kuybid and B.K. Kondratiev, who managed to gain invaluable experience during the year of work with the beam at the I-2. By the end of 1967, the output current of the I-100 LU proton beam was increased to 60 mA and beam injection was successfully ensured during the physical start-up of the IHEP proton synchrotron at 76 GeV. Later, the maximum pulsed beam current of 120 mA was recorded on the I-100. Shortly after the launch of the machine, all work on the I-100 was transferred to the IHEP staff. All our experience accumulated during the year of work with the I-2 beam, we passed on to them both orally and in the form of preprints, articles and detailed and “run-in” operational and other instructions.

For several years - before the construction of injectors in Brookhaven and Batavia (USA) - the proton LU I-100 was the largest the largest in the world (now in Europe); its parameters and the level of development of all its technological systems aroused deep respect abroad and, of course, increased the prestige of Soviet science and technology. The author of the physical design of the I-100 LU I.M. Kapchinsky became widely known in the world accelerator community. Prominent academicians A.P. Alexandrov and M.A. Markov was entitled "Engineering Masterpiece". A general view of the mounted I-100 accelerator in the injector hall is shown in Fig.6. Here is a live story by I.M. about the launch of the I-100.

The country highly appreciated the work of the developers of the I-100 LU-injector by awarding the State Prize of the USSR to 12 authors (including three from ITEP: I.M. Kapchinsky, N.V. Lazarev and V.K. Plotnikov), awarding the most distinguished workers with medals.

It is pleasant to remember the friendly, sometimes selfless work of specialists of all ranks, laboratory assistants and ordinary workers, united by a great common goal, who saw the attention and active support of our GKIAE and the USSR Academy of Sciences. It is a pity that now there are practically no financial and organizational conditions for the creation of large, world-class accelerator facilities in our country by the commonwealth of advanced domestic research centers.

It can be conditionally noted that this ended the first turbulent period in the development of the theory and technology of linear ion accelerators in our department. As a result, the issues of acceleration of very intense (near the Coulomb limit) beams with pulsed currents up to 250 mA, but with low average currents ~1–2 μA, were resolved. Part of the team continued to work on improving the performance of the I-2 injector: the stability of the pulsed beam in amplitude and time, reducing downtime to 0.5-1% or less (mainly due to the reconstruction of the RF system and ion source), as well as further modernization technological systems and expansion of possible areas of use of LU I-2. A long period of time has ended when I.M., being at the same time the head of both departments of linear accelerators at ITEP and IHEP, spent most of his time at IHEP.

It goes without saying that when the fever subsided, in addition to work related to the LUI-2 and I-100, it turned out possible in our team to develop directions. The universally recognized "Kapchinsky School" arose. A large number of scientific papers were published, many inventions were registered. A number of laboratory researchers (V.K. Plotnikov, V.A. Batalin, N.V. Lazarev, V.I. Bobylev, V.S. Skachkov, A.A. Kolomiets, A.M. Kozodaev, V. I. Pershin, A. A. Drozdovsky, A. I. Balabin) defended their dissertations. The number of scientific publications and inventions of our most active employees began to number many dozens (at present, their number for some employees has already exceeded 100-150, in subsequent years V.K. Plotnikov, and recently S.V. Plotnikov became doctors of physics. - S.A. Vysotsky, I.A. Vorobyov, TV Kulevoy). The Kapchinsky school has borne fruit not only at ITEP and IHEP, but also at a number of other accelerator centers in the country (JINR, INR, NIIEFA, KIPT, etc.).

In 1966, a book by I.M. Kapchinsky [14] on the theory of resonant IL ions (Fig. 7 above), written by him in connection with the work on the creation of injectors I-2 and I-100. His second book (Fig. 7 bottom left), which is an expanded version of the first book supplemented with new chapters, appeared in our country in 1982 [15].

In 1985 this book was translated into English and published in the USA [16] (Fig. 7 bottom right), which contributed to the world fame of our institute as one of the leading accelerator centers. In 1993, I.M. Kapchinsky was invited to the University of Maryland USA to give a course of lectures on the theory of LU, which he did shortly before his sudden death. In the same year, this course with drawings by the author (Fig. 8 above) was published in Los Alamos [17].

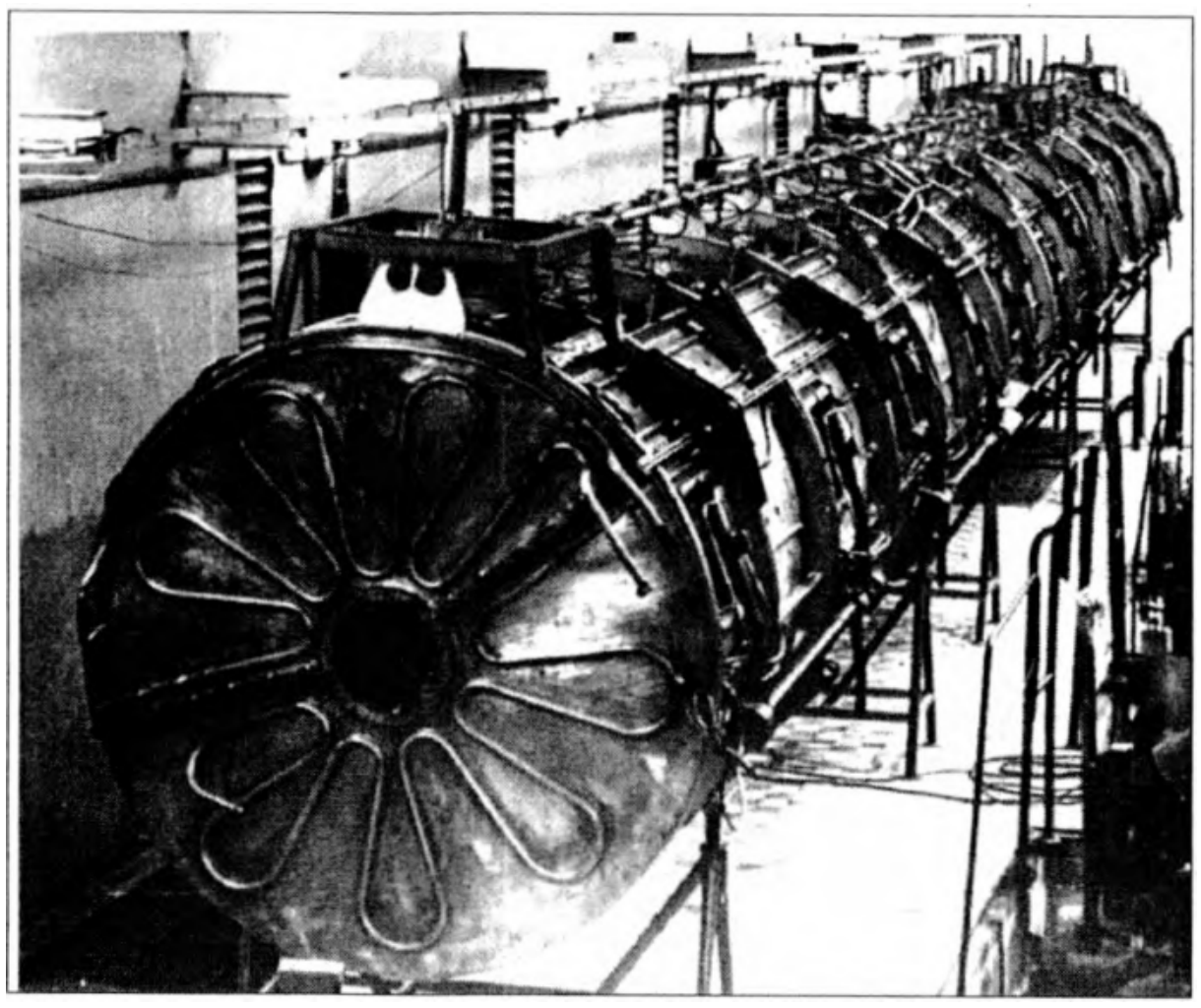

Fig.1. RF resonator I-2 before installation on the support beam.

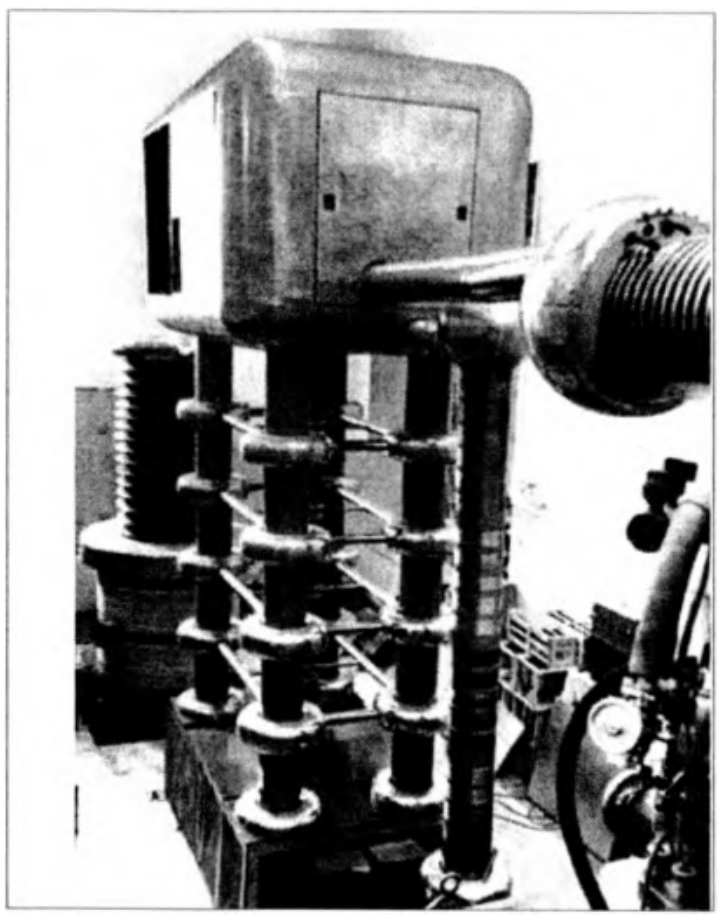

Fig.2. "Garage" for injector I-2. In the background is the IT-800 pulse transformer, at the top right is a part of the accelerating tube with an ion source in a spherical fairing.

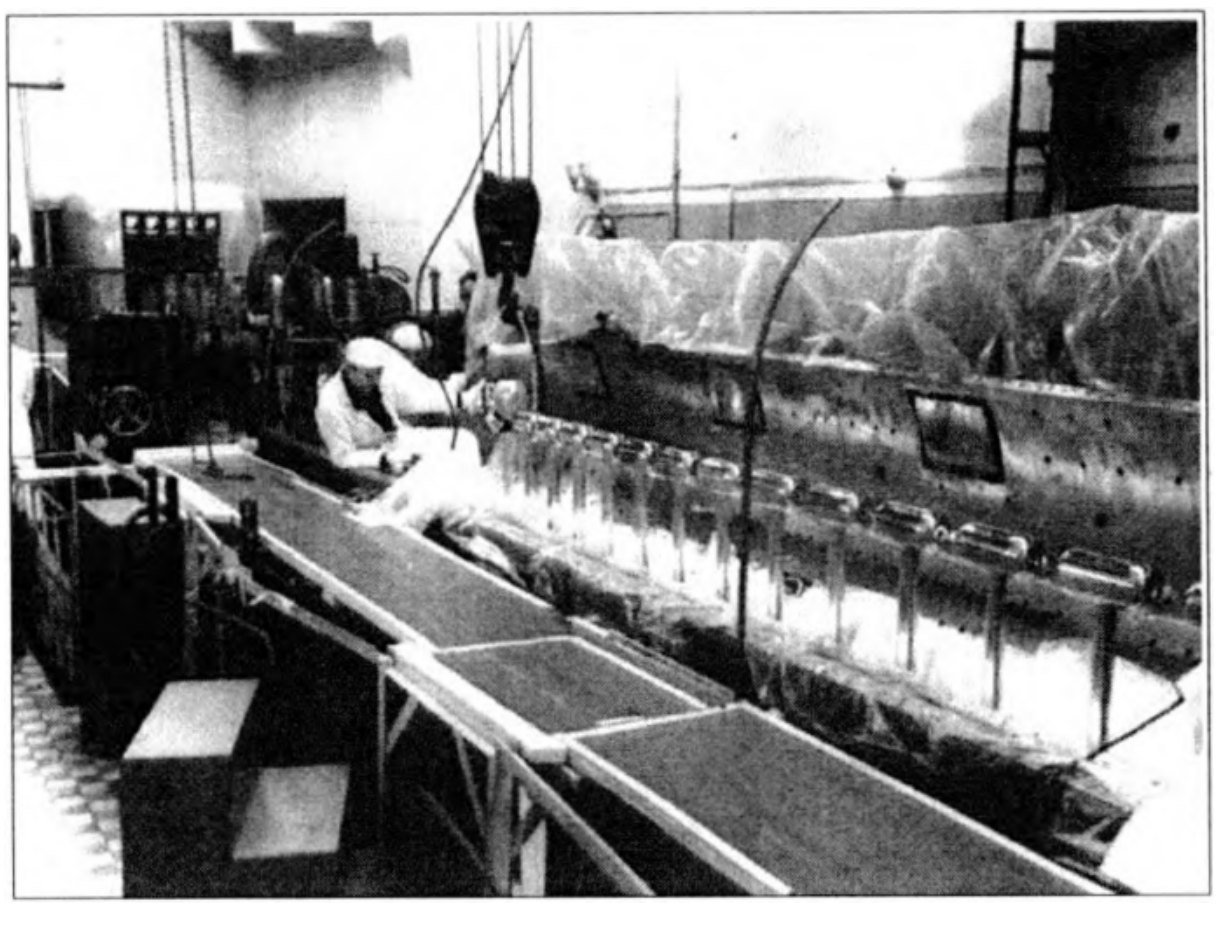

Fig.3. Precision adjustment of I-2 drift tubes.

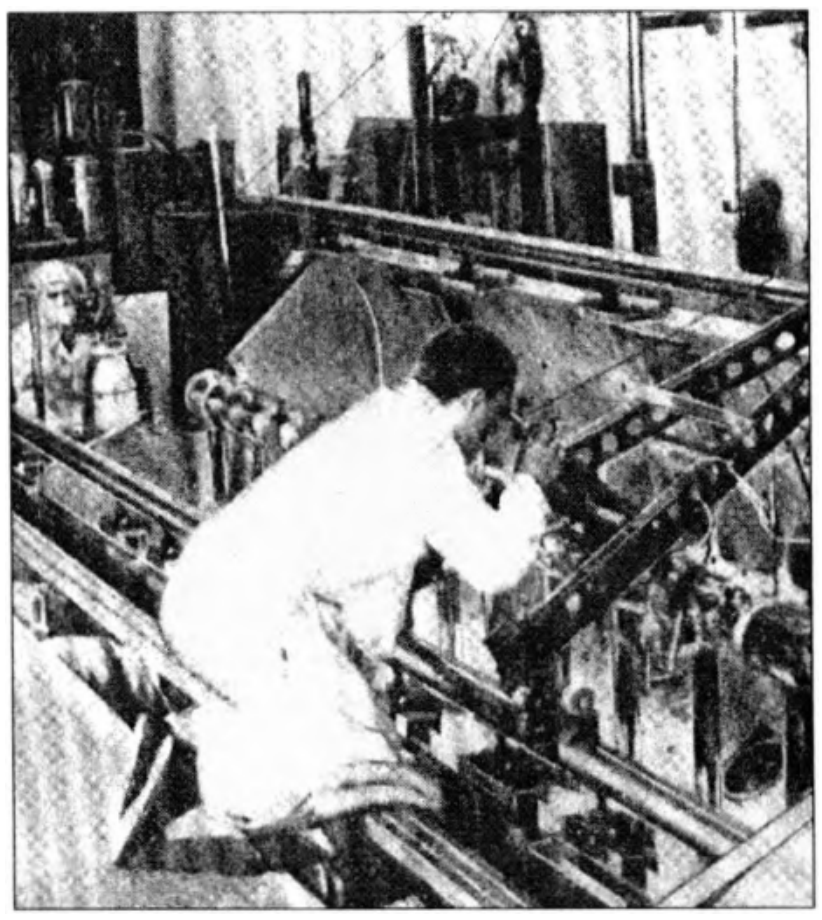

Fig.V.I. Edemsky is installing drift tubes in the open resonator I-2.

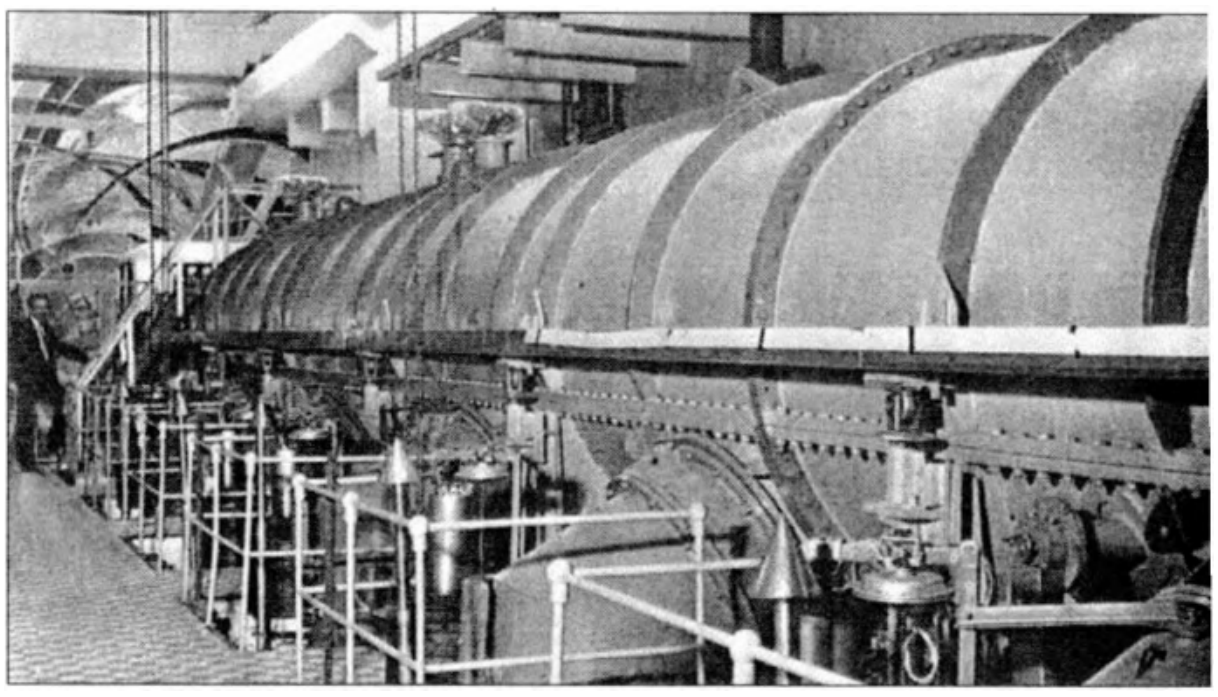

Fig.5. General view of the linear accelerator - injector I-2.

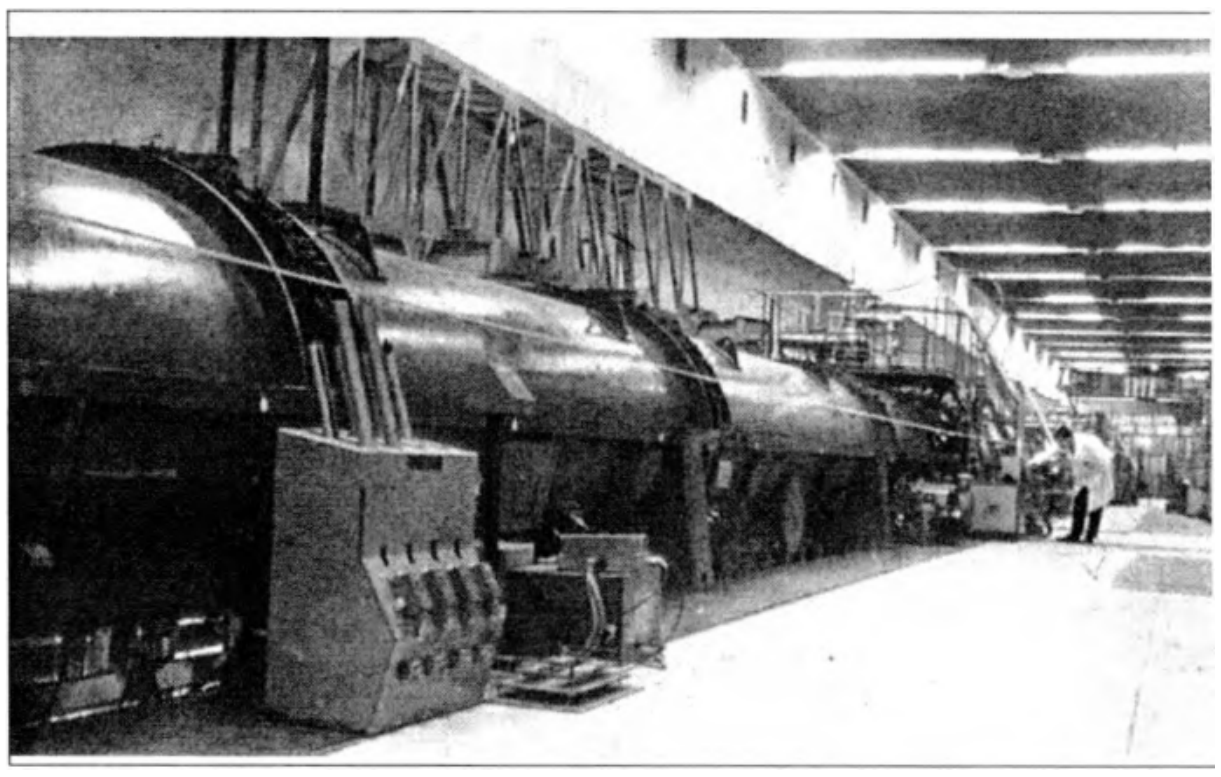

Fig.6. General view of the linear accelerator - injector I-100.

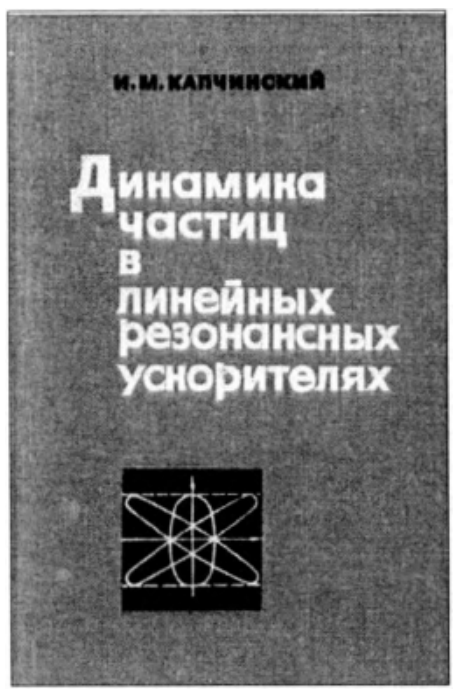

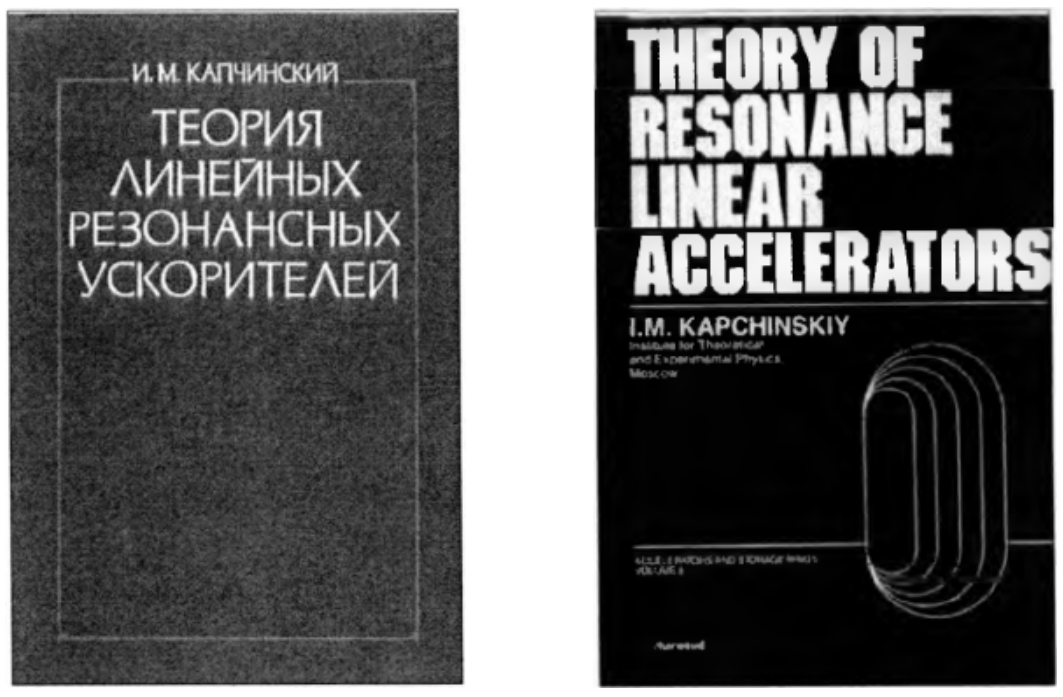

Fig.7. Books I. M. Kapchinsky on the theory of linear accelerators

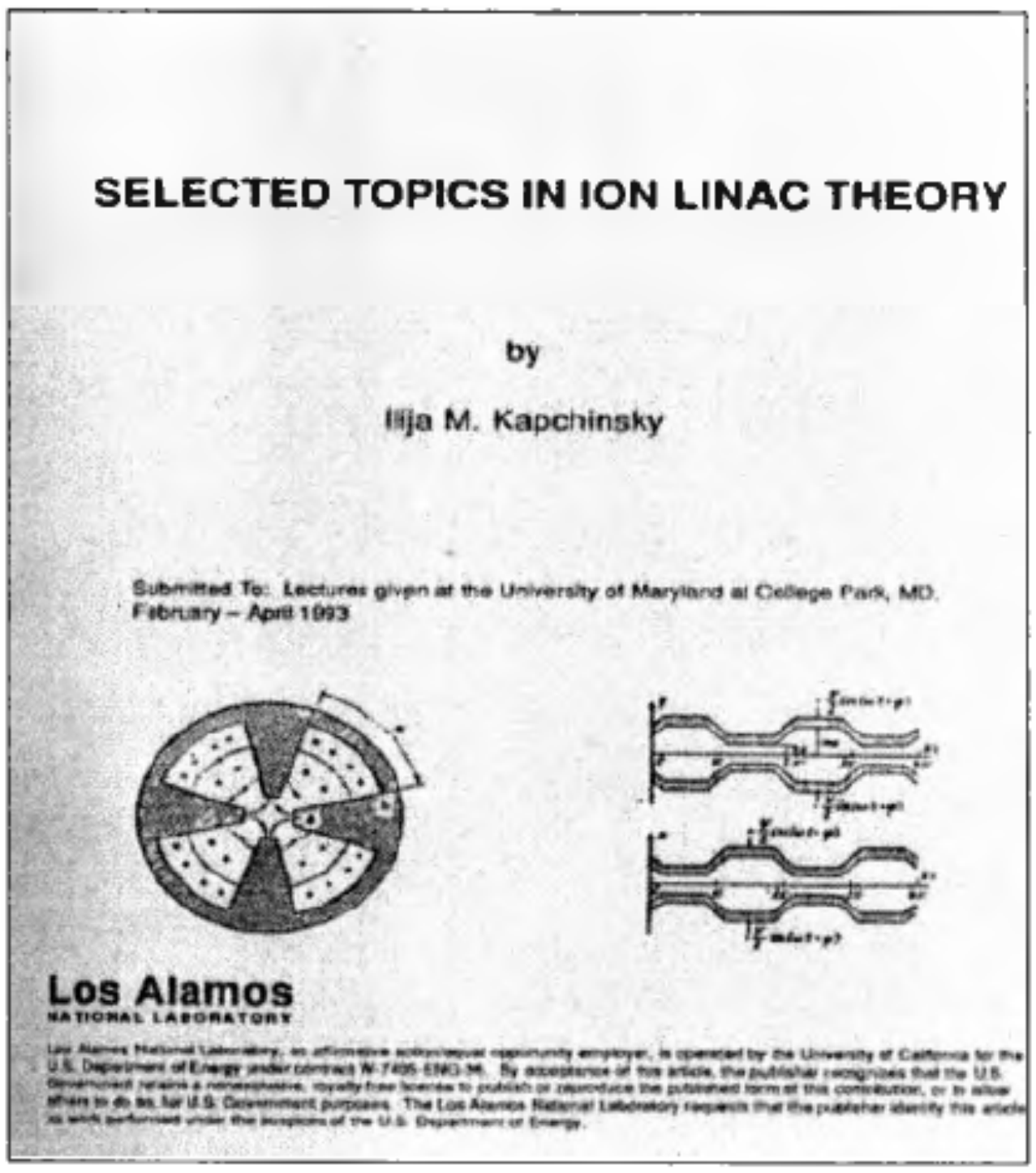

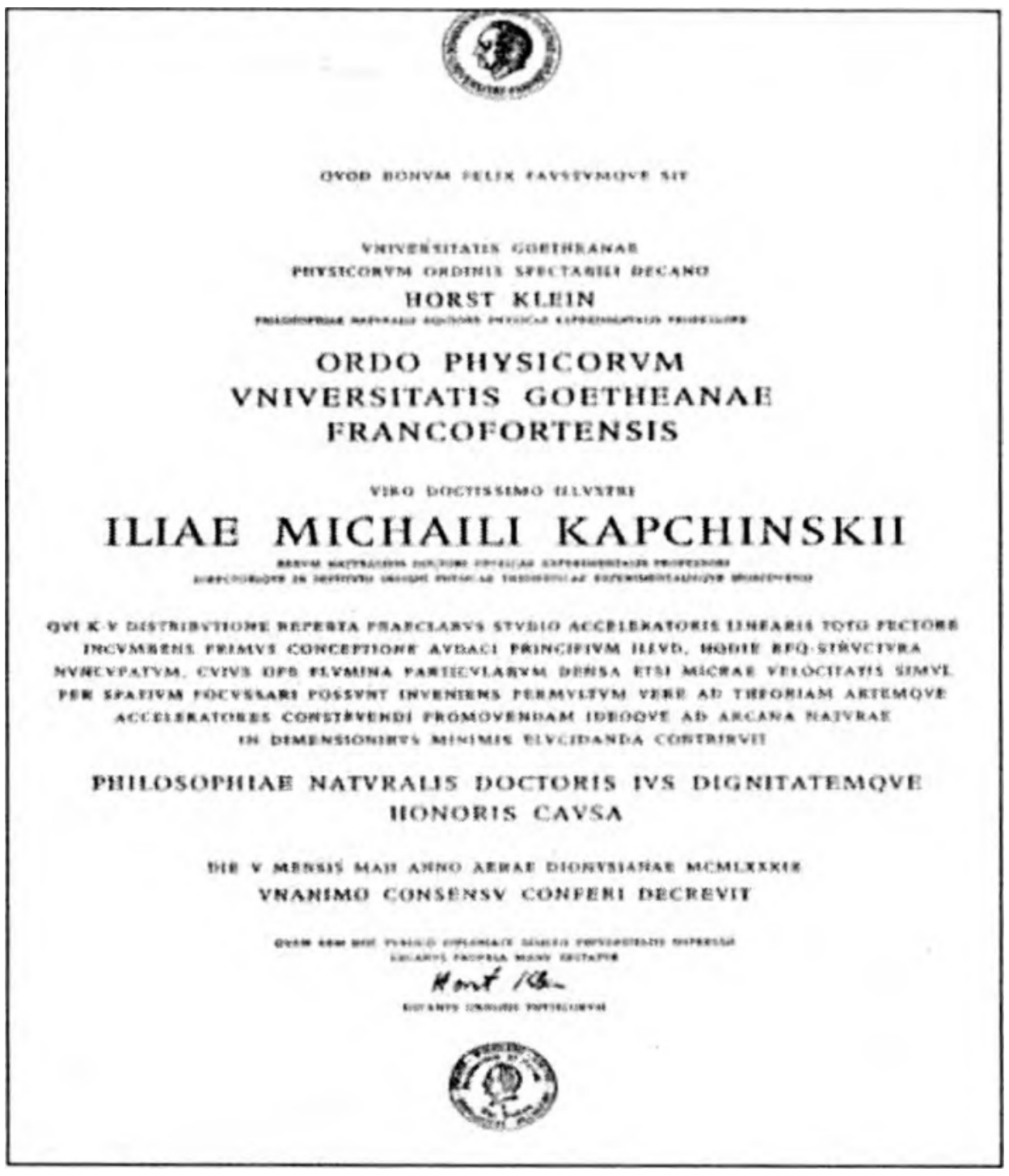

QVOD BONVM FELIX FAVSTVMQVE SIT

VNIVERSITATIS GOETHEANAE
PHYSICORVM ORDINIS SPECTABILI DECANO
HORST KLEIN

ORDO PHYSICORVM
VNIVERSITATIS GOETHEANAE
FRANCOFORTENSIS

VIRO DOCTISSIMO ILLVSTRI

ILIAE MICHAILI KAPCHINSKII

QVI K-V DISTRIBVTIONE REPERTA PRAECLARVS STVDIO ACCELERATORIS LINEARIS TOTO PECTORE
INCVMBENS PRIMVS CONCEPTIONE AVDACI PRINCIPIVM ILLVD, HODIE RFQ-STRVCTVRA
NVNCVPATVM, CVIVS OPE FLVMINA PARTICVLARVM DENSA ETSI MICRAE VELOCITATIS SIMVL
PER SPATIVM FOCVSSARI POSSVNT INVENIENS PERMVLTVM VERE AD THEORIAM ARTEMQVE
ACCELERATORES CONSTRVENDI PROMOVENDAM IDEOQVE AD ARCANA NATVRAE
IN DIMENSIONIBVS MINIMIS ELVCIDANDA CONTRIBVIT

PHILOSOPHIAE NATVRALIS DOCTORIS IVS DIGNITATEMQVE
HONORIS CAVSA

DIE V MENSIS MAII ANNO AERAE DIONYSIANAE MCMLXXXIX
VNANIMO CONSENSV CONFERI DECREVIT

Fig.8. Lectures by I. M. Kapchinsky at the University of Maryland and Diploma HONARIS CAUSA from Goethe University Frankfurt.

The most widely discussed in the literature scheme for generating neutrons with an energy of 14 MeV and such a high flux is a combination of a deuteron accelerator for an energy of 30–35 MeV at a continuous beam current of 100 mA and a flowing lithium target. The initial part of the accelerator for the electronuclear method of multiplication of nuclear fuel, which was seriously discussed in those years, requires even higher continuous beam currents - up to 300 mA. High-intensity ion accelerators with a beam current of about 100 mA are necessary for the transmutation of high-level long-lived nuclear power waste, as well as for solving many other important scientific and applied problems (neutron and neutrino generators, generator drivers for radioactive or "exotic" nuclei, etc.).

In addition to difficulties in ensuring the electrical strength of a high-voltage accelerating tube in the traditional scheme, there are others - to reduce the losses of accelerated particles, it is necessary to have a structure with a very high bandwidth, and this leads to an increase in the wavelength of the accelerating field. In four projects of neutron generators proposed in 1977 by Brookhaven and other laboratories in the USA, a wavelength of 6 m was adopted with a corresponding increase in all transverse dimensions of the structure and RF power losses. Due to the enormous cost of building and operating resonators with a diameter of more than 4 m and the open question of injection, these projects were rejected.

The key to solving the problem of accelerating large average currents lay in the invention by V.V. Vladimirsky, and M. Kapchinsky and V.A. Teplyakov of an accelerating structure with spatially homogeneous quadrupole focusing - POQF (SHQF – Spatially Homogeneous Quadrupole Focusing), proposed by them in 1968 [18]. A more general formula for this new phenomenon was later recognized as a discovery [19], for which I.M. Kapchinsky and V.A. Teplyakov was awarded the Lenin Prize (V.V. Vladimirsky had already been awarded the title of Lenin Prize laureate at that time (this award is not awarded twice) for the development of the project and the commissioning of the PS-70 synchrotron at IHEP in 1970).

The first estimates of the four-chamber resonator are given in the preprint by I.M. Kapchinsky and V.A. Teplyakov, published in 1969 at the ITEP [20], and in 1970 these estimates, as well as the limits of a possible decrease in the injection energy, were published by them in the PTE [21, 22].

In 1976 at ITEP I.M. began the development of structures with POKF in connection with the study of the project of a high-current LU for radiation materials science. Such a proposal was first published in the ITEP preprint [23] and at our conference on accelerators in Dubna in 1978 [24]. Around the same time, similar work began at Los Alamos, about which E.M. Pottmeyer made a presentation at the same conference [25]. Both scientific centers independently turned to the four-chamber resonator. The main considerations of the ITEP staff in favor of this solution, rather than a double H-resonator, as at IHEP, were associated with the possibility of significantly simplifying the problems of heat removal from the elements of the accelerating-focusing structure and with simpler measurements of the azimuthal field distribution.

All the above-mentioned advantages and rich possibilities of accelerator sections with POCF were carefully studied at the Los Alamos Laboratory in the USA, where the designation RPQ was introduced for this type of accelerators – Radio Frequency Quadrupole. About the successful launch in 1980 in LANL of its first POP structure – Proof-of-Principle at 640 keV and a pulsed proton beam current of 26 mA at a capture coefficient of 87%. LANL specialists immediately sent a very emotional telegram ("RFQ is alive and well…" etc.) to ITEP Director I.V. Chuvilo and I.M. Kapchinsky.

It must be admitted that the developers at Los Alamos have achieved impressive results. Numerous publications by Los Alamos staff in English on the theory and specific developments of nuclear weapons played a decisive role in spreading interest in these structures throughout the world. At present, many scientific centers in Russia, the USA, Germany, Japan, England, Switzerland (at CERN), France, Italy, Korea, China, India, etc., are developing linear ion accelerators based on NRA structures. Many of these structures are used regularly with the beam; they are integral parts of the largest accelerator facilities in a number of advanced laboratories in the world. The most complete analysis of accelerating structures with POCF (SHQF or RDQ) I.M. produced in his publications [26-28].

The next created under the leadership of I.M. at ITEP, the 3 MeV POKF accelerator was part of the 56 MeV pulsed proton prototype (PPI) under development for a high-current accelerator. The second part of the IPP represents the Alvarez section for an energy of 10 MeV. In the design of this section, Two innovations were changed and studied: the transition to a frequency of 297 MHz - doubled in relation to the section with POQF, which made it possible to study the issue of beam recapture and summation in the longitudinal phase space. The solution of this issue confirmed the practically chosen scheme of the heavy ion LU driver.

Another innovation proposed by I.M. at ITEP is the use of small quadrupole lenses with permanent magnets in drift tubes [29]. It is they that make it possible to manufacture drift tubes for subsequent sections of the Alvarez type at such a high (~300 MHz) operating frequency. In addition to improving the reliability of the LT, the obvious benefits of energy savings, the elimination of expensive and complex current stabilizers and the lens cooling system, permanent magnets have much greater radiation resistance than electromagnets with conventional winding insulation. V.S. Skachkov invented and developed at ITEP a completely new type of magnetic lenses [30]. Quadrupole rare-earth (REM) rod lenses of this type have a gradient of over 6 kG/cm with a drift tube aperture diameter of 20 mm. Studies have been carried out on the characteristics of various designs and technologies for manufacturing lenses with a pre-set or adjustable value of the magnetic field gradient, and the stability of this value over time. All these issues, as well as the choice of rare earth materials and other problems of introducing these new elements of accelerator technology, I.M. paid great attention. Over the following years, many of our publications were devoted to them (see, for example, [31, 32]), several inventions were formalized, extremely strong REM dipoles were created, which shows the pioneering nature of the developments of the department.

ITEP produced the first batch of REM lenses of 35 pieces, which were packed at NIIEFA into drift tubes of the first resonator LU ISTRA-36. This LU can be used in the ELYANG scheme, a high-power electronuclear neutron generator [33, 34], the creation of which, according to the completed project, in a new room next to the reactor was stopped due to the termination of funding.

Proposed by I.M. how the operating frequency of the Alvarez section of 297 MHz, which is optimal for LU, represents a new frequency range for domestic technology, since there are no megawatt power generators, or even suitable lamps for this frequency the industry does not release the frequency. Under the leadership of B.I. Polyakov (who by that time had transferred to ITEP from MRTI) and A.M. Raskopin in the department carried out, one might say, a daring attempt - to make the GI-27A1 lamp, designed, as already indicated, for a frequency of 150 MHz, work in a specially designed cavity resonator at a frequency of about 300 MHz. This work [35], carried out together with NPO Svetlana, was crowned with success, and now we have generators with a power of about 3 MW, which provides RF power to the adopted RL circuit.

At present, the development of high-intensity proton LAs is being actively continued in a number of scientific centers of the world. The main goal is to create an accelerator-driver for large physical installations. Apparently, in solving the global problems of mankind - improving the safety of nuclear power plants, mastering the thorium fuel cycle and transmuting long-lived waste from nuclear reactors - it is impossible to do without the use of such accelerators with missile defense structures. Numerous publications of works initiated by IM., in our department and in the reactor department of the Institute, are devoted to various aspects of these problems [36-40].

Contribution of I.M. in modern accelerator science is noted, in particular, by the fact that one of the most prominent American scientists in the field of accelerators, Professor of the University of Maryland M. Reiser devoted his fundamental monograph on the theory and calculation of charged particle beams [41] to two outstanding scientists who died at the same time - our I.M. Kapchinsky and American L.J. Laslett.

Here, many significant topics that were started thanks to the groundwork carried out under the guidance of I.M. (hope others will do this):
• the history of creation and fate of the heavy ion accelerator TIPR, the results of research carried out on it, the characteristics of ion sources of the ME\L/A type;
• studies carried out in the department on the use of carbon absorbers of lost particles and on the use of low-gap accelerating sections with individual RF power supply;

• launch of the FPF section and the results of its operation during the acceleration of helium nuclei at the I-2 LU;
• development and commissioning of the NRW section for the LU injector of the INR meson factory;
• development of work on the ISTRA-36 accelerator for the implementation of the ELYANG project;
• broad international cooperation, topics and results of the work of the department's employees abroad;
• development of new structures for nuclear reactors and a heavy ion accelerator on the problem of inertial thermonuclear fusion, in particular, for the ITEP-TVN project;
• works on the theory of accelerators, general physics works and DR-

Unfortunately, everything in our world has not only its beginning, but also its end. One of the most important achievements of I.M. - the active team of the department of linear accelerators created by him 15 years after the death of its founder gradually fell into decay, and in June 2008, by order of the ITEP director, the department was abolished, and the surviving scientists, highly qualified specialists and workers were transferred to the newly organized structural unit "ITEF Accelerator Complex ", adding to the depleted staff of the proton synchrotron.

It seems to me that if Ilya Mikhailovich were alive, he would develop in the department such work that would be consistent with the main topics of our industry, would find popular areas that lie in the area of the previous activity of our highly qualified team. Moreover, for many years he was a member of the Scientific and Technical Council, his opinions and proposals often contributed to the development of the right decisions of this body for the entire institute.

With the death of Ilya Mikhailovich Kapchinsky, a remarkable person, an internationally recognized scientist, ITEP suffered an irreparable loss. We, his closest employees, who have worked together with him for their best decades, mourn the loss of a leader, friend, teacher, but at the same time, we consider meeting with him and half a lifetime of joint work as happiness, we will carry the bright memory of him until the end of our days.

## ILYA MIKHAILOVICH AND ARTISTIC CREATIVITY

**V.A. Batalin**

In the autumn of 1964, Ilya Mikhailovich invited me to his office. Unexpectedly for me, the conversation went completely off the production topics.

- In January next year, - said I.M., - Academician A.L. Mints. The entire scientific community will congratulate him. We, too, should not be left behind.

I listened to him with increasing astonishment. What does this have to do with me? I had never seen the academician and knew very little about his activities. It turned out that this was only indirectly related to me. It turned out that I.M. decided to give A.L. Mints for the anniversary of his portrait. And, according to I.M., my wife, an artist, Ida Mikhailovna Egorkina, could write it. “But Ida is not a portrait painter,” I tried to object.

“I saw her good, living portraits,” I.M. said in response.

- These are watercolors, rather sketches, not portraits, and therefore alive, not dried. In addition, they are written from nature. Academician is unlikely to agree to pose.
- I have a lot of good and different photographs, - I.M. said. I tried to explain that Ida does not write from photographs, but I.M. said that he would invite her to some event with the participation of Mintz, where she could make some sketches.
- In general, invite her to me tomorrow, and I myself will talk to her, - said I.M.

Charming I.M., of course, persuaded Ida to take on a difficult job for her. When the portrait was ready I.M. looked at him for a long time and critically, and finally said:

- You got a portrait quite similar to A.L., but there is a significant flaw in it. In the portrait, Mintz's gaze is sharp and piercing. In fact, his gaze does not leave any definite impression.

I don’t remember how Ida coped with her sharp eyes, but apparently she coped, because on the anniversary the portrait was presented to Alexander Lvovich. I don't know his reaction either. Some time later I had to meet with A.L. and even get scolded by him. During the launch of the I-2 injector, after a successful acceleration at a low beam current, we decided to gradually increase the proton beam current. And at a fairly high current, the plastic screen, with which we observed beam focusing at the accelerator input, partially evaporated. Plastic fumes caught in the first accelerating gap caused breakdowns, and the launch program was thwarted.

Arriving the next day, A.L. terribly scolded me for a gross, in his opinion, mistake and was indignant at how such unprofessionalism could be allowed. Meanwhile, this screen was quite standard and was used on all accelerators of our institute. And the real blunder was that the beam current density was unusually high. And the screen could not stand it, despite the pulsed nature of the beam. Before the launch, we worked hard to increase the phase density of the beam current and achieved a record value for that time. When focusing on the input of the accelerator, the real beam current density turned out to be unacceptably high for plastic. Listening to A.L.'s scolding, I tried to look into his eyes. They really did not leave a definite impression behind the glasses. A.L. was especially worried about the need to open the casing of the accelerator. It seemed to me that the memory of the sad fate of the first casing weighed over him. Luckily, I didn't have to open the cover. The electrical strength of the gap was restored by simpler means.

Recently (in the summer of 2008) a TV program about the so-called. “Secret Physicists” was dedicated to A.L. Mints. They showed a large number of photographic portraits of A.L. On each of them, his gaze was sharp and piercing and left a very definite impression. Apparently, this is the specificity of photography. led by A. L. There is his bust, such a granite or bronze block. I thought that the pictorial portrait made by Ida would better correspond to his memory. As always, Ilya Mikhailovich looked far ahead.

Another work of art created on the initiative of I. M. is in Protvino. I.M. asked Ida to find a color solution for the hall of the I-100 accelerator. This time, the sketch made by Ida, apparently, was liked by everyone, since it was carried out.

## ILYA MIKHAILOVICH KAPCHINSKY - STROKE TO THE PORTRAIT

**A.M. Kozodaev**

Under the guidance of Ilya Mikhailovich Kapchinsky, I worked at the Institute of Theoretical and Experimental Physics for 35 years. We were with him in different age categories (he was almost 20 years older), so the contacts were more in the field of industrial life. But I was lucky enough to experience its unobtrusive beneficial effect. Communication with him taught me a lot and left a deep imprint in my memory.

1. Arrival at the laboratory

In 1958, being a very young man and working as a laboratory assistant at the Thermal Engineering Laboratory of the USSR Academy of Sciences (so it was then called ITEP), I began to look inside the institute for an occupation closer to the radio engineering that interested me. Someone advised to go to I.M. Kapchinsky (Fig. 1), he is gathering a team to laboratory No. 012 to build a linear proton accelerator, and there will be a lot of radio engineering. Came. A friendly man of short stature with a short haircut is sitting, smoking "Belomor". The conversation turned to the oscilloscope. How does it work, what controls its beam, what adjustments does it have? I gave general answers, realizing the paucity of my knowledge. After a ten-minute conversation, when I, realizing my complete ignorance, was ready to get up and leave, I.M. Kapchinsky, to my deep surprise, suddenly said that I would fit right in. Now I understand that Ilya Mikhailovich did not find out my knowledge (they simply did not exist), but my desire to receive it and work.

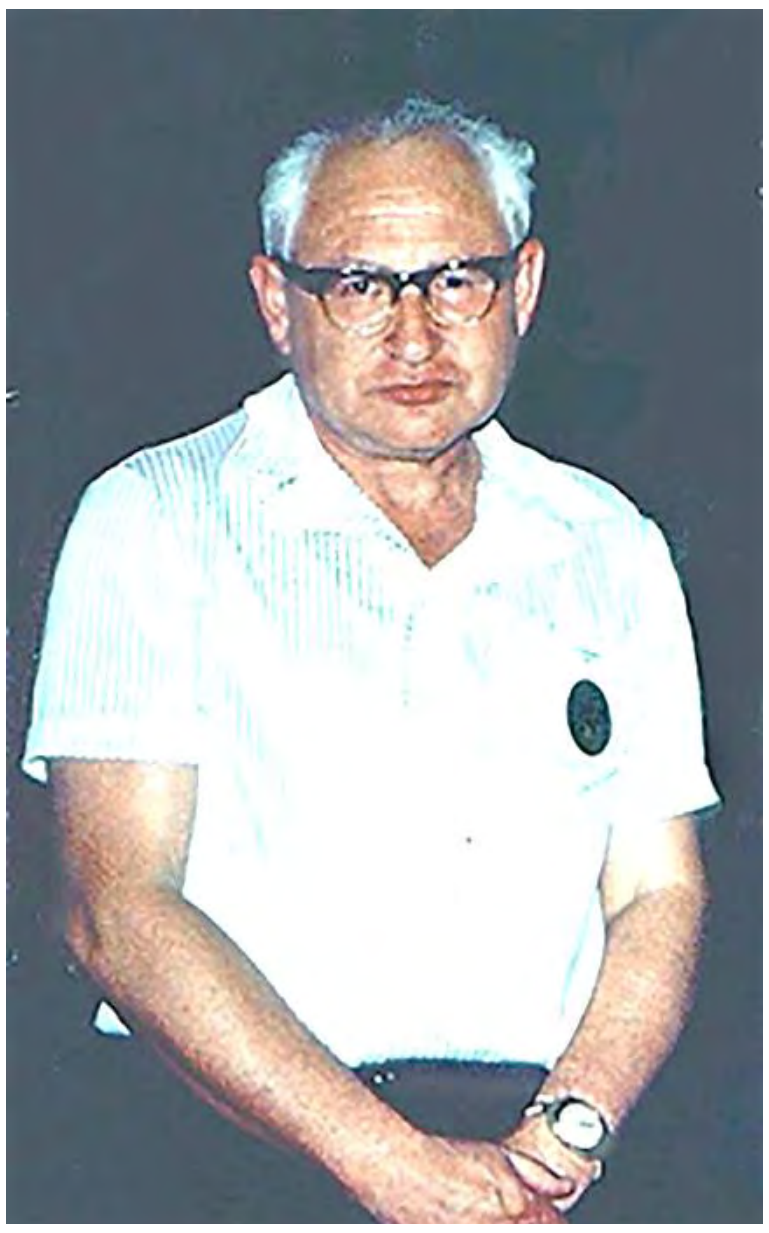

Fig. 1.
I.M. Kapchinsky.

There was a calm and very businesslike atmosphere in the laboratory. Everyone was ready to help me, the youngest and most inexperienced employee. N.V. La-zarev, to whose work I was connected, developed for the I-100 accelerator (which was to be built in Protvino near Serpukhov) quadrupole electromagnetic pulsed lenses and the ideology of their pulsed power supply. I had to deal with the manufacture of lenses, their power stand, and magnetic measurements. Once, when discussing the results obtained, Ilya Mikhailovich said: "Well, here you are, Sasha, and prepare a scientific and technical report on the work and the results obtained". I felt unprepared for such responsible work, and before N.V. Lazarev was uncomfortable, because everything was done under his leadership. When preparing and conducting magnetic measurements, it was necessary to move from general physical concepts to specific measured quantities. Ilya Mikhailovich quickly sketched out the relationship between magnetic induction, flux, electromotive force on two sheets of paper and showed how to obtain specific expressions for the number of turns of measuring harmonic coils, their sensitivity, expected voltage signals, etc. I was surprised and, I would say, inspired by the ability of Ilya Mikhailovich to turn abstract, intangible ideas into a tool for practical activity. The above report was prepared, but it required a fair amount of effort from me to penetrate into some sections of physics and mathematics, calculate, design and manufacture specific units, take measurements, analyze them, and formulate conclusions. I constantly received help from I.M. Kapchinsky, N.V. Lazarev, E.N. Daniltsev, S.V. Skachkov. After finishing the work, I myself felt that I had risen significantly above my original level. Later it became clear that this was one of the ways of Kapchinsky education of young specialists - to entrust them with work that requires more knowledge, more qualifications than those possessed by a young man.

Approximately at the same time, during the construction of the U-7 proton synchrotron at ITEP, magnetic measurements of ring magnets were carried out. Ilya Mikhailovich called me and said something like this: “A unique U-7 ring accelerator is being built at the institute, it will allow us to take a new step in understanding the structure of matter, etc. But there are difficulties. Temporarily, a specialist like you is needed to participate in round-clock magnetic measurements. Otherwise, everything will stop. I ask you to help”. After such words, it was impossible to refuse to jerk the measuring coil in the interpolar space of the magnets and record the readings of the device during night shifts, since it turned out that this work was the most important, but without me – well, nothing! This was another technique used by Ilya Mikhailovich when assigning uninteresting, routine, but necessary work to employees. On this subject, he sometimes recalled the parable about the construction of the Dome Cathedral in Riga, when a passerby asked one worker: “What are you doing?” – and received the answer: “Don't you see? I carry stones”. Another worker doing the same job replied: “Don’t you see? Building the Dome Cathedral!”

2. Launch of the I-2 accelerator

Ilya Mikhailovich held in his hands all the threads of control over the construction of the linear proton accelerator I-2 for an energy of 24.6 MeV. The accelerator included two cylindrical resonators operating at a frequency of 148.5 MHz and loaded with drift tubes with quadrupole electromagnetic lenses. A building for the accelerator was being built, equipment was being delivered and installed. Designers and specialists of related organizations often visited the site. As the time of launch approached, work in the I-2 building (Fig. 2) expanded.

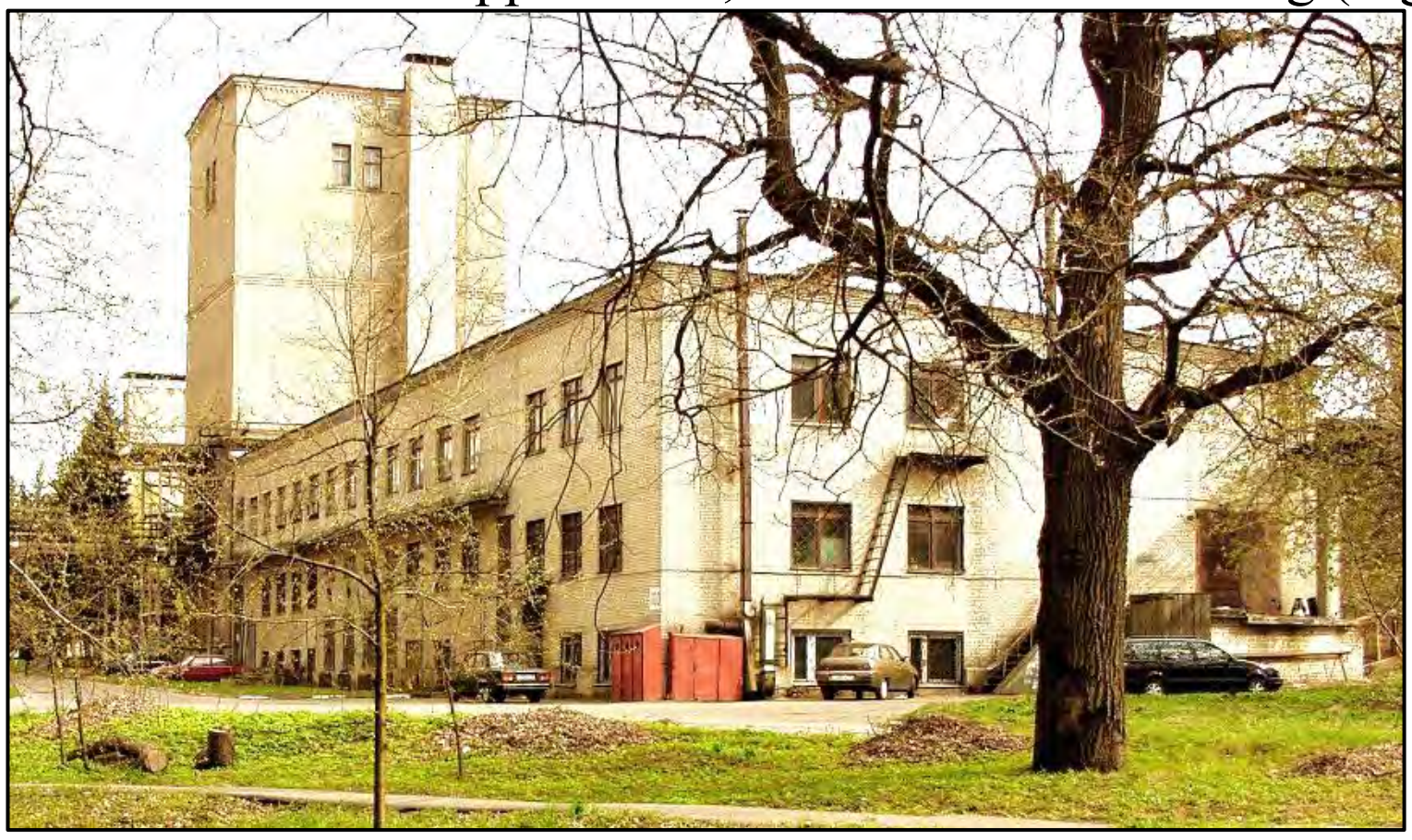

Fig. 2. Accelerator building I-2.

Mechanical installation of resonators and vacuum pumps, magnetic measurements of lenses, alignment, installation of high-frequency generators, adjustment of high-voltage equipment of the preinjector and ion source, power supply and cooling systems. All laboratory staff, together with colleagues from other institutes, tried to bring the moment of obtaining an accelerated beam closer.

Shortly before the start of adjustment work, Ilya Mikhailovich gave a course of lectures on the physics of particle acceleration to the leading staff of the laboratory. At this time, he was preparing his first book, Particle Dynamics in Linear Resonant Accelerators, which was published by Atomizdat in 1966. The desk reference document was a preprint by V.K. Plotnikov and I. M. Kapchinsky with the physical parameters of the I-2 accelerator. Weekly operational meetings were introduced, with the help of which Ilya Mikhailovich managed the preparations for launching the accelerator.

A dramatic event occurred during the pumping out of the first version of the vacuum casing, made of aluminum alloy. The casing could not withstand the pressure and crumpled. The instantaneous pressure difference was such that, being at that moment in a narrow passage through the concrete protection, I felt a strong air shock. Fortunately, there was no resonator in the casing. A huge black cloud hung over the head of the laboratory I.M. Kapchinsky and vacuum engineer V.V. Koloskov. K.N. Meshcheryakov – head of the Main Directorate of the State Committee for Atomic Energy – and V.M. Lupulov – the developer of the casing – accused them of "illiterate" pumping, although it soon became clear: the casing had not been calculated for stability. Deputy ITEP Director V.V. Vladimirsky took, to his credit, a resolute position in defense of the laboratory, and there were no serious consequences for people.

We returned to start-up work after the manufacture of another casing, made of stainless steel (Fig. 3). The physical launch took place calmly, without nervousness. It was November 2, 1966, on the nose of November 7 – the holiday of the October

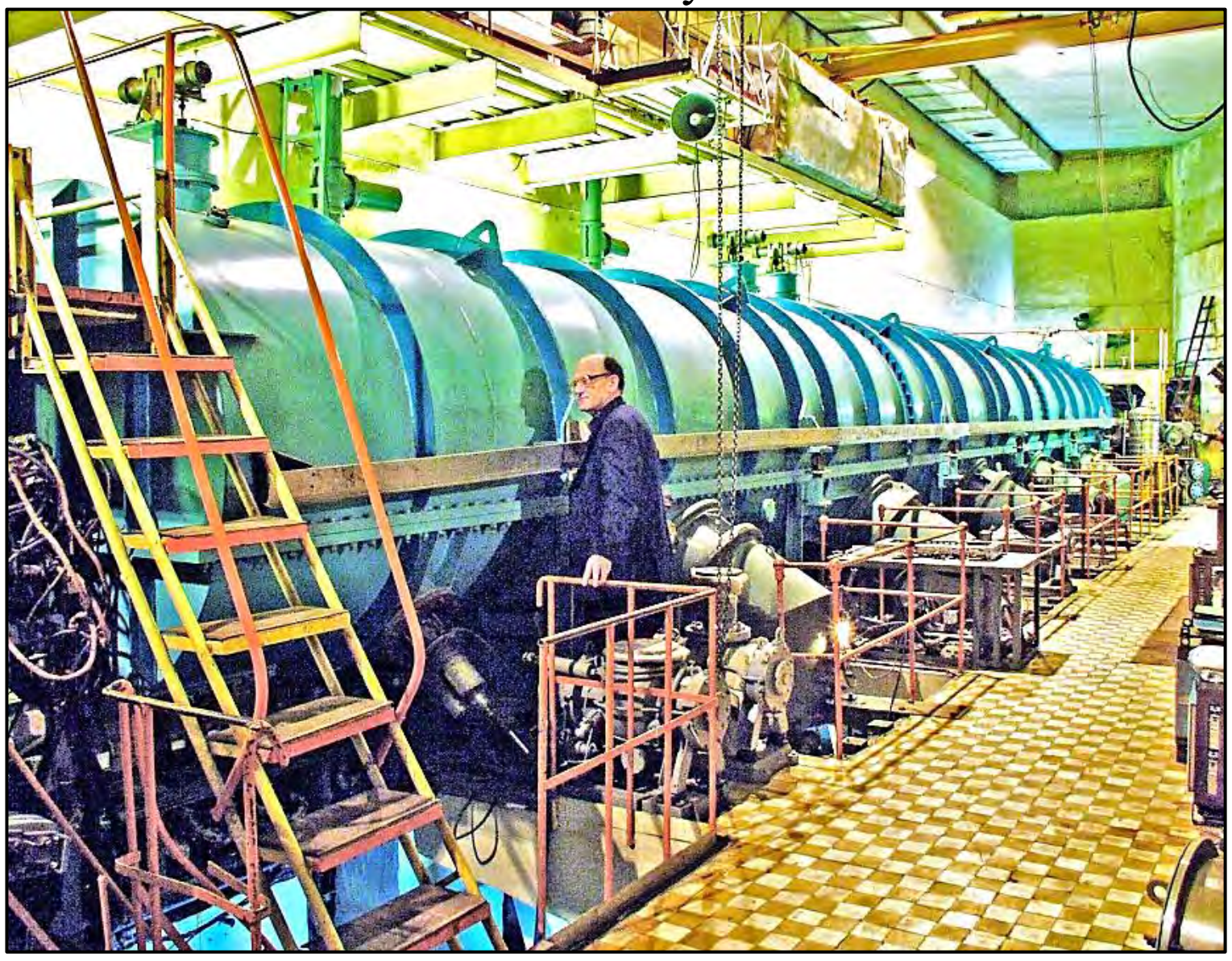

Fig. 3. Accelerator I-2.

Revolution. According to established tradition, a labor victory was needed. Everyone went about their business, my place was at the screen of the IO-4 oscilloscope, which recorded the beam current. Sometimes Ilya Mikhailovich walked behind the console, but he didn't pull anyone, he didn't interfere with anyone. The first signal on the IO-4 screen corresponded to the accelerated proton beam current of 7 mA. Immediately after that, something was optimized, tweaked, of course, without real measurements and analysis. At the end of the day, 30 mA was received (and after a few years of improvement, a record of 230 mA). Late in the evening, having played enough, we decided to stop there. Everyone was happy, N.V. Lazarev pulled out a bottle of alcohol from somewhere. But then it turned out that at the output of the second Alvarez resonator, the lens of the half-tube No. 235 related to the output ion conduit was switched on. Thus, the beam parameters obtained at the output could not be explained only by the action of the accelerator channel. For this reason, Ilya Mikhailovich flatly refused to participate in the celebration: in his understanding, physically the launch was not carried out cleanly, and there is nothing to celebrate yet. V.A. Batalin found a way out of this situation. He proposed to assume that not only the accelerator was launched, but also part of the ion pipeline. This eased the tension somewhat.

Reporting on the launch of the accelerator, the Pravda newspaper of November 5, 1966 wrote: "A good gift was made for the 49th anniversary of October". Information about the scientific and technical side of this event was placed in the journal "Instruments and Experimental Techniques" No. 5 for 1967, where 12 articles about the I-2 accelerator were printed at once. Yes, this was the first significant success of Ilya Mikhailovich at ITEP as a leader of the accelerator physicist and head of work, the success of the team he created.

## 3. Construction of the I-100 accelerator

Almost simultaneously with the construction of I-2, the leading specialists of the laboratory under the leadership of I.M. Kapchinsky participated in the creation of the I-100 – a linear proton accelerator with an energy of 100 MeV, which then served for a long time as an injector of the proton synchrotron at IHEP (Protvino).

Every week, Ilya Mikhailovich spent 1-2 days in Protvino and supervised the construction of the I-100 accelerator. Sometimes we, ITEP employees, also traveled with him. The I-100 accelerator is much larger than the I-2, both in terms of energy, and in length, and in terms of the amount of work. From the outside, it was clear how Ilya Mikhailovich, having arrived from the train, immediately screwed into the physical and technical problems of the construction of the I-100, laying out the tasks entangled in a ball on the shelves and placing them in order.

For the creation of the I-100 accelerator I.M. Kapchinsky, N.V. Lazarev and V.K. Plotnikov was awarded the USSR State Prize in 1970.

## 4. Improvement of the I-2 accelerator

During the trial operation of the I-2 accelerator at the end of 1966 and in 1967, as well as during the first months of its work at the proton synchrotron in 1967-1968, rich information was obtained on the physics and technology of the accelerator. In parallel with the operation of the machine, a process of its improvement was going on for several decades. Ilya Mikhailovich kept his finger on the pulse of this activity, while simultaneously conducting search work. In the style of I.M. Kapchinsky was to provide creative freedom to employees to solve specific problems. And if someone had a nonstandard proposal, he always supported it. Even if this suggestion might not be immediately useful. Without being involved in specific engineering matters himself, he easily entered into the details of ongoing developments and was often able to give good advice.

I remember, in 1969, we were engaged in reducing the rapid changes in the magnetic field of the analyzing magnet associated with the instability of the power supply voltage. The filtering of the supply voltage turned out to be very cumbersome and ineffective, since the inductance of the magnet and Foucault currents in its core repeatedly weakened the rapid instability of the field. Further reduction of the field instability required another way. We started developing active field noise suppression circuits. Once in the dining room at dinner, I shared this with Ilya Mikhailovich, not expecting to greatly interest him. However, he asked very carefully, pointed out a few significant points and said: “Evaluate everything on paper first. It costs nothing to correct a mistake on paper, and correcting what is done in hardware and getinax is expensive”. Estimates were made, magnetic field noise was effectively suppressed, and the principle “first evaluate on paper, and only then do” sunk into the head.

Another time, in the same canteen at dinner, the conversation turned to beam emittance and Liouville's theorem. I could not understand why the beam emittance could not be reduced. Ilya Mikhailovich, finishing the compote from a glass, said: “Imagine that metal balls are poured into this glass. There is free space between them, but it cannot be used, because the balls are solid and do not compress”.

Much attention to I.M. Kapchinsky paid attention to the completeness of the work, especially engineering and technical, which should ensure the reliable operation of the installations. Looking through the reports and documents of those years, I see with what completeness and thoroughness they were prepared. For each completed development, a report or description was issued with schematic diagrams, operating modes, test results, connection diagrams, operating instructions, a list of possible malfunctions, engineering and technical, which should ensure the reliable operation of the installations. Looking through the reports and documents of those years, I see with what completeness and thoroughness they were prepared.

Once Ilya Mikhailovich called me and after talking about current affairs he said: “Sasha, I categorically…”. His words were interrupted for a second, I thought feverishly what I had done, and he continued: "… advise you to write a dissertation on the basis of what has already been done”. It was believed that the dissertation should summarize what was already meaningful and done, and not be a fantasy on paper. In the laboratory, and then in the department, writing a dissertation was never an end in itself. New employees were told that they needed to work, delve into the problem, get real positive results, and then prepare a thesis.

Later, in 1985, there was a need to modernize the high voltage stabilization system on the forming lines of powerful pulse modulators of the I-2 RF power paths. Ilya Mikhailovich called me and in a very serious voice said something like this: “Sasha, you have an insanely lot of work, you need to rest. I-2 modulators are worn out and give an unacceptably long downtime. No one is able to modernize them in a short time, except for you and your fellows”. Such a disarming reception on the part of I.M. Kapchinsky was sometimes experienced by other employees. Work has begun. Through the efforts of Yu.B. Stasevich, A.R. Course, V.G. Kuzmichev, E.M. Kuznetsov and A.Yu. Lukashin, these systems were redesigned using the most modern solutions at that time and continue to serve.

I-2 is the first linac with hard focusing in Eurasia, which for some time was the world record holder in terms of accelerated current - 230 mA, on which for the first time joint-simultaneous operation on 3 beam extraction channels was mastered and a hard magnetic quadrupole was used in the particle focusing channel. I.M. Kap-chinsky was the chief designer of the unique physical installation, the developer, the head of the construction, operation and modernization. The I-2 accelerator, which has been successfully operating for 50 years, which is unprecedented in itself, can be safely called one of the monuments to the creative genius of Ilya Mikhailovich.

5. Collective acceleration methods

At the end of the sixties of the last century, interest arose in the accelerator world in the acceleration of electron rings "stuffed" with protons (collective acceleration method - CAM). Work began to develop in the Lawrence Laboratory at Berkeley (E.J. Lofgren, D. Keefe, A.W. Sessler) and at JINR (V.P. Sarantsev). It was expected that with the help of CAM it is possible to obtain high-intensity proton beams at an extremely high acceleration rate. The idea was accepted by I.M. Kap-chinsky. In April 1974, Laboratory No. 012 was transformed into Department No. 120 headed by I.M. Kapchinsky. As part of the department, laboratory No. 122 was created (headed by V.K. Plotnikov) with the task of conducting development and research on CAM.

At the founding meeting on October 1, 1970, I.M. Kapchinsky set before us the general (for the future) task of constructing an installation that includes a linear electron accelerator with an energy of 4-5 MeV and a pulsed current of 3-4 kA; a compressor in which the diameter of the ring should decrease by about an order of magnitude; a manipulator where adiabatic compression and saturation of the ring with protons should take place; a ring pipeline for the acceleration and focusing of the rings and a terminal plug for "shaking out" the accelerated protons. The output energy of protons was called 1 GeV, the average current was 10 mA.

Work boiled over. The calculation and physical part was conducted by I.M. Kap-chinsky and V.K. Plotnikov. At the institute, a preliminary version of the linear induction electron accelerator LIA-1.5 for an energy of 1.5 MeV was created. The main version of the LIA-5/5000 accelerator (5 MeV, 5000 A) was manufactured at the Scientific Research Institute of Electro-Physical Equipment (Leningrad). The department developed a number of other original units for the installation of the CAM.

It was clear that it would be necessary to switch large pulse powers in the range of durations from nano- to milliseconds. The limiting capabilities of powerful valve switching devices were clarified: thyratrons, ignitrons, tacitrons, thyristors, etc. I have prepared a large overview of the ultimate capabilities of these devices. I.M. Kapchinsky and N.V. Lazarev proposed to publish it in the form of a book in Atomizdat. To my doubts about the usefulness of such a publication, Ilya Mikhailovich noted that it would certainly be useful, since it would save specialists from a lot of rough work to find information and comparison of devices, he himself called the director of Atomizdat and agreed on the preparation of the publication. Here his understanding of the dependence of the success of a business not only on the course of the main direction, but also on the state of auxiliary affairs, was manifested.

Quite a lot of reports on the CAM were presented at the conference of those years. But their tone gradually passed from iridescent to indefinite. There were physical obstacles. At that time, I.M. Kapchinsky asked me to present considerations on the maximum technically achievable stability of currents and voltages, both direct and pulsed. I presented. It seemed to me that I was closer to real technology. However, Ilya Mikhailovich once again surprised me with his desire to penetrate the essence of the issue. Having found out that most of the signals in stabilization systems are based on measuring voltages, he rightly noted that a lot is determined by the achievable accuracy of these measurements.

Analyzing the physical processes of acceleration and realistically possible accuracy and stability of the processes of formation of fields, I.M. Kapchinsky came to the conclusion that CAM had no prospects. I think one should pay tribute to the courage of Ilya Mikhailovich, who himself presented his views to the director of the institute I.V. Chuvilo, who previously worked at JINR and believed in the success of CAM. The director took this with undisguised displeasure, but the work began to wane.

## 6. Accelerating RFQ structure

Proposed by V.V. Vladimirsky, I.M. Kapchinsky and V.A. Teplyakov, the accelerating structure with high-frequency spatially homogeneous quadrupole focusing (English name Radio Frequency Quadrupole – RFQ) made a revolution in the physics and technology of ion acceleration at low particle velocities and received wide distribution and general recognition in the accelerator laboratories of the world (Fig. 4).

The proposal was first publicly announced in August 1969 at the 7th International Conference on High-Energy Charged Particle Accelerators in Tsakhkadzor (near Yerevan) and, as it seemed to me, did not arouse much interest among specialists at that time. In subsequent publications, I.M. Kapchinsky and V.A. Teplyakov revealed the advantages of the RFQ structure (a significant decrease in the injection energy, an increase in the coefficient such as particle capture and beam intensity, simplicity of design, small transverse dimensions), but there was no social order for the implementation of the structure yet.

The situation changed when interest in the possibility of accelerating high-intensity proton (and deuteron) beams to generate powerful neutron fluxes and electronuclear fuel production increased sharply. In 1976, from a conference on proton linear accelerators (Canada, Chalk River), N.V. Lazarev brought informa-

Union of Soviet Socialist Republics | **DESCRIPTION of the invention to the copyright certificate** | **265312**

The authors of the invention **V.V. Vladimirsky, I.M. Kapchinsky and V.A. Teplyakov**

**ION LINEAR ACCELERATOR**

**Subject**

1. Ion linear accelerator containing a quadrupole focusing and accelerating system, characterized in that, in order to reduce the permissible lower level of injection energy, the focusing and accelerating systems are made in the form of a four-wire long line, in which the distance between two wires lying in the same plane with the axis of the accelerator varies along the length of the line as follows, that, where the distance between the wires of one pair is minimal, between the wires of the other is the maximum.

2. Accelerator as item 1, which differs in that the resonator is made in the form of a cylinder, divided by radial partitions into four identical chambers.

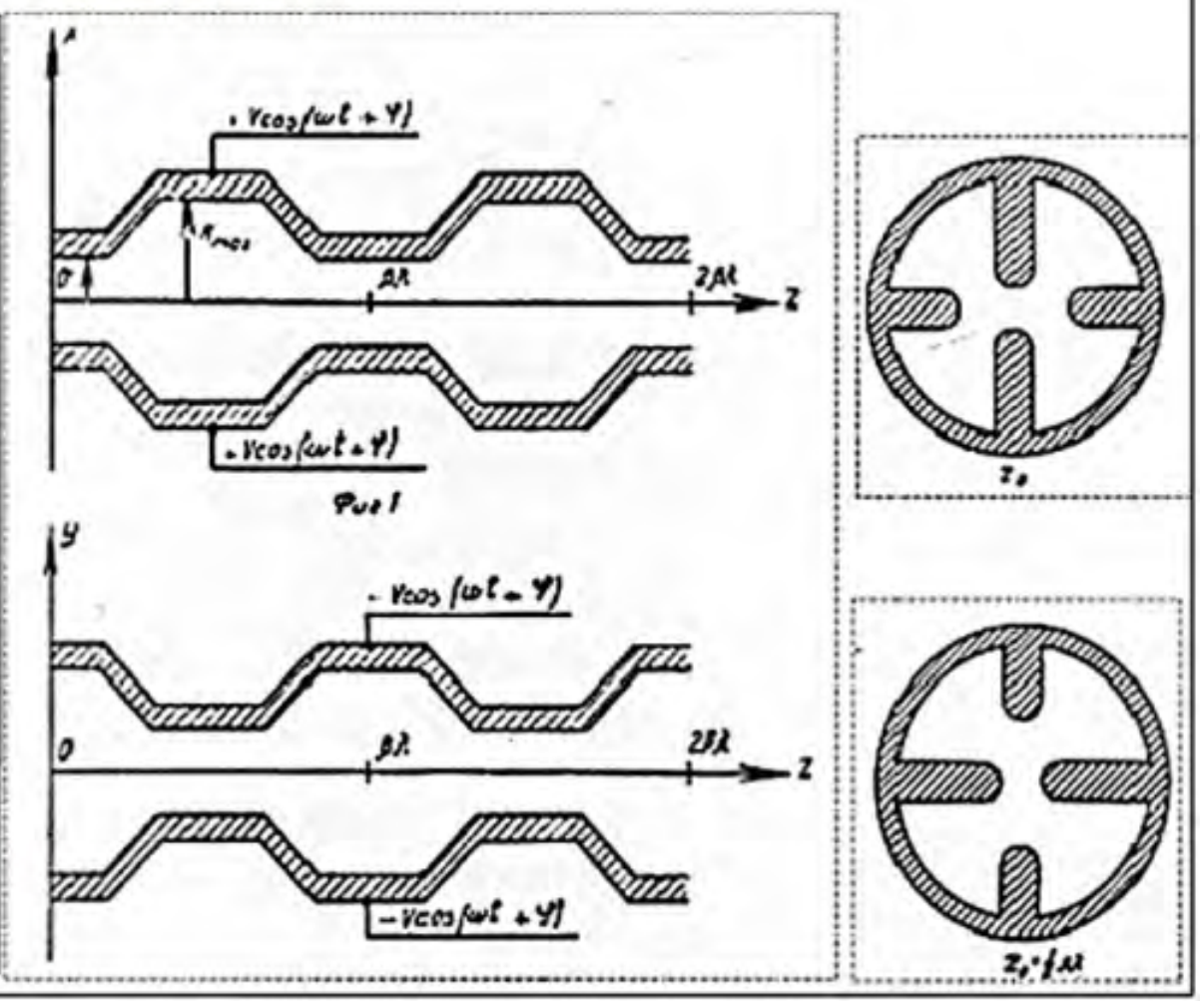

Fig. 4. Invention that revolutionized the technique of acceleration
at low particle velocities.

tion not only about the intensification of research work in the USA and Canada on the acceleration of average beam currents of 100-300 mA to an energy of ~ 1 GeV, but also information about the difficulties of a fundamental nature encountered in the Chalk River when trying to accelerate a continuous high-intensity beam to energy 800 keV. It was then that it became clear that only fhe RFQ structure could open the way to the acceleration of high-current beams! Its study and implementation began to engage in many laboratories around the world. In the USSR, such work was successfully carried out at ITEP and IHEP. The authors of the structure I.M. Kapchinsky and V.A. Teplyakov was awarded the Lenin Prize of the USSR for 1988.

## 7. Pulsed proton prototype of a high-current linear accelerator

Ilya Mikhailovich began to focus the forces of the department on solving the problems of developing a high-current accelerator. The structure of a high-current (for a beam current of 100-300 mA and an energy of 1 GeV) linear accelerator of protons (or deuterons) loomed, which, according to I.M. Kapchinsky, should consist of an electrostatic injector for an energy of 70-100 keV, the RFQ section for an output energy of about 3 MeV, Alvarez resonators up to an energy of 100- 150 MeV, and a system with washers and diaphragms proposed by V.G. Andreev, for higher energy.

It was decided to build the focusing channel of the Alvarez resonators using permanent magnets. On the initiative of I.M. Kapchinsky, the development of hard magnetic quadrupoles has already begun. By that time, the industry had mastered the production of permanent magnets with sufficiently strong and time-stable fields. Hard magnetic quadrupoles in the beam acceleration and transport channels could remove a number of technical problems in the development of high-power accelerators. To carry out developments and research in this new area for the department, a persistent specialist who could not back down in the face of difficulties was needed. I.M. Kapchinsky and N.V. Lazarev decided to involve V.S. Skachkov, especially since his father, S.V. Skachkov was an experienced magnetist and Vladimir Sergeevich could widely use his advice. The calculation was completely justified! V. Skachkov became perhaps the most knowledgeable specialist in Russia in the field of building individual elements and entire beam transport paths using hard magnetic materials. As a result, drift tubes with hard magnetic quadrupoles open into vacuum were created for the first time (Fig. 5). At the time when V.S. Skachkov switched to this direction, I jokingly told I.M. Kap-chinsky: “It will be difficult for me without Volodya's head”. And he, also jokingly, replied: “Grow your own”.

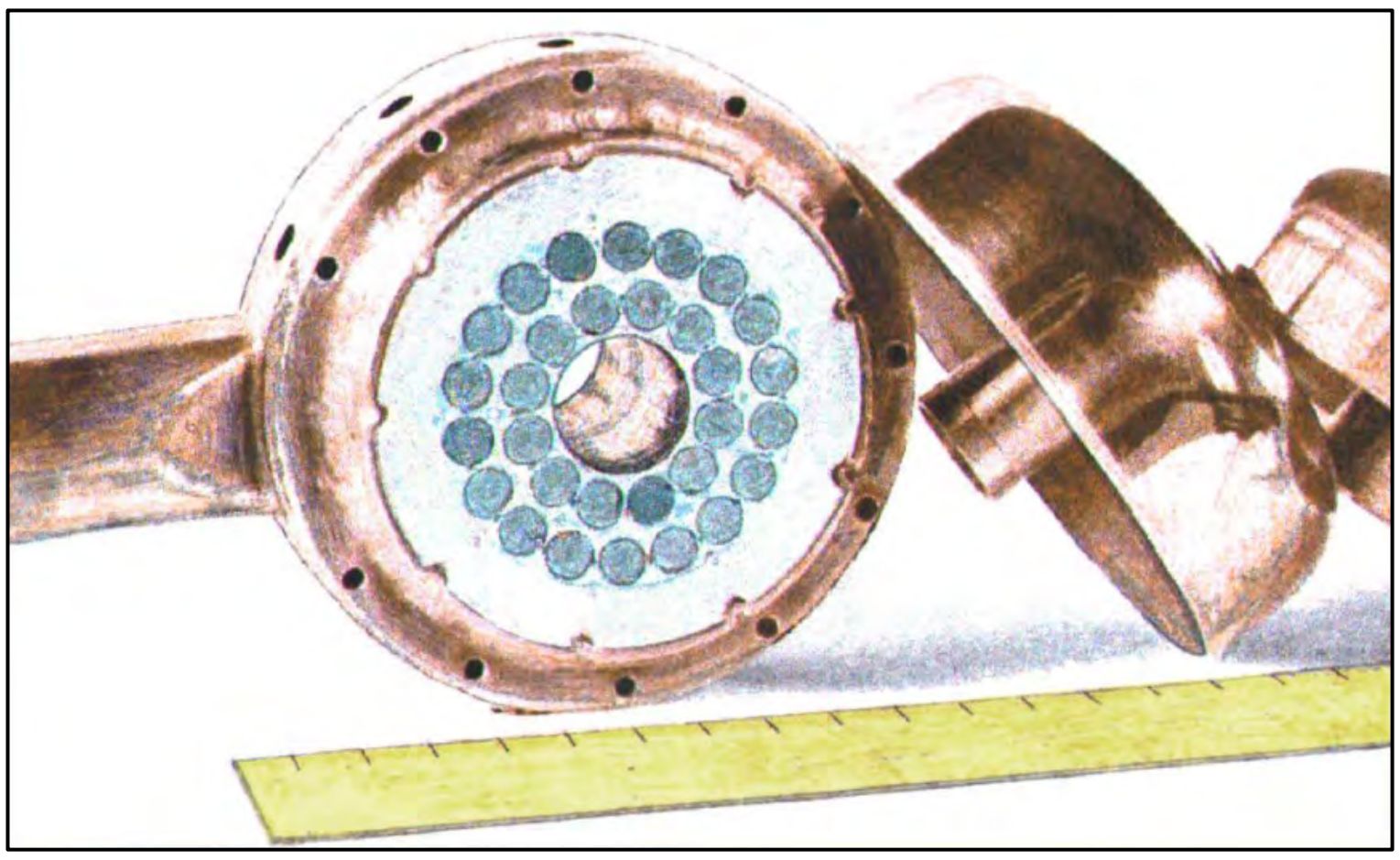

Fig. 5. Drift tube with a solid magnetic quadrupole.

The wavelength of the HF field in the resonators providing the main acceleration was chosen to be half as long as in the RFQ section. The features of the accelerator structure promised great advantages in accelerating strong beam currents, but at the same time Ilya Mikhailovich pointed out the importance of solving a number of not only physical, but also technical problems:

- creation of an electrostatic injector providing a beam of the required intensity;
- addressing the issue of lowering the level of radiation due to loss of particles;
- receipt and input into the resonators of huge RF powers, heat removal.

It was necessary to develop work on experimental confirmation of the effectiveness of the proposed solutions. In 1979 I.M. Kapchinsky prepared a physical justification for the design of a linear accelerator, which was based on the approaches mentioned above. The project pursued two goals: the construction of a new (instead of I-2) injector for the ITEP proton synchrotron and the study of this linac as a pulsed proton prototype of a high-current accelerator with a continuous beam. The prototype was supposed to consist of an initial part (the RFQ for an energy of 3 MeV, an RF field frequency of 148.5 MHz), a matching channel and one Alvarez resonator with an output energy of 10 MeV (frequency 297 MHz, in the focusing channel - magnetically hard quadrupoles). The design current of the beam in a pulse is 150-200 mA.

On January 1, 1980, as part of Department No. 120, Laboratory No. 123 was organized with the task of developing and building the prototype. I was assigned to lead it. Large strokes of the task were placed for all years from 1980 to 1985. In January 1980, Ilya Mikhailovich called us together and said that there was a great interest in the prototype from several directions. Based on its principles, it is possible to create a deuteron accelerator (100 mA, 35 MeV) for a high-intensity neutron generator, in which the director of the Institute I.V. Chuvilo is interested. The same block diagram of the accelerator is convenient for electronuclear breeding (300 mA, 1 GeV), which is of interest to the staff of the Reactor Department headed by Deputy Director V.V. Vladimirsky. Similar solutions can be used in the acceleration of heavy ions for work on pulsed thermonuclear fusion (average current 10 mA, 10 GeV/nucleus), which is in the area of interest of another deputy director V.G. Shevchenko. Therefore, work on the prototype must be clearly organized and carried out as intensively as possible.

The preparation of the prototype was not the only concern of the department of linear accelerators, which by that time had become one of the leaders in the institute. I.M. Kapchinsky also sought to advance in the areas of acceleration of heavy ions and high-current continuous proton beams. To work in these areas, a laboratory was formed headed by V.V. Kushin.

8. Preparation and launch of RFQ-1

On February 14, 1980, I submitted I.M. Kapchinsky, a network schedule for the creation of the first version of the initial part of the prototype (RFQ-1), which assumed the receipt of a beam accelerated to an energy of 3 MeV at the beginning of 1982 (which was realized!). And then it turned out that Ilya Mikhailovich, as one of the authors of RFQ, had just received a telegram from Los Alamos with congratulations on the occasion of a brilliant confirmation of the structure's efficiency! It turned out that in LANL they quickly created and carried out a physical launch of the RFQ section 1.15 m long at a frequency of 425 MHz. A pro-ton beam with a pulsed current of 15 mA was accelerated from an energy of 100 keV to 650 keV. The news was both happy and sad. The championship in the implementation of the bladed version of the RFQ was intercepted by the Americans. Some consolation was the fact that our accelerator was much more “serious”: protons had to be accelerated from 88 keV to 3 MeV at an order of magnitude higher beam current.

The RFQ-1 channel was formed by copper modulated electrodes of a four-wire line fed from a four-chamber high-frequency resonator (Fig. 6). The resonator was divided into 8 sections, had a total length of 4.7 m and was placed in a vacuum casing. The design and manufacture of the resonator was carried out under the direction of R.M. Vengrov.

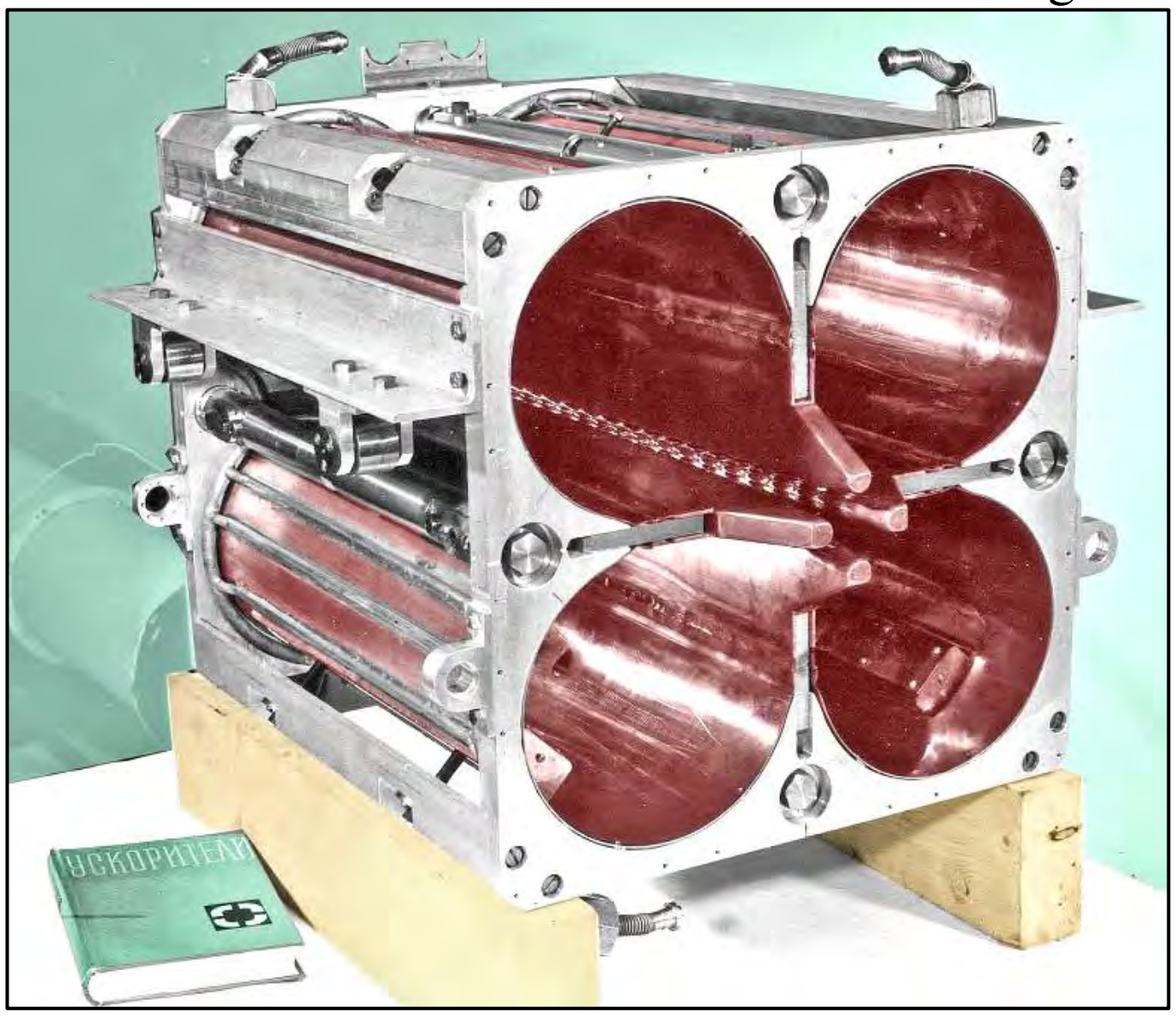

Fig. 6. Entrance section RFQ-1.

For the first time, an accelerated beam at the output of RFQ-1 (Fig. 7) was obtained on January 18, 1982, exactly at the time as planned. The fact of acceleration was recorded immediately after the HF field level rose to the calculated threshold value, and no extravagant surprise (unfortunately or fortunately?) did not occur.

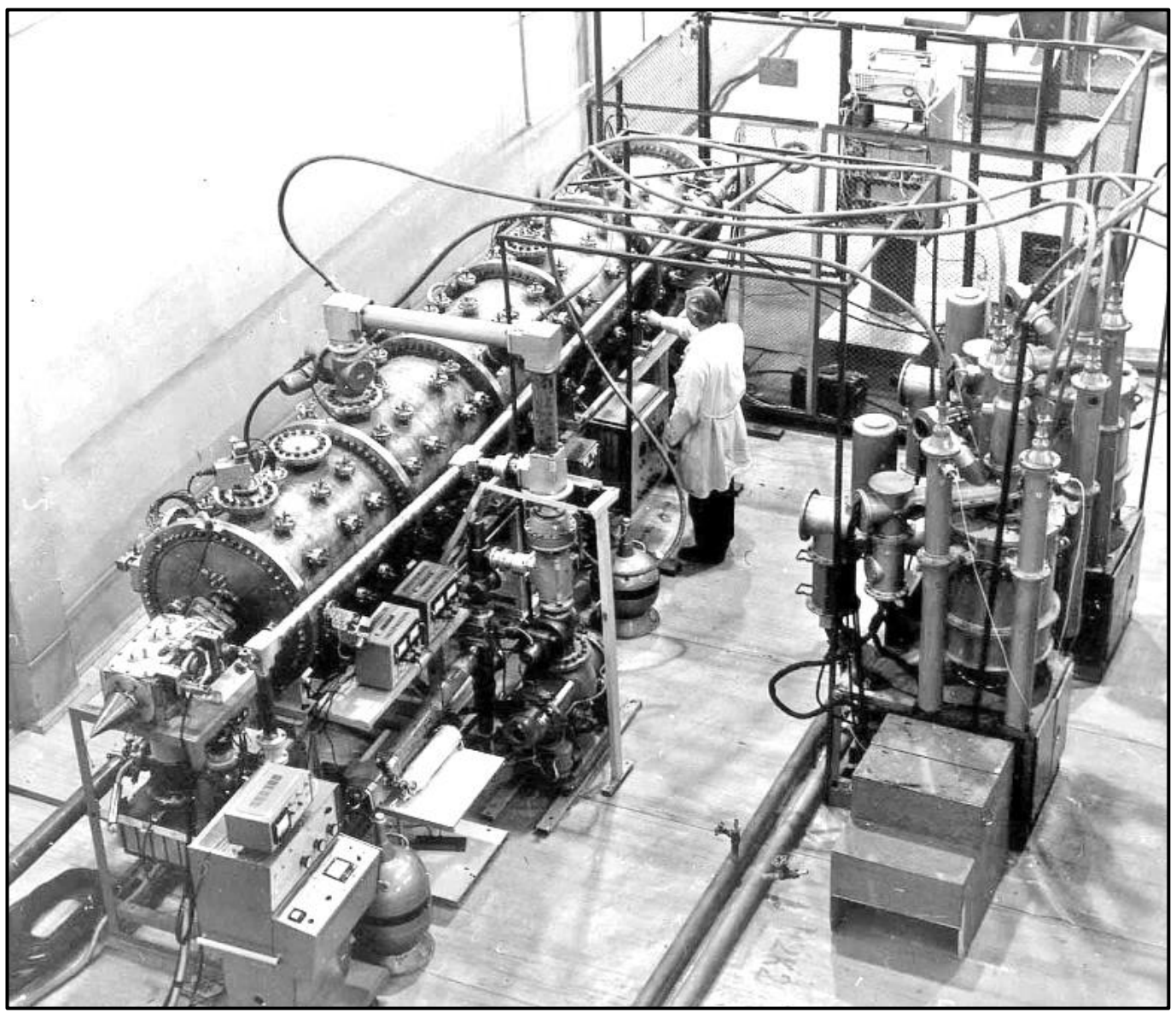

Fig. 7. Accelerator RFQ-1.
The first experimental dependences obtained at the facility confirmed the physical picture of the acceleration processes. The current of the proton beam, whose energy did not exceed 3 MeV, first increased with an increase in the HF field strength and then dropped almost to zero; the current of particles accelerated to an energy of 3 MeV, starting from the threshold value of the HF field, steadily increased (Fig. 8).

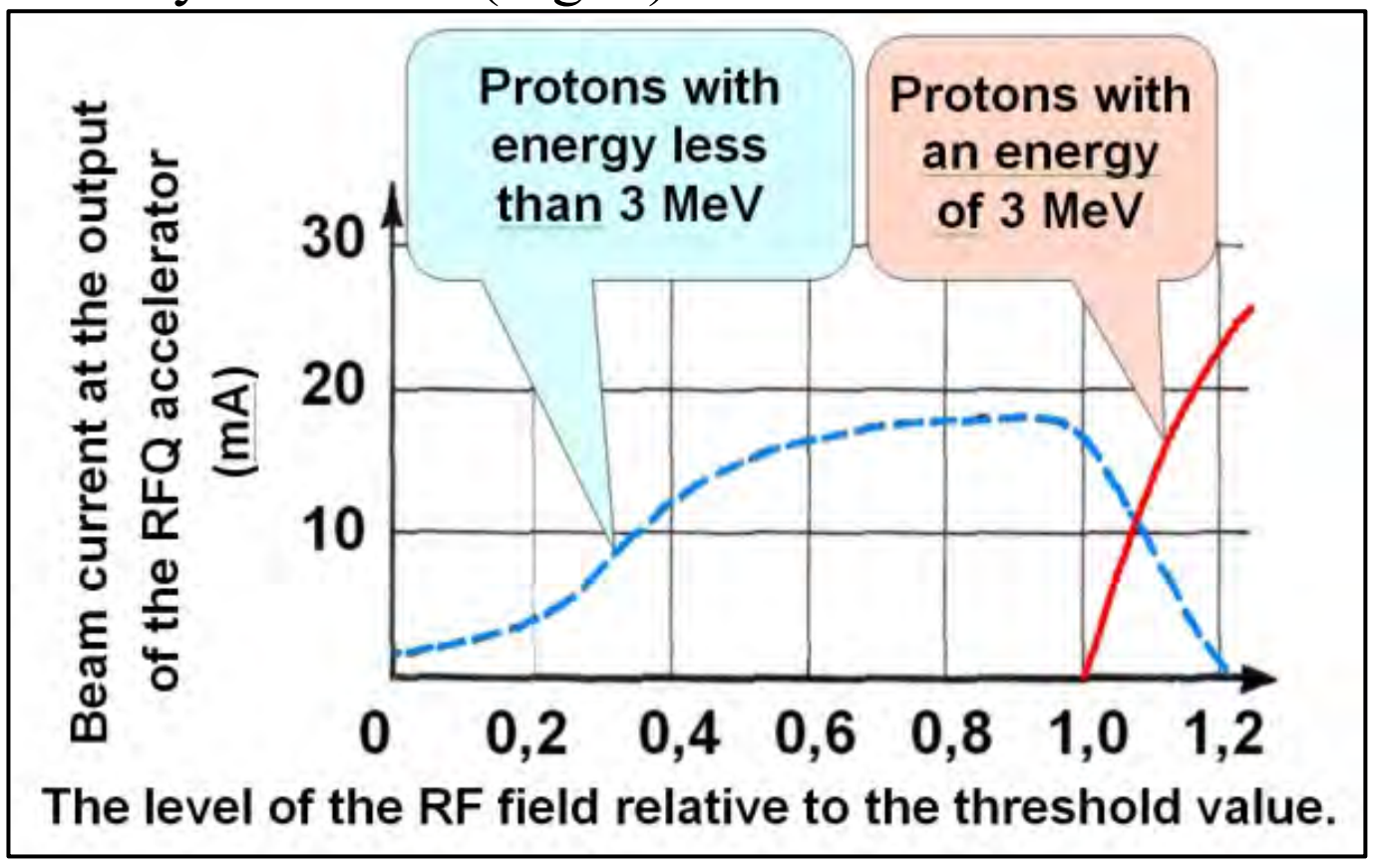

Fig. 8. The first dependencies obtained during the physical launch of RFQ-1.

All those preparing the launch of RFQ-1 experienced a certain sense of satisfaction: the accelerator with its technological systems was assembled and launched in just one year! But Ilya Mikhailovich took the successful launch for granted. Now, from the height of the past decades, it is clear that this, of course, was a great event - the first RFQ accelerator with the most convenient bladed version of the electrodes was launched in Russia (where it was proposed) and in Eurasia. At the PAC-83 conference (Santa Fe, USA), where N.V. Lazarev made a report on the launch of our RFQ-1, the Americans, being more informed about work on RFQ at Los Alamos, they reacted to the report with the exclamation: “Oh, Russian RFQ!” And Nikolai Vladimirovich reasoned with them by saying that there is only a Russian RFQ, proposed by V.V. Vladimirsky, I.M. Kapchinsky and V.A. Teplyakov, all other RFQ are the development of the Russian!

The maximum pulse current at the output of the RFQ-1 was obtained equal to 250 mA. The fraction of accelerated particles in it was 95%. In my opinion, the accelerated proton beam current of 237 mA is still the record for RFQ channels to this day. In the current range from 0 to 100 mA, there were practically no losses of particles in the channel. The obtained experimental data were close to the calculated ones.

9. Accelerator ISTRA-56

“We need to come up with a conditional name for the prototype, – Ilya Mikhailovich once told me. – Think, Sasha. It could be, for example, the name of a river...”. I was just preparing for a kayaking trip and was considering routes along the northern rivers Pinega, Mezen, Kuloy. Kuloy was the most inaccessible and most tempting river. Without hesitation, I suggested: “Kuloi!” Ilya Mikhailovich chewed his lips and said: “Somehow it doesn’t sound very harmonious. Here in the Kushin laboratory an accelerator is being developed with the code name of the Desna river near Moscow. It would be nice to come up with something similar for the prototype.” – “Well, then Istra”, – I answered. “This is good: related topics, related rivers,” – he concluded the conversation. Ilya Mikhailovich used an unobtrusive method of persuasion. He wanted and made it so that I expressed to them the previously made decision and took it as my own. It was a lesson in guiding people.

With the change in name, Ilya Mikhailovich also complicated the scheme of the accelerator. In the final version (ISTRA-56), the accelerator included a matching channel (MC) and three Alvarez resonators, which made up the main part (A-1, A-2, A-3). The beam current in the pulse was planned to be 150 mA. The average beam current could reach 500 μA. It was supposed to first place the accelerator in the experimental building and, at a low average current, begin work on the production of radionuclides. Large average beam currents, up to nominal 500 μA (and with the necessary power of technological systems - and up to 2 mA!) could be obtained in a specially constructed building.

The main parameters of the parts of the accelerator ISTRA-56.

| | RFQ | MC | A-1 | A-2 | A-3 |
|---|---|---|---|---|---|
| The output energy of the particles (MeV) | 3 | 3 | 10 | 36 | 56 |
| Focus type | RFQ | spatial-periodic | | | |
| Length (m) | 4.6 | 0.4 | 3.9 | 11.6 | 9.2 |
| Radio frequency (MHz) | 148.5 | 148.5 | 297 | 297 | 297 |
| Pulse power from the RF generator (MW) | 1.1 | 0.07 (buncher) | 1.7 | 5.6 | 4.6 |

With increasing interest in the acceleration of intense pulsed beams, attention was turned to the method of accumulating significant energy with its subsequent rapid transfer to the load. For energy storage, capacitive storages were most often used, however, the permissible energy density in them is 1-2 orders of magnitude lower than in inductive ones. Inductive storages seemed to be compact and very promising, although they had several problematic moments (large losses, commutation complexity, ponderomotive forces). In order to “feel” this direction, the development of an inductive storage was started with the official goal of using it in the power supply system of the RF path of the ISTRA-56 accelerator. I was afraid that I.M. Kapchinsky would not approve of this work, which could not be classified as a priority. However, he not only did not express disapproval, but, on the contrary, was interested in the progress of the work and supported them. As it became clear later, this was explained simply: Ilya Mikhailovich wanted to have explored possible technical ways to implement future plans. We took this as a lesson in science and technology policy. As a result, an original high-voltage modulator for an average power of 25 kW was developed, in which the properties of inductive and capacitive storage were optimally combined, efficiency was increased, and there was no high-voltage rectifier.

The design of accelerating resonators and the development of the main technological systems of ISTRA-56 were carried out on the basis of the calculated data obtained by I.M. Kapchinsky, A.I. Balabin, A.A. Kolomiets, I.M. Lipkin and others. However, Ilya Mikhailovich instructed me to collect all the information, edit and publish the preprint “Basic physical parameters of the ISTRA-56 proton linear accelerator” (in 2 parts). To me, who was engaged in the technical implementation of the project, and not the settlement work. His idea was understandable - so that I could better understand what we were doing and more clearly imagine the physics of the processes. So Ilya Mikhailovich, in addition to preparing a reference document, organized a kind of educational program for me.

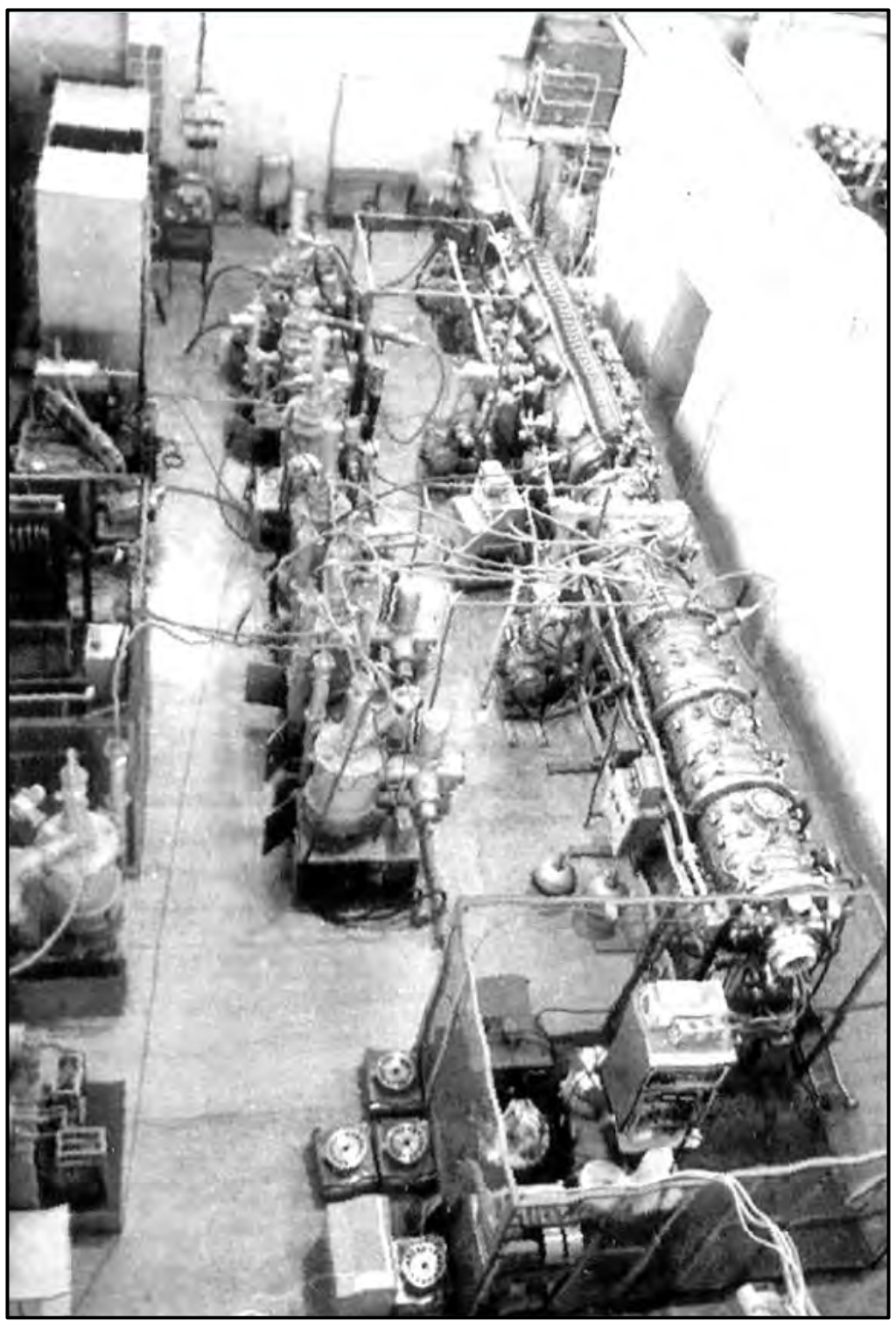

For the first time, the beam at the ISTRA-10 accelerator, which included RFQ-1, MC and A-1 (fig. 9), was accelerated to an energy of 10 MeV on September 19, 1989. The output pulse current was brought to about 100 mA with parameters approaching the calculated ones. The next sections of the accelerator were being prepared.

The first version of the initial part - RFQ-1 confirmed the correctness of physical representations and calculations, but its design was not rigid enough. With an aperture channel diameter of 15 mm, the displacement of individual electrodes reached a very large value of 1 mm, and the accelerated beam current was 80 mA! Such a large current value with a strongly deformed channel spoke of the enormous resilience of RFQ channels!

Physical ideas and technical solutions laid down by I.M. Kapchinsky and the team led by him into the ISTRA-56 accelerator are focused on the creation of high-power linear accelerators for use in nuclear energy to solve the problems of safety of energy production and transmutation of radioactive waste from nuclear power plants.

Ilya Mikhailovich sometimes said that he would like to see the fruits of his labors “still in this life”. Unfortunately, he didn't get to see everything. It was not possible to see that practically in every modern accelerator the initial part is built on the principles of RFQ, that generations of accelerator physicists study from his books...

10. Attitude towards people

Ilya Mikhailovich was characterized by an attentive, respectful attitude towards people. Once I met him on the way from the medical unit. He asked if I was ill. I answer that there is no temperature, but my throat hurts, weakness. The doctor said, will pass. "Which doctor saw you?" I gave the name of a doctor who did not enjoy authority among the staff. Then Ilya Mikhailovich began to explain that a sore throat could have very serious consequences and dragged me to his house to show his wife, who was known to be a good otolaryngologist. Naturally, I balked. But he was adamant. Fortunately, his wife Lyubov Mikhailovna did not detect a severe form of the disease.

Another time, Ilya Mikhailovich saw that I was carrying some weights along with everyone else at the subbotnik. He stopped work for a minute and deliberately (apparently more for those around him) reprimanded me that it was a crime with my severe myopia to lift weights, etc.

Once, for his birthday, I gave him a photograph that I had taken and pasted onto plywood. She seemed very interesting to me, which was hardly true. But Ilya Mikhailovich received her with animation, said that he would hang it on the wall, and started asking how to take care of her, which greatly embarrassed me.

On another occasion, being a guest and having tasted specially cooked meat, he loudly noted the high art of cooking sweet and sour. The hostess was very pleased.

During the adjustment work on the I-2 accelerator, the employees of our department and developers from another institute together prepared the automatic control equipment for operation. The equipment was located in high (more than 2 meters in height) racks. Nearby stood a safe (a little over a meter high) and a chair. We started a game, who will climb the high racks on these subjects faster. And when the lead engineer of the developers went upstairs, we quickly moved the safe away from the racks. The climber was running nervously at the top, scolding us and not knowing how to get down. And then we noticed Ilya Mikhailovich, who was standing at the door and saw everything! “What will happen?” – thought each of us. “Yes, – said Ilya Mikhailovich and called the “top climber” by name and patronymic, – before climbing, you had to think about how you would get down”. There were no other consequences. But, when necessary, Ilya Mikhailovich could be demanding, tough, and adamant.

On his desk, under glass, lay a paper with some rules of life. We were conquered by their wisdom and humanity. On fig. 10 shows some of them, written by Ilya Mikhailovich.

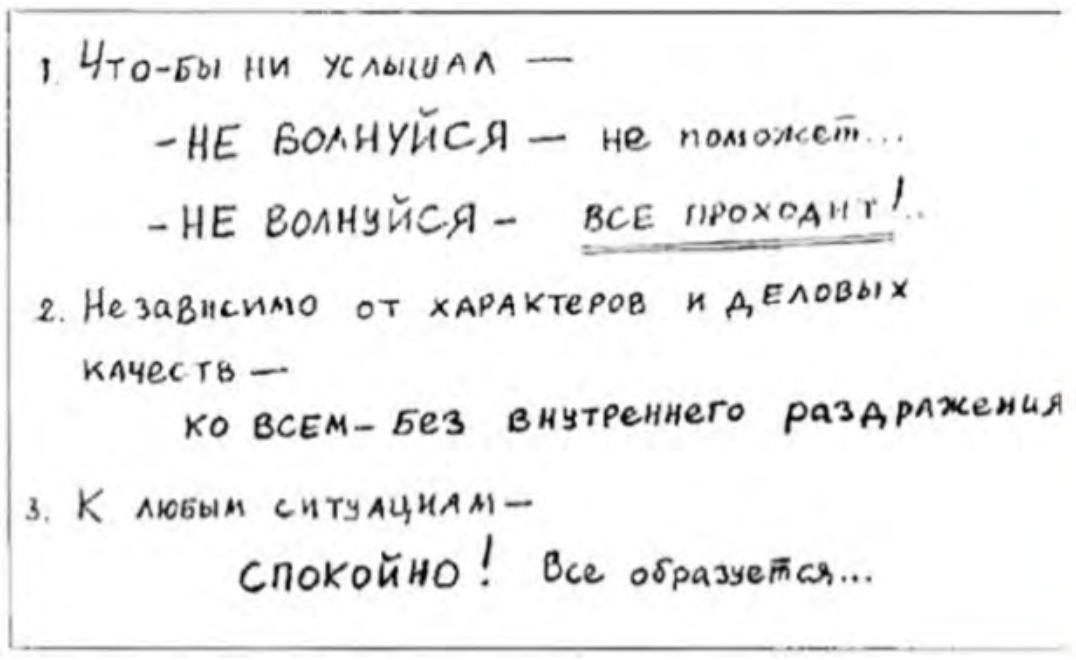
1. Что-бы ни услышал —
- НЕ ВОЛНУЙСЯ — не поможет...
- НЕ ВОЛНУЙСЯ — ВСЕ ПРОХОДИТ!..
2. Независимо от характеров и деловых качеств —
ко всем — без внутреннего раздражения
3. К любым ситуациям —
спокойно! Все образуется...

**Life Credo of Ilya Mikhailovich Kapchinsky**

1. **Whatever you hear –**
   **– Don’t get worried – it won’t help…**
   **– Don't worry – <u>everything passes!..</u>**
2. **Regardless of their personalities and business skills –**
   **– treat everyone without internal irritation**
3. **In any situation –**
   **KEEP CALM! Everything will work out…**

Fig. 10. Life Credo of Ilya Mikhailovich.

## 11. Epilogue

The first years of the last decade of the 20th century were, apparently, the most “turbid” time in the process of returning Russian society to the capitalist path. The pygmies fired on authorities and headquarters. In this rampant “democracy”, it was forgotten that Greek democracy had its own authority - Pericles. In our department, there were enough brawlers who publicly, at a meeting, made stupid accusations against the head of the department, and I.M. Kapchinsky had a hard time. Once, on the way from the dining room, he met R.M. Vengrov and me and said with a bitter smile:

"They are finishing me off, but you are silent". We babbled something about the "surprise and impudence of statements". Then came the nasty feeling of betrayal towards Ilya Mikhailovich.

Another time, I had to get a signature from the director under the order on bonuses for the temporary creative team for work on "ISTRA-10". The director did not share the idea of accelerating clusters, which was then discussed by some employees of the institute (including I.M. Kapchinsky), and made a noise: "Here, Ilya Mikhailovich ...". This was followed by expressions unworthy of the director and insulting to Ilya Mikhailovich. I said nothing: the fate of awards to several dozen people was being decided. The order was signed, but when I left the director's office and calmed down, I again experienced a feeling of betrayal. And I decided for myself - there will be no third time!

It didn't happen... Ilya Mikhailovich was invited to give a course of lectures at the University of Maryland (USA) and died there on May 2, 1993. And the feeling of guilt before him remained. Nothing in life can be left for later, "later" may not be. Everything must be done at once as conscience dictates, otherwise it will not give rest – this is the last lesson that I learned from communicating with Ilya Mikhailovich. It didn't happen.

Now our dissatisfaction with some decisions of I.M. Kapchinsky seems ridiculous. And he just knew better. With what joy I would today place the solution of many problems in his hands and enthusiastically carry out his instructions. But no, life requires my nerves.

Alas, everything passes, even life. Ilya Mikhailovich left behind a bright mark both as a scientist, and as an engineer, and as a person. Doctor of technical sciences, Professor, Laureate of Lenin and State Prize of the USSR, Honorary Doctor of the Goethe university Frankfurt (Germany), the author of many works on the theory of oscillations and pulse technology, on the physics of charged particle beams and accelerator technology, the scientific supervisor of the construction and commissioning of the I-2 and I-100 linear accelerators, the head of one of the leading departments of the ITEP, Ilya Mikhailovich was one of the largest authorities at the institute and will remain in our memory as a person of rare spiritual qualities, able to make strong-willed decisions and at the same time be very attentive to people.

The staff of the department of linear accelerators of the ITEP he brought up made a worthy contribution to the development of the physics and technology of linear accelerators of charged particles. People created the affairs and history of the department. People are very different, with their own strengths and weaknesses. Ilya Mikhailovich set the direction and style of activity. He managed to harness employees into a common cause and use their strengths, suppressing the weak ones. One thing is certain: the collective long-term scientific and technical activity of the department, by and large, served the progress of mankind. School I.M. Kapchinsky brought up talented specialists who enjoy a well-deserved reputation in the accelerator centers of the world.

## 32 YEARS OF WORK WITH ILYA MIKHAILOVICH IS A GIFT OF FATE

**R.P. Kuybida**

I entered the laboratory of Ilya Mikhailovich Kapchinsky in summer 1961, having previously worked in the Urals for 3 years as a young specialist, under the supervision of V.A. Teplyakov and B.K. Shembel. Laboratory of I.M. has just started; it was the stage of recruiting the first twenty employees.

At that time, the ITEP's main problem was the physical launch of the 7 GeV synchrotron, and of course, I was involved in this activity, and in the fall of that year have been enrolled in the laboratory. The works connected with LU I-2 construction have just begun.

For five years I participated in the development and testing on models of methods for precision measurements of geometric, magnetic, and other parameters of I-2 drift tubes, manufactured at the Elektrosila plant in Leningrad. I was already familiar with basic measurements procedures, but I had to solve a lot of new for me tasks under the supervision of Evgeny Nikolayevich Daniltsev and Vladimir Konstantinovich Plotnikov. For me, then a young student, they were great authorities, but I saw that they felt the same distance towards I.M. I.M. delved into and controlled the progress of many problems associated with the development of the first highcurrent linear accelerator of protons with hard focusing in our country both in his laboratory and in other institutes (primarily in RYAN and NIIEFA), as well as in factories and other organizations involved in the project. Simultaneously, he paid great attention to our work on development of test benches for magnetic measurements, for mechanical and vacuum testing of drift tubes, their certification, and methods for their alignment in resonators.

I.M. was always very attentive to the selection of persons for his team and their training. Simultaneously with the beginning of the installation of the technological equipment at the accelerator building area, I.M. organized a scientific seminar on the theory of linear accelerators for his team. Soon, at the request of Academician A.L. Mintz, director of RYAN (now MRTI), he repeated the lectures in another institute. Later the lecture course on the initiative of B.I. Bondarev has been published as RYAN preprint. After that I.M. converted it into monograph and publish the book with a green cover, which, as I understand, was the first fundamental work on the of linear resonant accelerators theory.

In 1984, I worked at the “Atom to the World” exhibition in the Yugoslav city of Maribor and got into a lot of trouble when a person, who had an internship at CERN, visited the exhibition. He saw a book by I.M. and began to ask it very insistently. In order to hand over the book I had to get a permission at our embassy in Belgrade.

I.M. knew how to use each employee in the most rational way. So, I.M. entrusted V.I. Edemsky, a radio engineer at the Institute of Chemical Physics, to supervise the installation of all technological equipment at the accelerator. This led to his approval for the position of the mechanics service head, which he carried out successfully until his retirement, constantly showing himself as very creative designer.

I.M. always addressed employees by first names, even when the laboratory staff exceeded 100 people, which made an impression on the young. He retained the form of address by name or by name and patronymic almost forever from the moment they met. So, even thirty years later he addressed me as Slava.

I remember for the rest of my life the next episode. Being already the control service head of the I-2, I repeatedly reported to the operational meetings about all difficulties that could lead to an occurrence of emergency. And when this event happened, I.M. reprimanded me, and to my reasoned objection that I warned about this, he replied:

– Slava, you did not convey this to my consciousness.

Once, in cold weather, I was impressed by a picture in I.M. office. He was sitting at the table, throwing a winter coat over his shoulders and writing something, a Belomorkanal cigarette was smoking in his left hand while near his nose he holds a handkerchief. That was the time when electric heaters were not available even for leaders.

It amazed me that he always wrote cleanly without blots and treated paper very carefully, remembering that during World War II, as a student at Moscow State University, he had to write on every piece of paper that came to hand. If a document was not supposed to be transferred somewhere, then he wrote on the reverse pages of the computer listings that were available in a large number.

I.M. knew how to listen and hear the opinions of his employees, but what struck me most of all was with what attention he listened to me when I told him about the development of personal computers and software, as well as about the state of these affairs at ITEP.

Traditionally, all new devices for calculations, starting with electromechanical calculators first came to I.M., and then moved further along the employees. The first electronic calculator DVK-2 came to I.M., but its usage did not give desired results. When I.M. received the first real personal computer PC-286 he ordered a program to increase the bit capacity to improve the accuracy of calculations. He quickly begun to program using BASIC language that allowed him to reduce significantly requirement for machine time on large computers of BESM-6 series.

I.M. supported the idea of computers implementation in measuring and control systems of the linear accelerator, but at that time available to us computers did not allow to reduce such processes complexity because of their low reliability. When our division obtained new MicroVAX computers, I.M. accepted without discussions a decision to open them for common use, explaining that any division should contribute into increasing the institute resources.

It seemed to me that I have grown up in I.M. mind when one day I refused to give him advice on a matter of principle. We were returning from Protvino, where I.M. supervised the construction of the I-100 LU, whereas I was involved in the control of incoming drift tubes. In one of them a defect was found in the quadrupole lens winding. I.M. asked me if I should order a duplicate of just this tube, or duplicate the whole set of tubes. I replied that this question is out of my competence.

He was always strictly followed the established rules, but once he rejoiced that in our country rules may be not mandatory. (The crane operator lifted the oil tank to repair the transformer IT-800, violating the crane limitation).

Two high-current linacs with DTL structure, developed under the direction of I.M., are operating until now as proton synchrotron injectors. The first – at ITEP, since 1966, at 25 MeV and the second – at IHEP, since 1967 at 100 MeV. Numerous ion accelerators in the world based on the RFQ structure demonstrate wide possibilities of their use for solving various tasks in the nuclear power industry. In particular, within the framework of the research program for inertial thermonuclear fusion, a prototype of an RFQ for initial part of high-current linac was built to accelerate heavy ions in range up to uranium. The modernization of the RFQ structure with magnetic coupling windows proposed by V.A. Andreev was successfully implemented in the TIPr accelerator at ITEP and today is widely distributed throughout the world. After a successful launch, the solution was subsequently used for many years for experiments with heavy plasma and for the study of materials for nuclear reactors.

A talented person is talented in everything. I didn't know that I.M. dabbled in writing poetry. Once a poetry evening was held in the institute club, at which employees read their poems. One of the authors, I.M. Lipkin was a member of the Linac department. I.M. told him in my presence that he was very sorry that he did not know about this evening in advance. I thought then that he, probably, could also present his poems. I soon forgot about this, but later I was happy to learn that we can get acquainted with samples of I.M. work in this collection of memoirs.

What a pity that I.M., a scientific leader in the field of ion high-intensity linear accelerators, generally recognized in the country and the world, passed away so suddenly and early. Many promising and very important works started by him remained unfinished.

Unfortunately, there was no one at the institute who was equal to I.M. in terms of talent and experience, who could continue the work of the ITEP department of linear accelerators, which did so much for the development of the theory and technology of ion linear accelerators.

I feel grief from the loss, but I am grateful to fate for meeting Ilya Mikhailovich and for three whole decades of interesting and fruitful work with him.

## IN MEMORY OF THE SUPERVISOR

**A.A. Kolomiets**

I have to confess that the proposal of N.V. Lazarev to take part in the release of a collection of memoirs about I.M. Kapchinsky put me in a difficult position. The fact is that due to the difference in age and position in the department, I never entered the circle of those close to I.M. personally, did not participate in discussions of any scientific programs and plans, nor in meetings on current activities. So to tell the world about any interesting details of the work of I.M. as the head of the department is not possible for me.

In the same way, I cannot talk about the private life of I.M., since my informal communication with him was limited to attending a couple of birthday parties. Although even such fleeting contacts revealed to me many unexpected things in him. For example, I was quite surprised when I.M. said that in his youth he was a desperate football fan and did not miss a single match of his favorite team.

These considerations seemed to me sufficient reason to evade the duties of contributing as an historiographer of I.M., but N.V. told me that it is impossible for a person who has worked in the department for 40 years can not to say anything reasonable about his head. Being ashamed of such an argument, I realized that I had no other way than the beaten path of most memoirists, who, regardless of the name in the title of their work, write about themselves, their loved ones.

So, I appeared in the department of linear accelerators (LU) of ITEP at the end of 1966 for undergraduate practice. The personnel department dealt with me quite unexpectedly by issuing a pass for a week and advice to find a suitable place for myself. With that I went to explore ITEP. The LU department was not the first one I visited. But this department immediately attracted my attention by the atmosphere of general enthusiasm for its work. After getting acquainted with several more laboratories, I chose the LU department and I never regretted my choice while the department was headed by I.M.

The festive atmosphere in the department when I appeared there was explained by the fact that the I-2 linear accelerator had just been successfully launched, which completed the very large work of the entire team in its development and construction. Now the stage of experimental research and adjustment of the accelerator, reaching the design parameters and putting it into routine operation was ahead. A significant part of this work was carried out by a group led by V.A. Batalin, in which I started working at ITEP. It was a great success for me to join the work at this stage and as part of this particular group.

The work consisted in achieving the design parameters of the I-2 LU and was quite understandable. All that was required was to learn how to obtain a high-brightness beam and eliminate all reasons for the decrease in brightness in initial beam formation system and in LEBT. Since computer simulation of beam motion was in its infancy at the time, the work to optimize beam parameters was carried out solely on the basis of experimental activities. Appropriate measurement techniques were developed, the required measuring devices were manufactured, the beam parameters were measured, and the design of the ion source and focusing elements was corrected based on the data obtained. As a result of this activity, an original design of the LU I-2 ion source was developed, the dynamics of the beam in the accelerating tube was significantly improved, and the design of the LEBT was significantly changed.

It is clear that participation in such work as part of a highly qualified group gave yesterday's student a unique opportunity to quickly learn the basics of the physics of charged particle beams and acquire some skills in experimental work. It should be noted that I.M.'s book, which then also existed in the form of a series of lectures on the fundamentals of the theory of linear resonant accelerators, was an indispensable tool for these works.

I recall with particular pleasure the atmosphere in which the work in the group took place. We worked together, everyone was interested in achieving the set goals. Both the proposals of the group leader and the newcomer were equally carefully discussed. It seemed to me then that this is the usual working environment of a scientific institution, which arises by itself. But now I understand that such an atmosphere depends, first of all, on the scale and style of work of the supervisor. In our case, I.M. controlled and directed the work, in no way restraining the initiative of the participants and, moreover, in every possible way supporting and creating the necessary conditions for its successful advancement. But what I.M. strictly kept under his personal control was the preparation of publications. Not a single printed work of any level could leave the department without being personally thoroughly checked by him both in terms of content and design. This had two important implications. Firstly, it brought up a certain culture of preparing publications among young employees, and secondly, it worked for the constant growth of the authority of both I.M. himself and the department he headed.

Of course, our group was no exception in the department. Similar work was carried out in other groups. As a result, in a short time, the operational parameters of the I-2 linac were brought to a level exceeding the design ones, and the number of dissertations defended in the department was constantly increasing.

After the successful commissioning of the I-2, I.M. was able to start implementing his ideas on the development of intensive ion linear accelerators. Of course, the accelerating structure with spatially uniform quadrupole focusing (worldwide known as RFQ) proposed by I.M. together with V.V. Vladimirsky and V.A. Teplyakov was a key element of high current accelerators. The construction of the RFQ accelerator was started by in ITEP itself, in contrast to I-2, in the construction of which several organizations took part.

The complexity of the tasks that needed to be solved in the process of work was determined by their novelty. Almost everything, from methods for calculating the geometry of electrodes to resonant structures, had to be developed from scratch. New laboratories and groups appeared in the department, which were responsible for the development of all systems of the new accelerator facility. During this period, I.M. published the main works on the theory of accelerating structures using focusing by a high-frequency electric field. In addition, there have been many publications by the staff of the department on issues related to the development of the RFQ structures.

I.M. always understood the importance of using computer methods to build intense linacs. He initiated these works during the creation of the I-2. Despite the fact that the low productivity of the machines available at that time and the insufficient development of computational methods did not allow obtaining sufficiently reliable results, he constantly made efforts to expand the use of computer methods in the work of the department. The department has always had wide access to ITEP's computing facilities. A number of employees under the leadership of I.M. began to create programs required for RFQ structures development. In addition, at his request, ITEP mathematicians began developing a program that simulates the dynamics of a beam of charged particles in an accelerator, taking into account the action of space charge forces. On behalf of I.M. I interacted with mathematicians in the development of this program. This was extremely important for me personally, as it made it possible to learn modern programming methods and the use of numerical simulation to solve specific problems related to the development of high current linacs. After the construction of an RFQ accelerator at ITEP, I got the opportunity to conduct an experimental study of its operation. Accompanying the experiments with computer simulations showed, on the one hand, the reliability of the results obtained by computer simulations, and, on the other hand, the very great usefulness of such an approach for studying the behavior of an intense beam in an accelerating and focusing structures.

Upon completion of this work, I.M. literally forced me to formalize its results in the form of a PhD thesis.

If what is written above causes the impression of some special attention to I.M. to my person, then this is nothing more than an aberration of the presentation from the first person. In fact, quite a few employees of the department who have carried out important work, obtained significant results, and defended dissertations can tell something similar. This was a consequence of the general creative atmosphere in the department.

Works by I.M. during that period did not go unnoticed by the world accelerator community. Moreover, his authority has reached an extremely high level. To the greatest extent, the employees of Los Alamos (LANL) contributed to this, who were the first to evaluate the significance of the accelerating structure with spatially uniform quadrupole focusing, began their own work in this direction and widely disseminated information about the new structure calling it RFQ.

This contributed to the emergence of extensive international contacts between the ITEP Linac Department and the world's accelerator centers. Visits to ITEP by delegations from the most famous accelerator centers have become commonplace. Numerous memorandums of intent were signed to organize joint work. The closest and broadest cooperation was planned with LANL, within the framework of which the joint development and construction of a linac based on a new accelerating structure was envisaged. It is hard to even imagine what an impetus this cooperation would give, the access of the department's employees to modern technologies and equipment, the development of the department's capabilities, and, possibly, of all domestic accelerator technology.

Unfortunately, these plans were not destined to materialize. All contacts with LANL were interrupted by the US government after the entry of Soviet troops into Afghanistan.

Despite this unfortunate circumstance, the authority of I M. abroad continued to grow steadily. He got the opportunity to visit many accelerator centers in Europe, Frankfurt University awarded him an honorary title, his book was published in the USA. For I.M., these trips were by no means tourism. Each of his visits to any accelerator center was the beginning of a new long-term cooperation. Very many employees of the department took part in joint work with INFN(Italy), GSI (Germany) and others. contributed to the growth of the qualifications of the specialists of the ITEP Linac department.

However, international cooperation was considered by I.M. as something useful, but secondary. More important for him was the implementation of his own plans for the development of intense proton and heavy ion accelerators to solve the problems of nuclear and thermonuclear energy production. Unfortunately, the situation changed after the unexpected collapse of the USSR. The almost complete cessation of funding for domestic accelerators, no matter how large on a global scale, and, in general, the reduction in funding for science, led to the fact that the main mode of existence of the ITEP Linac department was participation in international projects under the auspices of the ISTC and similar organizations. The authority of I. M. was clearly manifested in the fact that one of the first projects of the ISTC provided for the joint development of ATW by the forces of LANL and the ITEP Accelerator Facility for Nuclear Power.

I.M. left us in 1993. The department's attempts to organize work without I. M. finally confirmed that the scientific team cannot exist without a pronounced leader - the Scientist. In fact, all the time until the disbandment of the department in 2008, its activities were reduced to more or less successful use of the potential accumulated under I.M.

In conclusion, I can only express my gratitude to fate, which made it possible to work in a team headed by Ilya Mikhailovich Kapchinsky, to whom I owe all the achievements in my professional activities.

## "WHY IS IT GROWING?"

**V. S. Skachkov**

In June 1974, I, a MEPhI graduate, defended my graduation project and, however, a little later, I became an engineer-physicist in ITEP Laboratory No. 12. At that time, work was underway there to improve the technological systems of the I-2 linear accelerator, the injector of the U-7 proton synchrotron, and valuable experience had already been accumulated in the development of automated installations and several “smart” electronic devices were created that combined high electrical voltages and pulsed currents. I also took part in several developments, each time trying to offer a new, original solution. Particularly well proven is the “groped” method of forming a flat top of a current pulse in an inductive load. My affairs were going well, since the development of this direction was quite clearly and quite naturally drawn as part of the work on my dissertation. Soon, in the fall of 1974, I wrote an application for admission to the ITEP graduate school. I doubted whether it would be appropriate to indicate my position next to the signature, because I was still a radio electrician of the 5th category! Doubts turned out to be in vain: it was common for that era and people not to attach much importance to nothing decisive formalities.

All this time, as well as over the next almost 20 years, my work was easy and interesting. Many employees of the laboratory, including myself, went about their business with enthusiasm, the working day often reached 10 or more hours, but no one noticed this. Pretty soon, an understanding came of why such a business atmosphere arose in the laboratory, where the source that fueled it set the vector of movement for the entire scientific department, determined and supported the steady growth of the positive derivative in new developments. Such was the situation at the time when an unexpected and sharp turn took place in my life. N.V. Lazarev (at that time the chief engineer of the linear accelerator division) somehow called me to his office and said that I.M. Kapchinsky proposes to reorient my scientific activity in connection with the idea that arose to create a permanent magnet focusing channel for a linear ion accelerator. I accepted the offer without hesitation, since it was easy to guess that in this direction a vast field untouched by anyone for theoretical and experimental activity opens in front of me. It was precisely such a task, to the solution of which at first you do not know how to approach, that I most of all expected.

From that moment on, our meetings with Ilya Mikhailovich became regular. He was keenly interested in the promotion of the first ordinary variant of quadrupoles, the basis for the calculation of which was laid by him a year before I was involved in this problem. Frankly, this design was so classical that I always resisted calling it my development, although Ilya Mikhailovich convinced me of this. I don’t know if there was some subtle intent of my leader, but such thoughts arise because he had an enviable knowledge of the psychology of people. Perhaps he was taking advantage of my disdain for honors, especially when it comes to microscopically small accomplishments. Once it seemed that, due to a completely trifling result, I was recommended to make a publication, and I expressed my assessment: "The mountain gave birth to a mouse." Ilya Mikhailovich agreed that time, and this made me happy. From the very first day, I started looking for a design in which the field sources would be distributed in the most rational way.

How pleasant it was to hear during the entire period of development of the theory of a new design simple words of approval and recognition of the significance of my numerous findings! With what tact and benevolence Ilya Mikhailovich spoke about my very risky extrapolations, often built on a very meager evidence base! At the same time, how calmly, without unnecessary pathos and pomposity, he reacted when I finally reported to him about the proof of the central theorem, obtained at the turn of 1975 and 1976! In relation to myself, I often say jokingly: a physicist is not a profession, but a diagnosis. Meetings with people like Ilya Mikhailovich convince me that physicists are a special breed of people. It gives them great satisfaction to notice and visually explain the inner, deeply hidden, often seemingly contradictory connections in the diverse world around us. Then even a person who by nature has very modest physical data becomes a titan pioneer in science.

At one of the Ph.D. thesis defenses at MEPhI, where the director of one of the plants, it seemed to me, spoke very convincingly about his developments in methods for the industrial production of large resonators for linear ion accelerators. The same could not be said about the subsequent speeches, which at such meetings are usually highly formalized, and listening to them quickly becomes boring. At the stage of assessing the results achieved by the applicant, the floor was given to Ilya Mikhailovich. With one small remark, he immediately enlivened the meeting, captured the attention of the members of the dissertation council who were present.

– ...Here, someone has already said that it is more difficult to build proton accelerators than electronic ones. And I would say that they are 1836 times more difficult to build than electronic ones.

Of course, waves of doubt always roam the defenses, which you can't always discern behind the smart words of speaking scientists, but after such remarks, unanimity grows in the ranks of judges, which determines a successful outcome for the applicant.

Whatever the rank of the scientific and production conferences that took place with my participation, I do not remember a case when anyone seriously objected to Ilya Mikhailovich. In matters that make up modern knowledge about accelerator science and technology, his authority was indisputable. He freely operated on the main theorems and fundamental dependencies describing the behavior of beams of accelerated charged particles, and the amount of information that he kept in his memory was simply shocking with its colossal size. Quite by accident, I happened to find out where its moving border passes. Here is how it was. Having received permission to enter for the next report, I stepped forward and closed the door of office No. 40 behind me. I was getting ready to unfold my materials, when suddenly, unexpectedly for me, Ilya Mikhailovich somehow said in an undertone:

– Volodya, I found a mistake in you.

Ilya Mikhailovich handed me a sheet of paper showing the rule for the distribution of magnetization in a quadrupole lens. For 2-3 seconds I looked at the drawing and, making sure that everything depicted on it was familiar to me and made by my hand, I answered just as calmly:

– There is no mistake, Ilya Mikhailovich.

– Well, how about the orientation of the magnetization is not measured from the abscissa axis?! – objected Ilya Mikhailovich.

– Yes, – I answered, – the original $2\varphi$ rule is written in unit vectors of a cylindrical coordinate system, and in the Cartesian system, the rotation speed of this vector is one more.

For some time I interpreted this episode as a small test for the speed of orientation of the applicant in his own works, it seemed very improbable that Ilya Mikhailovich could not know this rule by heart. Overwhelmed by doubts, I decided to find out. It was only necessary to make a significant pause filled with thoughts during the next report before writing down the rule for changing the field in a quadrupole lens with chalk on a blackboard. In contrast to the commonly used formula, which Ilya Mikhailovich certainly knew by heart, in my special case of the position of the poles in space, the dependences of the field components on the coordinates are reversed! Soon the right opportunity presented itself.

At that very climactic moment, I raised my hand with the chalk to my forehead and thoughtfully said:

– What should be multiplied by? By $x$ or $y$? – And, feeling that Ilya Mikhailovich's attention was definitely captured, I resolutely uttered the finished phrase: – By $x$!

For a moment I expected objections, expressed with more or less anger, but there was no movement behind me. I turned around, as if to show that I was looking for support for my rightness, and looked at the teacher... Ilya Mikhailovich smiled benevolently. It seemed to me that this is how he expresses his attitude towards a student who "fell into a mess", and I continued:

– Yes, yes, in this coordinate system it will be exactly like that.

This time Ilya Mikhailovich thought. He took a sheet of paper and a pen and said with some degree of bewilderment:

– Volodya, you took me by surprise.

My demonstration of readiness to immediately present evidence immediately met with strong disagreement, and a minute later a formula appeared at Ilya Mikhailovich's hand, written out on the blackboard.

– Well, of course, because in an electric quadrupole, it is precisely this dependence that is observed! – He concluded as a sign that everything is now falling into place.

When a conversation about Ilya Mikhailovich comes up in various companies, the theme of his greatness as a major figure and organizer of science almost always arises by itself. No one has ever argued about this, but only cited various instructive stories of witnesses. For me, the power of the incomprehensible magic of Ilya Mikhailovich manifested itself in one discussion that took place in the mid-80s in the office of director I.V. Chuvilo. Shortly before this meeting, I was warned that Academician G.N. Flerov intend to visit institute and that I should be ready. If I remember correctly, I was called when Ivan Vasilievich and Ilya Mikhailovich had already told about the main achievements in the development of a high-current linear accelerator prototype, and I was mostly silent during the conversation. Unexpectedly, Georgy Nikolayevich expressed his high appreciation of the development of the original permanent magnet quadrupole lens at ITEP, after which Ilya Mikhailovich replied:

– So we will give the author the State Prize!

After these words, the participants in the discussion apparently felt that the subject of the conversation had been exhausted, and its results satisfied everyone. Unlike them, I felt dumbfounded, and, thank God, that after that I was not given the floor – embarrassment could happen.

Often there were cases in which Ilya Mikhailovich seemed unprotected. Once I ran to the institute dining room and saw him in line. He had apparently not yet fully recovered from his cold, which betrayed his voice. An elderly woman standing behind him noticed this and began to shame Ilya Mikhailovich, not limiting herself to impartial expressions.

– How did you manage to appear in a crowded place, because you are sick! – Without a shadow of embarrassment, she blurted out, clearly counting on the support of others.

After a brief hesitation, Ilya Mikhailovich, without hiding his resentment, answered:

– This morning my wife told me that I am no longer contagious.

Such argumentation disarmed the attacking side, and that immediately stopped their attacks. This saved her and me – barely holding back, I boiled like a kettle.

Dinner at the same table with Ilya Mikhailovich was always very interesting, but fraught with danger. He ate quickly, although he actively supported the conversation on any topic, and to keep up with him succeeded with great effort. Often and, it seemed to me, excessively he salted food. Much later, I heard that doctors seriously associated this preference with heart disease.

After the death of Ilya Mikhailovich, it seemed to me that the ITEP accelerator sector was empty. And not only because the star of the highest magnitude in the world died out untimely – everyone felt this huge loss, but also because from that difficult moment there was no person with whom each of us could discuss any physical idea. Just like that, come up and talk, knowing for sure that you will be listened to with attention and get an accurate physical representation of the subject of your search as completely as it is known at the moment. But doubts did not leave him. Once, at the end of a planned conversation with me, he, peering at some American article, thought aloud:

– Everyone vied with each other saying that the beam emittance is increasing ... Why is it growing? ... – And saying goodbye to me, he again plunged into the interrupted process. This is how I remember Ilya Mikhailovich.

## THE BIG IS SEEN IN THE DISTANCE...

**V.A. Bomko**

National Scientific Center "Kharkov Institute of Physics and Technology"

July 5, 2009 will be the 90th anniversary of Ilya Mikhailovich Kapchinsky, which, in fact, is a celebration of a large number of his students and colleagues in research work, dispersed over many countries and continents. His ideas have captured the minds of everyone who has anything to do with the problems of creating linear accelerators for heavy particles. In a short period of time, calculated from the moment he entered the Institute of Theoretical and Experimental Physics, starting in 1958, Ilya Mikhailovich not only comprehended all the achievements known at that time in the field of physics and technology of linear accelerators, but soon became its generally recognized leader and organizer . Of the whole variety of ideas put forward by him and developments carried out by him on the problem of linear accelerators, Ilya Mikhailovich used only a part of them as the basis of his famous monographs, the first of which was called "Dynamics of particles in linear resonant accelerators." A distinctive feature of this book is the presentation of only our own deep developments on the dynamics of high-current beams in a linear accelerator during their acceleration in a structure with drift tubes. We have obtained a detailed theoretical substantiation of a device that ensures the radial and phase stability of bunches of particles accelerated in an external high-frequency electric field, taking into account their own Coulomb forces. He was the first to use the method of microcanonical phase distribution of particles in a bunch and proposed many other innovations in the approach to solving the problem of beam dynamics. The book immediately became a rarity and it was impossible to acquire it. I remember in 1966 I was very lucky: while on vacation in Mariupol, I accidentally saw one copy of it on the shelf of a bookstore. Only in a place where there were no specialists in this profile, this book could be delayed. And the second book of Ilya Mikhailovich "Theory of linear accelerators", published in 1982, turned out to be completely inaccessible, and one could get acquainted with its contents only in a special library.

Here I want to note that I got acquainted with the scientific developments of Ilya Minilovich back in 1963 on the basis of materials received at the Physicotechnical Institute of the Academy of Sciences of the Ukrainian SSR, as the Kharkov Institute of Physics and Technology was then called, ITEP reports "Physical substantiation of the project of a linear proton accelerator for an energy of 100 MeV" and "Physical substantiation of the I-2 project". There, along with a deep theoretical consideration of the dynamics of beam bunches during acceleration, a comprehensive justification was given for the choice of the main parameters of the largest accelerator in the world at that time - the injector of the proton synchrotron at an energy of 76 GeV, as well as the linear accelerator of protons at an energy of 25 MeV, the proton synchrotron injector at energy 10 GeV. The development of these machines was carried out at that time by several leading research, design and production organizations in the country. The choice of such quantities as the specific acceleration, the wavelength of the accelerating field, the magnitude of the synchronous phase with quadrupole focusing, the number of resonators, the gap coefficient, the multiplicity of the acceleration period, the parameters of the hard-focusing channel was substantiated. The dynamics of the beam and the design characteristics of the accelerators were discussed: the phase incursion and the frequency of radial and longitudinal oscillations, the channel capacity and the limiting beam current resulting from this; the influence of space charge on longitudinal oscillations, the choice of "idle" gaps. The issue of tolerances for systematic and random errors, the choice of apertures of drift tubes were discussed especially carefully.

It should be noted that, since 1955, at the Physicotechnical Institute of the Academy of Sciences of the Ukrainian SSR, Kharkov, work has been carried out on theoretical and experimental studies of the method of beam focusing using quadrupoles. Theoretical developments were carried out by Ya.B. Fainber and A.A. Sharshanov. The design of electrostatic and magnetic quadrupoles was created. As a result, already in 1960, a quadrupole focusing system was implemented in the drift tubes of the 5.5 MeV linear proton accelerator in operation, which was originally launched on grid focusing. This hard focusing system was structurally imperfect and hastily executed, but even in this form it showed higher efficiency. At that time, we did not have the task of creating an accelerator of national importance, as was the case at ITEP.

Ilya Mikhailovich in the early 60s led the development of large accelerators I-2 and I-100, not only in terms of physical justification, but also served as curator of the design and manufacture of their main components. Many participants in the implementation of these projects noted his scrupulousness in relation to the accuracy of the execution of products. He strictly demanded compliance with the specified tolerances. This was especially true of the I-2 accelerating structure. It seemed to us that in this respect there was some excess in the rigidity of the requirements. Indeed, although the energy of protons at the input of the linear accelerator at an injector potential of 700 kV provided acceptable sizes of drift tubes for placing magnetic quadrupoles in them with a realistically feasible magnetic field gradient, Ilya Mikhailovich chose the path of using a two-fold drift with a cell length of 2 VA., although this meant an almost 2-fold increase in the length of the first resonator. In addition, only in this accelerator, along with the main windings of the quadrupole poles, additional windings were installed to ensure the correction of the magnitude and distribution of magnetic field lines during the setup and operation of the accelerator, despite the fact that everything was designed for a constant power supply. Ultimately, this complication paid off. As noted more than once, the reliability of maintaining the parameters of the accelerated beam at the required level turned out to be the highest for more than 40 years.

I will not dwell in detail on the scientific achievements of the team led by Ilya Mikhailovich, which he created, nurtured and constantly aimed at the accomplishment of major tasks of a scientific and technical nature. The creative energy of this team made it possible not only to create perfect installations of exceptional importance, but also to significantly raise the level of world accelerator technology. Ilya Mikhailovich did not confine himself to the scientific interests of his department; he generously shared his achievements on a global scale. A.A. Kolomiets, in his article on the participation of the department of linear accelerators in international cooperation, notes that the employees of the department published more than 120 reports in the Proceedings of various International Conferences. In this connection, I recall our conversation with Ilya Mikhailovich after my report at the All-Union Conference on Charged Particle Accelerators in Dubna, where I noted that the method of tuning invented by us and formation of a uniform distribution of the accelerating field of an interdigital accelerating structure, reinvented in Japan 10 years after it was introduced in our main | section of the linear accelerator of multiply charged ions, LUMZI, without reference to our work. Ilya Mikhailovich asked: - "Where did you publish your developments?" - I answered that they were repeatedly published in the Proceedings of the Conference. - "Then it's clear, the Japanese don't read them."

Since the 1960s, I have often visited ITEP. Almost every time there was at least a brief meeting with him in his office. I remember such a conversation. Being familiar with the materials of the physical substantiation of the choice of the main parameters of the I-2, I expressed concern about the calculated value of the accelerating field strength at the gaps between the drift tubes, reaching 130 and even in some gaps - 150 kV/cm. I shared my experience that at the same operating wavelength in our currently operating accelerators LUP-20 and LUMZI, the field in the gaps does not exceed 100 kV/cm, breakdowns occur higher. The time was such that the I-2 systems were already in the process of installation. Ilya Mikhailovich became interested in my remark, began to ask in detail about the geometrical parameters of the drift tubes, about the vacuum and other details. Then, when I appeared at ITEP after the launch of I-2, he said with a smile: “You scared me then.” The point is that in KIPT accelerators pumping to high vacuum was carried out by high-power diffusion pumps developed at the institute (20 or 40 thousand l/sec) with water traps of oil vapors, which led to a significant migration of oil vapors into the working volume of the resonator. This was the reason that even at a sufficiently high vacuum, the limiting level of the field strength in the gaps decreased. In the I-2 accelerator, cryogenic cooling of Louvre traps was used in the traps, and then a transition was made to oil-free pumps.

I happened to take part in the International Conference on High Energy Particle Accelerators, which was held in Tsaghkadzor, Armenia. There, Ilya Mikhailovich delivered his famous report on the structure of the POKF, which subsequently had a decisive influence on the course of development of the technology of linear accelerators. More than 300 foreign specialists in accelerators took part in this conference hospitality abounded. Every morning, six large tourist buses drove under the hotel buildings, which took away the conference participants for the whole day on excursions to the ancient sights that Armenia is so rich in. The temptation was very great. But on the day on which the report of Ilya Mikhailovich was listed? the huge conference room was packed.

Ilya Mikhailovich took an active part in regular seminars on linear accelerators of charged particles, organized by the KIPT every two years (in 2007, the 20th International Seminar on linear accelerators was held in Alushta). At the same time, Ilya Mikhailovich often visited our department and got acquainted with our developments of interdigitated accelerating structures and the method of smooth control of the energy of accelerated particles and stabilization of the distribution of the accelerating field. I remember meeting with him, especially on the two-day trips to the Donets to the region of the university biological station in Gaidary that took place after the end of the seminar.

Ilya Mikhailovich combined the quality of a world-class scientist with human simplicity and accessibility. He showed great interest in the people around him. When I was on business trips to ITEP, I often met him in the institute's canteen, where one could talk about serious matters at the same table.

I am extremely grateful to him for agreeing to oppose my doctoral dissertation, which was defended in 1984 at the MEPhI Academic Council. It was a great honor for me, since not every applicant for this degree could receive such consent. It so happened that the members of the Academic Council had to meet twice on this occasion. The fact is that the opponent, in addition to I.M. Kapchinsky and B.P. Murina was N.P. Sobenin, who, as it turned out two days before my defense, had no right to oppose, since he was the secretary of this Council. Having learned this, he urgently achieved in the VAK, of which he was a member of the Council, the transfer of powers of the secretary of the MEPhI Academic Council to another member of the Council, who, on the night before my defense, was taken by ambulance to the hospital. In the morning, when the Council met in full force, it was announced that, for the reasons indicated, my defense was postponed. (We can say that the reason for this incident were three factors: the defense should take place on the ball on Monday, February 13th of a leap year). The real defense of my dissertation took place at the meeting of the MEPhI Academic Council on April 9, 1984 and went "without a hitch." I still keep Ilya Mikhailovich's review, as well as the minutes of the Council meetings.

Years have passed since Ilya Mikhailovich Kapchinsky, the great scientist who made a revolution in accelerator science, a prominent organizer of it, who raised a first-class team that continues to multiply achievements in science and technology, left us. His students are alone in their native ITEF, while others are in many large centers of Russia and abroad.

## I.M. KAPCHINSKY IN IHEP

**V. A. Teplyakov**
Institute for High Energy Physics, Protvino

I met I.M. Kapchinsky in absentia. Since 1950, I began working at the Institute of Chemical Physics of the USSR Academy of Sciences under Academician N.N. Semyonov, in the laboratory of Boris Konstantinovich Shembel. We took part in the development of a linear proton accelerator. B.K. he explained to me, a laboratory assistant: protons are accelerated by a high-frequency electric field in the gaps between the drift tubes. When the field becomes retarding, the protons fly inside the drift tubes. This introductory conversation helped me understand the report of A.I. Akhiezer and Ya.B. Feinberg. For a long time I did not suspect that there are other institutions and organizations that are also studying the linear accelerator. I experimentally studied the dependence of the frequency of a short cylindrical resonator (compartment) on the dimensions of the drift tube.

In 1952, the first All-Union Conference on accelerators took place. B.K. Shembel reported on the method developed by us for the engineering calculation of a linear accelerator. I did not suspect how many people in different organizations build accelerators, many jointly create powerful accelerators in Dubna, there is the A.L. Mints and the institute where V.V. Vladimirsky. Me B.K. said that Vladimirsky had expressed an interesting idea: to focus protons outside the drift tubes!

Now it seems strange that we did not know anything about colleagues. The leaders knew something, but they did not tell the laboratory assistants about these secrets.
When we started publishing our articles in the PTE, they were subject to review by the ITEP. The reviews were kind. And only later did I learn that Ilya Mikhailovich Kapchinsky was the reviewer of a number of our articles.

I remember that Ilya Mikhailovich and I first met in Moscow. I came from the Urals to ITEP to see V.V. Vladimirsky. I wanted to know his opinion on the possibility of focusing with an accelerating field. V.V. said: "We watched it - nothing good will come of it." But the main question for V.V. I had a continuous mode accelerator. And then, me V.V. reprimanded: “I will never do this myself and I will not support you!” It's good that I found Ilya Mikhailonich and we talked about focusing with an accelerating field. Wow was the first friendly conversation with a literate person. After all, my comrades considered me a dreamer and did not take me seriously. It is unpleasant when no one tries to understand your timid guesses.

When I returned from the Urals to Moscow, Ilya Mikhailovich invited me to the Institute for High Energy Physics, where he was also head of the Injector Department. It was October 1966. The department accepted the linear accelerator for installation, commissioning and subsequent operation. Our joint work with the employees of the Radio Engineering Institute and those seconded from other organizations went well. This was the organizing role of I.M. Kapchinsky and chief engineer of the department Sergey Aleksandrovich Ilyevsky. They skillfully planned the work and clearly carried out the planned. But at that time, builders were working, the accelerator with its numerous systems was being mounted, the installation of drift tubes in the accelerator resonators was being completed. Work was in full swing, but there was no fuss. It was interesting for me to watch how two experienced leaders at the end of each week summed up what had been done and figured out a plan for the next week. I.M. went home to Moscow for the weekend. And our work continues...

Ilya Mikhailovich instructed me to understand the operation of individual systems of the accelerator, to identify and eliminate shortcomings. And I did not come to Protvino alone: my friends, the laboratory staff, came with me.

I sought to get a lathe in the laboratory. To this, Ilyevsky said: "We need to ensure the operation of the accelerator and for this there is a workshop of the department." I had to seek support from I.M. He agreed with me: I was allowed to do some experiments in the laboratory. We raised a small lathe to the fourth floor. Tolya Sokolov was especially pleased with the machine. Now he had his own job. Later everyone was convinced, and S.A. also that the machine really fell into good hands. Soon, a proton gun, which we brought from the Urals, was restored next to the machine. Experience in the development of various proton sources and their operation allowed Vitaly Vasilyevich Nizhegorodtsev to obtain the first protons accelerated to 700 keV already in early 1967. Installation and adjustment of the I-100 accelerator went well. We pretty quickly eliminated the cause of illegal proton screening in the second resonator. Vladimir Konstantinovich Plotnikov took part in this. He was invited to IHEP by Ilya Mikhailovich. (I.M. wanted to involve his employees from ITEP to participate in the launch of the I-100.)

We successfully met the deadlines for launching the accelerator by November 1967. It was a sunny, warm August. There are many mushrooms in the forest. On Thursday I invited I.M.to visit nature. He willingly agreed and we, as we were dressed at work, without changing clothes, went to the forest. Let's go in my car in the direction of Tarusa. I haven't been to those places yet. The location was convenient to leave the car. After wandering through the forest and picking up a small basket of whites, we returned to my one-room "bachelor" apartment. I.M. said he knew how to cook mushrooms with mashed potatoes. I did not have oil for the planned dish. The shops were no longer open. Saved by a neighbor. Dinner was hearty (just put the potatoes in the pan).

The next morning I went to I.M. He asked me: how am I doing? I didn't really understand the question. (Everything was in order at work.) I.M. took out a simple pencil from his pocket. - "I could have lost him, bending down for a mushroom," he said. I looked at him in bewilderment.
- “And what did they drop?”
- “Wallet.”
- “And in it?”

It turned out that there was a passport in the wallet, important information, a receipt for paying for the monument for 8 thousand, money. The monument - a marble slab - should be received next week. After standing silently, I left the office. At noon, I again went to I.M.; he's with S.A. discusseing some business issues.

- “Can we take a ride?” I asked.

And we drove into the forest, realizing the hopelessness of our trip. There was nothing to loose.

The starting point of our yesterday's journey was easy to find. Again, the eyes were white. There was no basket. I strung them on a broken twig. I have prepared several such “skewers”. He folded them under conspicuous bushes.

Surprisingly, I found signs of yesterday's path. We crossed an overgrown ravine, and on the other side we managed to find the necessary signs. We walked near the edge of the ravine. I moved a little to the right, and I.M. continued to move along the coast.

“We didn’t go there,” I said.

- “I found it”! - answered I.M.

And we took a course to the car. I didn't see any more mushrooms.

The handwriting of Ilya Mikhailovich, as a theoretician, I already guessed well. I once managed to observe when he enthusiastically worked. He then developed the "theory of crankshafts." I was surprised that I.M. writes out transformations in formulas very slowly. And I knew how fast he writes articles. Watching, I realized: I.M. I'm sure he's not wrong. However, he liked, if possible, to illustrate (for himself) the theory with some numerical example.

“Now, with the advent of computers, it is convenient to do this,” I remarked.

- “I'm very afraid that young people who quickly master computers will not master not only mental arithmetic, but also physics and mathematics”, - he answered.

One morning, going to work, we met with I.M. at the very entrance to the injector building.
- “And how to set the potentials on the modulated electrodes of the CNC?” - he asked.

- “In a four-chamber resonator of the magnetron type”, - I answered.

(Viktor Borisovich Stepanov then conducted experiments with a model of a four-chamber resonator. He tried to excite a rotating wave in it).

Ilya Mikhailovich liked the idea of a four-chamber resonator.
- “How simple!” - he said.

Two evenings later, in Protvino, he wrote the theory of a four-chamber resonator. Naturally, he would like to quickly compare the natural frequencies of the quadrupole and dipole modes of oscillation with his theory. Soon such an opportunity appeared. V. B. Stepanov recalled: Ilya Mikhailovich was very surprised - the measured frequencies differed greatly from the calculated ones (the end regions of the resonator introduced a large detuning).

On weekends, Ilya Mikhailovich and I often went to Moscow together (our families lived almost nearby). On the way, we had enough time to talk, and not only about accelerators. I.M. was an interesting conversationalist! Once we got into a conversation, and he said that he wanted to do literary work. I think he could do it! I treated and still treat Ilya Mikhailovich with great respect as an accelerator-theorist.

I was particularly impressed by his work on the microcanonical charge distribution. It was this work that inspired accelerators all over the world to increase the current of accelerated protons in linear accelerators by several times and bring it to the limit value.

Under the leadership of Ilya Mikhailovich Kapchinsky, two of the largest linear proton accelerators in the USSR were built at ITEP and IHEP. They are still in operation. Under his leadership, the ITEP created NCCH (Initial Part of the Accelerator) with HFC-(High-Frequency Quadrupole Focusing). Many of Ilya Mikhailovich's employees became candidates and doctors of sciences.

**List of scientific papers**

# СПИСОК НАУЧНЫХ ТРУДОВ

Список научных трудов Капчинского Ильи Михайловича

| NN п/п | Наименование научного труда | Печатный или рукописный | Издательство, журнал (N, стр., год) или N авторского свидетельства | Колич. печатн. листов | Фамилии соавторов работы |
|---|---|---|---|---|---|
| 1 | 2 | 3 | 4 | 5 | 6 |
| 1 | Нелинейные искажения в RC-генераторах синусоидальных колебаний | печатн. | ЖТФ, том XVI, вып. 8, 1946, с. 893 | 1,5 л | |
| 2 | ~~[illegible]~~ Высокочастотный модулятор | | Авт. свид. ~~[illegible]~~ №16018 с приоритетом 10 окт. 1950. | | |
| 3 | Теория RC-генераторов в области гармонического режима (кандидатская диссертация) | рукопись | МЭИ, 1947 | 5 л | |
| 4 | Методы теории колебаний в радиотехнике (монография) | печатн. | Госэнергоиздат 1954 | 22 л | |
| 5 | Физическое обоснование проекта линейного ускорителя протонов И-2. | рукопись | Отчет ИТЭФ АН, 1958, Москва. | 1 л | |

| 1 | 2 | 3 | 4 | 5 | 6 |
|---|---|---|---|---|---|
| 6 | Методы теории колебаний в радиотехнике | печатн. | Сёко сюпен-ся, Токио, 1959 (на японском языке) | 22 л | |
| 7 | Limitations of proton beam current in a strong focusing linear accelerator associated with the beam space charge. | печатн. | Proc. of the Intern. Conf. on high-energy accelerators and instrumentation. CERN, 1959, p. 274 | 1,5 л | [illegible] V. V. Vladimirsky |
| 8 | Нагрузка ускоряющего поля в линейном ускорителе-инжекторе током протонного пучка | рукопись | Отчет ИТЭФ АН, 1960, Москва | 0,8 л | |
| 9 | Физическое обоснование проекта линейного ускорителя протонов на энергию 100 МэВ | рукопись | Отчет ИТЭФ АН, 1960, Москва | 3 л | |
| 10 | Теория и расчет фокусировки пучков с большим пространственным зарядом в [illegible] х жестко-фокусирую[illegible] (докторс-кая диссер[illegible] | рукопись | РТИ АН, 1962, Москва. | 12 л | |

| 1 | | 3 | 4 | 5 | 6 |
|---|---|---|---|---|---|
| 11 | О токов инжекции в протонном синхро= троне с жесткой фокусировкой | печати. | Атомная энергия, том 13, вып. 3, 1962, с. 235 | 0,6л | |
| 12 | Взаимодействие тока пучка с ускоряющим полем в линейном ускорителе протонов | печати. | Препринт ИТЭФ АН №106, 1962, Москва. | 2л | |
| 13 | Магнитные квадрупольные линзы для линейного ускорителя с про= летными трубками | печати. | ПТЭ №3, 1963, с. 15 | 0,5л | В.К. Плотников |
| 14 | Расплывание пучка заряженных частиц | печати. | Радиотехника и электроника, №6, 1963, с. 985 | 0,5л | |
| 15 | Влияние пространственного заряда на фазовые колебания частиц в ионном линейном ускорителе | печати. | Труды Международной конференции по ускорителям, Дубна, 1963, с. 906 | 0,8л | А.С. Кронрод |
| 16 | Методы теории колебаний в | | Пекин, 1963, | 22л | |

| 1 | 2 | 3 | 4 | 5 | 6 |
|---|---|---|---|---|---|
| 17 | О проекте инжектора для протонного синхротрона на 70 ГэВ | печатн. | Труды Международной конференции по ускорителям, Дубна, 1963, с. 906 | 0,8л | В.Г. Кульман, И.В. Лазарев, Б.П. Мурин, И.Х. Невяжский, В.К. Плотников, Б.И. Полязков |
| 18 | Влияние пространственного заряда на фазовые колебания частиц в ионном линейном ускорителе | печатн. | ПТЭ №3, 1964, с. 26 | 0,8л | А.С. Кронрод |
| 19 | Динамика частиц в линейных резонансных ускорителях | печатн. | Выпуск РТИ АН, НДТ-641, 1964 | 14л | |
| 20 | Основные физические параметры линейного ускорителя И-2 | печатн. | Препринт ИТЭФ-389 [illegible] 1965, Москва | 2л | В.К. Плотников |
| 21 | Динамика частиц в линейных резонансных ускорителях (монография) | печатн. | Атомиздат, 1966 | 20л | |
| 22 | Запуск линейного ускорителя протонов с жесткой фокусировкой на энергию 25 МэВ | печатн. | Атомная энергия, Том 22, вып. 3, 1967, с. 239 | 0,2л. | В.А. Батанин, В.Г. Кульман, И.В. Лазарев, Б.П. Мурин, И.Х. Невяжский, В.К. Плотников, Б.И. Полязков, А.И. Солнышков |
| 23 | Расчетные значения физических параметров линейного ускорителя И-100. | печатн. | Препринт ИФВЭ, ИНЖ 67-38, Серпухов, 1967 | 2л | А.П. Мальцев, В.К. Плотников |

| 1 | 2 | 3 | 4 | 5 | 6 |
|---|---|---|---|---|---|
| 24 | Линейный ускоритель протонов И-2 на энергию 25 МэВ | печатн. | ПТЭ №5, 1967, с. 9 | 0,5 л | В.А. Баталин, Е.Н. Данильцев, А.А. Жданко и др. |
| 25 | Линейный ускоритель протонов И-2 I. Характеристики продольного и поперечного движения частиц | печатн. | ПТЭ №5, 1967, с. 12 | 0,5л | В.К. Плотников |
| 26 | Линейный ускоритель протонов И-2 III. Фокусирующая система | печатн. | ПТЭ №5, 1967, с. 28 | 0,5л. | Р.М. Венгров, Н.Д. Веселов, Г.В. Веселова и др. |
| 27 | Линейный ускоритель протонов И-2 VIII. Наблюдения за поведением фундамента | печатн. | ПТЭ №5, 1967, с.52 | 0,3л | Н.И. Парфёнов, Е.Н. Данильцев, М.В. Щелканов. |
| 28 | Линейный ускоритель протонов И-2 XI. Наладка режима ускорения и параметры пучка | печатн. | ПТЭ №5, 1967, с. 65 | 0,5л | В.А.Баталин, И.В. Лазарев, В.К. Плотников, Б.И. Поляков |
| 29 | Требования к предварительным каскадам ускорения, связанные с повышением интенсивности пучка в протонном синхротроне на 70 ГэВ. | печатн. | Препринт ИФВЭ ИНЖ 68-14-К, Серпухов, 1968. | 1л | |
| 30 | Экспериментальное изучение динамики протонов в жесткофокусирующем линейном ускорителе. | печатн. | Атомная энергия, Том 27, вып. 5, 1969, с. 428 | 0,5л | В.А. Баталин В.К. Плотников |

| 1 | 2 | 3 | 4 | 5 | 6 |
|---|---|---|---|---|---|
| 31 | 25-MeV Proton Linac | печатн. | Proc. of the 6-th Intern. Conf. on high-energy accelerators, Cambrige, USA, 1967, p. A 30 - A31 | 0,3л | N.V. Lazarev, V.K. Plotnikov, V.G. Kulman, B.P. Murin et al. |
| 32 | Adjustment of acceleration conditions and beam parametars of I-2 linear accelerator. | печатн. | —"—, p. A1 - A7 | 0,5л | V.A. Batalin, N.V. Lazarev, V.K. Plotnikov, B.I. Poljakov |
| 33 | Injector for the Serpukhov PS: start-up works | печатн. | —"—, p. 255 | 0,3л | A.L. Mintz, I.K. Nevjazhsy et al |
| 34 | Motion of accelerated bunches taking into account the proper field | печатн. | —"—, p. A148 | 0,3л | D.G. Koshkarev |
| 35 | Продольное кулоновское расталкивание ионов в линейном ускорителе при предельно высоких значениях фазовой плотности пучка | печатн. | Атомная энергия, т. 25, вып. 2, 1968, с. 104 | 0,3л | |
| 36 | Методика определения расчетного режима ускорения в инжекторе протонного синхротрона ИФВЭ | печатн. | Препринт ИФВЭ ИНЖ 68-49-К, Серпухов, 1968 | 0,4л | А.Г. Афонин, С.А. Ильевский, Л. М. Потименкова, В. А. Тепляков |

| 1 | 2 | 3 | 4 | 5 | 6 |
|---|---|---|---|---|---|
| 37 | Параметры линейного ускорителя И-100 при ускорении пучков высокой интенсивности | печатн. | Препринт ИФВЭ ИНЖ 68-63, Серпухов, 1968. | 0,6л | С. А. Ильевский, Б. И. Поляков, Л. М. Попиненкова и др. |
| 38 | Изучение поперечных колебаний пучка в жесткофокусирующем канале методом выключения линз | печатн. | Препринт ИФВЭ ИНЖ 69-3 Серпухов, 1969 | 0,5л | В.К. Плотников Э.Д. Деренко |
| 39 | Линейный ускоритель ионов с пространственно-однородной жесткой фокусировкой | печатн. | ПТЭ № 2, 1970 с. 19 | 0,5л | В. А. Тепляков |
| 40 | Проводка пучка в линейном ускорителе протонов на энергию 100 МэВ | печатн. | Препринт ИФВЭ ИНЖ 69-9 Серпухов, 1969 | 0,5л | В.А. Баталин, С.А. Ильевский, В.В. Нижегородцев, В.К. Плотников и др. |
| 41 | О возможности снижения энергии инжекции и повышения предельного тока в ионном линейном ускорителе | печатн. | ПТЭ № 4, 1970 с. 17 | 0,3л | В. А. Тепляков |
| 42 | Проектные соображения о реконструкции протонного синхротрона ИТЭФ | печатн. | Препринт ИТЭФ-611 [illegible] ч. III, Москва, 1968 | 0,3л | Л. Л. Гольдин Д. Г. Кошкарев |
| 43 | Перевод протонного синхротрона ИТЭФ на работу от инжектора — линейного ускорителя на 25 МэВ | печатн. | ПТЭ № 6, 1969 с. 14 | 0,3л | Л. З. Барабаш, В.А. Баталин, И.А. Веселов и др. |

| 1 | 2 | 3 | 4 | 5 | 6 |
|---|---|---|---|---|---|
| 44 | Исследование спектра пучка на выходе линейного ускорителя ИТЭФ. | печатн. | Труды Всесоюзного совещания по ускорителям заряженных частиц т. I (Москва, 1968) 1970, с. 551 | 0,3л | В. А. Баталин<br>В. К. Плотников |
| 45 | О проекте линейного ускорителя протонов с пониженной энергией инжекции и высокой интенсивностью пучка | печатн. | Международная Конференция по ускорителям частиц высоких энергий, т. I, Ереван, 1970, с. 153 | 0,2л | А. П. Мальцев<br>В. А. Тепляков |
| 46 | Ускорение протонных пучков высокой интенсивности в инжекторе Серпуховского синхротрона | печатн. | —"— с. 157 | 0,2л | С. А. Ильевский, В. Г. Кульман, А. Г. Лонидзе, Б. П. Мурин, и др. ~~Б. И. Поляков, А. М. Пономаренко, В. А. Тепляков~~, ~~В. Г. Тишин~~ |
| 47 | Пуск секции линейного ускорителя с высокочастотной квадрупольной фокусировкой. | печатн. | Атомная энергия, т. 34, вып. 1, 1973, с. 56 | 0,25л | С. А. Ильевский, Г. В. Кузнецов, А. П. Мальцев, К. Г. Мирзоев и др. |
| 48 | Экспериментальные и теоретические исследования динамики пучка в линейном ускорителе И-2 и разработка новых узлов ускорителя (1970) | печатн. | Препринт ИТЭФ-894 ~~[illegible]~~ Москва, 1972 1971. | 2,5л | В. А. Баталин, В. И. Бобылев, Е. Н. Данильцев, ~~А. В. Карцев~~, ~~А. М. Козодаев, Б. В. Колосков~~, ~~Р. П. Куйбида, И. В. Лазарев~~, ~~В. К. Плотников~~, ~~В. И. Эдемский~~ и др |

| 1 | 2 | 3 | 4 | 5 | 6 |
|---|---|---|---|---|---|
| 49 | Новые разработки и усовершенствования технологических систем линейного ускорителя И-2 | печатн. | Труды Второго Всесоюзного совещания по ускорителям заряженных частиц, т. I (Москва, 1970), 1972, с. 64 | 0,2 л | В.А. Баталин, В.И. Бобылев, Е.И. Денисьцев и др. |
| 50 | Линейный ускоритель ионов | | Авт. свид. № 265312 с приоритетом от 25 окт. 1968 | | В.В. Владимирский, В.А. Тепляков |
| 51 | Линейный ускоритель ионов с высокочастотной жесткой фокусировкой. Динамика пучка в секциях с пространственно-однородной фокусировкой. Часть I | печатн. | Препринт ИФВЭ ИНЖ 72-19, Серпухов, 1972 | 1 л | |
| 52 | Линейный ускоритель ионов с высокочастотной жесткой фокусировкой. Динамика пучка в секциях с пространственно-однородной фокусировкой. Часть II | печатн. | Препринт ИФВЭ ИНЖ 72-30 Серпухов, 1972 | 1 л | |
| 53 | Экспериментальные и теоретические исследования динамики пучка в линейном ускорителе И-2 и разработка новых узлов ускорителя (1971) | печатн. | Препринт ИТЭФ-960 Москва, 1972 | 2,5 л | В.А. Баталин, Н.В. Лазарев и др. |

| 1 | 2 | 3 | 4 | 5 | 6 |
|---|---|---|---|---|---|
| 54 | Ускоряющая система | | Авт. свид. №317 350 с приоритетом от 11 мая 1970 | | В. А. Тепляков<br>В. Б. Степанов |
| 55 | The experience of big pulsed current acceleration on the linac I-2 | печатн. | Proc. of the 1972 proton linear accelerator Confer. Los Alamos, USA, p. 275 | 0,5л | V. A. Batalin,<br>N. V. Lazarev et al |
| 56 | Многоцелевое использование протонного линейного ускорителя И-2. | печатн. | Атомная энергия, Том 33, вып. 5, 1972, с. 938 | 0,2л | В. А. Баталин,<br>В. И. Бобылев,<br>Е. И. Данильцев и др. |
| 57 | Увеличение интенсивности пучка в инжекторе протонного синхротрона ИТЭФ | печатн. | ПТЭ №5, 1972, с. 17 | 0,2л | В. А. Баталин, А. А. Коломиец, Б. К. Кондратьев, Р. П. Куйбида |
| 58 | Увеличение интенсивности пучка протонного линейного ускорителя | печатн. | ПТЭ №1, 1973, с. 15<br>Труды 3[го] Всесоюзного совещания по ускорителям заряженных частиц, т. I (Москва, 1972), 1973, с. 275 | 0,3л | В. А. Баталин, А. А. Коломиец, ~~Б. К. Кондратьев, Р. П. Куйбида, В. С. Кузнецов, И. П. Кузнецова, Г. И. Трубников~~ и др |

| 1 | 2 | 3 | 4 | 5 | 6 |
|---|---|---|---|---|---|
| 59 | Оценка возможности создания коллективного ускорителя с многократным использованием электронных колец | печатн. | Препринт ИТЭФ-63 [redacted] Москва, 1973 | 1л | |
| 60 | О выборе закона модуляции электронов по энергии для подавления азимутальной когерентной неустойчивости кольцевого пучка | печатн. | Препринт ИТЭФ-80 [redacted] Москва, 1973 | 1,5л | П. Р. Зенкевич |
| 61 | Некоторые исследования работы линейного ускорителя И-2 при выходном токе пучка до 200 мА. Разработка новых технологических узлов (1972) | печатн. | Препринт ИТЭФ-106 [redacted] Москва, 1973 | 2,5л | В. А. Баталин, В. И. Бобылев и др. |
| 62 | О коллективном сильноточном ускорителе | печатн. | Атомная энергия, т. 37, вып. 3, 1974, с. 223 | 0,4л | И. В. Чувило, Д. Г. Кошкарев, П. Р. Зенкевич, В. К. Плотников, Ю. Г. Глобенко |
| 63 | Коллективные методы ускорения и развитие работ этого направления в ИТЭФ. | печатн. | Препринт ИТЭФ-60 [redacted] Москва, 1974 | 0,6л | П. Р. Зенкевич В. К. Плотников И. В. Чувило |

| 1 | 2 | 3 | 4 | 5 | 6 |
|---|---|---|---|---|---|
| 64 | Модуляция протонного пучка линейного ускорителя по энергии | печатн. | Атомная энергия, т. 37, вып. 5, 1974, с. 393 | 0,4л | В. И. Бобылев |
| 65 | ITEP experimental assembly for ERA collective method investigations. | печатн. | Proc. of the Int. Conf. on high-energy accelerators, Stanford, 1974, p. 314 | 0,4л | I. V. Chuvilo, V. K. Plotnikov, R. M. Vengrov |
| 66 | Разработка новых узлов и эксплуатация линейного ускорителя И-2 в 1973 г. | печатн. | Препринт ИТЭФ-10 [redacted], Москва, 1975 | 3,25л | И. В. Лазарев, В. А. Батанин, В. И. Бобылев и др. |
| 67 | О силе, удерживающей ионы в ускоряемом электрон-ионном кольце | печатн. | Препринт ИТЭФ-12 [redacted], Москва, 1975 | 0,8л | |
| 68 | Предложение по программе сооружения ускорителей протонов на сверхвысокие энергии | печатн. | Препринт ИТЭФ-11 [redacted], Москва, 1975 | 0,8л | В. В. Владимирский, Б. Л. Иоффе и др. |
| 69 | О возможности использования постоянных магнитов в жесткофокусирующих каналах ионных линейных ускорителей | печатн. | Препринт ИТЭФ-78 [redacted], Москва, 1975 | 0,5л | И. В. Лазарев |
| 70 | Некоторые исследования, разработки и опыт эксплуатации линейного ускорителя И-2 в 1974 году. | печатн. | Препринт ИТЭФ-86 [redacted], Москва, 1975 | 3л | В. А. Батанин, В. И. Бобылев и др. |

| 1 | 2 | 3 | 4 | 5 | 6 |
|---|---|---|---|---|---|
| 71 | Развитие теории и техники линейных резонансных ускорителей ионов | печатн. | Труды IV Всес. Совещ. по ускорителям заряженных частиц, т. I, Москва, 1975, с. 114 | 1 л | |
| 72 | Реконструкция линейного ускорителя протонов И-2 | печатн. | Труды IV Всес. Совещ. по ускорителям заряженных частиц, т. I, Москва, 1975, с. 167 | 0,3 л | В.А. Баталин, В.И. Бобылев, Н.В. Лазарев и др. |
| 73 | Источник ионов | | Авт. свид. N 528 815 с приоритетом от 29 дек. 1975 | | В.А. Баталин, Ю.А. Быковский, М.А. Кожушнер, Ю.П. Козырев, И.И. Левинтов |
| 74 | Опыт работы инжектора И-2 на три канала вывода пучка | печатн. | Вопросы атомной науки и техники, вып 2(3), Харьков, 1976, с. 3 | 0,3 л | В.А. Баталин, В.И. Бобылев, Н.В. Лазарев и др. |
| 75 | О наблюдении продольных когерентных колебаний частиц по измерению распределения поля вдоль резонаторов | печатн. | Вопросы атомной науки и техники, вып. 2 (3), Харьков, 1976, с. 33 | 0,1 л | В.И. Бобылев, А.А. Коломиец, Р.П. Куйбида |
| 76 | Расчет энергетики цилиндрического резонатора, нагруженного трубками дрейфа. | печатн. | Препринт ИТЭФ-15 [illegible] Москва, 1977 | 1 л | И.С. Илларионова, М.А. Ольшанецкий |

| 1 | 2 | 3 | 4 | 5 | 6 |
|---|---|---|---|---|---|
| 77 | О работах по исследованию электронных конц в ИТЭФ | печатн. | Труды IV Всес. Совещ. по ускорителям заряженных частиц, т. I, Москва, 1975, с. 91 | 0,3л | Р.М. Вентров, А.А. Вроздовский, В.В. Куренин, В.К. Плотников, Н.А. Попова, И.В. Чувило |
| 78 | Энергетика цилиндрического резонатора, нагруженного трубками дрейфа. | печатн. | Препринт ИТЭФ-16 Москва, 1977 | 1,5л | И.С. Илларионова, М.А. Ольшанецкий |
| 79 | Design and development of permanent magnet quadrupoles for ion linacs | печатн. | Proc. of the Linac Conf., Chalk River, 1976, p. 350 | 0,3л | A.M. Kozodaev, N.V. Lazarev, V.S. Skachkov |
| 80 | О проекте линейного ускорителя для испытательного нейтронного генератора | печатн. | ПТЭ №4, 1977, с. 23 | 0,3л | |
| 81 | Применение линейного ускорителя И-2 для прикладных исследований | печатн. | Доклады Второго Всес. Совещ. по применению ускорителей в народном хозяйстве (1975), Ленинград, 1976, с. 249 | 0,3л | В.А. Баталин, О.М. Гудков и др. |
| 82 | Схема интенсивного нейтронного генератора на 14 МэВ с двухчастотным линейным ускорителем дейтронов. Предложение ИТЭФ. | печатн. | Препринт ИТЭФ-118 Москва, 1977. | 2л | Б.Л. Иоффе, Н.В. Лазарев, А.Д. Леонгардт, И.В. Чувило, Р.Г. Васильков. |

| 1 | 2 | 3 | 4 | 5 | 6 |
|---|---|---|---|---|---|
| 88 | К вопросу о расширении фазового объема протонного пучка в линейном ускорителе | печатн. | Труды VI Всес. Совещ. по ускорителям заряж. частиц, т. I (Дубна, 1978), 1979, с. 256 | 0,3 л | Р. П. Кульбида |
| 89 | The linear accelerator structures with space-uniform quadrupole focusing | печатн. | IEEE Transactions on NS, NS-26, No3, June 1979, p. 3462 | 0,5 л | N.V. Lazarev |
| 90 | Проблемы сооружения сильноточных линейных ускорителей ионов | печатн. | Сб. „Проблемы ускорителей заряженных частиц" Дубна, 1980, с. 162 | 1 л | |
| 91 | Исследование четырехкамерного H-резонатора. | печатн. | Препринт ИТЭФ-49 Москва, 1980 | 1,5 л | М.Л. Гусев, Е.Н. Данильцев, С.Б. Угаров, В.И. Эдемский [redacted] |
| 92 | Сильноточные линейные ускорители ионов | печатн. | Успехи физ. наук, том 132, вып. 4, 1980, с. 639 | 2,5 л | |
| 93 | Исследования ИТЭФ по тяжелоионному термоядерному синтезу | печатн. | Препринт ИТЭФ-64, Москва, 1981 | 0,3 л | П.Р. Зенкевич, В.С. Имшенник, Д.Г. Кошкарев, В.Г. Шевченко |
| 94 | Устройство для увеличения мощности при генерации сгустков тяжелых ускоренных ионов | | Авт. свид. N 824786 с приоритетом от 19 февраля 1980. | | В. К. Плотников |

| 1 | 2 | 3 | 4 | 5 | 6 |
|---|---|---|---|---|---|
| 95 | О максимальной напряженности поля в структуре с пространственно-однородной квадрупольной фокусировкой. | печатн. | Препринт ИТЭФ-3 Москва, 1982. | 0,2 л | А.И. Балабин |
| 96 | Ускорительный комплекс для инерционного термоядерного синтеза на тяжелых ионах. | печатн. | Препринт ИТЭФ-143 Москва, 1980 | 0,3 л | П.Р. Зенкевич<br>Д.Г. Кошкарев |
| 97 | The pulsed proton prototype of a high current ion linac | печатн. | Proc. of the Linac Conf., Santa Fe, USA, 1981, p. 92 | 0,3 л | R.M. Vengrov<br>E.N. Daniltzev<br>A.M. Kozodaev et al |
| 98 | Ускоряющая система линейного ускорителя | | Авт. свид. N 890957 с приоритетом от 25 июля 1980. | | Р.М. Венгров<br>Е.Н. Данильцев |
| 99 | The launching of a 3-MeV proton RFQ linac | печатн. | IEEE Transactions on Nuclear Science, NS-30, No 4, part II, 1983, p. 3579 | 0,3 л | V.A. Andreev<br>V.S. Artemov et al |
| 100 | Параметры и особенности сильноточного линейного ускорителя для имитационных исследований радиационного повреждения материалов термоядерного реактора | печатн. | Труды VIII Всес. Совещ. по ускорителям заряженных частиц, т. II (Протвино, 1982), 1983 с. 99 | 0,5 л | |

| 1 | 2 | 3 | 4 | 5 | 6 |
|---|---|---|---|---|---|
| 101 | Ускорение пучка в начальной части импульсного протонного прототипа линейного ускорителя ионов. | печатн. | Труды VIII Всес. Совещ. по ускорителям заряж. частиц, т. II, (Протвино, 1982), 1983, с. 3 | 0,2л | В.А. Андреев, В.С. Артемов, В.И. Бобылев и др |
| 102 | Исследования в ИТЭФ по применению тяжелоионных пучков для УТС | печатн. | Труды VIII Всес. Совещ. по ускорителям заряж. частиц, т. I (Протвино, 1982), 1983, с. 92 | 0,5л | П.Р. Зенкевич, В.С. Имшенник, Д.Г. Кошкарев, В.Г. Шевченко |
| 103 | Разработка и запуск линейного протонного ускорителя с пространственно-однородной квадрупольной фокусировкой в ИТЭФ | печатн. | Вопросы атомной науки и техники. Серия: техника физического эксперимента, вып. 3(15), Харьков, 1983, с. 3 | 0,5л | А.И. Балабин, Р.М. Венгров, Е.И. Даниельцев, А.М. Коздаев и др |
| 104 | О выборе профиля электродов согласующего раструба на входе линейного ускорителя с пространственно-однородной квадрупольной фокусировкой | печатн. | Вопросы атомной науки и техники. Серия: техника физического эксперимента, вып. 3(15), Харьков, 1983, с. 39 | 0,25л | А.И. Балабин, И.М. Липкин |
| 105 | History of RFQ development | печатн. | Proc. of the 1984 Linac Conf., Darmstadt, FRG, GSI-84-11 p. 43 | 0,3л | |

| 1 | 2 | 3 | 4 | 5 | 6 |
|---|---|---|---|---|---|
| 106 | Физическое обоснование установки тяжелоионного термоядерного синтеза | рукопись | отчет ИТЭФ У 97114, Москва, 1984 | 11л | В.С. Имшенник, Д.Г. Кошкарев, В.Г. Шевченко |
| 107 | Система ВЧ питания четырехкамерного резонатора для ускорителя ионов | | Авт. свид. N1128756 с приоритетом от 2 апреля 1982. | | В.И. Бобылев, Р.М. Венгров и др. |
| 108 | Ускоряющая структура для линейных ускорителей ионов | | Авт. свид. N 1160915 с приоритетом от 6 января 1984. | | В.В. Кушин |
| 109 | К вопросу о распределениях плотности заряда при автофазировке пучков в линейных ускорителях | печатн. | Препринт ИТЭФ-151, Москва, 1985 | 0,4л | И.М. Липкин |
| 110 | Теория линейных резонансных ускорителей (монография) | печатн. | Энергоиздат, Москва, 1982. | 15л | |
| 111 | Линейные ускорители тяжелых малозарядных ионов для инерционного термоядерного синтеза | печатн. | Вопросы атомной науки и техники. Серия: Техника физического эксперимента, вып. 2 (23), Харьков, 1985, с. 10 | 0,5л | В.В. Кушин, И.В. Лазарев |

| 1 | 2 | 3 | 4 | 5 | 6 |
|---|---|---|---|---|---|
| 112 | Ускоряющая структура с пространственно-однородной квадрупольной фокусировкой для тяжелых ионов на частоте 6 МГц | печатн. | Труды IX Всес. Совещ. по ускорителям заряж. частиц, том I (Дубна, 1984), 1985, с. 218 | 0,3л | Е.Н. Данильцев; А.Ю. Дзюн, А.Б. Зарубин и др. |
| 113 | Опыт использования несвнопольных квадрупольных линз с постоянными магнитами на линейном ускорителе И-2. | печатн. | Труды IX Всес. Совещ. по ускорителям заряж. частиц, том II, (Дубна, 1984), 1985 с. 56 | 0,3л | В.С. Скачков, В.С. Артемов и др. |
| 114 | Исследование возможности получения ионов гелия в дуоплазматроне с холодным катодом. | печатн. | Вопросы атомной науки и техники. Серия: техника физического эксперимента, вып. 1 (22), Харьков, 1985, с. 48 | 0,2л | В.А. Баталин, Ю.Н. Волков, Б.К. Кондратьев и др. |
| 115 | Theory of Resonance Linear Accelerators | печатн. | Harwood academic publishers, London-Paris-New York, 1985 | 15л | |
| 116 | О согласовании пучка с пространственно-однородным квадрупольным каналом | печатн. | ЖТФ, том 55, вып 3, 1985, с. 586 | 0,3л | А.И. Балабин, В.С. Кабанов, В.В. Кушин, и др. |

| 1 | 2 | 3 | 4 | 5 | 6 |
|---|---|---|---|---|---|
| 117 | Об аппроксимациях критерия Килпатрика | печатн. | ПТЭ № 1, 1986 с. 33 | 0,2 л | |
| 118 | О некоторых результатах работ по проблеме управляемого термоядерного синтеза на тяжелых ионах в ИТЭФ | печатн. | Препринт ИТЭФ-133 Москва, 1985 | 0,5 л | П. Р. Зенкевич, В.С. Ищенко, Д.Г. Кошкарев, Н.Д. Гурзов, В.Г. Шевченко |
| 119 | Исследование характеристик пучка в электронной пушке ускорителя ЛИУ-5/5000 с десорбционным катодом. | печатн. | Вопросы атомной науки и техники. Серия: техника физического эксперимента, вып 3(24), Харьков, 1985, с. 15 | 0,3 л | В.В. Зиновьев, С.В. Мартынов и др. ~~[illegible] Ю.В. Павлинов, В.И. Першин, В.К. Плотников, Н.Я. Попова~~ |
| 120 | Численное исследование шестимерного согласования пучка в линейном ускорителе с пространственно-однородной квадрупольной фокусировкой. | печатн. | Вопросы атомной науки и техники. Серия: техника физического эксперимента, вып. 3(24), Харьков, 1985, с. 54 | 0,3 л | А. И. Балабин, В.С. Кебанов, В.В. Кушин, И. М. Липкин |
| 121 | RF linac for heavy ion fusion driver | печатн | Proc. of the 1986 Linear Accel. Conf. SLAC-Report-303, 1986, p. 318 | 0,5 л | V.V. Kushin, N.V. Lazarev, V.G. Shevchenko et al |

| 1 | 2 | 3 | 4 | 5 | 6 |
|---|---|---|---|---|---|
| 122 | RF linac for heavy ion fusion driver | печати | AIP Conf. Proc. 152, Heavy ion inertial fusion. Washington DC, 1986, p. 49 | 0,5л | V.V. Kushin, N.V. Lazarev, V.G. Shevchenko et al |
| 123 | Пути улучшения параметров линейного индукционного ускорителя ЛИУ-5/5000. | печатн. | Препринт ИТЭФ-163 Москва, 1987 | 1,75л | В.И. Бобылев, Ю.Я. Лапицкий, В.К. Плотников, И.В. Чувило |
| 124 | Поперечное согласование пучка в ускорителе с пространственно-однородной квадрупольной фокусировкой | печатн. | Труды X Всес. Совещ. по ускорителям заряженных частиц, т. I (Дубна, 1986), 1987, с. 403 | 0,5л | А.И. Балабин, И.А. Воробьев, ~~А.М. Козодаев, А.А. Козлинец,~~ И.В. Шахров и др. |
| 125 | Ускорение ионов $He^{2+}$ в линейном протонном ускорителе И-2 до энергии 24 МэВ/ядро. | печатн. | Вопросы атомной науки и техники. Серия: техника физического эксперимента, вып. 4 (35), Харьков, 1987, с. 3 | 0,2л | В.С. Артемов, В.А. Баталин, В.В. Кушин, Н.В Лазарев и др. |
| 126 | Ускорение ионов $Xe^{2+}$ в первой секции линейного ускорителя для установки инерционного УТС на пучке тяжелых ионов | печатн. | Труды XIII Междунар. Конф. по ускорителям частиц высоких энергий, (Новосибирск, 1986), т. I, 1987, с. 237 | 0,3л | В.С. Артемов, В.А. Баталин, Е.И. Данильцев и др. |

| 1 | 2 | 3 | 4 | 5 | 6 |
|---|---|---|---|---|---|
| 127 | Фокусирующий канал ускорителя протонов на энергию 10 МэВ | печатн. | Вопросы атомной науки и техники. Серия: техника физического эксперимента, вып. 1 (36), Харьков, 1988, с. 14 | 0,1л | Р.М. Венгров<br>Р.П. Куйбида<br>Н.В. Лазарев и др. |
| 128 | Наладка режима ускорения ионов гелия в протонном линейном ускорителе И-2 | печатн. | Препринт ИТЭФ-165-88 Москва, 1988. | 1л | В.С. Артемов, В.А. Баталин, Ю.Е. Корчагин и др. |
| 129 | Экспериментальное изучение динамики пучка ионов гелия в линейном ускорителе И-2 | печатн. | Препринт ИТЭФ-166-88 Москва, 1988. | 0,5л | Р.А. Романовский<br>В.С. Столбунов |
| 130 | Развитие линейных ускорителей ионов с высокочастотной квадрупольной фокусировкой | печатн. | Труды XI Всес. Совещ. по ускорителям заряженных частиц, т. I, (Дубна, 1988), 1989, с. 37 | 0,6л | В.А. Тепляков |
| 131 | Сооружение в ИТЭФ линейного ускорителя протонов „Истра-56" | печатн. | Труды XI Всес. Совещ. по ускорителям заряженных частиц, т. I, (Дубна, 1988), 1989, с. 245 | 0,3л | В.А. Андреев, В.С. Артемов, А.И. Балабин, Р.М. Венгров и др. |

| 1 | 2 | 3 | 4 | 5 | 6 |
|---|---|---|---|---|---|
| 132 | Основные физические параметры протонного линейного ускорителя „Истра-56" (часть I) | печатн. | Препринт ИТЭФ-157-89 Москва, 1989 | 2л | А.И. Балабин, А.М. Козодаев, А.А. Коломиец, И.М. Лыткин, С.Б. Угаров |
| 133 | Основные физические параметры протонного линейного ускорителя „Истра-56" (часть II) | печатн. | Препринт ИТЭФ-158-89 Москва, 1989 | 0,8л | И.А. Воробьев, А.М. Козодаев, А.А. Коломиец, И.М. Лыткин |
| 134 | Пробный ввод ВЧ мощности в резонатор Альвареца на частоте 297 МГц. | печатн. | Вопросы атомной науки и техники. Серия: Ядерно-физические исследования, вып. 5(5), Харьков, 1989, с. 10 | 0,3л | В.С. Артемов, Р.М. Венгров, А.М. Козодаев и др. |
| 135 | Увеличение предельной частоты лампы ГИ-27АМ при создании импульсного генератора на 300 МГц с мощностью 3,5 МВт для питания линейного ускорителя протонов | печатн. | II Всес. Совещ. по новым методам ускорения заряженных частиц. Нор-Амберд, Армения, (аннотации докладов) 1989, с.8 | | В.А. Звягинцев, Н.В. Лазарев и др. |
| 136 | Подготовка и пробное включение второго варианта начальной части ускорителя „Истра-56" | рукопись | Отчет ИТЭФ № 663 Москва, 1989 | 2л | А.М. Козодаев, В.В. Кушин и др. |

| 1 | 2 | 3 | 4 | 5 | 6 |
|---|---|---|---|---|---|
| 137 | Разработка методики юстировки линейного ускорителя „Истра-10" | рукопись | Отчет ИТЭФ № 565 Москва, 1988 | 1л | А.М. Козодаев, Р.М. Венгров и др |
| 138 | ~~Технология изготовления трубок дрейфа для первого~~ резонатора на энергию 10МэВ | рукопись | Отчет ИТЭФ № 697 Москва, 1989 | 5л | В.В. Куракин, В.С. Скачков К.И. Бурцев и др. |
| 139 | Drift tubes for focusing channel of ion linear accelerator | печатн. | Proc. of the 1989 IEEE Particle Accel. Confer., vol. 2, Chicago, 1989, p.1073 | 0,25л | [illegible] V.S. Skachkov, A.M. Kozodaev, V.V. Kurakin et al |
| 140 | The 6 MHz RFQ linac for [illegible] HIF driver | печатн. | Proc. of the 1987 IEEE Particle Accel. Confer., vol. 1, Washington DC, 1987, p. 388 | 0,3л | V.S. Artemov, A.I. Balabin et al |
| 141 | Acceleration of Key ions in the APF structure and the linac I-2 for injection into 10 GeV synchrotron | печатн. | Proc. of the 1988 Linac Conf., CEBAF-Report-89-001, Newport News, 1989, p 146 | 0,2л | I.V. Chuvilo, V.S. Artemov et al. |
| 142 | Подготовка и проведение физического пуска линейного протонного ускорителя „Истра-10" | рукопись | Отчет ИТЭФ № 719 Москва, 1990 | | В.А. Андреев, В.С. Артемов, В.И. Беляев и др. |

| 1 | 2 | 3 | 4 | 5 | 6 |
|---|---|---|---|---|---|
| 143 | К выбору метода адиабатической группировки пучков в линейном ускорителе с пространственно-однородной квадрупольной фокусировкой | печатн. | Препринт ИТЭФ-64-90 Москва, 1990 | 0,3л | |

/Капчинский И.М./

Ученый секретарь ИТЭФ /Терехов Ю.В./

| | | | | |
|---|---|---|---|---|
| 144 | Пространственно-однородная квадрупольная фокусифовка | Печ. | Физическая энциклопедия. Том 4, с. 154-155. М. Научное изд. "Большая Российская энциклопедия", 1994. | И.М. Капчинский (*Статья была написана и направлена автором в редакцию ФЭ в начале 1993 года*). |
| 145 | Selected Topics in Ion Linac Theory. | | Los Alamos, 1993, LA-UR-93-4192. | И.М. Капчинский (*Этот курс лекций, прочитанный автором в Мэрилендском Университете США, был издан в Лос-Аламосе в конце 1993 года*). |
| 146 | Linear Accelerator for Investigation of Radiation Damages in Fusion Reactor Materials | Печ., 5 стр. | Workshop on Intense Neutron Sources for Fusion Materials Testing, EAA, IAE (I.V. Kurchatov Inst.) July 6-8, 1993, | I.M. Kapchinsky, I.V. Chuvilo A.M. Kozodaev,, A.A. Drozdovsky, N.V. Lazarev, V.K. Plotnikov, I.A. Vorobjov |
| 147 | Linear Accelerator for Plutonium Conversion and Transmutation of NPP Wastes | Печ. | Proc. of the Particle Accelerator Conf., Washington, May 17-20, 1993, p.1675-1680. | I.M. Kapchinsky, I.V. Chuvilo, A.A. Kolomiets, N.V. Lazarev, I.M. Lipkiin, V.K. Plotnikov, I.A. Vorobjov |
| 148 | Circular Permanent Magnet Quadrupoles for Higher Frequency and Higher Shunt Impedance Linacs | Печ. | Proc. of the 3rd EPAC, Berlin, March 24-28, 1992, Vol.2, p. 1400-1402. | V.S. Skachkov, A.V. Selin, I.M. Kapchinsky, N.V.Lazarev |

| | | | |
|---|---|---|---|
| 144 | Spacially Uniform Quadrupole Focusing | Physical encyclopedia. Volume 4, p. 154-155. M. Scientific ed. "Great Russian Encyclopedia", 1994. | I.M. Kapchinsky (The article was written and sent by the author to the editors of the Faculty of Economics in early 1993. |

| | | | | |
|---|---|---|---|---|
| 149 | Linear Accelerators for Transmutation of NPP Wastes | Печ. | Preprint ITEP-100-92, M., 1992, то же в Трудах Рабочего совещания по ускорителям для трансмутации по проекту ATW/ABC Los Alamos, November. 16-20, 1992 | I.M. Kapchinsky, A.A. Kolomiets, N.V. Lazarev, I.M. Lipkiin, I.V. Chuvilo |
| 150 | Permanent Magnet Small-Size Quadrupole Lenses for Ion Linear Accelerators | Печ. | Proc. of the IEEE Transactions on Magnetics, January 1992, Vol. NS 28, No.1, p.531-533. | .M. Kapchinsky, N.V.Lazarev, E.A. Levashova, A.P. Preobraginsky, A.V. Selin, V.S. Skachkov |
| 151 | Радиационно-стойкие квадрупольные линзы на постоянных магнитах | Печ. | Сб. Аннотаций XII Международной конференции по магнитной технологии, Ленинград, 1991г. с.201. | I.M. Kapchinsky, N.V.Lazarev, E.A. Levashova, A.P. Preobraginsky, A.V. Selin, V.S. Skachkov |
| 152 | Сильноточные ионные ЛУ для трансмутации долгоживущих отходов | Печ. | Proc. of the 1990 Proton Linear Accelerator Conf., p.778-781. | П.П. Благоволин, И.М. Капчинский, Н.В. Лазарев, И.В. Чувило |
| 153 | Подготовка и проведение физического пуска протонного ЛУ Истра-10 | Печ. | XII Всесоюзное совещание по ускорителям заряженных частиц. Том 2, с. 57-59, Дубна, 1992. | В.А. Андреев, В.С. Артемов, Р.М. Венгров, И.М. Капчинский, А.М. Козодаев, А.А. Коломиец, Н.В. Лазарев и др. |

151 Radiation-resistant permanent magnet quadrupole lenses, Abstracts of the XII International Conference on Magnetic Technology, Leningrad, 1991. p.201.

152 High-current ion beams for transmutation of long-lived waste.

153 Preparation and conduct of the physical launch of the proton LA Istra-10, XII All-Union Conference on Charged Particle Accelerators. Volume 2, p. 57-59, Dubna. 1992.

| | | | | |
|---|---|---|---|---|
| 154 | Запуск линейного ускорителя протонов Истра-10 | Печ. | Proc. of the 1990 Proton Linear Accelerator Conf., p.782-784. | В.А. Андреев, В.С. Артемов, Р.М. Венгров, И.М. Капчинский, А.М. Козодаев, А.А. Коломиец, Н.В.Лазарев и др. |
| 155 | Подготовка, предварительная настройка и пробное ВЧ возбуждение первого резонатора основной части ускорителя Истра-56 | Рук. на 40 стр. | Отчет ИТЭФ № 582, М., ИТЭФ, 1988 г. | И.М. Капчинский, А.М. Козодаев, В.В. Кушин, Н.В.Лазарев, Р.М. Венгров, В.А. Андреев, А.А. Коломиец, А.М. Раскопин, В.С. Артемов, В.С. Косяк и др. |
| 156 | Разработка новых элементов и исследование динамики пучка установки НЧУ-1 – первого варианта начальной части импульсного протонного прототипа сильноточного ЛУ | Рук. на 179 стр. | Отчет ИТЭФ № 556, ИТЭФ, М., ИТЭФ, 1988 г. | И.М. Капчинский, А.М. Козодаев, В.В. Кушин, Н.В.Лазарев, Р.М. Венгров, В.А. Андреев, А.А. Коломиец, Ю.Б. Стасевич, А.М. Вишневский и др. |
| 157 | Экспериментальные работы на ускорителе тяжелых малозарядных ионов ТИПр-1 в 1988 г. | Рук. на 16 стр. | Отчет ИТЭФ о НИР № 4.08.033, М., 1988 г. | В.А. Баталин, А.Б. Зарубин, И.М. Капчинский, В.В. Кушин, Н.В. Лазарев |

154 Launch of the linear proton accelerator Istra-10
155 Training, preliminary tuning and trial RF excitation of the first resonator of the main part of the Istra-5 accelerator 6. on page 40 ITEP Report No. 582, Moscow, ITEP, 1988
156 Development of new elements and study of the beam dynamics of the NChU-1 facility - the first version of the initial part of the pulsed proton prototype of a high-current LA. on page 179 ITEP Report No. 556, ITEP, Moscow, ITEP, 1988
157 Experimental work at the TIPR-1 heavy low-charge ion accelerator in 1988. on page 16 ITEP Research Report No. 4.08.033, Issue 1988

## Afterword

In this short collection, dedicated to the memory of Ilya Mikhailovich Kapchinsky, our remarkable leader and elder friend, the main results of his extensive scientific, pedagogical and engineering activities are only very superficially covered. According to all 150 published scientific works of I.M. it is very difficult to make any serious review.

His life was cut short as if by a shot quite suddenly. Numerous folders remained, both with already completed and with just started work. I.M. achieved a lot - world fame and recognition as a leader in an important direction in accelerator science and technology. A number of his publications, starting with the very first, prepared in 1959 jointly with V.V. Vladimirsky to the International Conference at CERN, all four monographs, as well as articles in the journals PTE, UFN, AE, etc., in the Physical Encyclopedia, are a kind of intellectual wealth open to all those interested in this field of knowledge. His works are by no means outdated; they continue to be used in mastering the theory of resonant linear ion accelerators and in developing projects for more powerful and advanced machines of this class. There is no doubt that if Ilya Mikhailovich's life had lasted up to the present time, we would have become witnesses of his new scientific insights and technical achievements.

N.V. Lazarev

## Photos and Documents

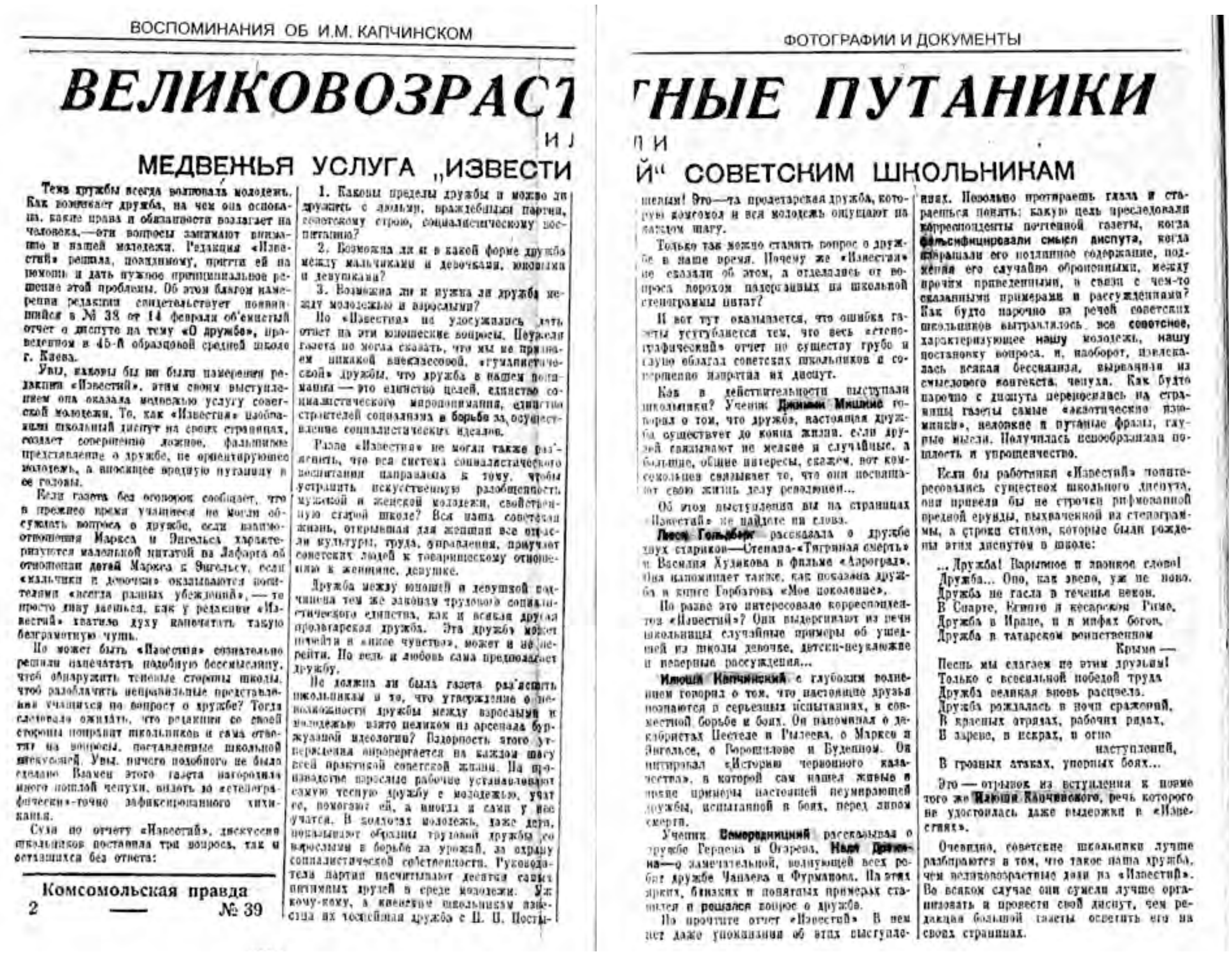

ВОСПОМИНАНИЯ ОБ И.М. КАПЧИНСКОМ

ФОТОГРАФИИ И ДОКУМЕНТЫ

# ВЕЛИКОВОЗРАСТНЫЕ ПУТАНИКИ

## МЕДВЕЖЬЯ УСЛУГА „ИЗВЕСТИЙ" СОВЕТСКИМ ШКОЛЬНИКАМ

Тема дружбы всегда волновала молодежь. Как возникает дружба, на чем она основана, какие права и обязанности возлагает на человека,—эти вопросы занимают внимание и нашей молодежи. Редакция «Известий» решила, повидимому, притти ей на помощь и дать нужное принципиальное решение этой проблемы. Об этом благом намерении редакции свидетельствует появившийся в № 38 от 14 февраля об'емистый отчет о диспуте на тему «О дружбе», проведенном в 45-й образцовой средней школе г. Киева.

Увы, каковы бы ни были намерения редакции «Известий», этим своим выступлением она оказала медвежью услугу советской молодежи. То, как «Известия» изобразили школьный диспут на своих страницах, создает совершенно ложное, фальшивое представление о дружбе, не ориентирующее молодежь, а вносящее вредную путаницу в ее головы.

Если газета без оговорок сообщает, что в прежнее время учащиеся не могли обсуждать вопроса о дружбе, если взаимоотношения Маркса и Энгельса характеризуются маленькой цитатой из Лафарга об отношении детей Маркса к Энгельсу, если «мальчики и девочки» оказываются носителями «всегда разных убеждений»,—то просто диву даешься, как у редакции «Известий» хватило духу напечатать такую безграмотную чушь.

Но может быть «Известия» сознательно решили напечатать подобную бессмыслицу, чтоб обнаружить теневые стороны школы, чтоб разоблачить неправильные представления учащихся по вопросу о дружбе? Тогда следовало ожидать, что редакция со своей стороны поправит школьников и сама ответит на вопросы, поставленные школьной дискуссией. Увы, ничего подобного не было сделано. Взамен этого газета нагородила много пошлой чепухи, видать за «стенографически»-точно зафиксированного хихиканья.

Судя по отчету «Известий», дискуссия школьников поставила три вопроса, так и оставшихся без ответа:

1. Каковы пределы дружбы и можно ли дружить с людьми, враждебными партии, советскому строю, социалистическому воспитанию?

2. Возможна ли и в какой форме дружба между мальчиками и девочками, юношами и девушками?

3. Возможна ли и нужна ли дружба между молодежью и взрослыми?

Но «Известия» не удосужились дать ответ на эти юношеские вопросы. Неужели газета не могла сказать, что мы не признаем никакой внеклассовой, «гуманистической» дружбы, что дружба в нашем понимании — это единство целей, единство социалистического миропонимания, единство строителей социализма в борьбе за осуществление социалистических идеалов.

Разве «Известия» не могли также раз'яснить, что вся система социалистического воспитания направлена к тому, чтобы устранить искусственную разобщенность мужской и женской молодежи, свойственную старой школе? Вся наша советская жизнь, открывшая для женщин все отрасли культуры, труда, управления, приучает советских людей к товарищескому отношению к женщине, девушке.

Дружба между юношей и девушкой подчинена тем же законам трудового социалистического единства, как и всякая другая пролетарская дружба. Эта дружба может перейти в «иное чувство», может и не перейти. Но ведь и любовь сама предполагает дружбу.

Не должна ли была газета раз'яснить школьникам и то, что утверждение о невозможности дружбы между взрослыми и молодежью взято целиком из арсенала буржуазной идеологии? Вздорность этого утверждения опровергается на каждом шагу всей практикой советской жизни. На производстве взрослые рабочие устанавливают самую тесную дружбу с молодежью, учат ее, помогают ей, а иногда и сами у нее учатся. В колхозах молодежь, даже дети, показывают образцы трудовой дружбы со взрослыми в борьбе за урожай, за охрану социалистической собственности. Руководители партии насчитывают десятки самых интимных друзей в среде молодежи. Уж кому-кому, а киевским школьникам известна их теснейшая дружба с П. П. Постышевым! Это—та пролетарская дружба, которую комсомол и вся молодежь ощущают на каждом шагу.

Только так можно ставить вопрос о дружбе в наше время. Почему же «Известия» не сказали об этом, а отделались от вопроса ворохом надерганных из школьной стенограммы цитат?

И вот тут оказывается, что ошибка газеты усугубляется тем, что весь «стенографический» отчет по существу грубо и глупо оболгал советских школьников и совершенно извратил их диспут.

Как в действительности выступали школьники? Ученик Джимми Мишине говорил о том, что дружба, настоящая дружба существует до конца жизни, если друзей связывают не мелкие и случайные, а большие, общие интересы, скажем, вот комсомольцев связывает то, что они посвящают свою жизнь делу революции...

Об этом выступлении вы на страницах «Известий» не найдете ни слова.

Леся Гольдберг рассказала о дружбе двух стариков—Степана-«Тигриная смерть» и Василия Худякова в фильме «Аэроград». Она напоминает также, как показана дружба в книге Горбатова «Мое поколение».

Но разве это интересовало корреспондентов «Известий»? Они выдергивают из речи школьницы случайные примеры об ушедшей из школы девочке, детски-неуклюжие и неверные рассуждения...

Илюша Капчинский с глубоким волнением говорил о том, что настоящие друзья познаются в серьезных испытаниях, в совместной борьбе и боях. Он напоминал о декабристах Пестеле и Рылееве, о Марксе и Энгельсе, о Ворошилове и Буденном. Он цитировал «Историю червонного казачества», в которой сам нашел живые и яркие примеры настоящей неумирающей дружбы, испытанной в боях, перед лицом смерти.

Ученик Самородницкий рассказывал о дружбе Герцена и Огарева, Надя Дрягина—о замечательной, волнующей всех ребят дружбе Чапаева и Фурманова. На этих ярких, близких и понятных примерах ставился и решался вопрос о дружбе.

Но прочтите отчет «Известий». В нем нет даже упоминания об этих выступлениях. Невольно протираешь глаза и стараешься понять: какую цель преследовали корреспонденты почтенной газеты, когда фальсифицировали смысл диспута, когда извращали его подлинное содержание, подменяя его случайно оброненными, между прочим приведенными, в связи с чем-то сказанными примерами и рассуждениями? Как будто нарочно из речей советских школьников вытравлялось все советское, характеризующее нашу молодежь, нашу постановку вопроса, и, наоборот, извлекалась всякая бессвязная, вырванная из смыслового контекста, чепуха. Как будто нарочно с диспута переносились на страницы газеты самые «аэкзотические изюминки», неловкие и путаные фразы, глупые мысли. Получилась несообразная пошлость и упрощенчество.

Если бы работники «Известий» поинтересовались существом школьного диспута, они привели бы не строчки рифмованной вредной ерунды, выхваченной из стенограммы, а строки стихов, которые были рождены этим диспутом в школе:

...Дружба! Варывное и звонкое слово!
Дружба... Оно, как звено, уж не ново.
Дружба не гасла в теченье веков.
В Спарте, Египте и кесарском Риме,
Дружба в Иране, и в мифах богов,
Дружба в татарском воинственном
Крыме —
Песнь мы слагаем не этим друзьям!
Только с всесильной победой труда
Дружба великая вновь расцвела.
Дружба рождалась в ночи сражений,
В красных отрядах, рабочих рядах,
В зареве, в искрах, в огне
наступлений,
В грозных атаках, упорных боях...

Это — отрывок из вступления к поэме того же Илюши Капчинского, речь которого не удостоилась даже выдержки в «Известиях».

Очевидно, советские школьники лучше разбираются в том, что такое наша дружба, чем великовозрастные дяди из «Известий». Во всяком случае они сумели лучше организовать и провести свой диспут, чем редакция большой газеты осветить его на своих страницах.

Комсомольская правда
2 — № 39

## Список дисциплин по выбору

| № | Наименование дисциплин | Семестр | Число часов |
|---|---|---|---|
| | **Кафедра теоретической физики и оптики** | | |
| | Квантовая теория или | 7—9 | 160 |
| | Высшая оптика | 7—9 | 160 |
| | **Кафедра колебаний** | | |
| | Теория колебаний или | 7—9 | 160 |
| | Теория звука | 7—9 | 160 |
| | **Кафедра молекулярных и тепловых явлений** | | |
| | Молекулярные и тепловые явления или | 7—9 | 160 |
| | Физика горения | 7—9 | 160 |
| | **Кафедра магнетизма** | | |
| | Общая теория магнетизма | 7—9 | 160 |
| | **Кафедра рентгено-структурного анализа** | | |
| | Структурная металлофизика | 7—9 | 160 |
| | **Кафедра электрических явлений в газах** | | |
| | Электрические явления в газах и их применение | 7—9 | 160 |
| | **Кафедра технической физики** | | |
| | Специальный курс | 7—9 | 160 |
| | **Кафедра математики** | | |
| | Специальный семинар по математической физике | 7—9 | 160 |
| | **Список факультативных дисциплин** | | |
| 1 | Векторный анализ | | |
| 2 | Теория вероятностей | | |

| № | Наименование дисциплин | Семестр | Число часов |
|---|---|---|---|
| 3 | Топология | | |
| 4 | Тензорный анализ | | |
| 5 | Интегральные уравнения | | |
| 6 | Математическая обработка результатов наблюдений | | |
| 7 | Вариационное исчисление | | |
| 8 | Новые методы математич. теории теплопроводности | | |
| 9 | Теория диффузии | | |
| 10 | Физико-химический анализ | | |
| 11 | Спектроскопия рентгеновских лучей | | |
| 12 | Физика твердого тела | | |
| 13 | Отдельные главы по методике рентгено-структурного анализа | | |
| 14 | Кристаллография | | |
| 15 | Теория диэлектриков | | |
| 16 | Магнитные материалы и проводники | | |
| 17 | Электромагнитные процессы в металлах | | |
| 18 | Учение об элементарных актах | | |
| 19 | Фазовые превращения в сплавах | | |
| 20 | Современные воззрения на турбулентное течение | | |
| 21 | Теория подобия и конвективный теплообмен | | |
| 22 | Дополнительные главы по акустике | | |
| 23 | Излучение и распространение радиоволн | | |
| 24 | Специальный семинар по различным вопросам физики | | |
| 25 | Электронная оптика | | |
| 26 | Коронный разряд и его технические применения | | |
| 27 | Физические основы радиотехники | | |
| 28 | Теория относительности | | |
| 29 | Нелинейная теория колебаний | | |
| 30 | Прикладная термодинамика | | |
| 31 | Термоионные явления | | |
| 32 | Переменные токи | | |
| 33 | Квантовая теория ферромагнетизма | | |
| 34 | Магнитный анализ металлов | | |
| | **Курсовые проекты или курсовые работы** | | |
| | Курсовая работа | 8—10 | 624 |

## Распределение курсовых экзаменов и зачетов по семестрам

| № | Наименование дисциплин | СЕМЕСТРЫ Экзамен | Зачет |
|---|---|---|---|
| 1 | Политическая экономия | 2.3 | — |
| 2 | Диалектический и исторический материализм | 6 | — |
| 3 | Ленинизм | 8 | — |
| 4 | Педагогика | 7 | — |
| 5 | Педагогическая практика | — | 8 |
| 6 | Методика преподавания физики | 7 | — |
| 7 | Иностранный язык | 2, 4, 6 | — |
| 8 | Военная подготовка | — | 4 |
| 9 | Физическая культура | — | 2, 4 |
| 10 | История физики | 10 | — |
| 11 | Астрономия | 1 | — |
| 12 | Дифференциальное и интегральное исчисление | 1, 3 | — |

| № | Наименование дисциплин | СЕМЕСТРЫ Экз. | Зачет |
|---|---|---|---|
| 13 | Аналитическая геометрия | 2 | — |
| 14 | Высшая алгебра | 1 | — |
| 15 | Диференциальная геометрия | 3 | — |
| 16 | Диференциальные уравнения | 4 | — |
| 17 | Методы математической физики | 6 | — |
| 18 | Физика (общая) | 2, 3, 4 | 2, 3, 4 |
| 19 | Химия | 5 | 5 |
| 20 | Теоретическая механика | 5 | — |
| 21 | Термодинамика | 5 | — |
| 22 | Статистическая физика | 7 | — |
| 23 | Электродинамика | 8 | — |
| 24 | Квантовая физика | 9 | — |
| 25 | Специальные дисциплины | 9 | — |
| 26 | Специальный практикум | — | 6, 7 |
| 27 | Специальная работа | — | 10 |

## Сводные данные по бюджету времени (в шестидневках)

| КУРС | Теорет. обуч. | Экзаменац. сессия | Учебная практика | Производств. практика | Госуд. экз. и дипл. проект | Каникулы | ВСЕГО |
|---|---|---|---|---|---|---|---|
| I | 42 | 6 | — | — | — | 12 | 60 |
| II | 42 | 6 | — | — | — | 12 | 60 |
| III | 40 | 8 | — | — | — | 12 | 60 |
| IV | 40 | 8 | — | — | — | 12 | 60 |
| V | 82 | 6 | — | — | 10 | 7 | 55 |
| ВСЕГО | 196 | 34 | — | — | 10 | 55 | 295 |

Учебный план. Физфак МГУ.
1938. Оборот

Справка

[illegible] батальоном Краснопресненского района [illegible] в том, что он состоял бойцом [illegible] батальона и с сего числа из батальона отчислен.

Начальник [illegible] батальона - капитан
[illegible]

27 июля [illegible]
№ 96

Справка из истребительного батальона. 27 июля 1941.

„ЗА ДОБЛЕСТНЫЙ ТРУД
В ВЕЛИКОЙ ОТЕЧЕСТВЕННОЙ ВОЙНЕ
1941—1945 гг."

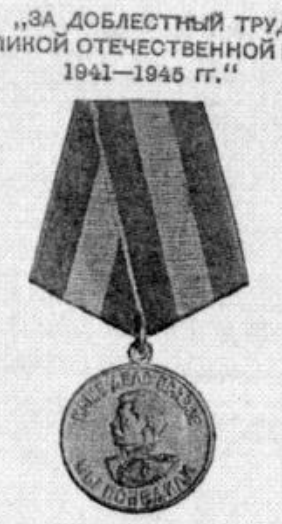

Р № 252842

УДОСТОВЕРЕНИЕ

За доблестный и самоотверженный труд в период Великой Отечественной войны
Капчинский
Илья Михайлович

указом ПРЕЗИДИУМА ВЕРХОВНОГО СОВЕТА СССР от 6 июня 1945 года награжден медалью
„ЗА ДОБЛЕСТНЫЙ ТРУД В ВЕЛИКОЙ ОТЕЧЕСТВЕННОЙ ВОЙНЕ 1941—1945 гг."
От имени ПРЕЗИДИУМА ВЕРХОВНОГО СОВЕТА СССР медаль вручена
7 мая 1946 г.
(должность и подпись лица, вручившего медаль)
м. п.

Удостоверение к медали
«За доблестный труд в Великой Отечественной войне». 1946.

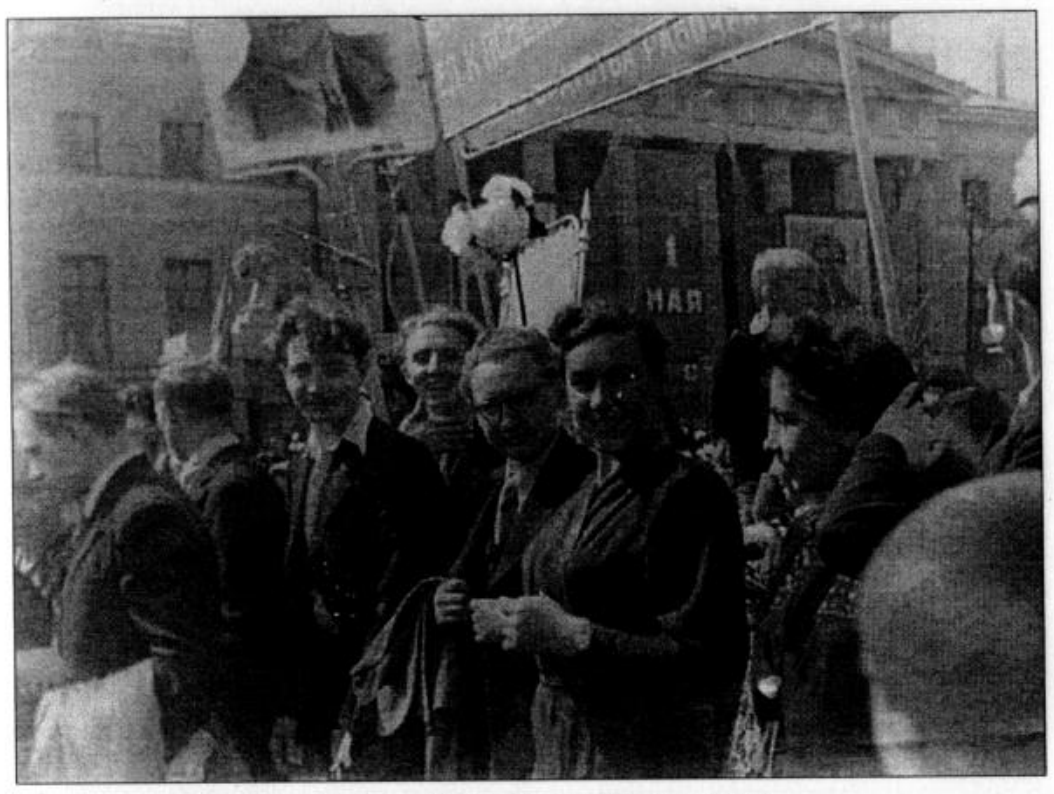

Первомайская демонстрация, 1949.
Слева - Р.М. Венгров, справа - И.Б. Андреева.

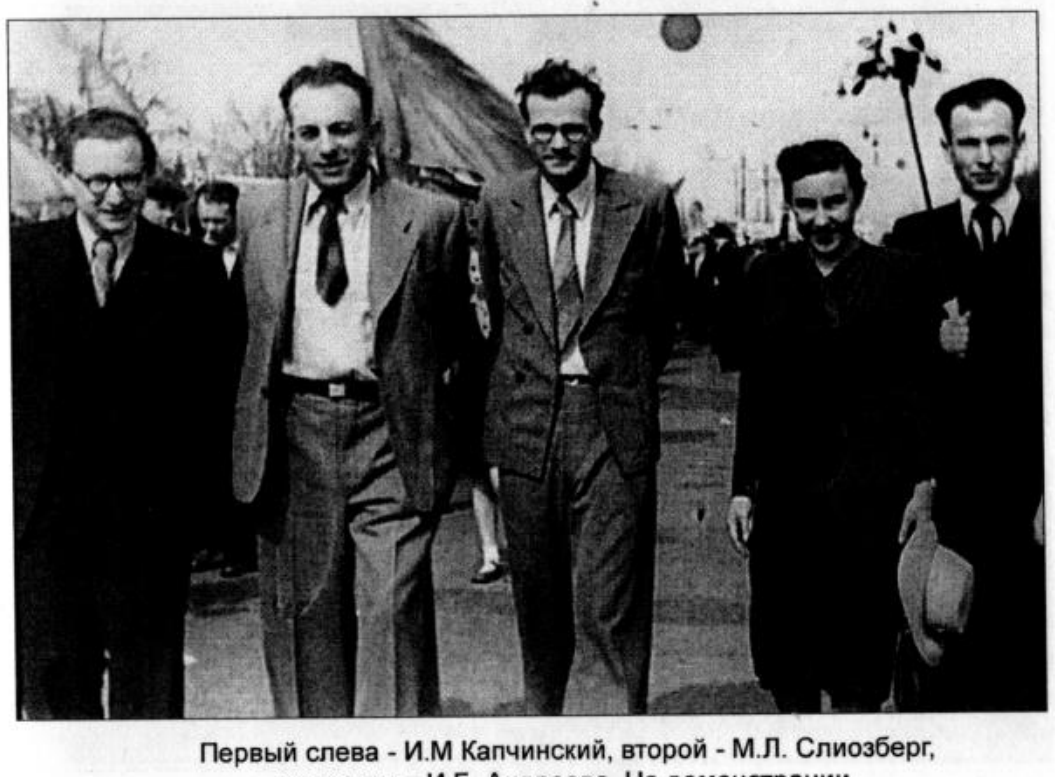

Первый слева - И.М Капчинский, второй - М.Л. Слиозберг, четвертая - И.Б. Андреева. На демонстрации.

Top left:Information from the fighter battalion. July 27, 1941.

Bottom left: CERTIFICATE
For valiant and selfless LABOR DURING THE GREAT PATRIOTIC WAR
Kapchinsky, Ilya Mikhailovitch
PRESIDIUM OF THE SUPREME COUNCIL OF THE USSR June 1945 AWARDED WITH A MEDAL "FOR VALIANT LABOR IN THE GREAT PATRIOTIC WAR 1941-1945"
From the "penalty of the PRESIDIUM OF THE SUPERIOR COUNCIL of the USSR.
Certificate for the medal "For Valiant Labor in the Great Patriotic War". 1946.

Top right: May Day demonstration, 1949. Left - R.M. Vengrov, on the right - I.B. Andreeva

Bottom right: First from the left - I M Kapchinsky, second - M L. Sliozberg, fourth - I.B. Andreeva. At the demonstration.

ВОСПОМИНАНИЯ ОБ И.М. КАПЧИНСКОМ

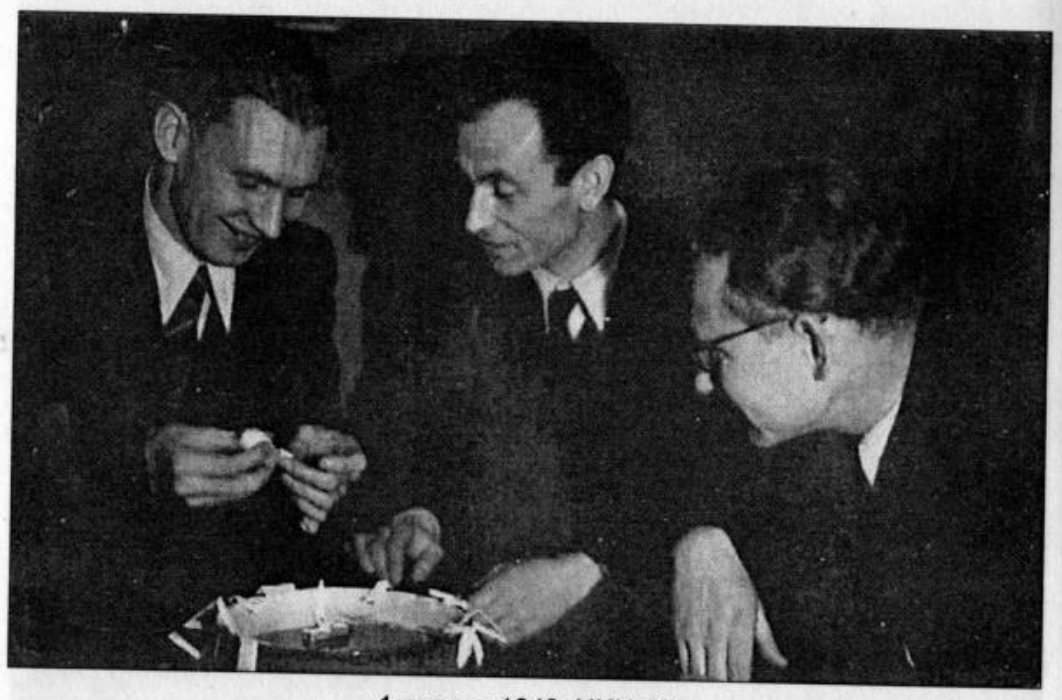

1 января 1949. НИИ-20

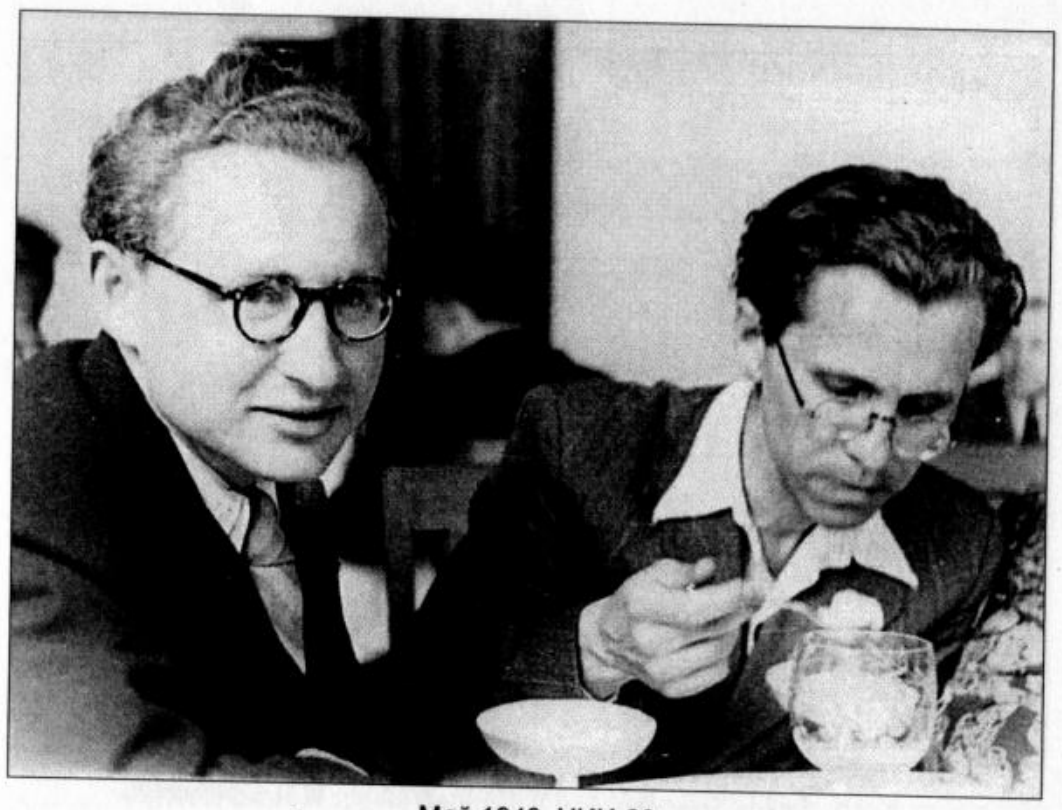

Май 1949. НИИ-20

ФОТОГРАФИИ И ДОКУМЕНТЫ

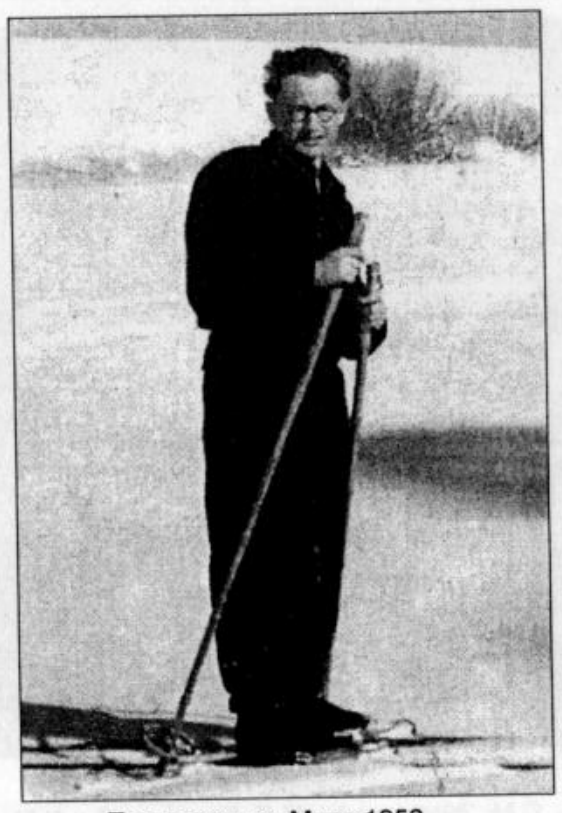

Подмосковье. Март 1952.

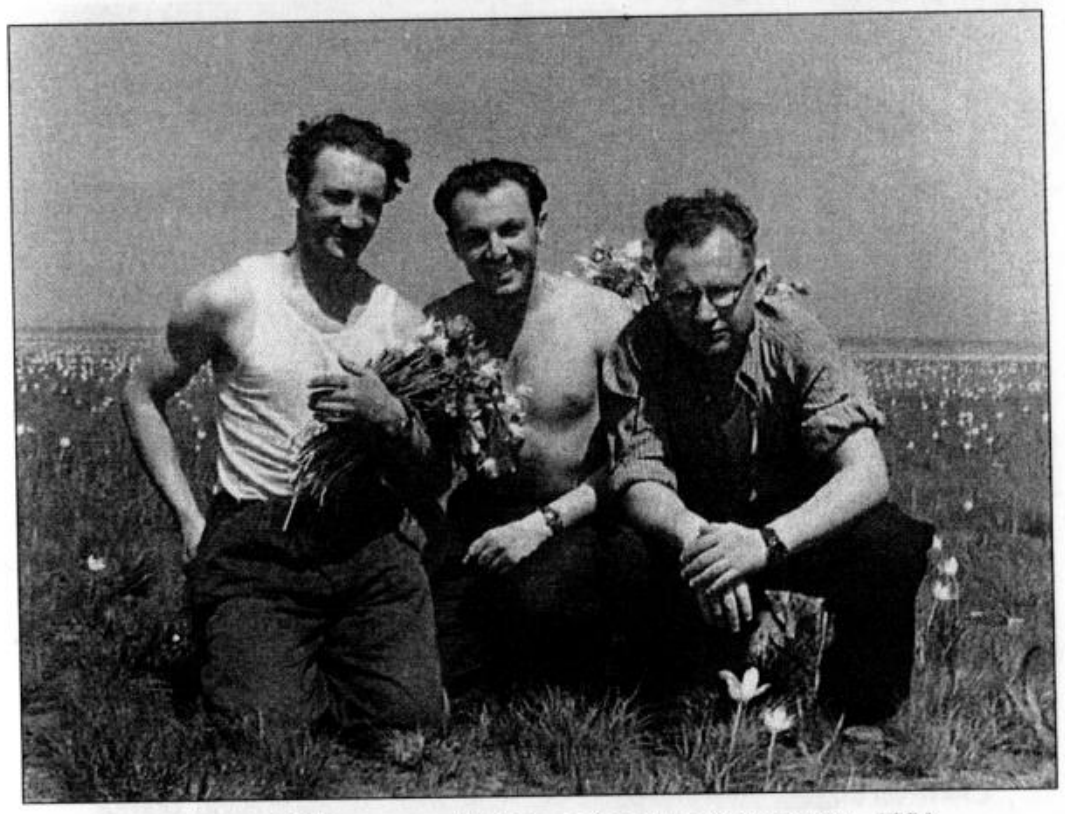

Слева - Р.М.Венгров и М.Л.Слиозберг. Полигон Донгуз. 1954

Top right: January 1949.

Bottom left: May 1949.

Top right: Moscow region. March 1952.

Bottom right: Left - R.M.Vengrov and M.L.Sliozberg. Landfill Donguz. 1954

ВОСПОМИНАНИЯ ОБ И.М. КАПЧИНСКОМ

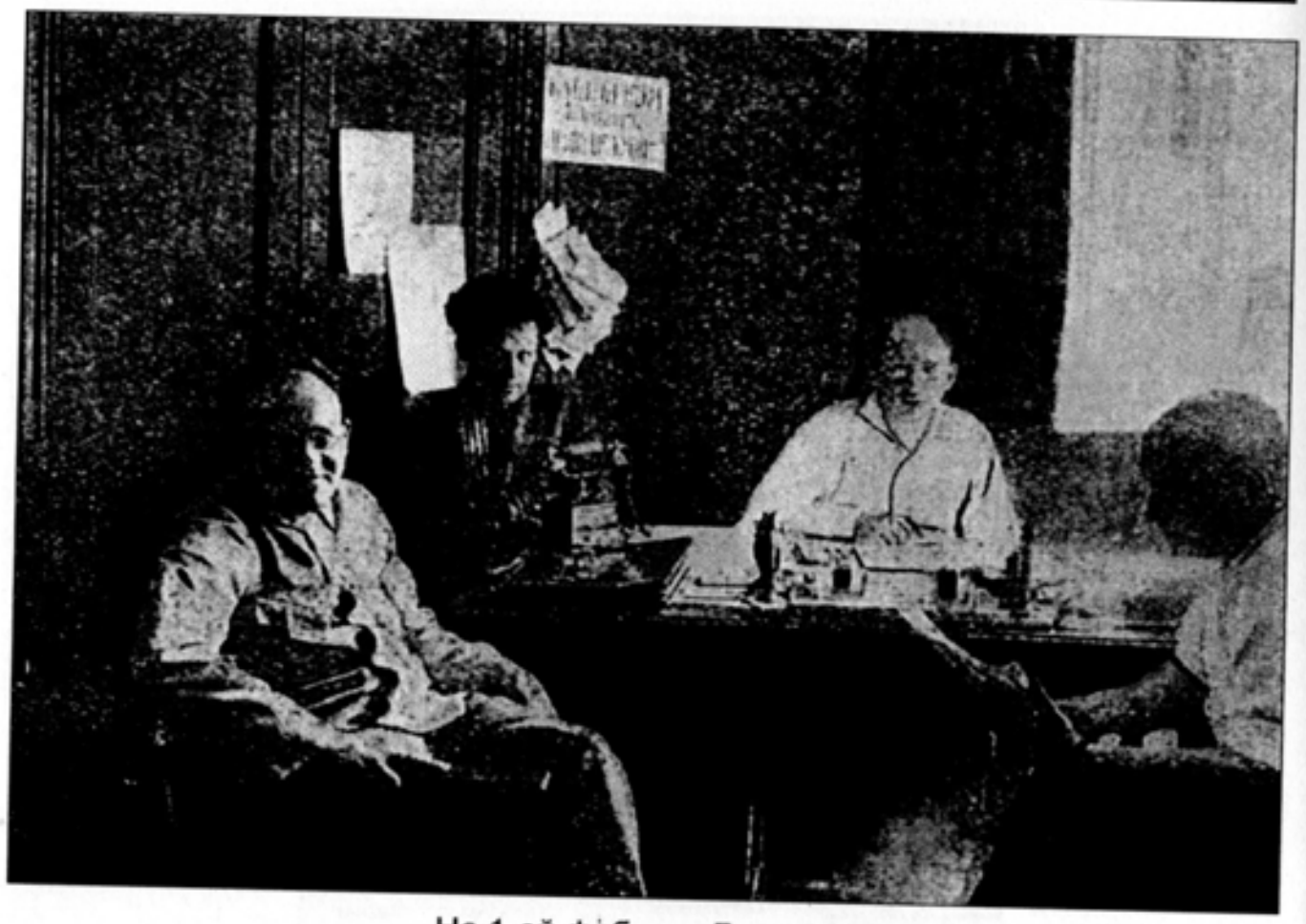

На 1-ой фабрике Госкино.
И.Э. Бабель, С.М. Эйзенштейн, М.Я. Капчинский и Я.М. Блиох.
Журнал «Советское кино», №2, 1926

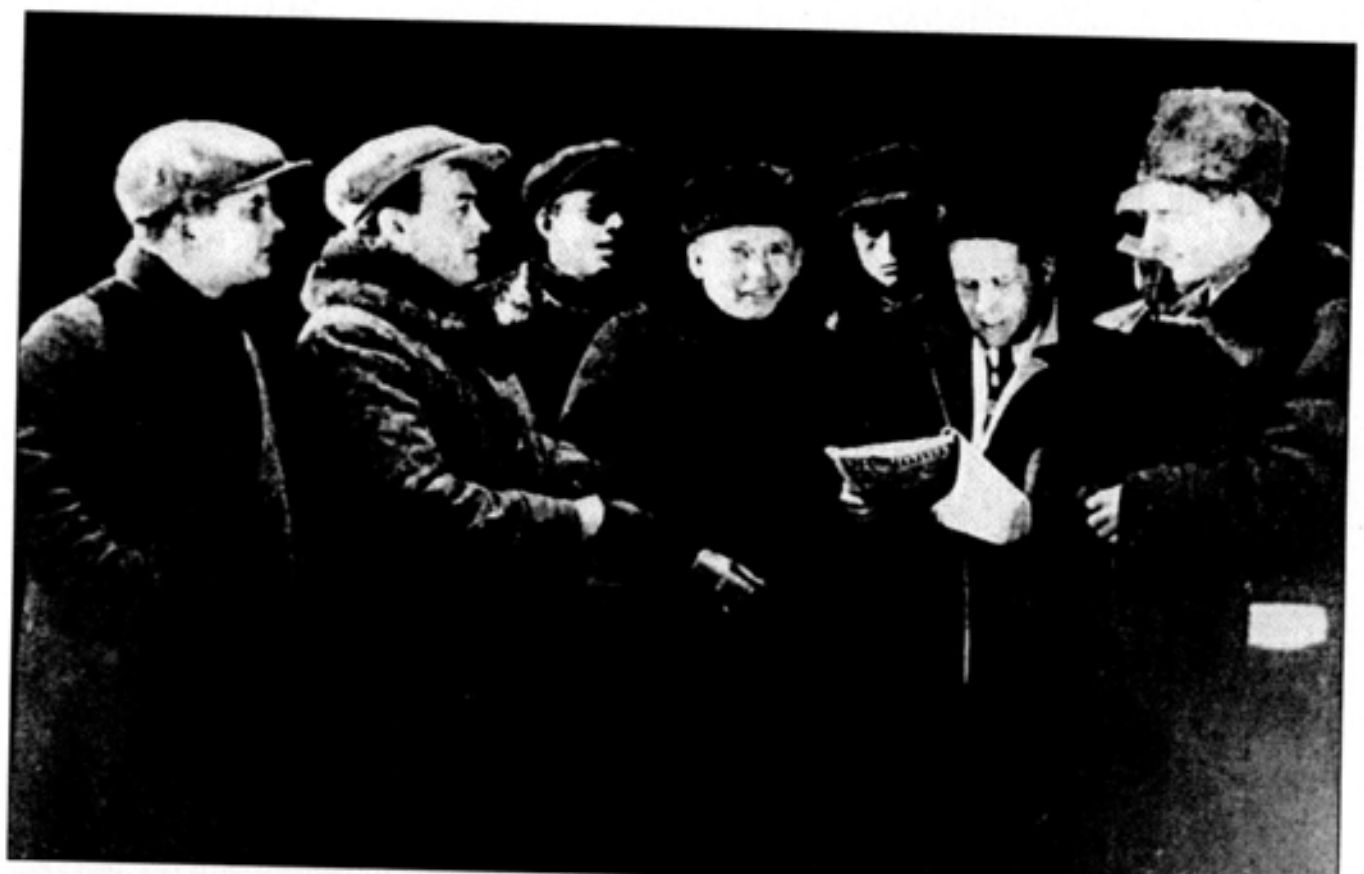

Съемочная группа к/ф «Броненосец Потемкин».
Слева направо: М.М. Штраух, Я.М. Блиох, М.С. Гоморов,
М.Я. Капчинский, А.И. Левшин, С.М. Эйзенштейн, Э.К. Тиссэ.
Снято на второй день утром после показа фильма в Большом Театре
24 декабря 1925 г. делегатам 14-го съезда ВКП(б).

ФОТОГРАФИИ И ДОКУМЕНТЫ

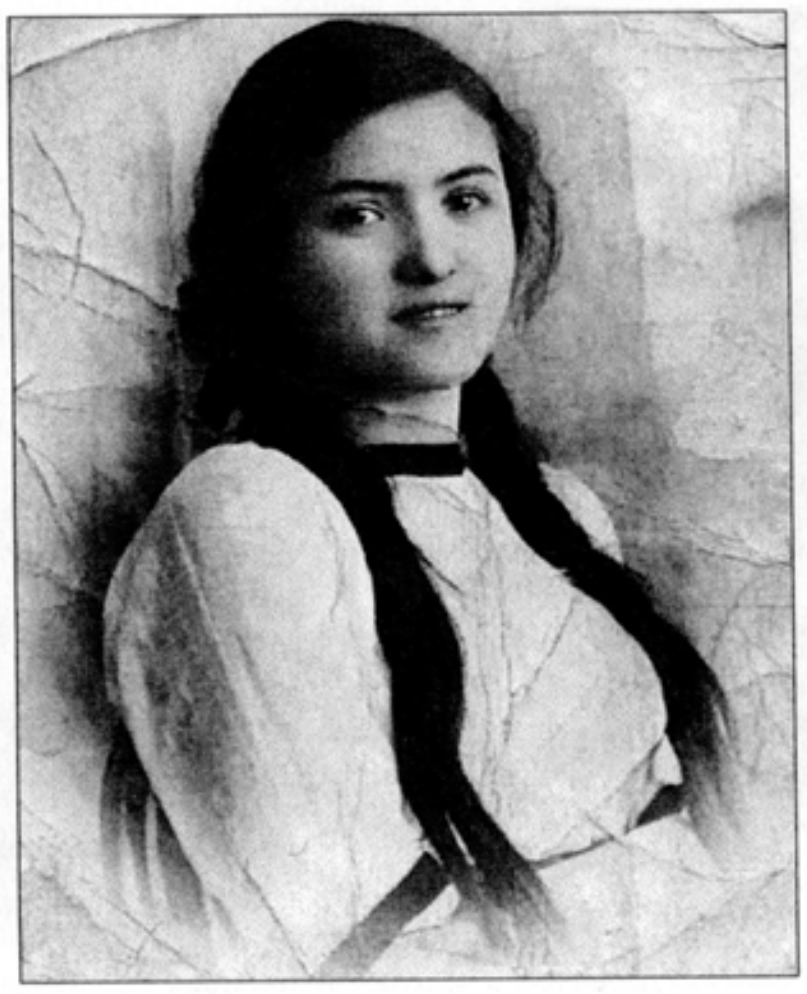

С.Г. Капчинская. Надпись на обороте:
«Въ память любимому другу и товарищу отъ Сарры
19/VI 19»..

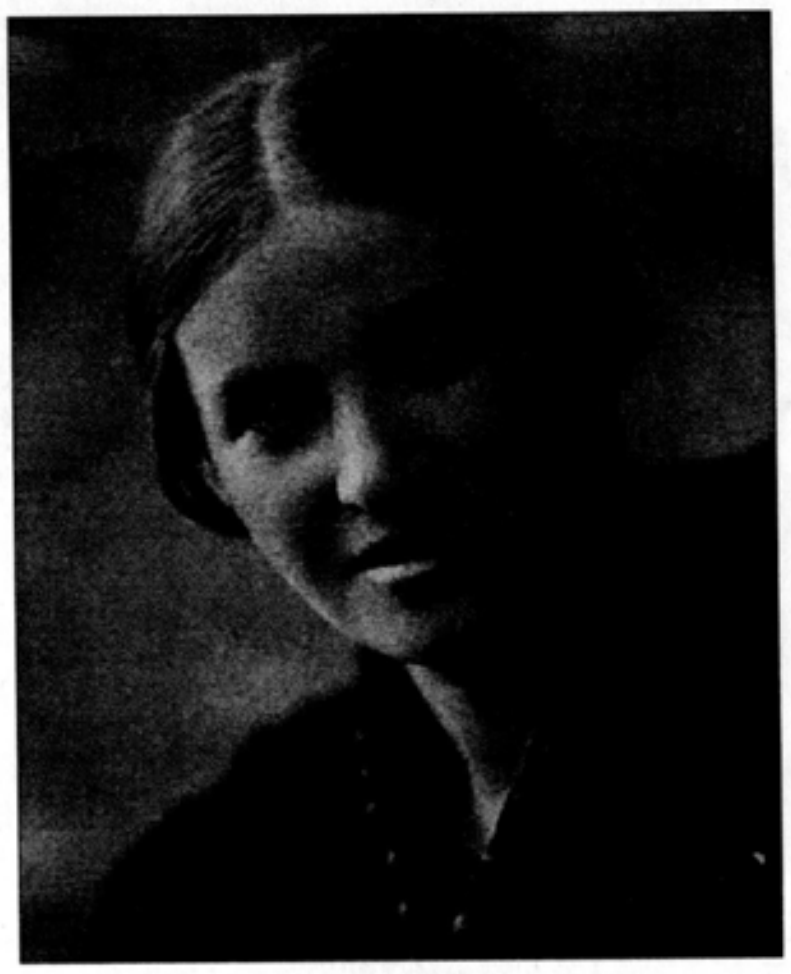

С.Г. Капчинская, 1924

Top left: At the 1st Goskino factory. I.E. Babel, S.M. Eisenstein, M. Ya. Kapchinsky and Ya M Bliokh. Magazine "Soviet Cinema", No. 2, 1926

Bottom left: Film crew of the movie "Battleship Potemkin". From left to right: M M. Strauch, I M. Bliokh, M.S. Gomorov, M.Ya. Kapchinsky, A.I. Levshin, S.M. Eisenstein, E K. Tisse.
Taken on the second day in the morning after the film was shown at the Bolshoi Theater on December 24, 1925 to the delegates of the 14th Congress of the CPSU(b).

Top right: S.G. Kapchinskaya. Inscription on the back:
"In memory of a beloved friend and comrade from Sarah 19L / 1 19"..

Bottom right: S.G. Kapchinskaya, 1924

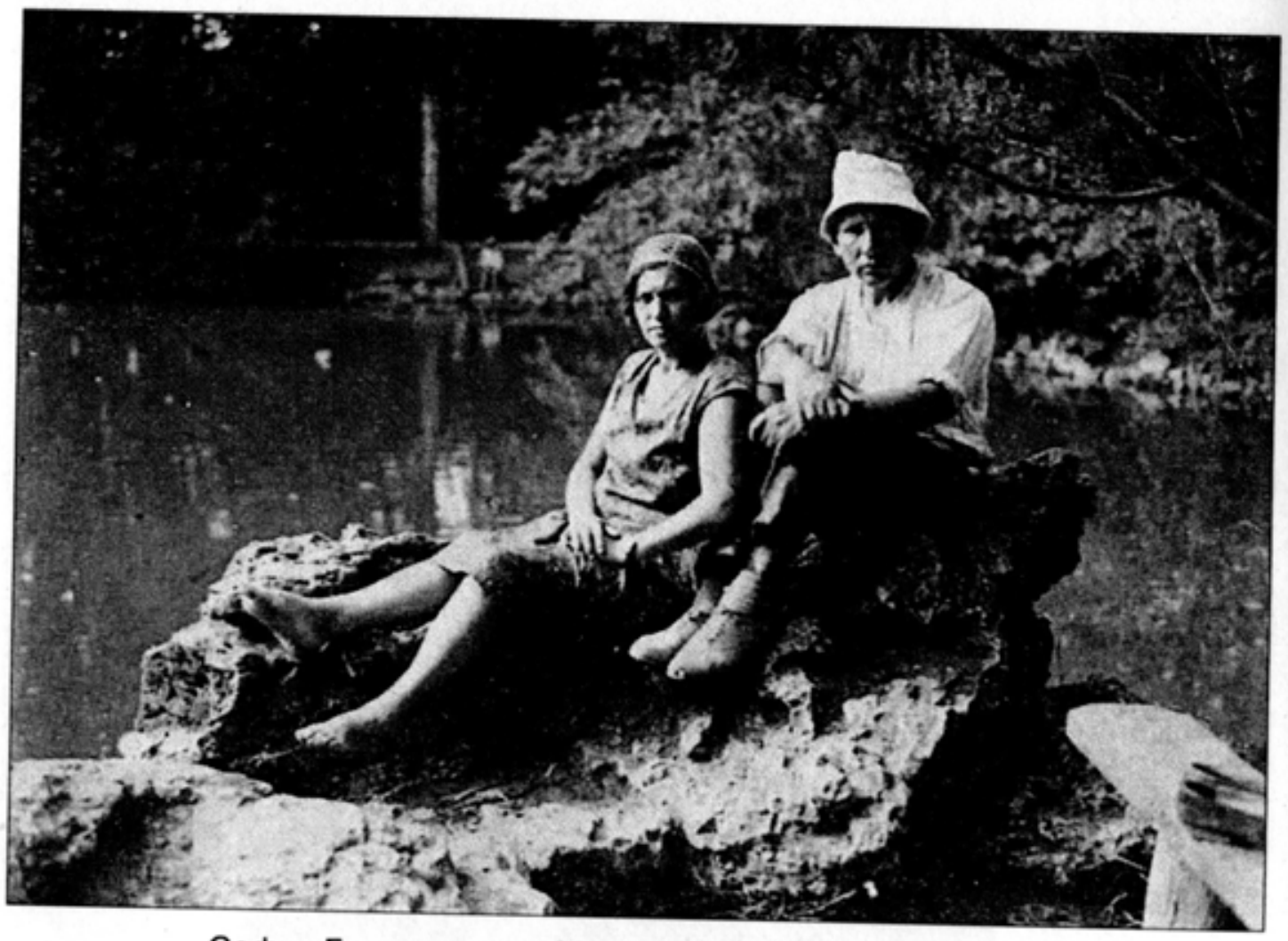

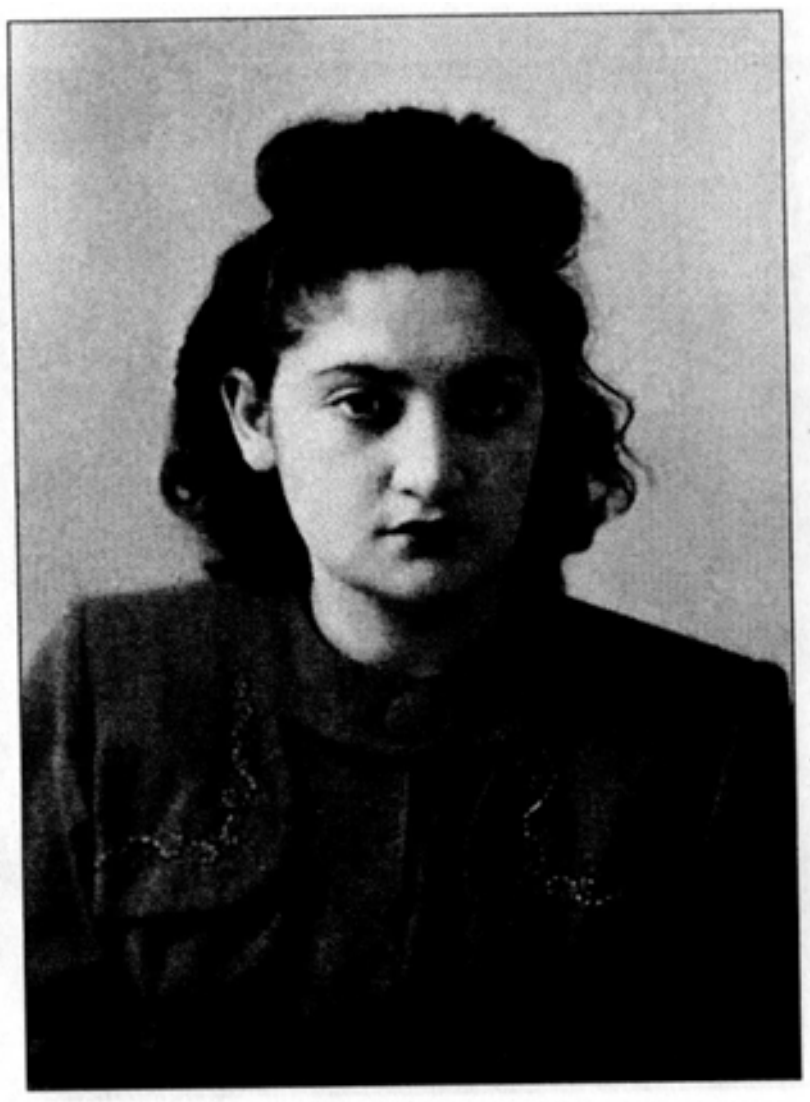

Софья Григорьевна и Михаил Яковлевич Капчинские.
Г. Цхалтубо (Грузия). Ориентировочно - 1920-ые гг.

Л.М. Капчинская. 28 апреля 1952

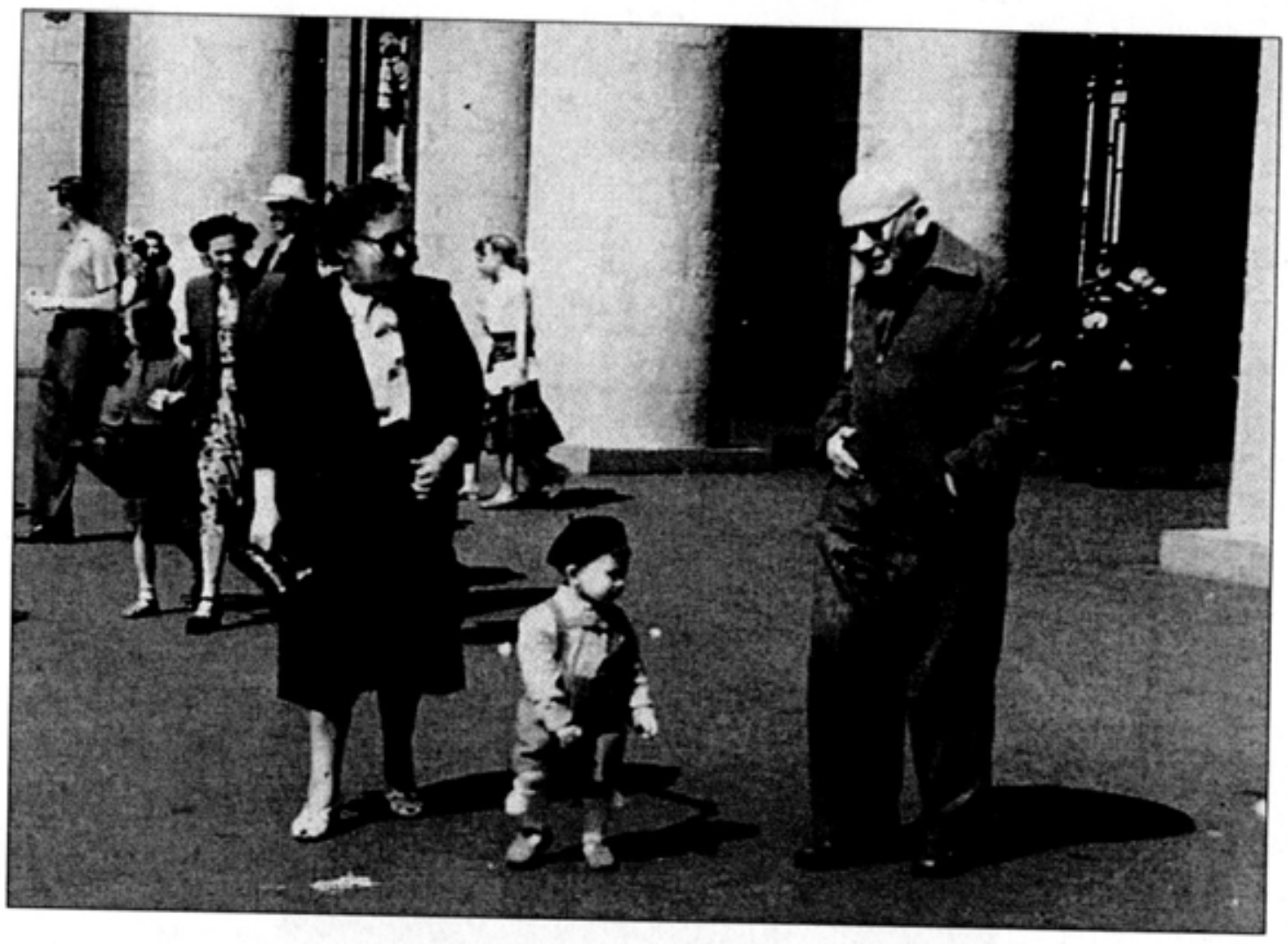

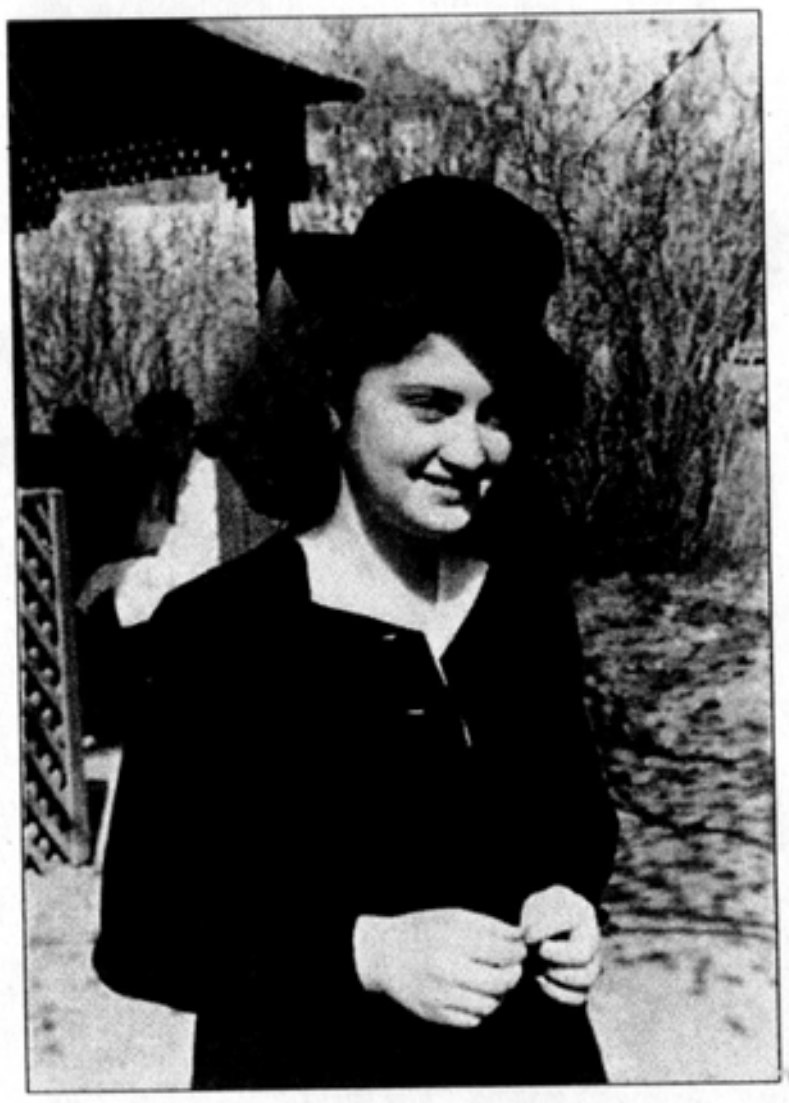

Софья Григорьевна и Михаил Яковлевич Капчинские.
12 мая 1957.

Л.М. Капчинская. Апрель 1952.
√ курс. Первая Советская больница (г. Саратов),

Top left: Sofia Grigoryevna and Mikhail Yakovlevich Kapchinsky. G Tskhaltubo (Georgia). Estimated 1920s.

Bottom left: Sofia Grigoryevna and Mikhail Yakovlevich Kapchinsky. May 12, 1957.

Top right: L.M. Kapchinskaya. April 28, 1952

Bottom right: .M. Kapchinskaya. April 1952 V course. First Soviet Hospital (Saratov), surgery.

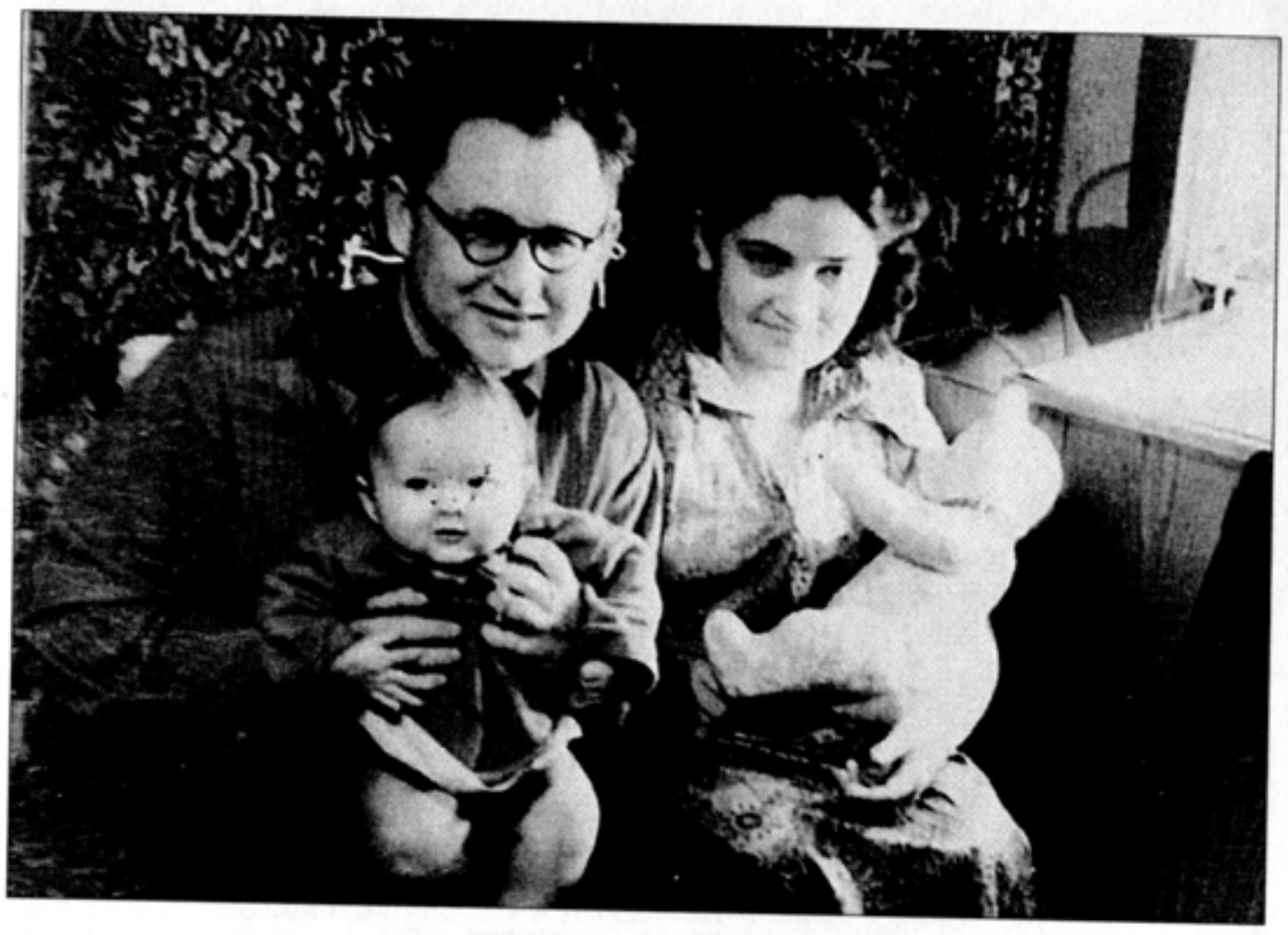

Москва. 20 января 1956

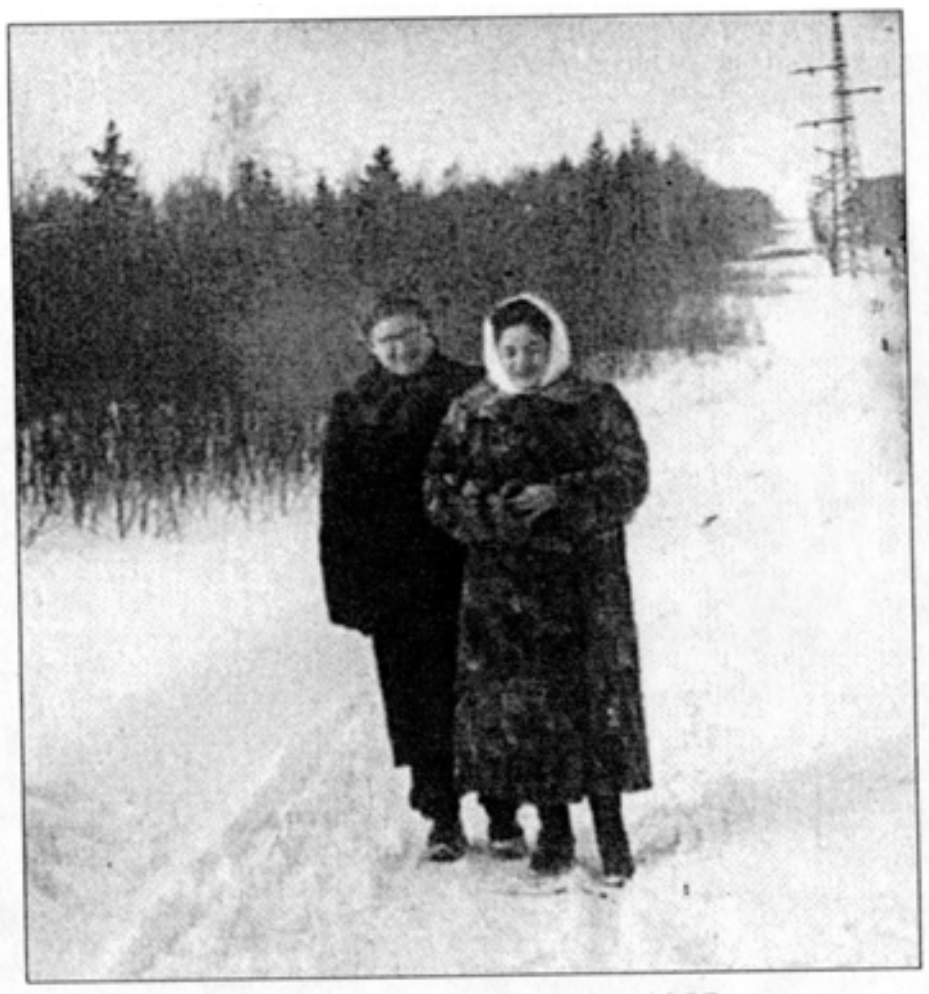

Красная Пахра. Февраль 1955.

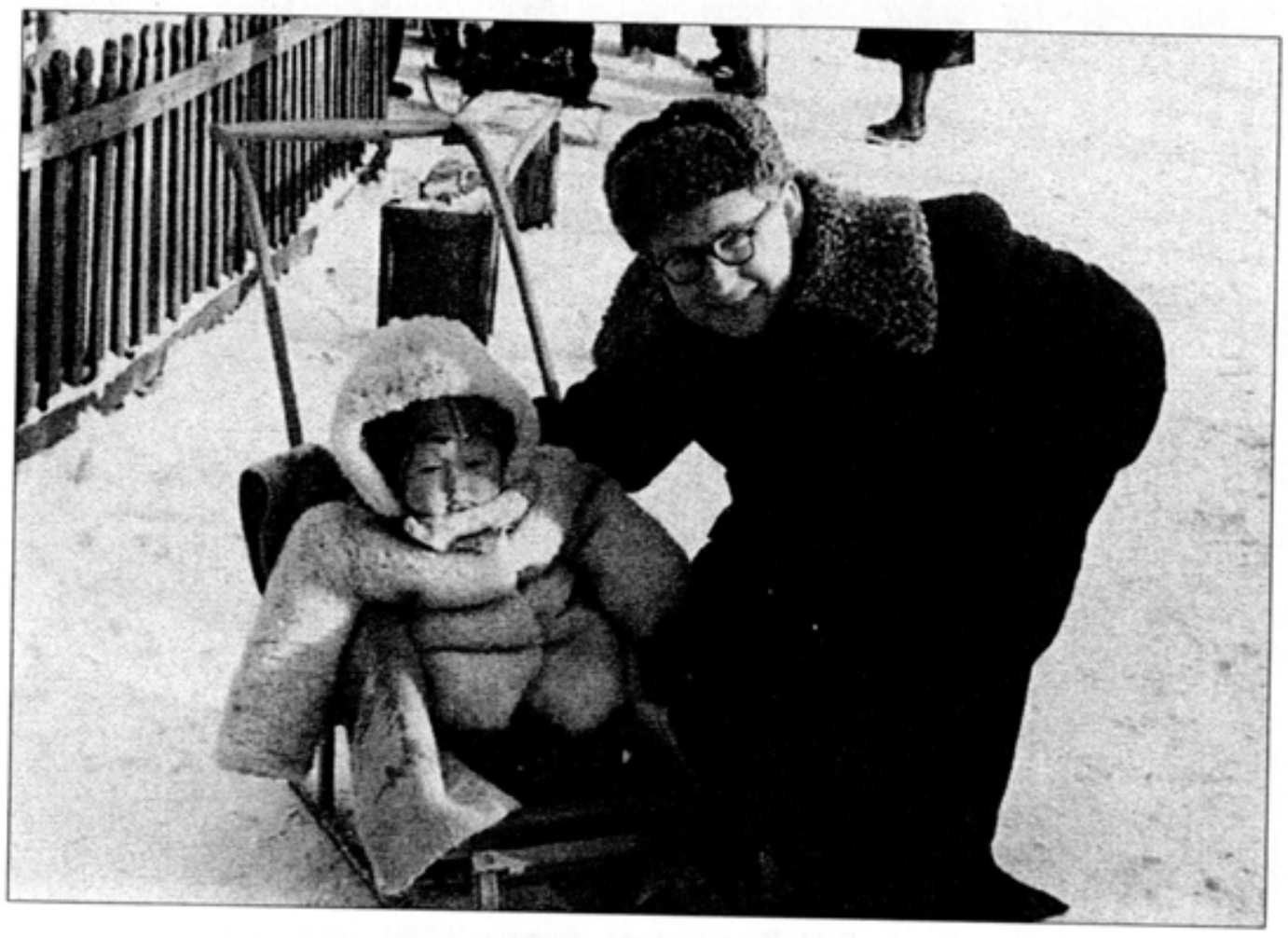

Москва. 18 ноября 1956

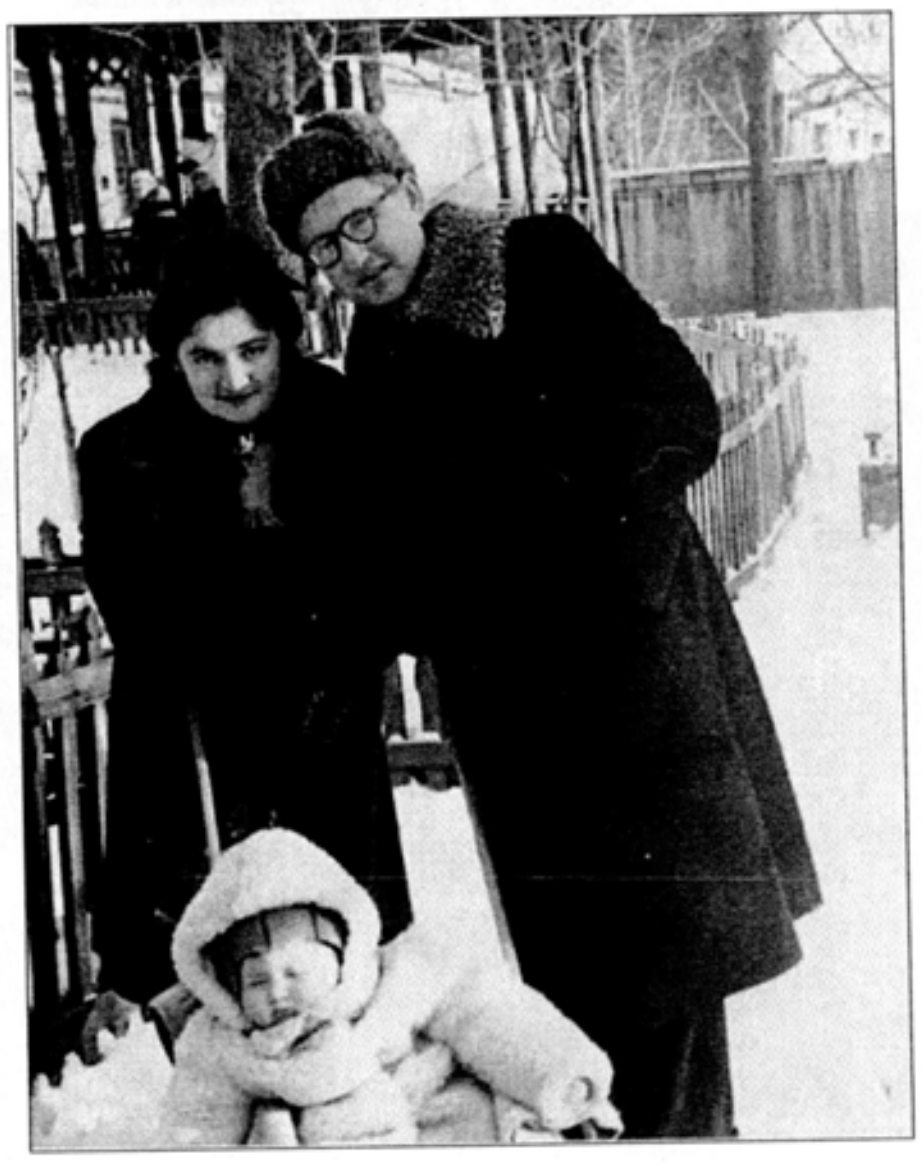

Москва. 18 ноября 1956

Top left: Moscow. January 20, 1956

Bottom left: Moscow. November 18, 1956

Top right: Red Pakhra. February 1955

Bottom right: Moscow. November 18, 1956

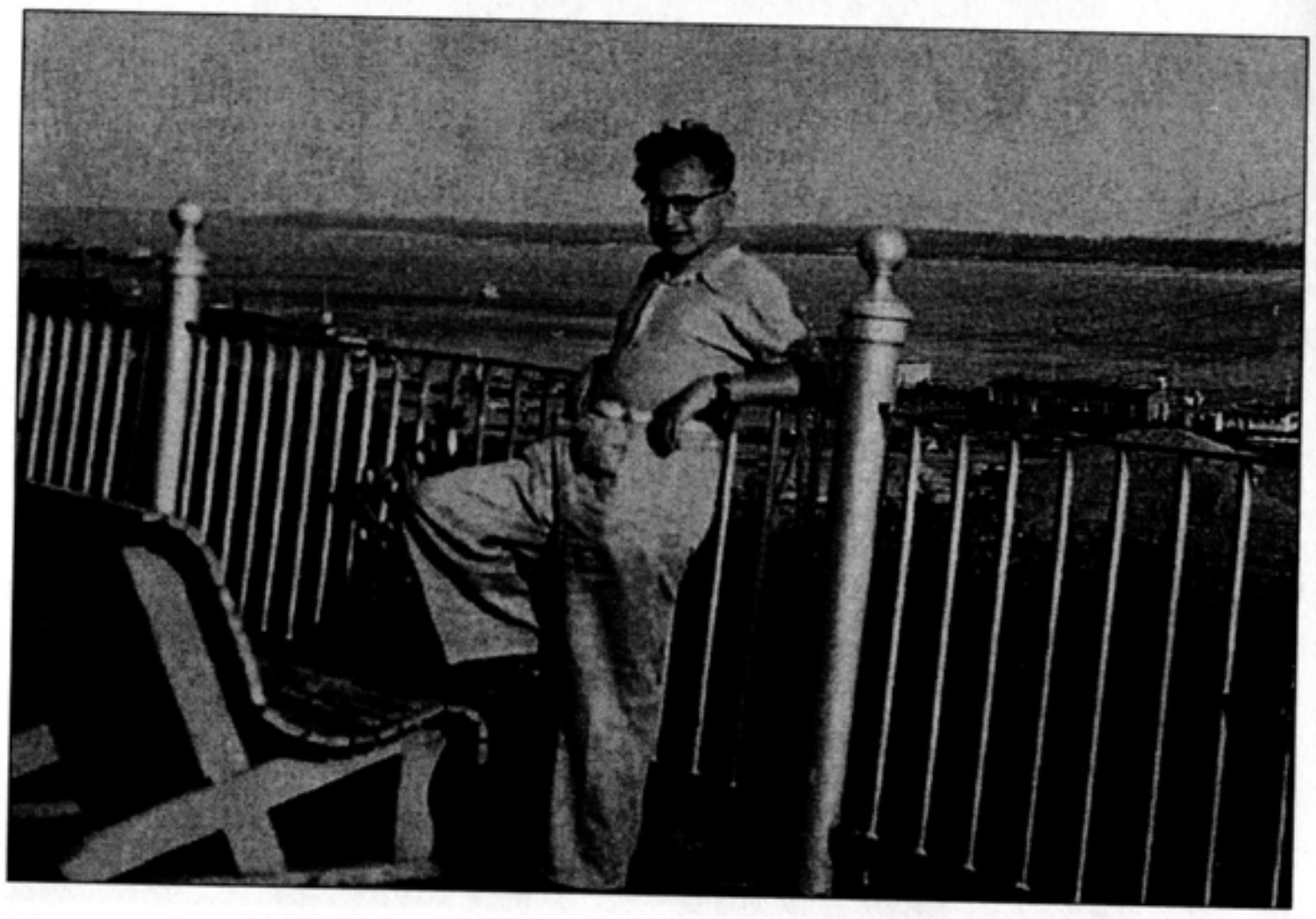

Саратов. 1954 или 55

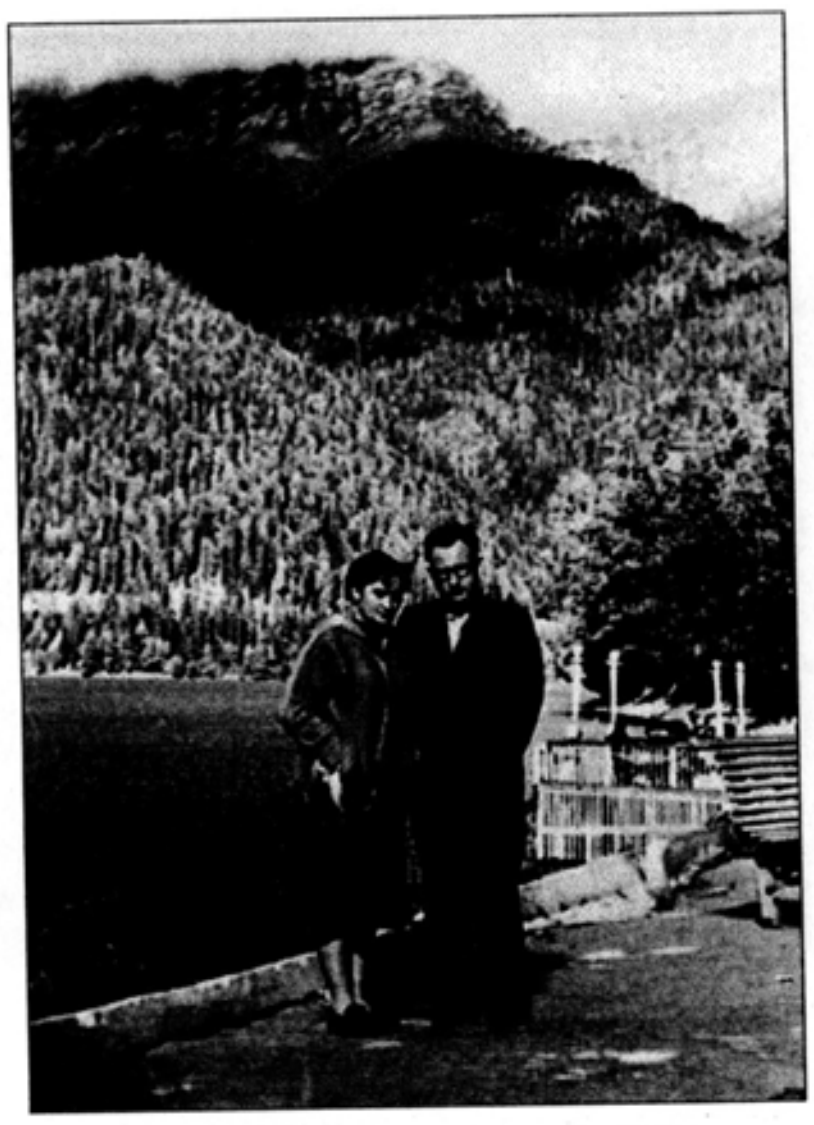

Оз. Рица (Абхазия). 1957

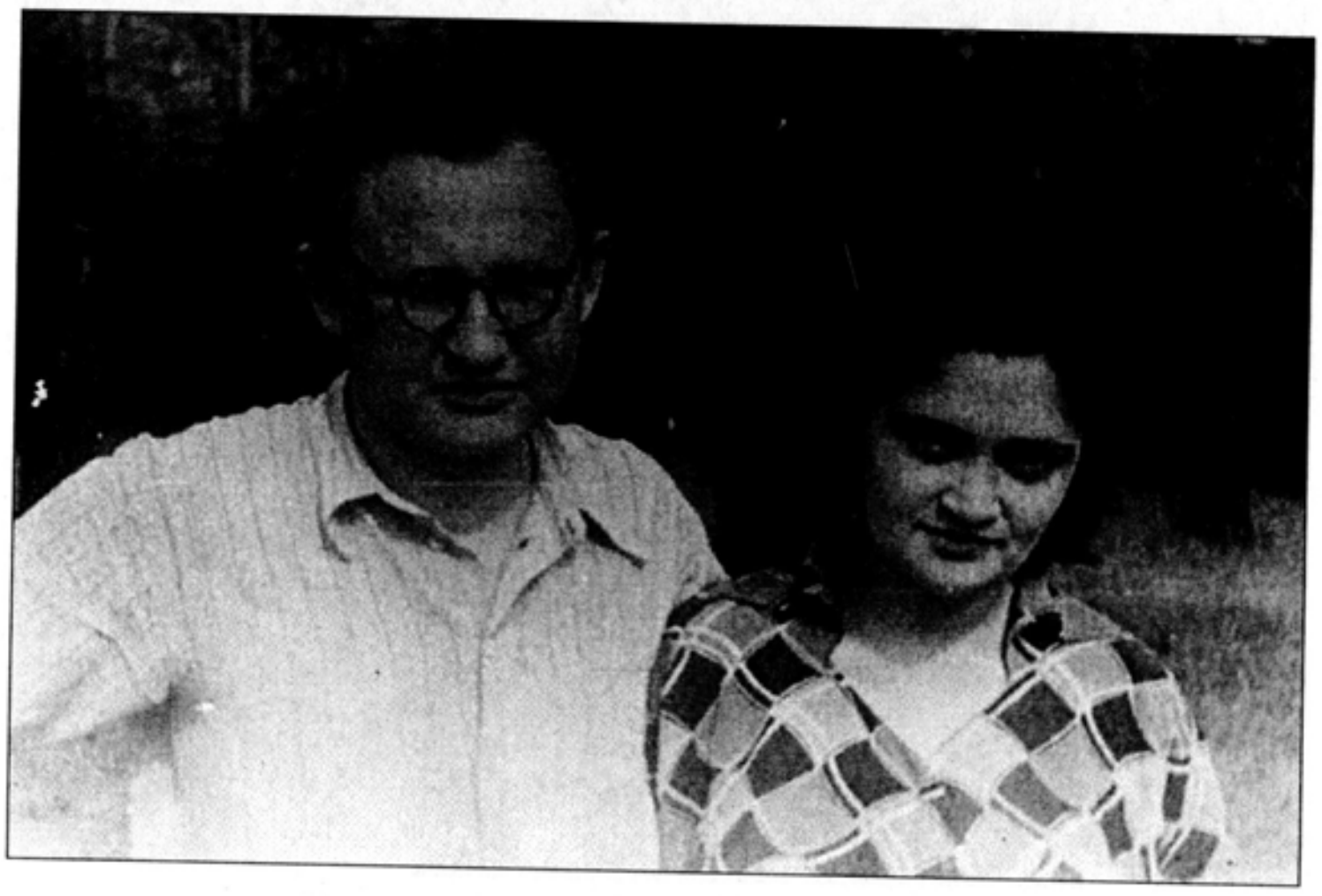

Подмосковье, дача Венгровых. Июнь 1954.

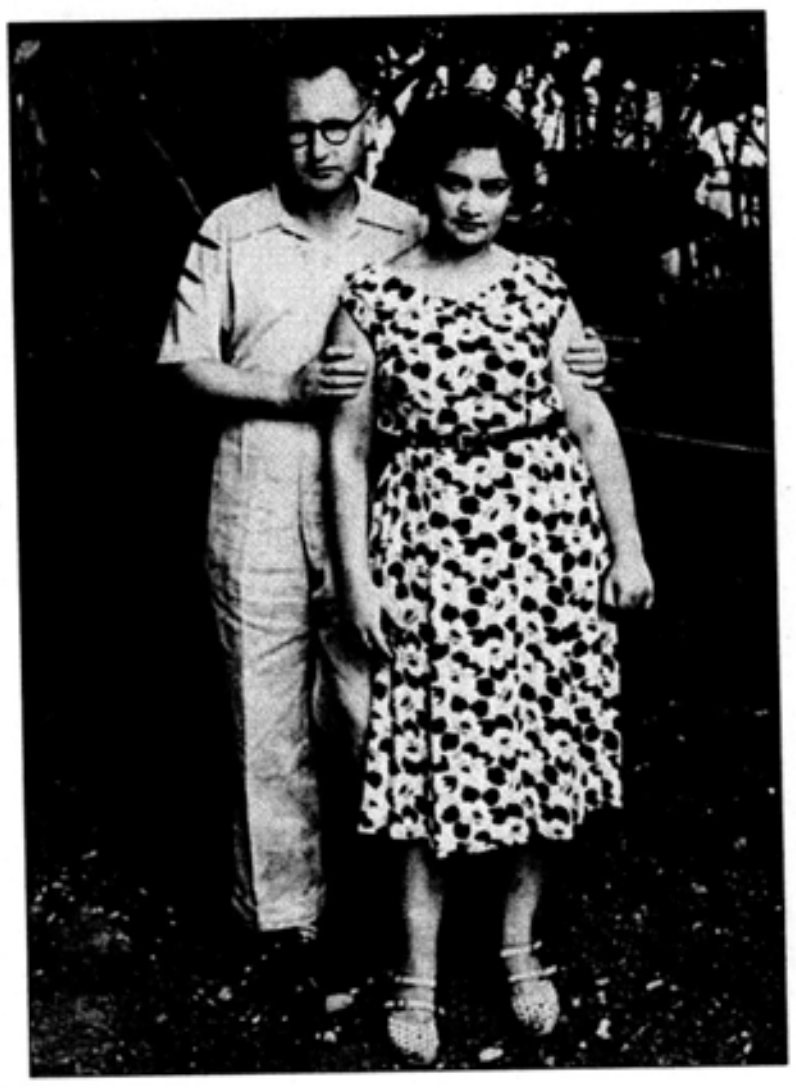

Сухуми. октябрь 1957

Top left:  Saratov, July 1955.

Bottom left: Moscow region, dacha of the Hungarians. June 1954

Top right: Oz. Ritsa (Abkhazia). 1957

Bottom right: Sukhumi, October 1957

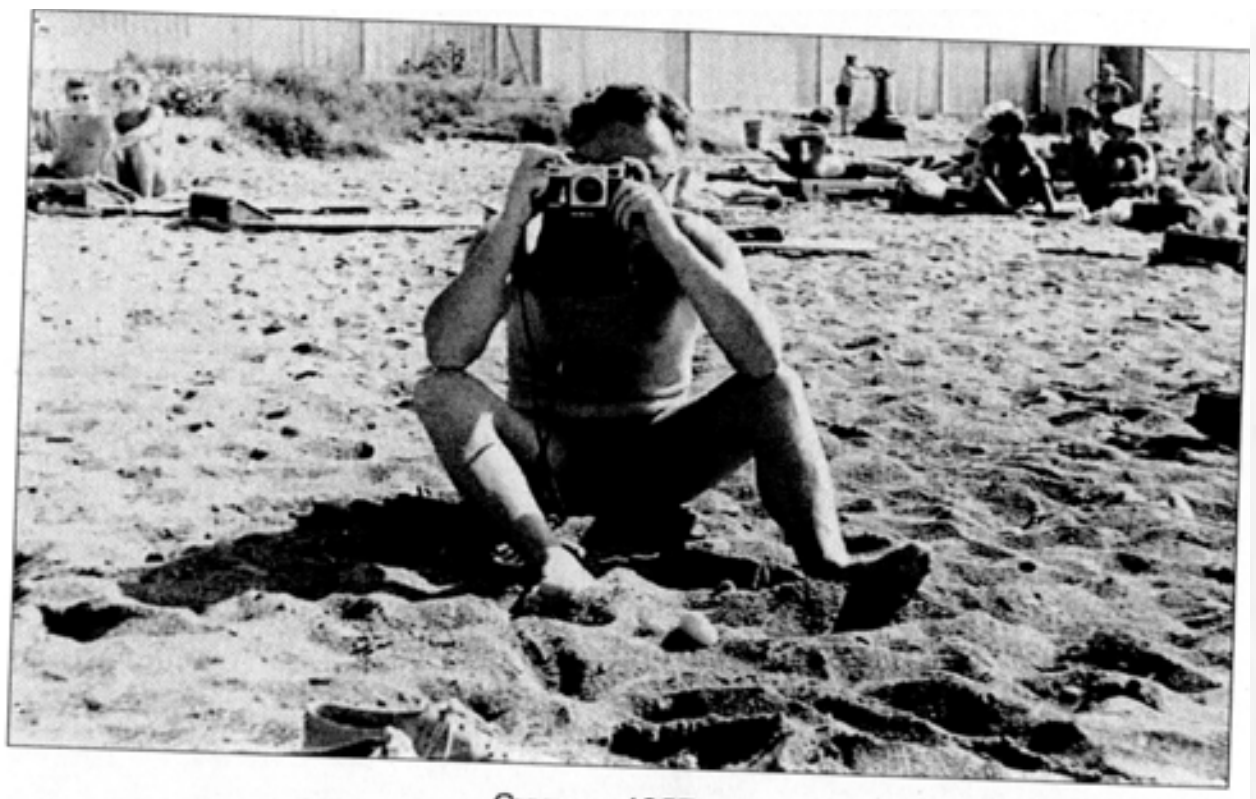

Сухуми. 1957

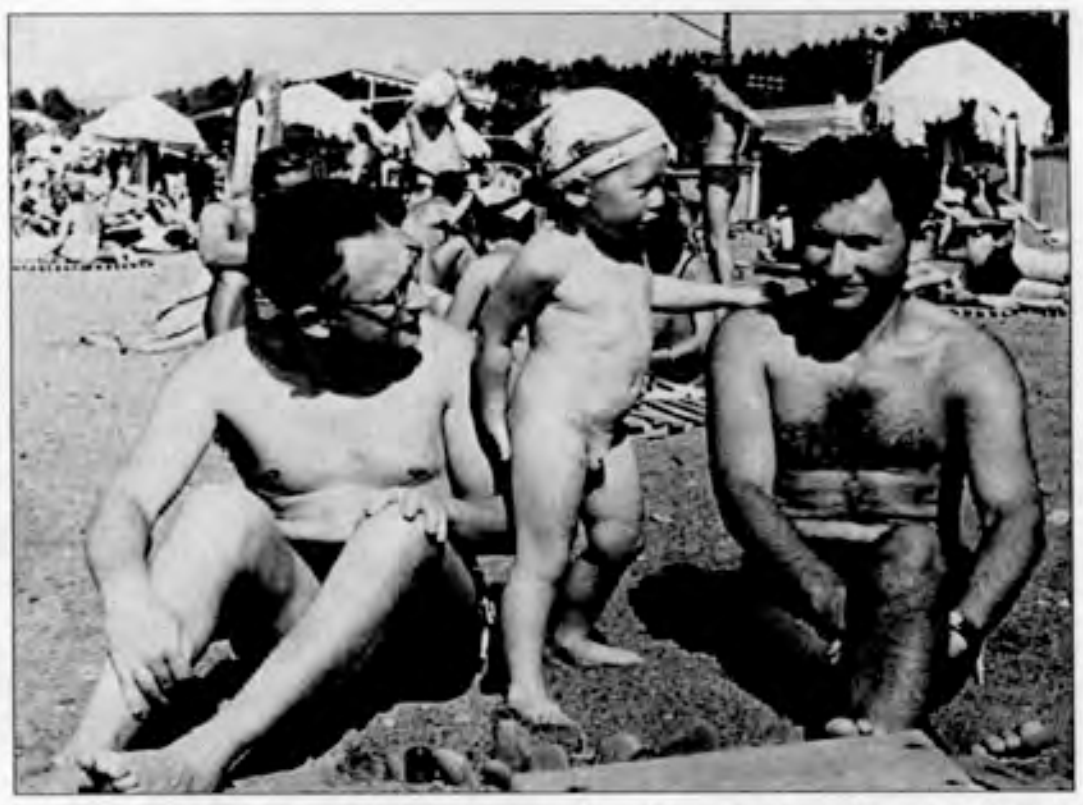

Сухуми, октябрь 1958. Справа - Лев Михайлович Капчинский.

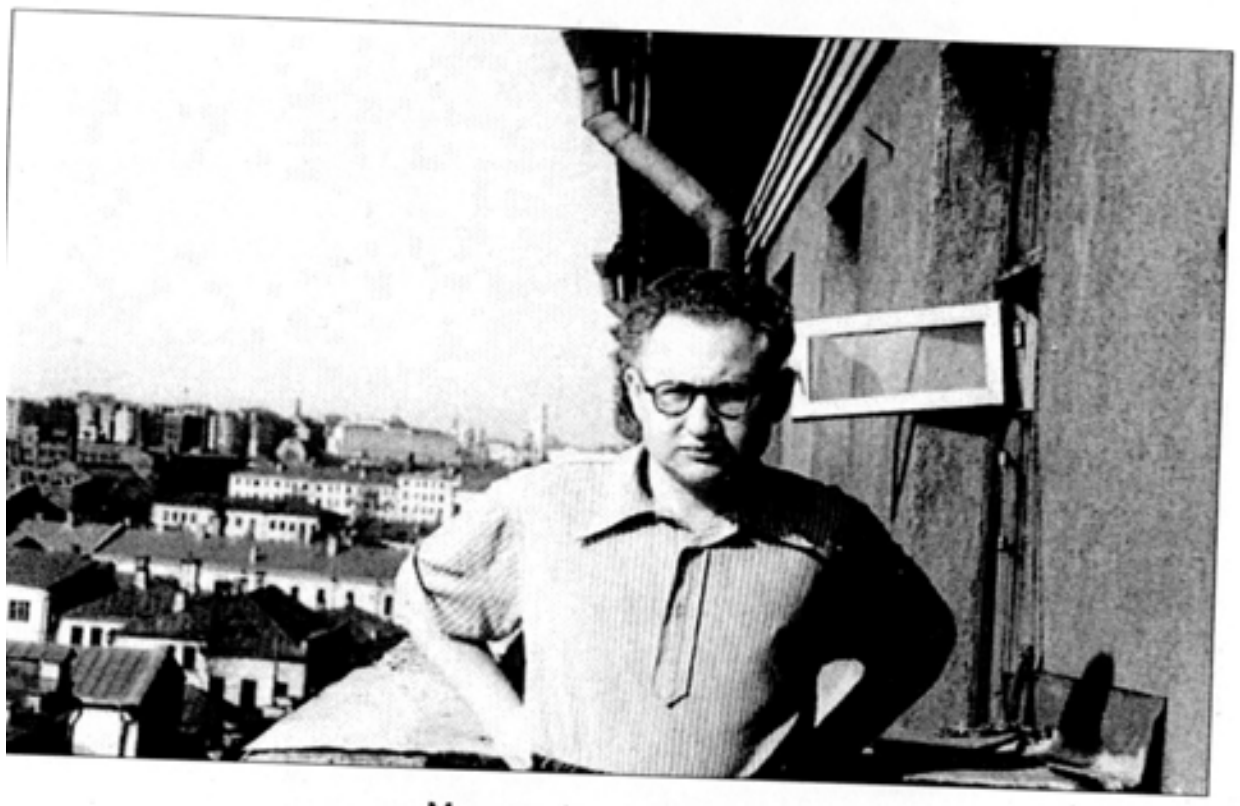

Москва. 1 мая 1957

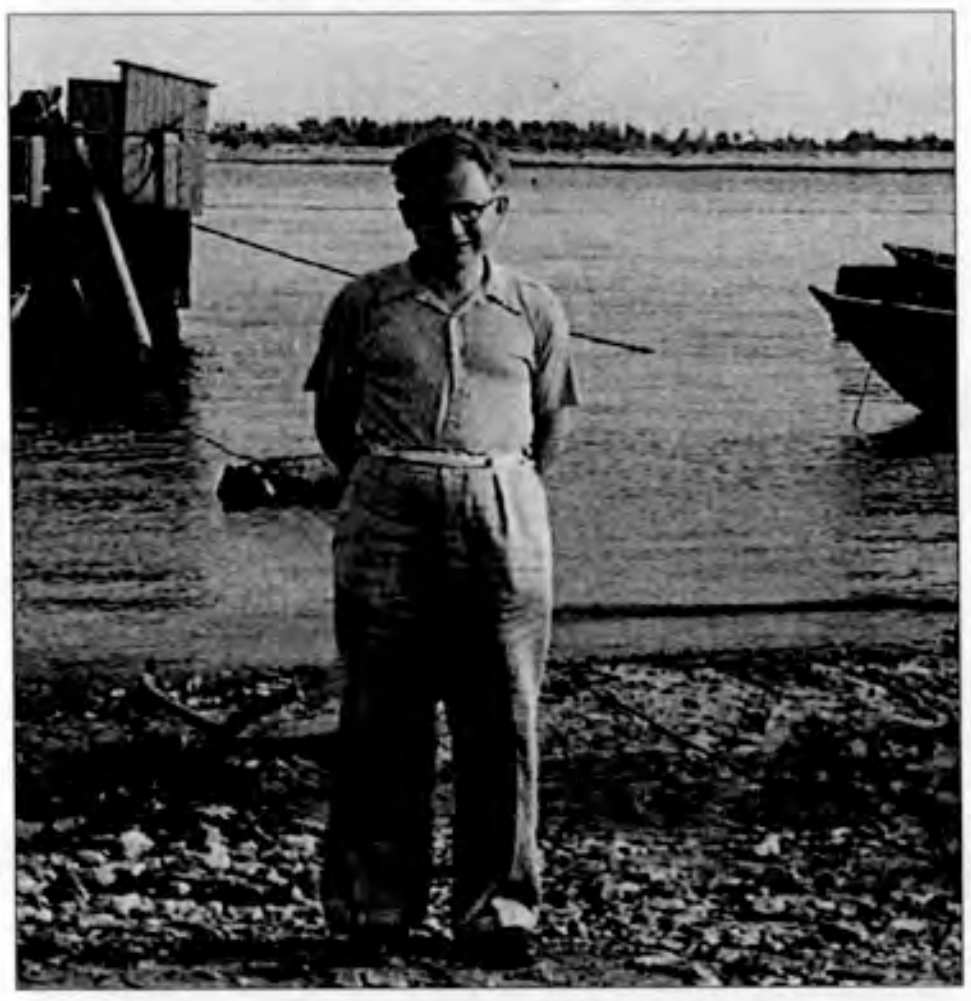

Саратов. 1958

Top left: Sukhumi. 1957

Bottom left: Moscow 1 May 1957

Top right: Sukhumi, October 1958. On the right - Lev Mikhailovich Kapchinsky

Bottom right: Saratov. 1958

Moscow region, dacha of the Hungarians. June 1954

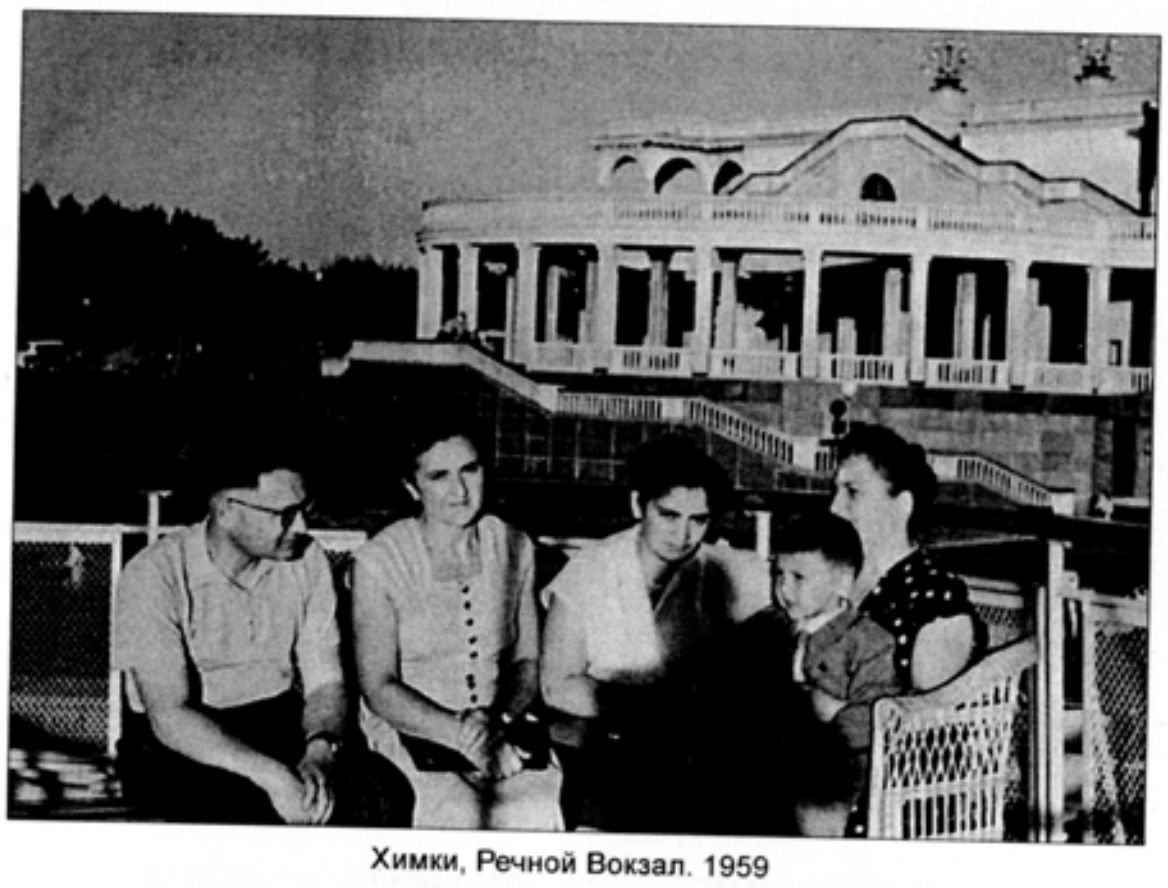

Химки, Речной Вокзал. 1959

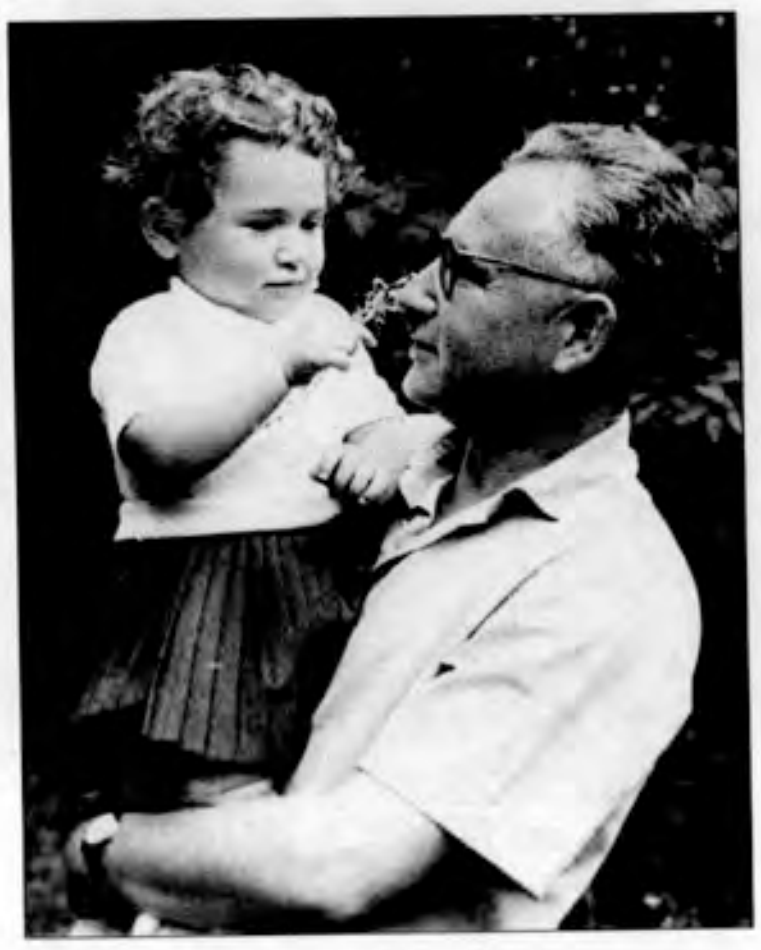

С дочерью. Август 1963

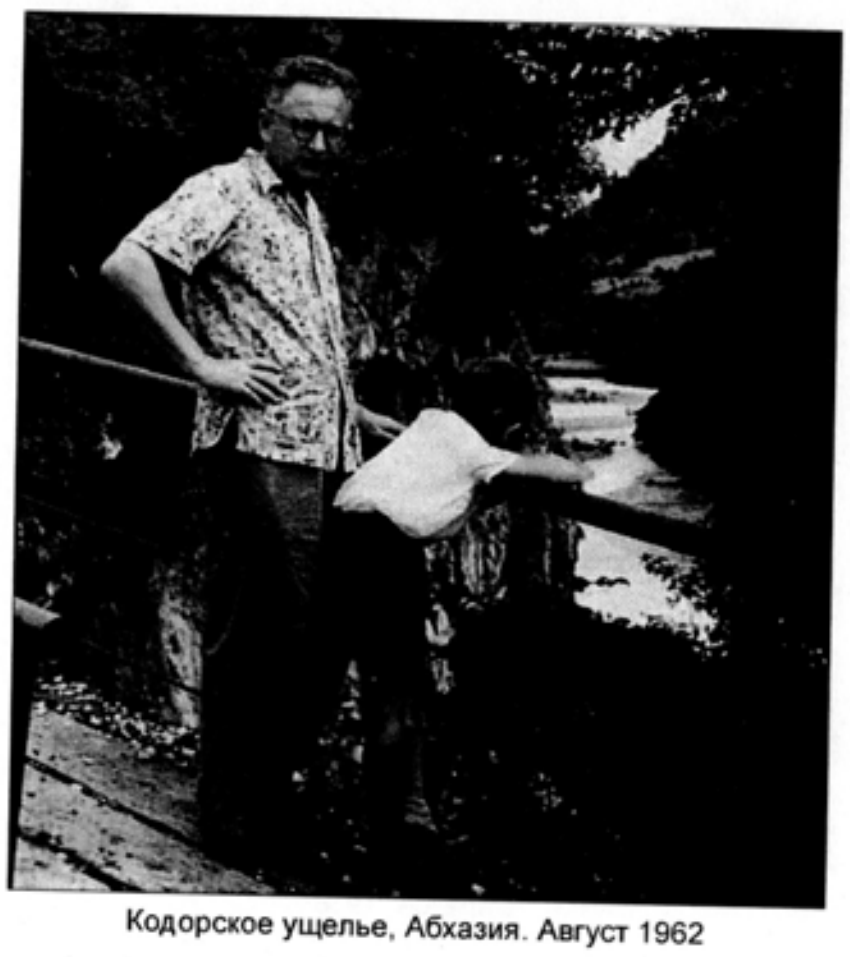

Кодорское ущелье, Абхазия. Август 1962

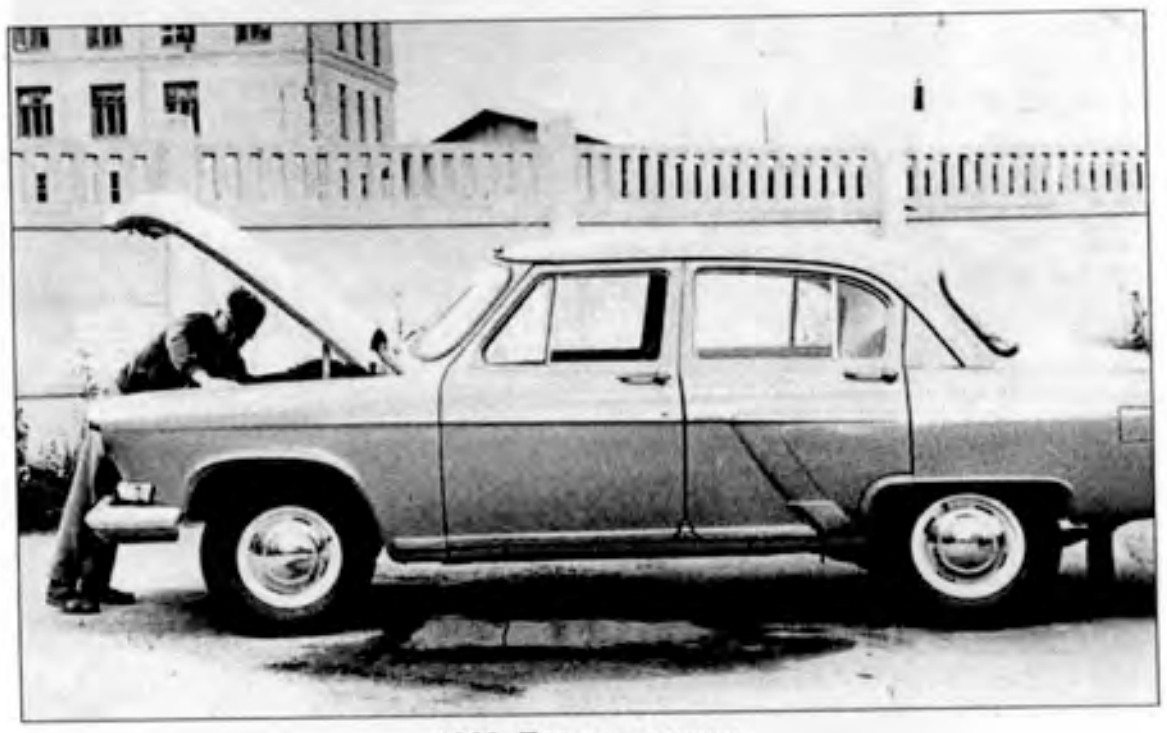

1965. Первая машина

Top left: Khimki, River Station. 1959

Bottom left: Kodori Gorge, Abkhazia. August 1962

Top right: With daughter. August 1963

Bottom right: 1965. First car

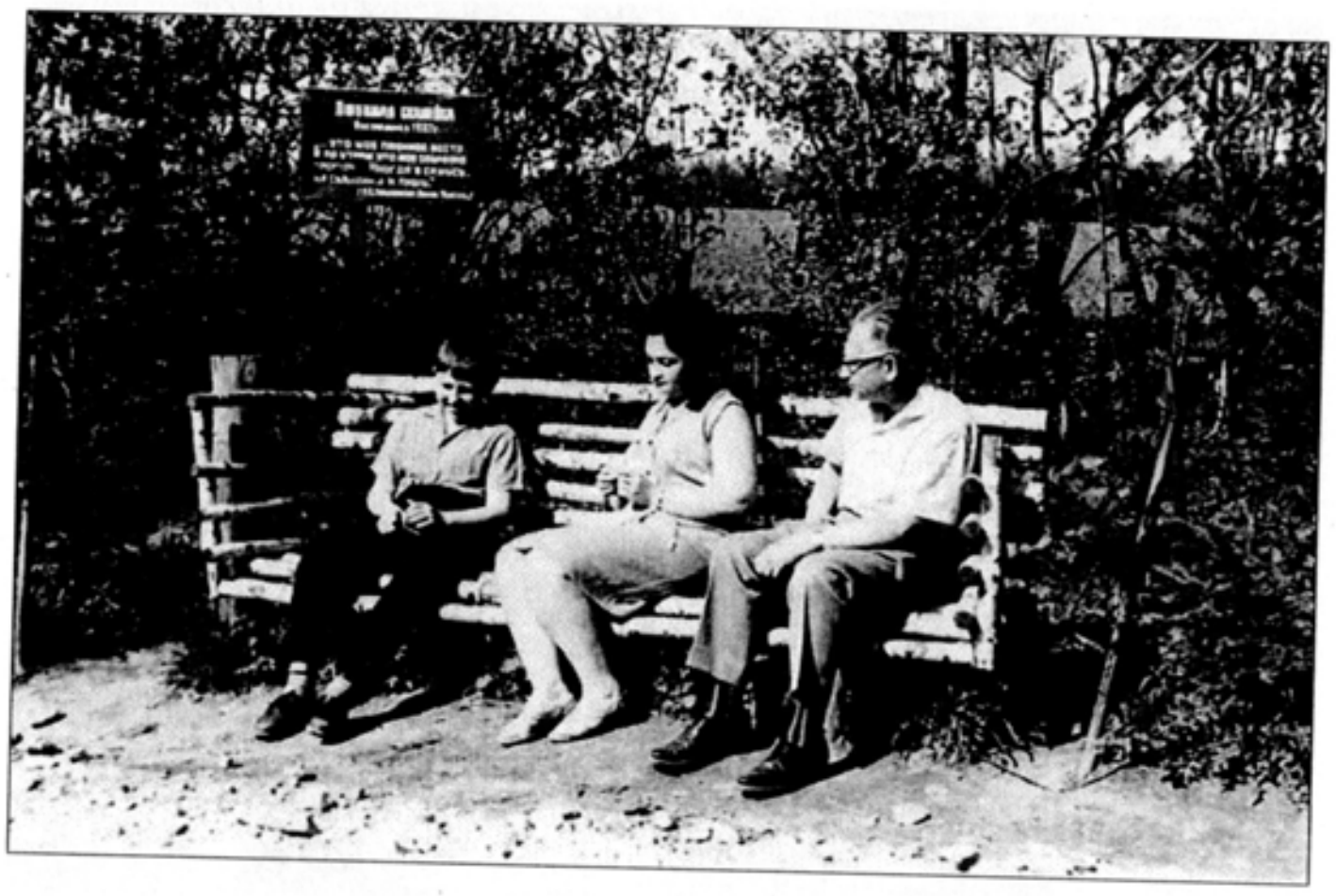

Ясная Поляна. 1967

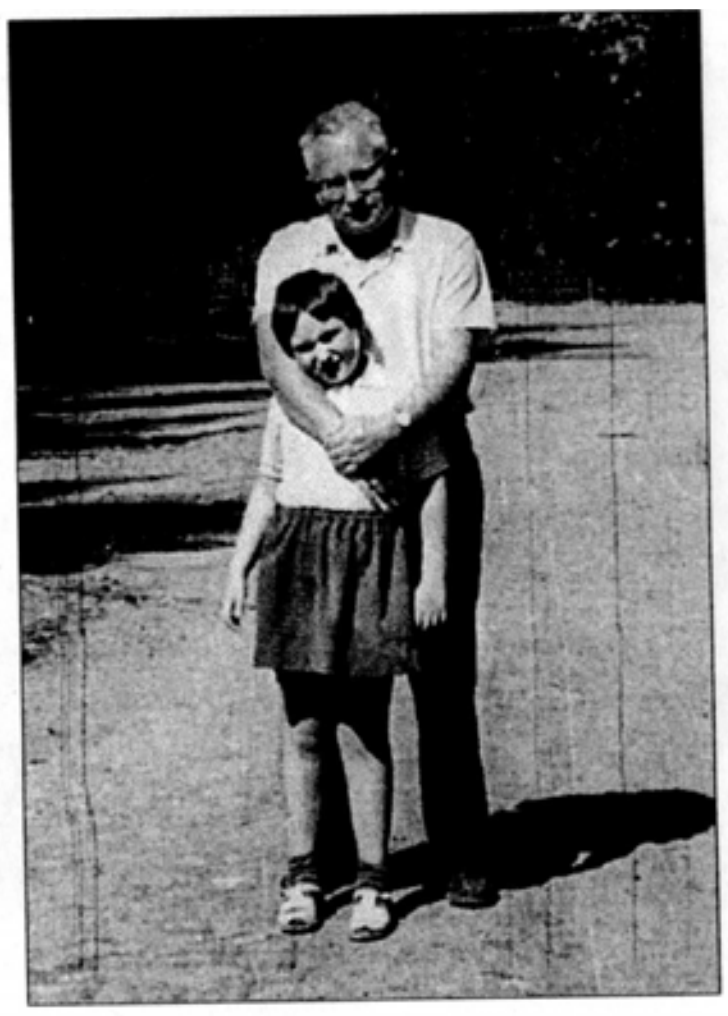

С дочерью. Протвино, 1967

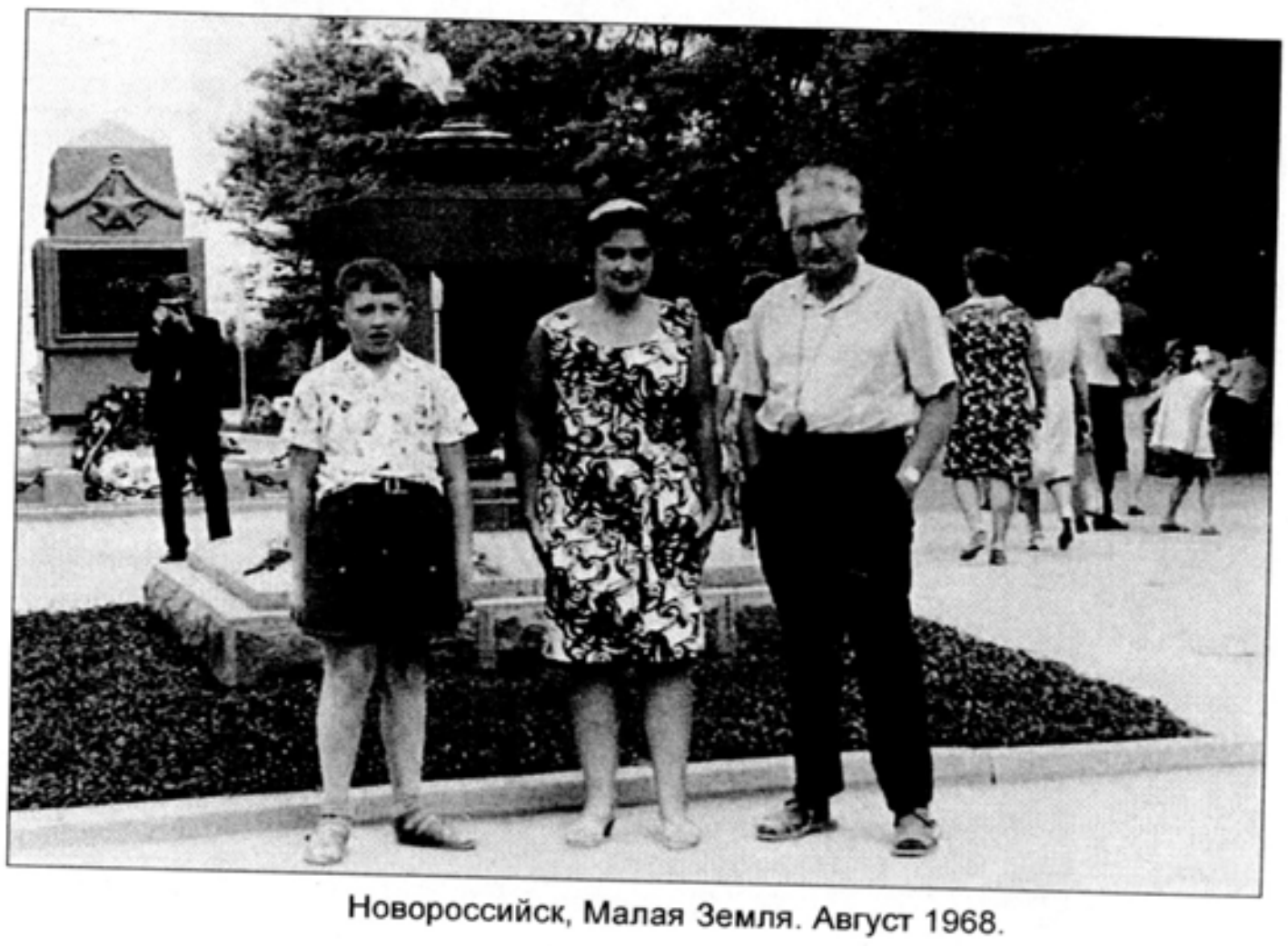

Новороссийск, Малая Земля. Август 1968.

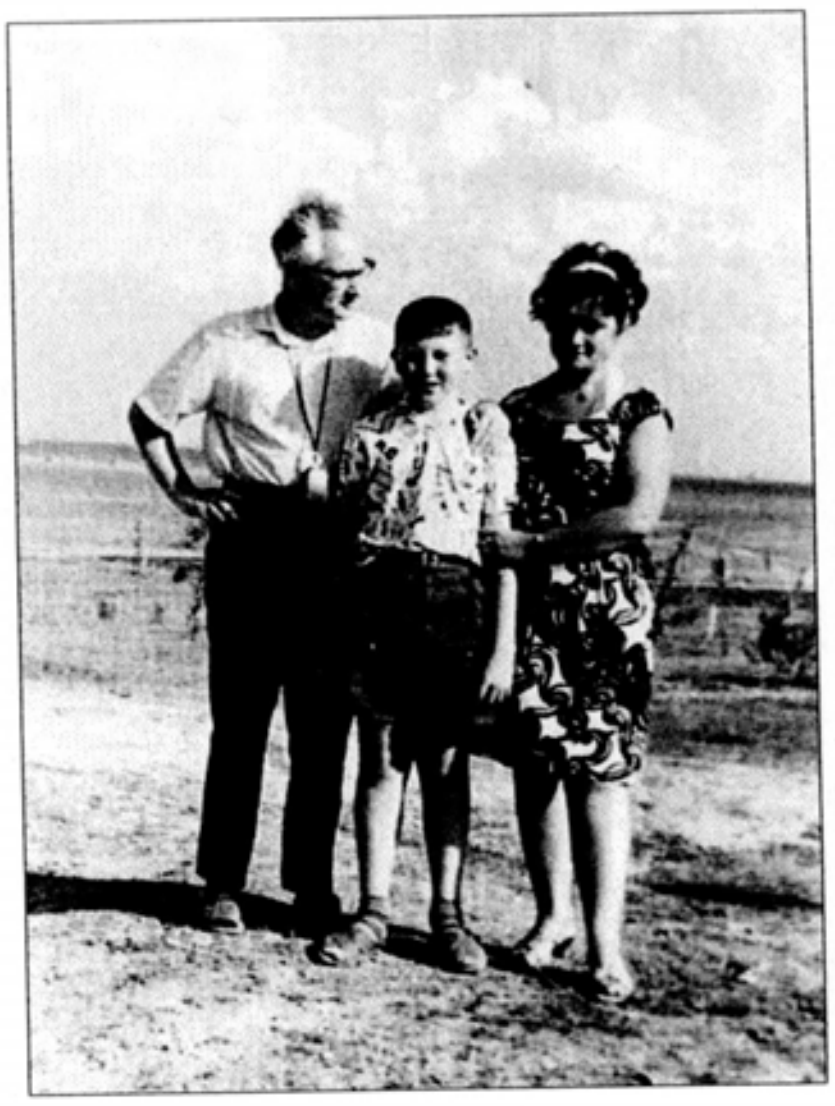

Новороссийск. Август 1968

Top left: Yasnaya Polyana. 1967

Bottom left: Novorossiysk, Malaya Zemlya. August 1968.

Top right: With daughter. Protvino, 1967

Bottom right: Novorossiysk. August 1968

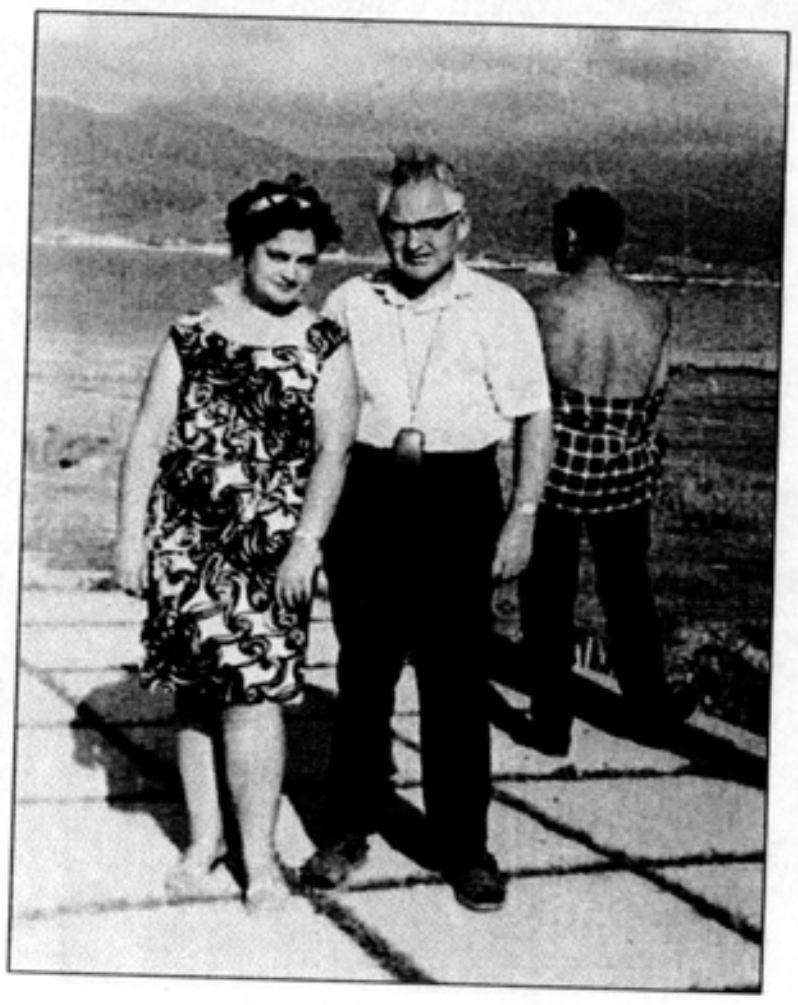

Новороссийск, август 1968

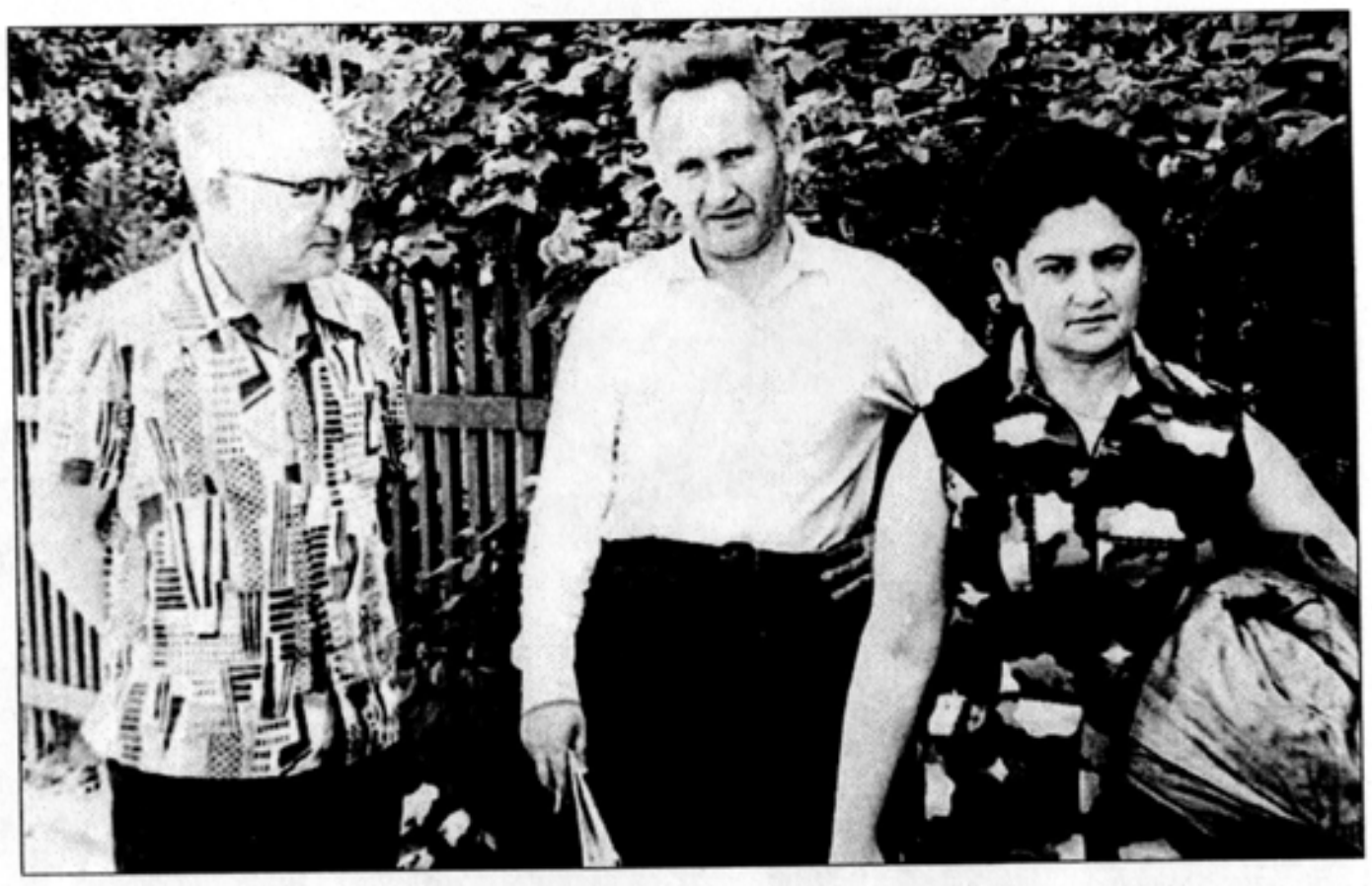

Москва, 1972. В центре - Лев Михайлович Капчинский.

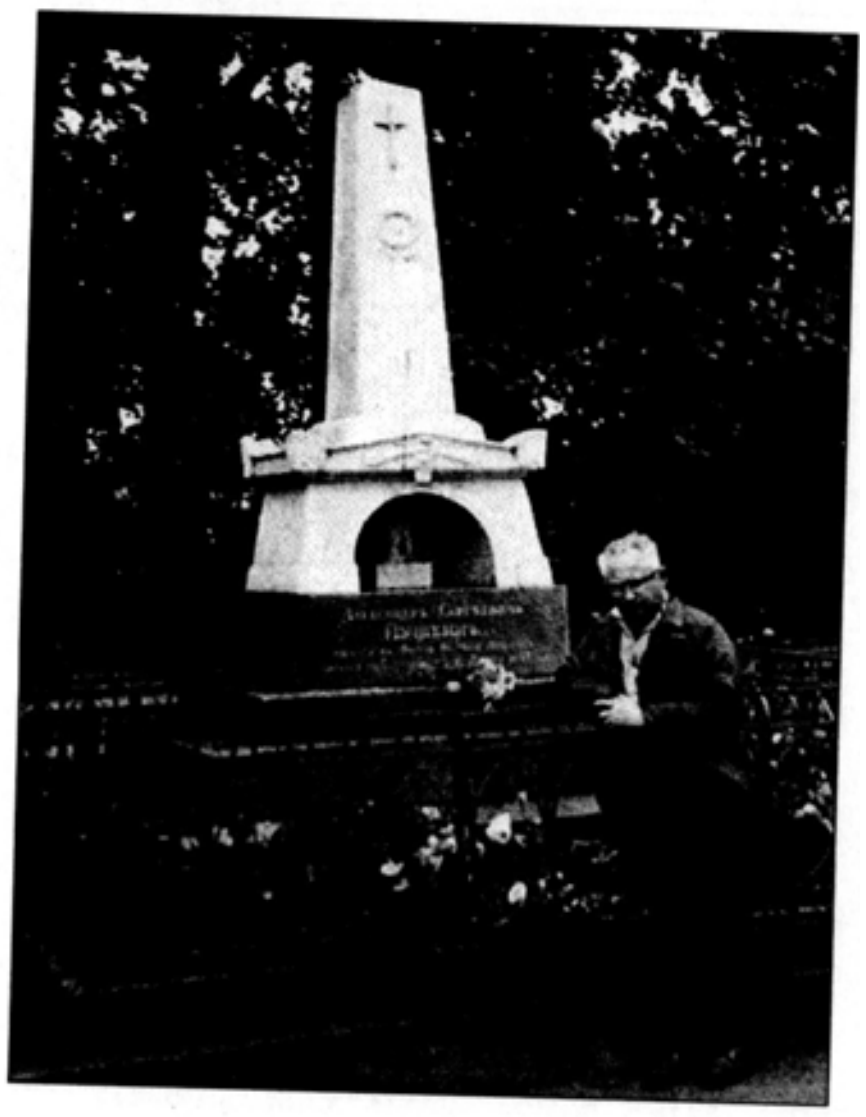

Святогорский монастырь, у могилы А.С. Пушкина. 1!

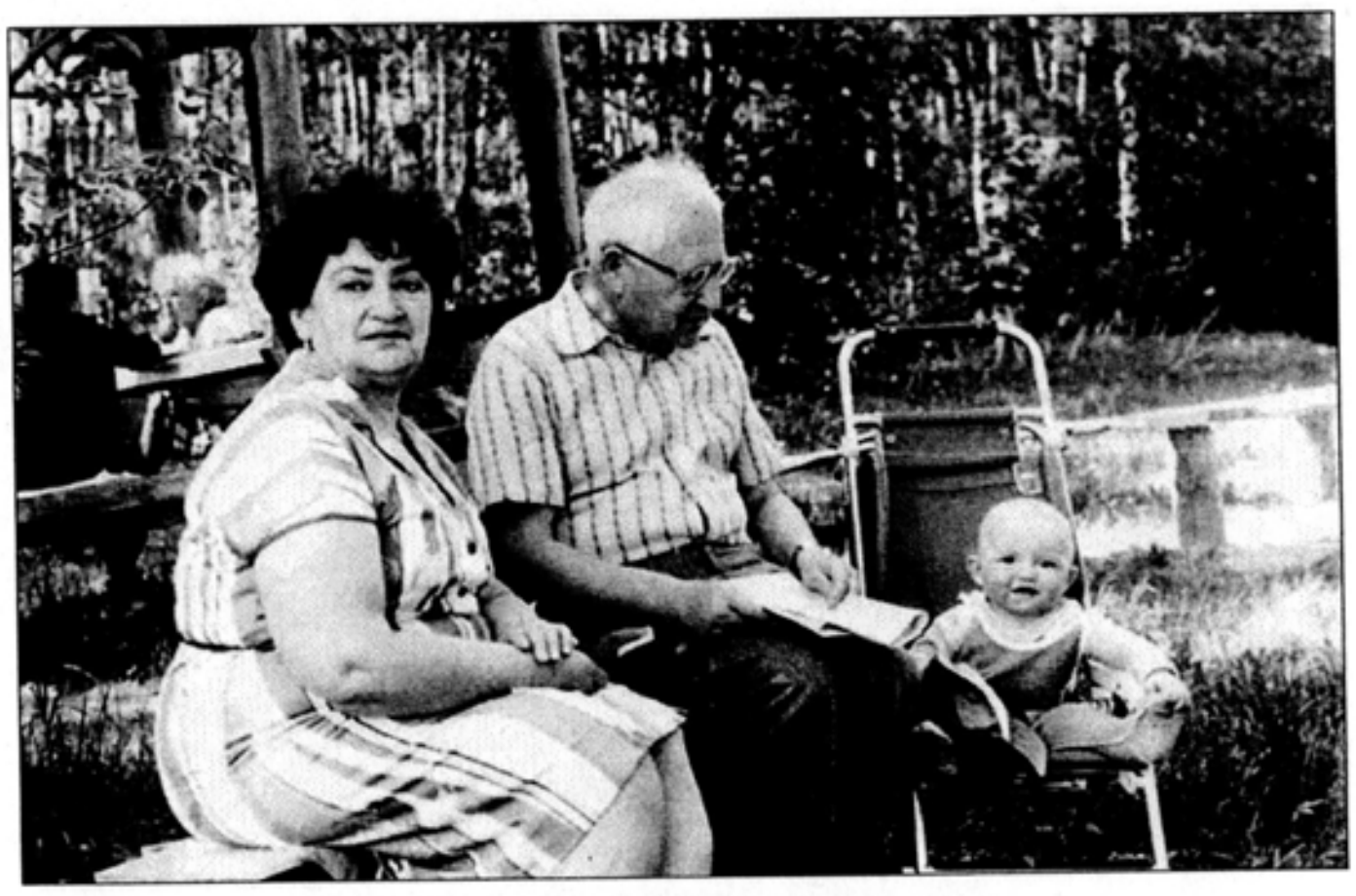

С внуком. Июнь 1988

Top left: Novorossiysk, August 1968

Bottom left: Svyatogorsky Monastery, at the grave of A.S. Pushkin. 1969

Top right: Moscow, 1972. In the center - Lev Mikhailovich Kapchinsky.

Bottom right: With grandson. June 1988

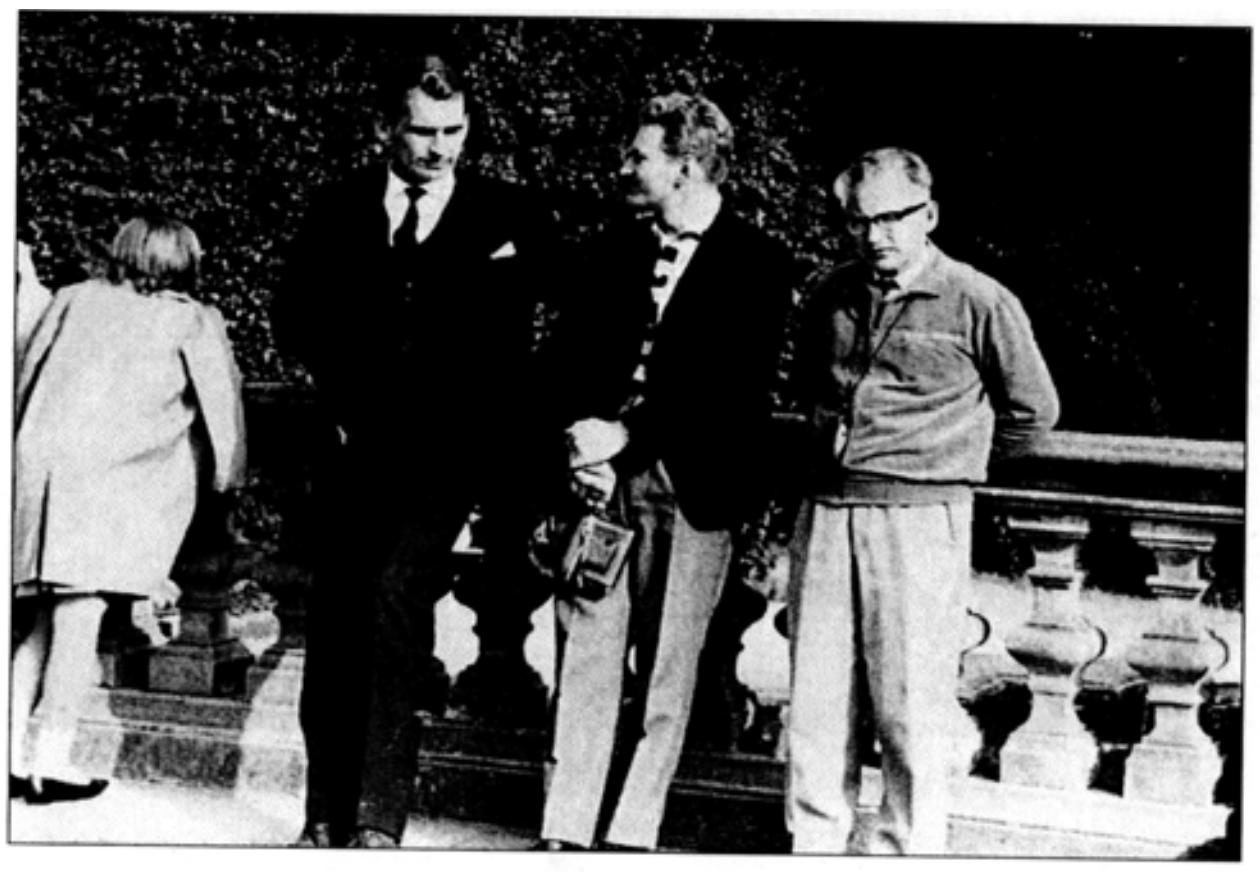

Краков, 1967. В центре - В.К. Плотников.

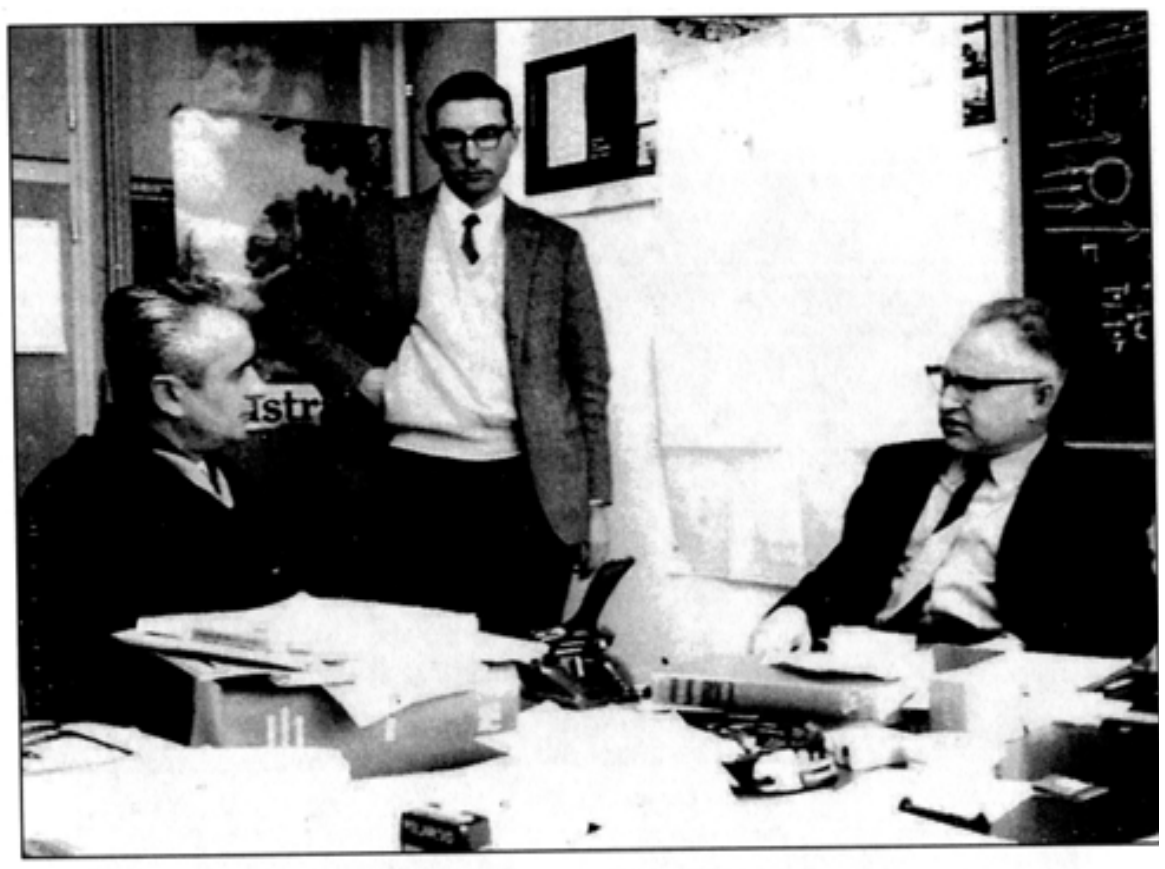

ЦЕРН, Женева. Март 1968.
В центре - сотрудник ЦЕРНа, слева - С.А. Ильевский.

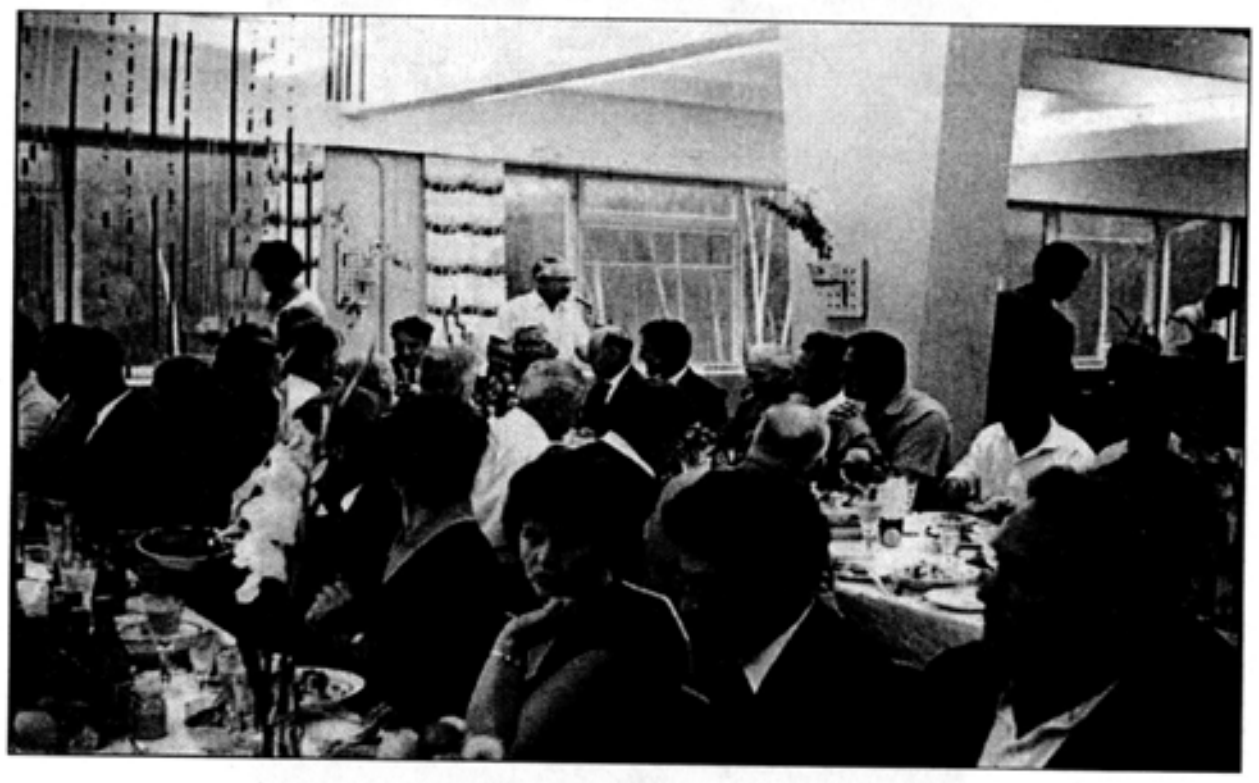

Протвино. Банкет, посвященный пуску И-100. 1967.

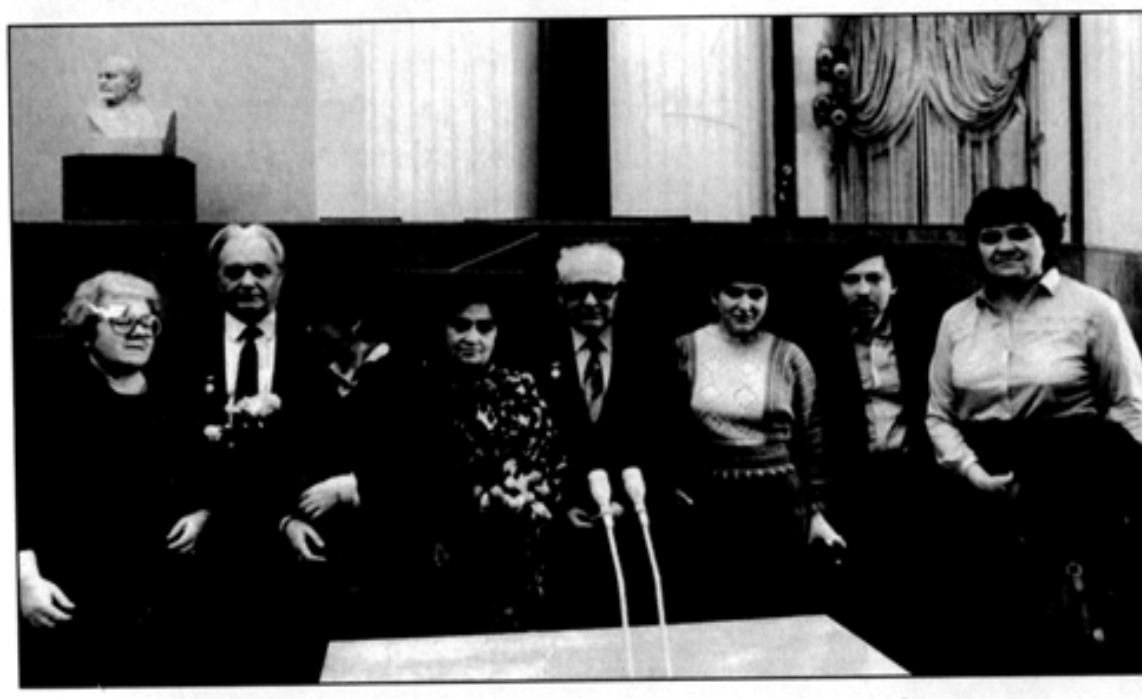

Екатерининский зал Кремля, 13 мая 1988.
Вручение Ленинской премии. Слева - В.А. Тепляков с семьей.

Top left: Krakow, 1967 In the center - V.K. Plotnikov.

Bottom left: Protvino. Banquet dedicated to the launch of the I-100. 1967.

Top right: CERN, Geneva. March 1968. In the center - a CERN employee, on the left - S.A. Ilyevsky.

Bottom right: Catherine's Hall of the Kremlin, May 13, 1988.
Presentation of the Lenin Prize. Left - V.A. Teplyakov with his family.

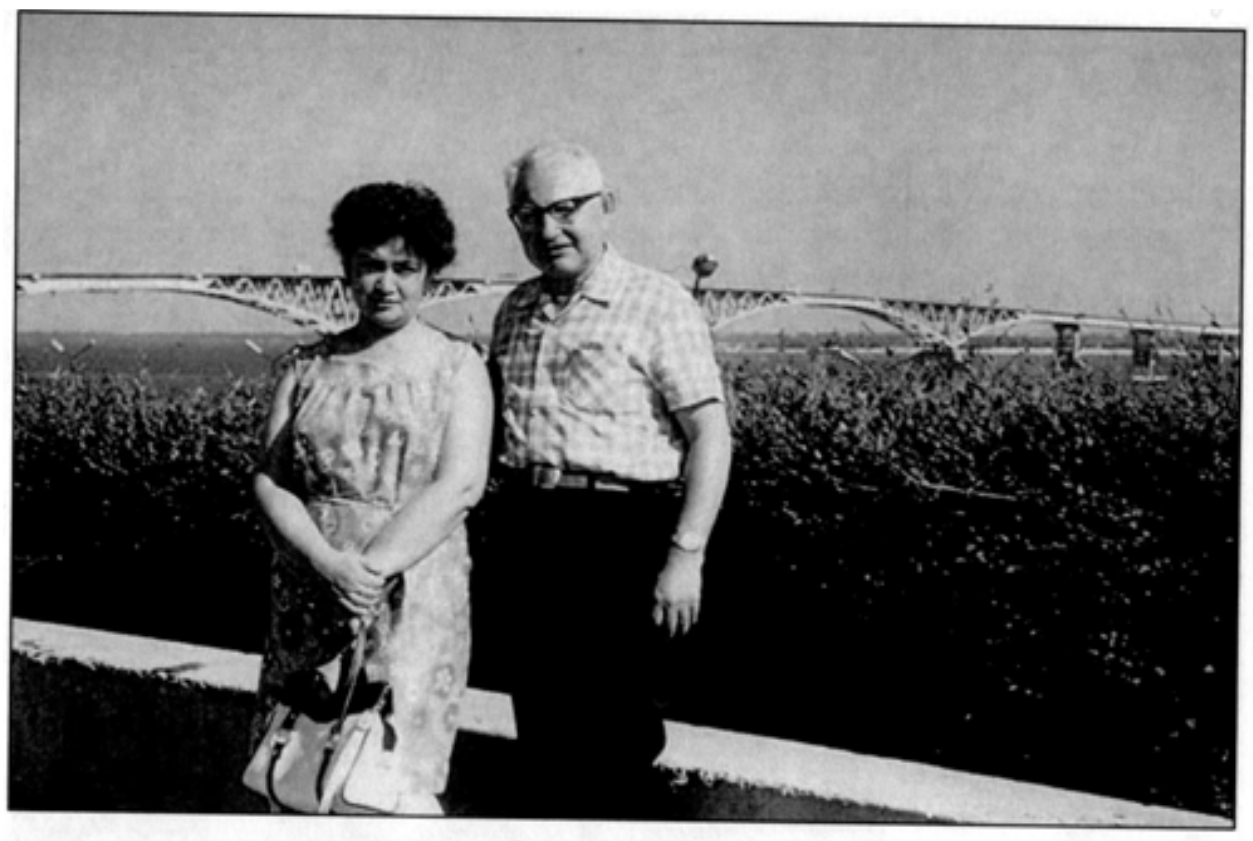

Саратов, 1976

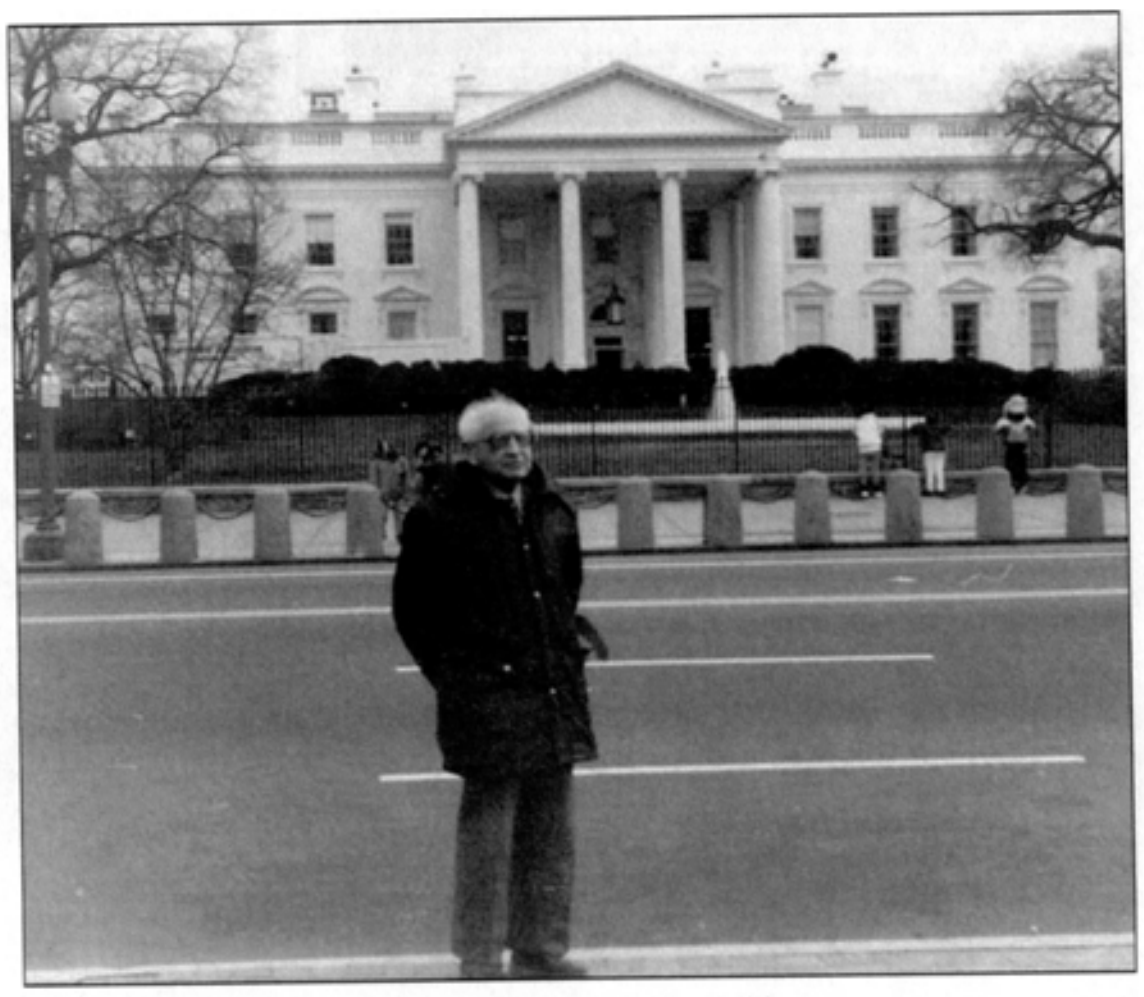

Вашингтон, февраль 1993

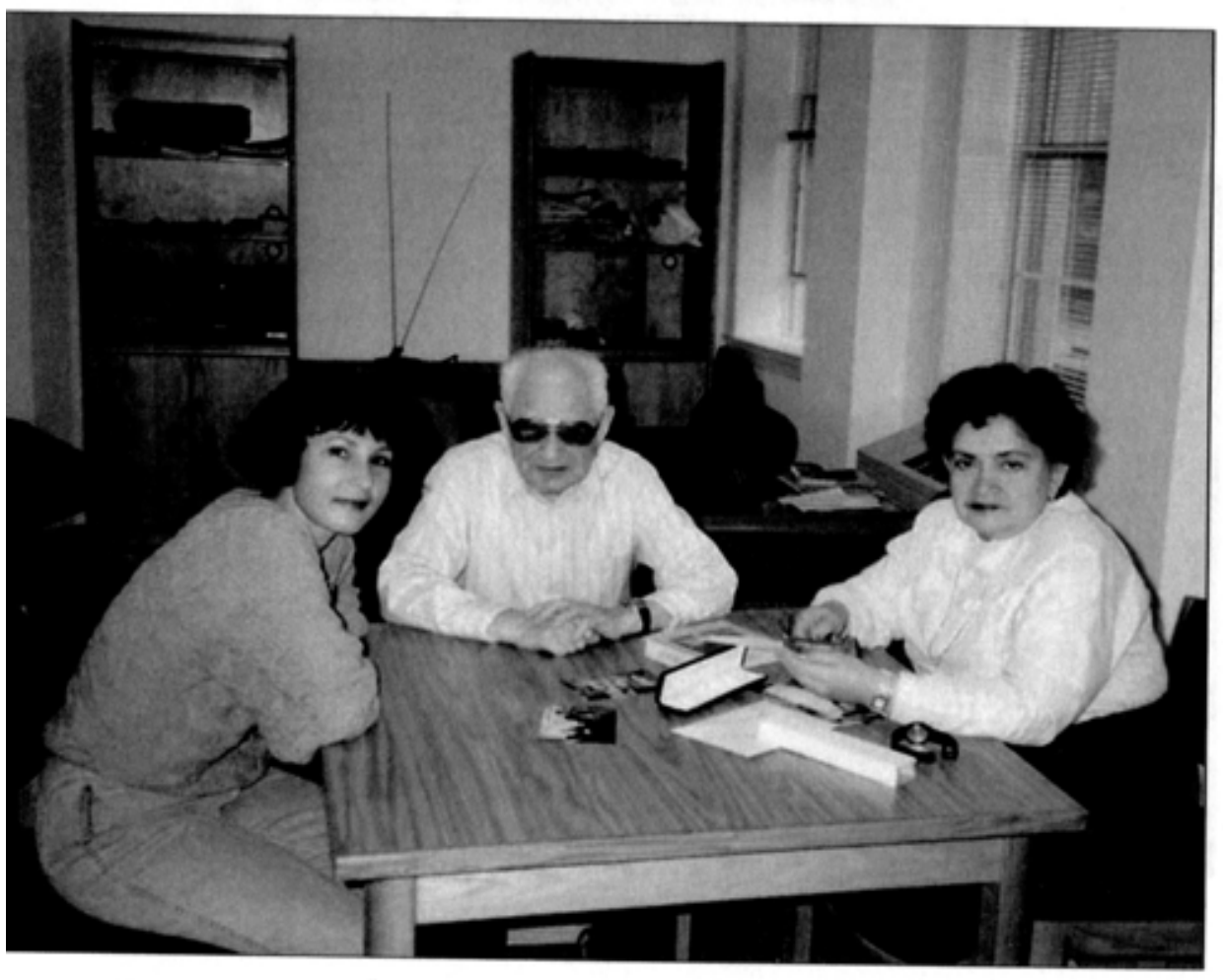

Вашингтон, в комнате кампуса Мэрилендского университета, 1993

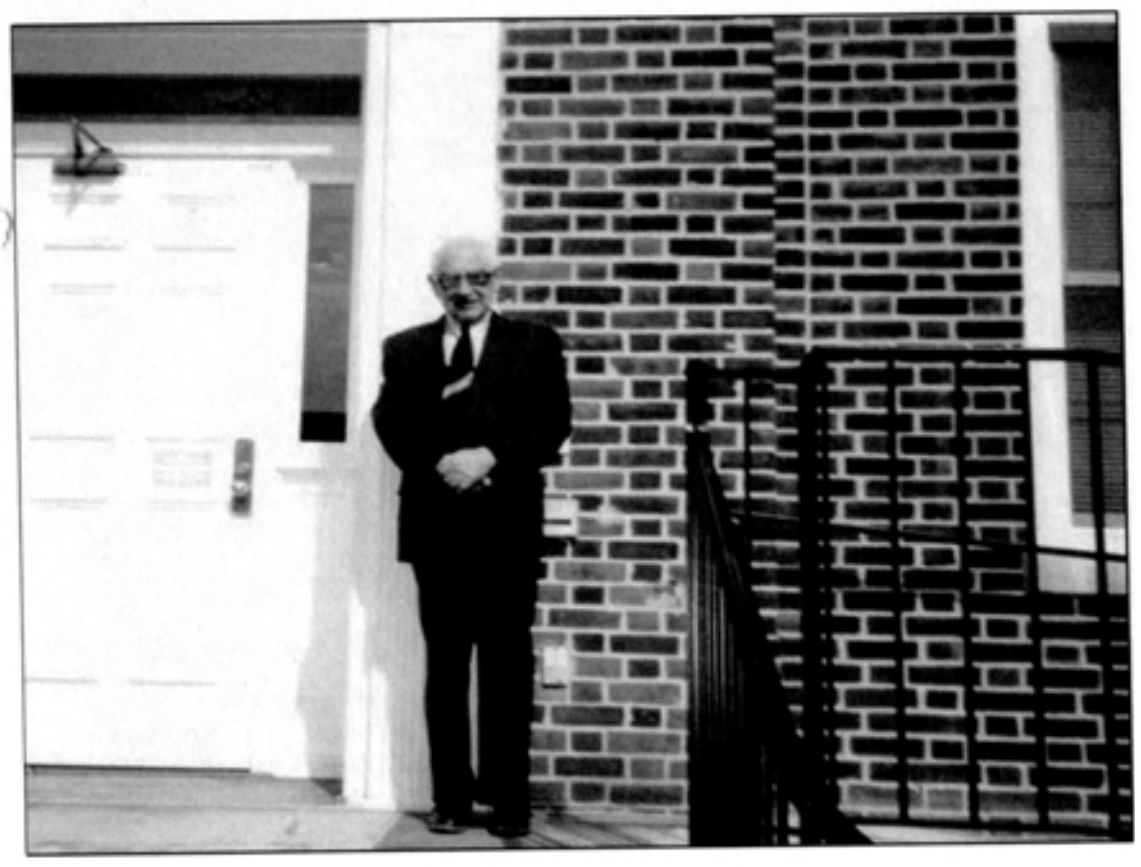

Вашингтон, кампус Мэрилендского университета, 1993

Top left: Saratov, 1976

Bottom left: Washington, in a room on the University of Maryland campus, 1993

Top right: Washington, February 1993

Bottom right: Washington, University of Maryland campus, 1993

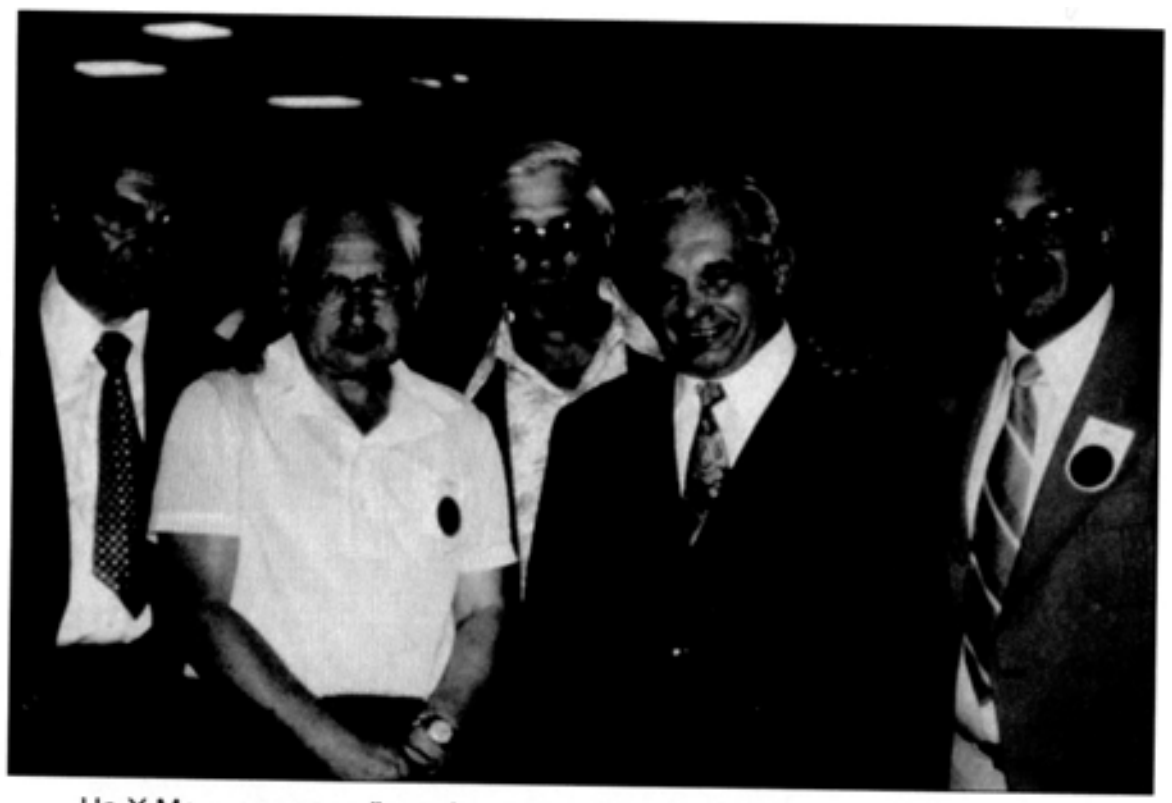

На X Международной конференции по ускорителям высоких энергий. Протвино, июль 1977. Слева направо: Н.В. Лазарев, И.М. Капчинский, В.Г. Андреев (МРТИ), В.А. Тепляков, Дон Свенсон (Лос Аламос).

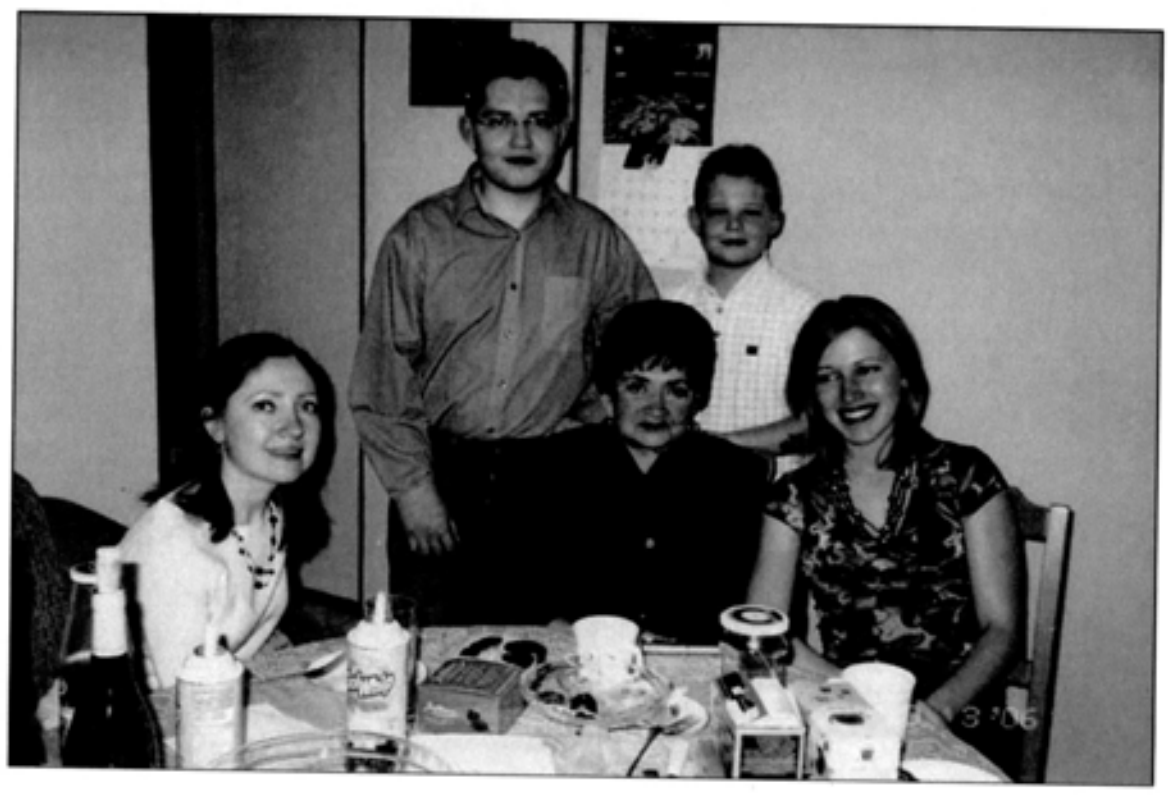

Внуки и внучки. 2006

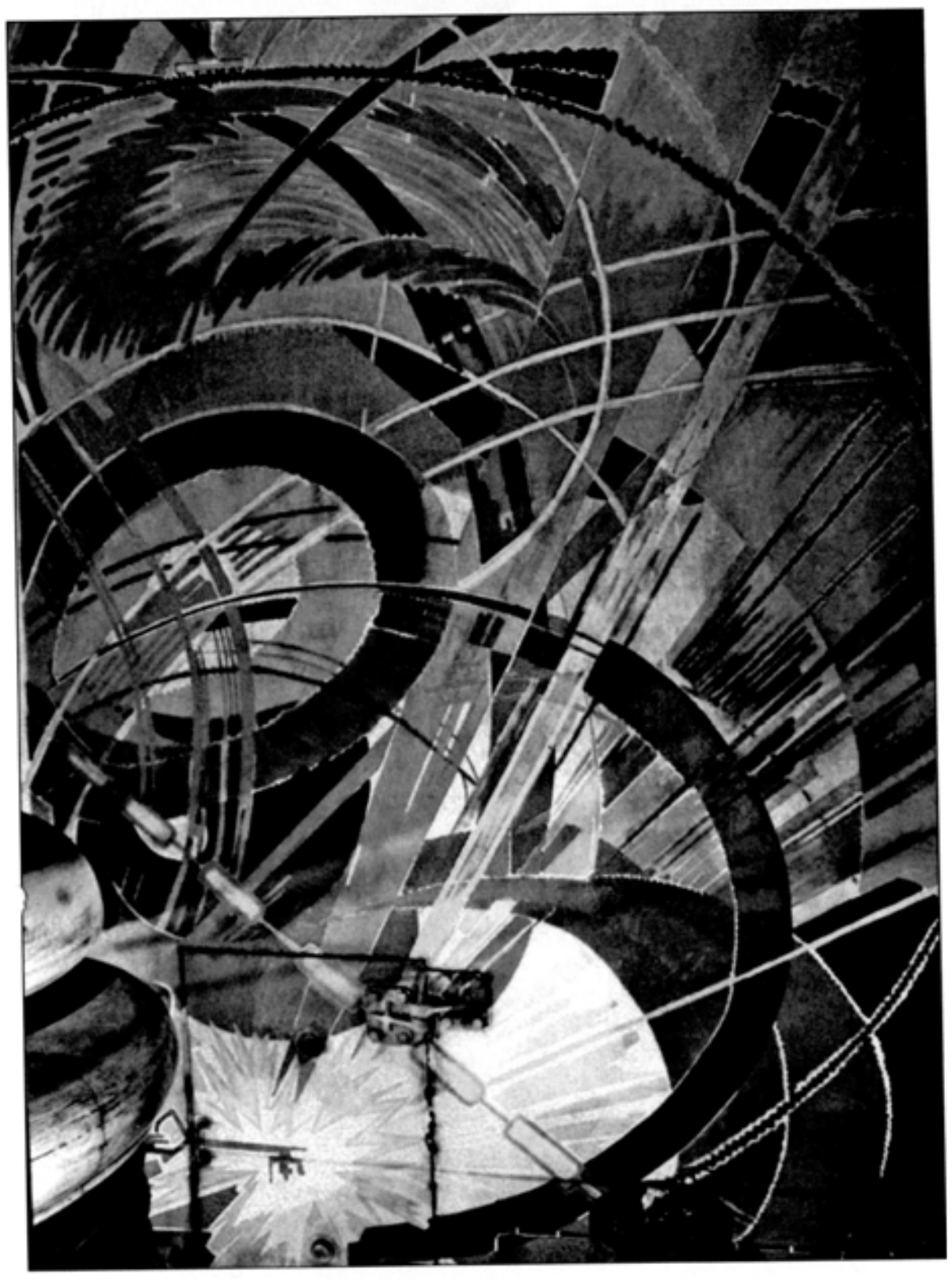

Панно в зале форинжектора, Протвино. Работа Иды Егоркиной-Баталиной.

Top left: At the X International Conference on High Energy Accelerators. Protvino, July 1977. From left to right: N.V. Lazarev, I M. Kapchinsky, V.G. Andreev (MRTI), V.A. Teplyakov, Don Swenson (Los Alamos).

Bottom left: Grandchildren and granddaughters. 2006

Right: Panel in the pre-injector room, Protvino. The work of Ida Egorkina-Batalina.

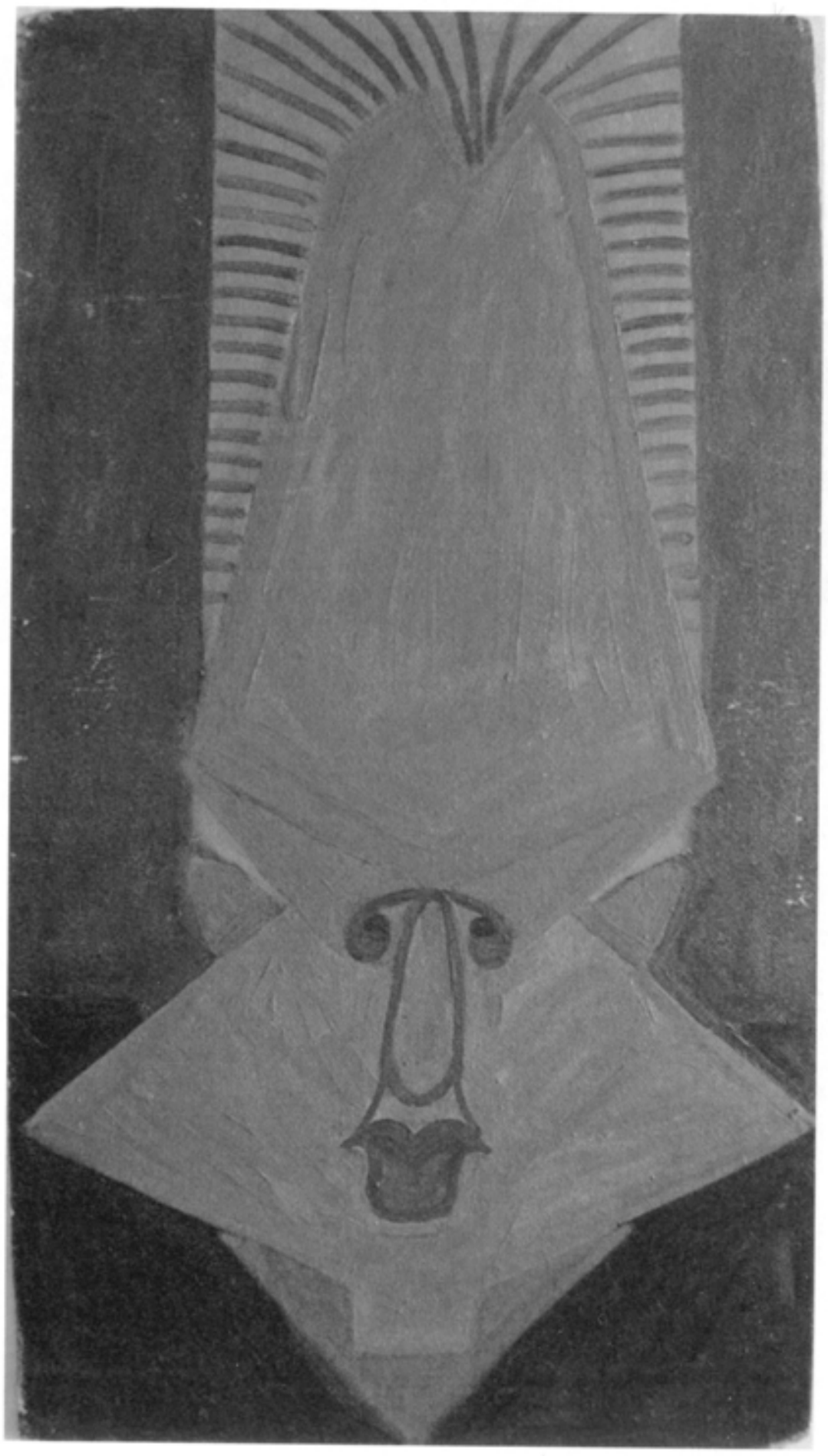

А. Степанова. Эскиз к портрету И.М. Капчинского. 1968

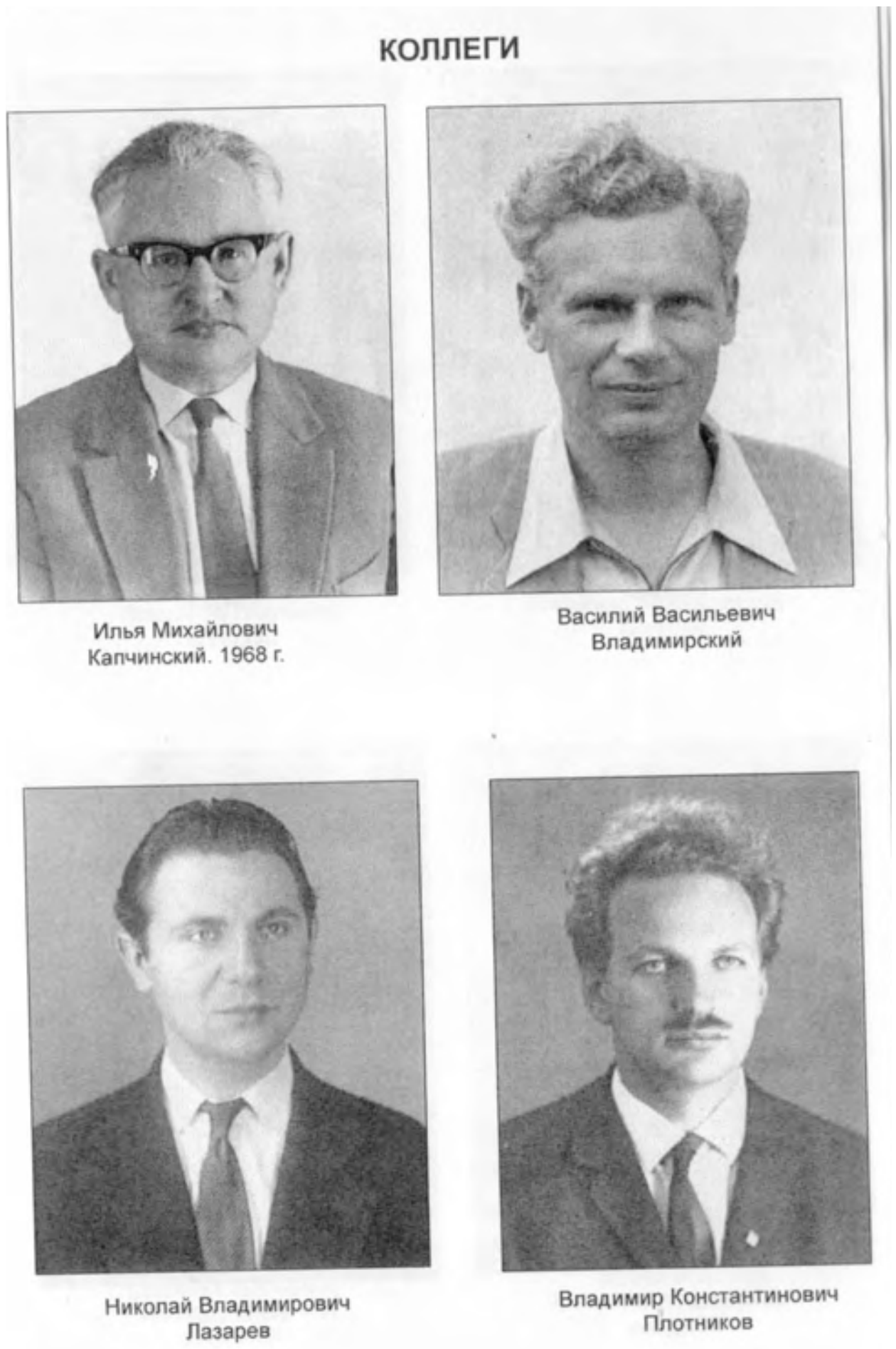

Left: Panel in the pre-injector room, Protvino. The work of Ida Egorkina-Batalina.

Right – top left: A. Stepanova. Sketch for the portrait of I.M. Kapchinsky. 1968

Bottom left: Nikolai Vladimirovich Lazarev

Top right: Vasily Vasilievich Vladimirsky

Bottom right: Vladimir Konstantinovich Plotnikov

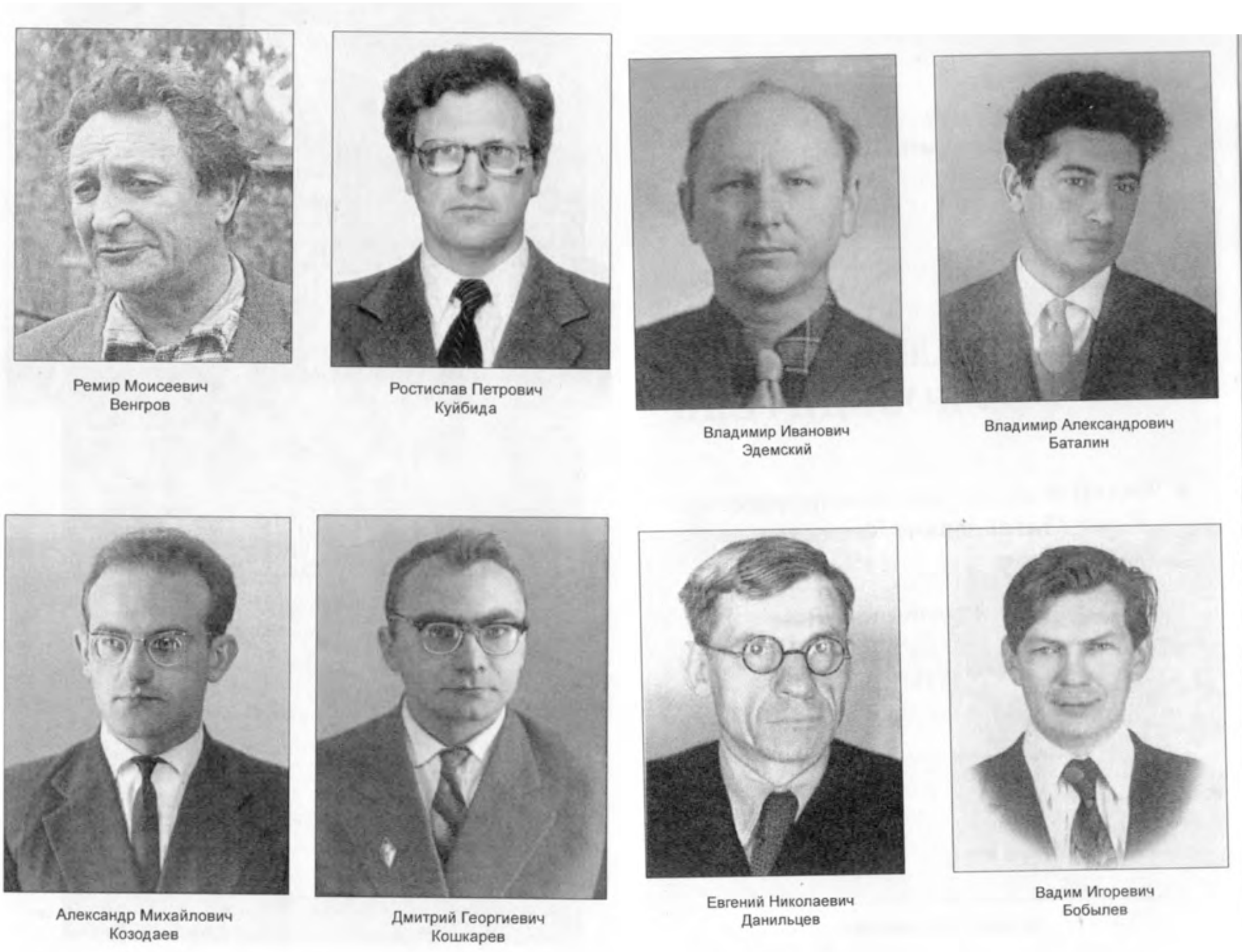

Remir Moiseevich Vengrov; Rostislav Petrovich Kuybida; Vladimir Ivanovich Edemsky; Vladimir Alexandrovich Batalin

Alexander Mikhailovich Kozodaev; Dmitry Georgievich Koshkarev; Evgeny Nikolaevich Daniltsev; Vadim Igorevich Bobylev

Институт Теоретической и Экспериментальной Физики

5 –09

# УЧЕНЫЙ, УЧИТЕЛЬ, РУКОВОДИТЕЛЬ

## К 90-летию со дня рождения профессора Ильи Михайловича Капчинского

### Сборник воспоминаний

### Часть I

Москва 2009

Institute of Theoretical and Experimental Physics, 5-09

SCIENTIST,

TEACHER,

SUPERVISOR

To the 90th anniversary of the birth of Professor Ilya Mikhailovich Kapchinsky

Collection of memories Part I

Moscow 2009, ITEP preprint. June 2009

$$V_k = \gamma \frac{\beta\lambda}{S}\left(\frac{a}{\lambda}\right)^2 \frac{v_f}{\mu}\mu\lambda$$

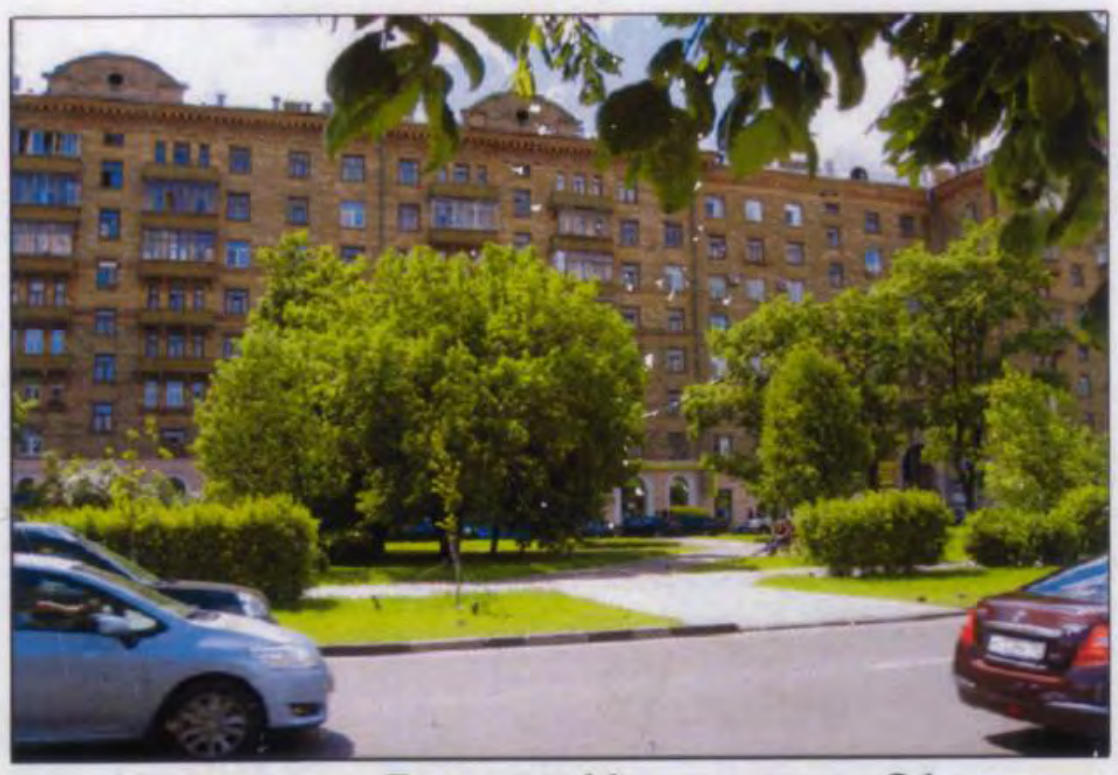

Москва, ул. Дмитрия Ульянова, д. 24

$$R_c = \sqrt{\frac{r_a^2}{2\mu_0^2} + \sqrt{\left(\frac{r_a^2}{2\mu_0^2}\right)^2 + \left(\frac{F_0}{\mu_0}\right)^2}}$$

$$I_{\lim} = \frac{1}{2} j_0 V_k \left[1 - (V_k / V_x)^2\right]$$

$$I(x, y, p_x, p_y, z) = a_{11}(z) \cdot x^2 + a_{12}(z) \cdot y^2 + a_{21}(z) \cdot p_x^2 + a_{22}(z) \cdot p_y^2,$$

$$U(x, y, z) = -\frac{\rho(z)}{4\varepsilon_0} \cdot \left[x^2 + y^2 + \frac{a_x - a_y}{a_x + a_y} \cdot (x^2 - y^2)\right],$$

$$\rho(x, y, z) = n_0 \cdot \iint f(x, y, p_x, p_y, z) dp_x dp_y.$$

Что-бы ни услышал —
-НЕ ВОЛНУЙСЯ — не поможет...
-НЕ ВОЛНУЙСЯ - ВСЕ ПРОХОДИТ!..

$$V_k = \gamma \frac{\beta\lambda}{S}\left(\frac{a}{\lambda}\right)^2 \frac{v_f}{\mu}\mu\lambda$$

$$I_{\lim} = \frac{1}{2} j_0 V_k \left[1 - (V_0 / V_k)^2\right]$$

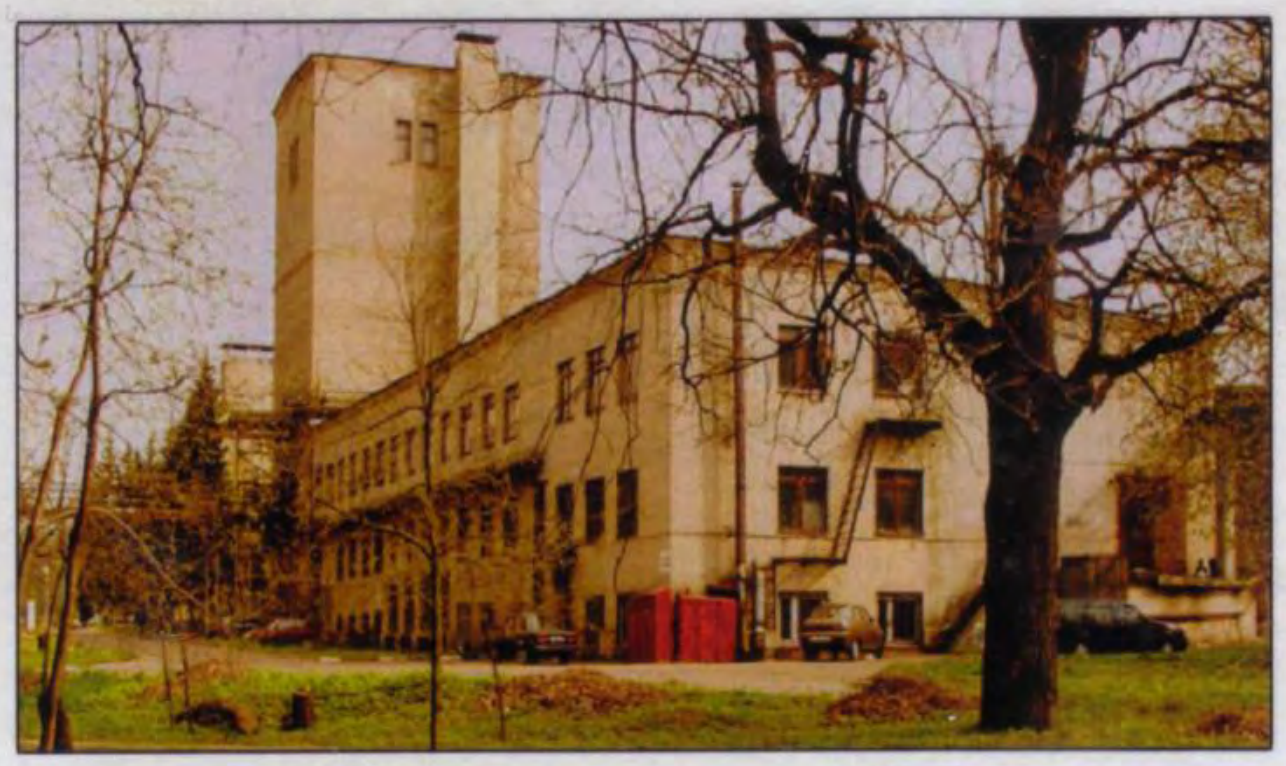

Ускорительный корпус ИТЭФ

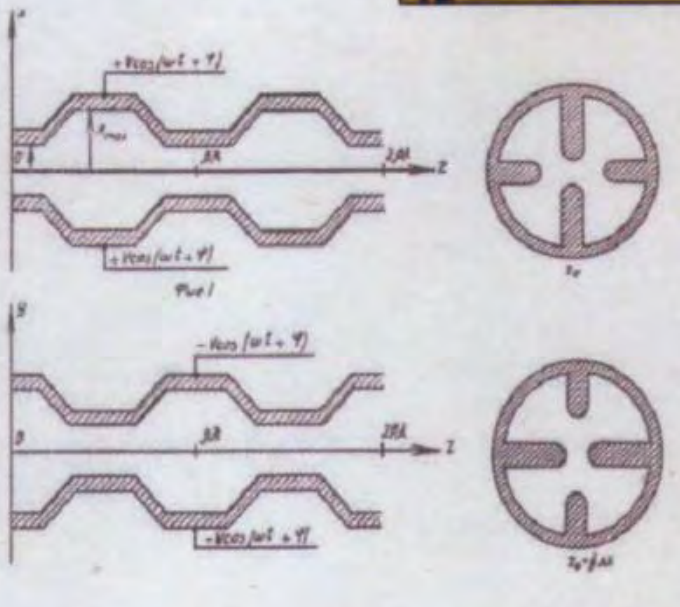

$$V_k = \gamma \frac{\beta\lambda}{S}\left(\frac{a}{\lambda}\right)^2 \frac{v_f}{\mu}\mu\lambda$$

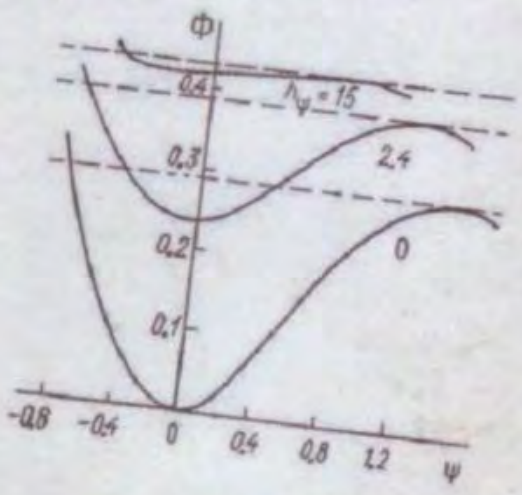